PhD in Astrophysics

# Numerical simulations of astrophysical dynamos and applications to giant planets

Albert Elias-López

September 2025

Supervisors:
Daniele Viganò
Fabio del Sordo

Tutor:
Lluís Font Guiteras

Committee members:
Maarit Korpi-Lagg
Ankit Barik
Nina-Elisabeth Nèmec

INSTITUT DE CIÈNCIES DE L'ESPAI

CSIC Consejo Superior de Investigaciones Científicas

IEEC

Universitat Autònoma de Barcelona

## Acknowledgements

First and foremost, I would like to thank my supervisors, Daniele Viganò and Fabio del Sordo, for their unwavering guidance and support throughout this research. Their scientific expertise, experience with computing proposals, writing advice, and motivational spirit have provided constant direction during these PhD years. I am deeply grateful for this opportunity.

I am also thankful to my colleagues at ICE-CSIC for fostering such an inclusive and intellectually stimulating work environment. In particular, Clàudia and Simran, the PhD students in my group, with whom I'm confident, will have a friendship that will last a lifetime. I extend my gratitude to Clara, Dani, Celsa, Abu, Michele, Dane, Antia, Simone, Pedro, Victor, Maider, Rebeca, Cristina, Maryam, Luís, Guiosue, Pau, and many others whose names I may forget. My time at Flatiron also brought me meaningful connections: thank you to Lucas, Devina, Hiroka, Tom, Tobin, Shaker, Vicente, Samuele, Ulf, Don, Antonio, Lukas, Moha, Azin, and François.

Many people have contributed to my learning journey. I thank Axel Brandenburg and Thomas Gastine for their initial guidance with the public MHD codes, the *Pencil Code* and *MagIC*, respectively. Axel also hosted the first conference I attended, where I gave my first talk and, memorably, had my laptop stolen. Eva, thank you for hosting me at the Scuola Normale and for your insightful ideas on the vorticity project. Alexis, your advice on *MagIC* and your thoughtful feedback on the cold Jupiter project were invaluable. A very special thank you goes to Matteo Cantiello, who could be considered my third supervisor, for your mentorship, for helping me join the CCA Pre-doc program, and for your pivotal role in the hot Jupiter project. Living in New York has undoubtedly left a lasting impression on me.

To the friends who have carried me through different stages of life: from my bachelor's (Elena, Natanael, and Èric), from Saldes (Adrià, Marta, Irene, and Irene), from the choir (Aleix, Núria, Xavi, Èric, Sílvia, Laura, and Laura), from Xivarri (Gemma, Gemma, Quel, Paula, and Laia), and many others (Elle, Laia, Sergi), thank you for the joy, distraction, and support. Music has been a constant source of strength throughout these years. I'm deeply grateful to two wonderful people, musicians and mentors: Rafael, my guitar teacher, and Joel, my friend and choir director.

Finally, my deepest thanks go to Carme, Jordi, and Marta, my parents and sister. So much of who I am comes from you. Thank you for the puzzles since childhood, for the sibling rivalry (obviously very healthy), for the countless tuppers, and most of all, for the endless love.

Looking back on these years, and now feeling profoundly wise, I will borrow the words of one of the greats: *”People who think they know everything are a great annoyance to those of us who do.”* - Isaac Asimov. Though, truthfully, most days my self-awareness makes me identify with another: *“Two things are infinite: the universe and human stupidity; and I'm not sure about the universe.”* - Albert Einstein. And so I won't end with two contradictory quotes, but with the best advice from a wise fish: *“Just keep swimming.”*.

# Abstract

## Numerical simulations of astrophysical dynamos and applications to giant planets

Magnetic fields are ubiquitous in astrophysical systems, from small objects like planets to large-scale environments such as the intracluster medium, and they play a central role in governing the dynamics of these systems. The magnetic diffusion timescale is usually much smaller than the evolutionary timescales. Thus, long-lived fields generated in the past cannot usually explain the current magnetization of planets, stars, and galaxies. Self-sustaining dynamos provide instead the most robust framework for explaining this ubiquitous magnetization. In astrophysics, the *dynamo* refers to the process that converts the kinetic energy of conducting fluids into magnetic energy. Mathematically, this theory relies on the combination of Maxwell's equations and fluid dynamics: magnetohydrodynamics (MHD). Numerical simulations are fundamental for solving these equations and for understanding dynamo processes.

This thesis is divided into two parts, both of which involve numerical self-excited dynamos simulations applied to different astrophysical scenarios: the interstellar medium (ISM) and the interiors of gas giant planets. In the first part, using 3D MHD simulations in a periodic box with the *Pencil Code*, I study magnetic field growth from irrotational subsonic expansion flows, a simplified model for supernova-driven flows in the ISM. These curl-free flows approximate stellar explosions and winds, the ISM's main energy sources, and bring the fluid to a turbulent state.

The second part of the thesis focuses on planetary dynamos. I give an overview of planetary magnetic fields and their numerical modeling via spherical shell convection dynamos. Despite the growing number of confirmed exoplanets, their magnetic fields remain elusive, but they could be characterized by the detection of coherent radio emission from the current and next-generation low-frequency radio interferometers. Understanding planetary dynamos is another ingredient needed for studying planetary interiors, evolution, habitability, and atmospheric dynamics. Using a series of 3D spherical shell dynamo simulations, I explore the evolution of magnetic fields in cold gas giants. I use the *MagIC* code and couple it with 1D thermodynamic profiles obtained with planetary evolution models produced with the *MESA* code. I simulate dynamos at various stages of planetary evolution and then obtain a slow decay in field strength ($\propto t^{-0.3}$), consistent with previously proposed scaling laws. The results show a transition from multipolar to dipolar regimes and provide evolutionary trends for the dimensionless dynamo numbers.

Finally, I examine dynamos in hot Jupiters (HJs), highly irradiated gas giants that may host the strongest magnetic fields, making them prime radio emission targets. I use 1D models to study how internal convective structures and rotation affect dynamo behavior. Most HJs remain in a fast-rotating regime ($Ro \lesssim 0.1$), but high-mass, distant HJs can shift into different regimes. When heating is confined to outer layers, convection within the dynamo region is suppressed, weakening the magnetic field strength predictions and the chances of observational detectability, which might explain the lack of definitive magnetic field measurements by multiple radio campaigns performed in the last two decades.

**Keywords: Magnetohydrodynamics (MHD) - Magnetic fields – Computational methods – Numerical dynamos - Instabilities - Turbulence - Interstellar medium - Galactic magnetic fields - Planetary magnetic fields - Gas giant evolution - Planetary dynamo evolution - Dipolarity transition - Hot Jupiter inflation - Rossby number - Hot Jupiter dynamo - Radio emission**

**Abbreviations**

BSC: Barcelona Supercomputing Center

CCA: Center for Computational Astrophysics

CMB: Core-mantle boundary (not cosmic microwave background!)

ECM: Electron cyclotron maser

ECMI: Electron cyclotron maser instability

HD: Hydrodynamics

HJ: Hot Jupiter

IAC balance: magnetic, Archimedean, and Coriolis balance

LSD: Large-scale dynamo

MAC balance: magnetic, Archimedean, and Coriolis balance

MHD: Magnetohydrodynamics

MLT: Mixing length theory

RCB: Radiative convective boundary

RES: Red Española de Supercomputación

SN: Supernova explosions

SSD: Small-scale dynamo

SSL: Stably stratified layer

# Contents

# 1

# Introduction

Magnetic fields are present in nearly all astrophysical environments, spanning a wide range of physical spatial scales. From relatively small objects, such as planets, sunspots, or accretion disks, to the largest scales, defined by galactic intercluster media, non-negligible magnetism is present. It often plays a crucial role in shaping the dynamics of these systems.

Permanent magnetization, analogous to the ferromagnetism in some metals on Earth, cannot be considered responsible for such fields. The primary reason is that, in most astrophysical scenarios, magnetic fields are embedded in fluids (e.g., plasma, gas) rather than solids over a broad range of temperatures and pressures; therefore, the proper framework for studying them is to couple fluid dynamics with Maxwell's equations. Even in the rare cases where solid matter is present, the Curie temperatures (above which a material becomes paramagnetic instead of ferromagnetic) are typically exceeded. Consequently, the magnetization of fluids and plasmas outside laboratory conditions requires an explanation within fluid dynamics and plasma physics. There are three most plausible mechanisms for explaining the existence of astrophysical magnetic fields:

1. The magnetic diffusion timescale is significantly longer than the evolutionary timescale of the system, preventing the decay of an initial primordial field, as described by the fossil field theory (Cowling, 1945). This might hold for some cases, such as some magnetized intermediate-mass (1.5-6 $M_\odot$, where $M_\odot$ is the solar mass) main sequence Ap/Bp stars (Hidalgo et al., 2025), which are seen to have intense, large-scale, mostly dipolar, axisymmetric, and, apparently static fields (Aurière et al., 2007), white dwarfs (Ferrario et al. (2015), at least during a part of their life), and neutron stars (Goldreich and Julian, 1969; Pons and Viganò, 2019), where a pre-existing stellar magnetic field can be strongly amplified during the post-main-sequence phase core collapse. In these cases, the field is created and then it decays slowly enough to be relevant for long timescales. However, this scenario cannot be applied to most astrophysical objects (e.g., low-mass stars, planets, stellar formation, galactic fields), for which magnetic fields are complex and, in many cases, evolve on relatively short timescales (in human terms).

2. External magnetic fields induce secondary currents in a smaller system, generating a local magnetic field. This mechanism may contribute to the magnetic properties of particular objects; however, its applicability is limited and requires specific conditions. A typical example is the variation of Jupiter's magnetic field near Europa, which is caused by the induced currents generated by the interaction of the Jovian magnetic field with Europa's postulated salty liquid ocean (Kivelson et al., 1999; Khurana et al., 2020; Keller et al., 2024).

3. A self-sustaining dynamo, which is arguably the most interesting one and the main object of this thesis, where I study in particular the interstellar medium and planetary interior scenarios. In these cases, the magnetic field must be continuously generated to compensate for its dissipation, as the evolutionary timescales of the objects exceed the magnetic diffusion timescales by far.

The word *dynamo* encompasses the processes during which kinetic energy is converted into electric or magnetic energy, but it has a slightly different connotation between astrophysics and engineering. For daily life applications, this word is commonly associated with the wired, coiled structures that convert mechanical kinetic energy into electrical current in the presence of an external magnetic field, typically provided by a magnet. In astrophysics, the dynamo refers specifically to the third case above, where kinetic energy is provided by the movement of electrically conducting fluids, which are simultaneously the source of the magnetic fields. One such definition can be found, for example, in Andrew D. Gilbert's Handbook of Mathematical Fluid Dynamics (2003): *Dynamo theory concerns the generation of magnetic field from the flow of an electrically conducting fluid, relevant to the magnetic fields of the Earth, Sun, planets, stars, and galaxies.* Both physical scenarios rely on Maxwell's equations for their mathematical description. However, in astrophysics and plasma physics, electromagnetism is intricately intertwined with the high complexity of fluid dynamics, leading to the formulation of magnetohydrodynamics (MHD).

The thesis is divided into two parts, reflecting the different nature of the main topics I have been involved in. Although galactic and planetary magnetic fields have vastly different length scales, both rely on numerical modeling to understand their dynamo mechanisms. The reason is that laboratory experiments at comparable astrophysical scales or regimes are not feasible. Moreover, although the MHD equations have been theoretically used to study specific aspects under ideal assumptions, such as dynamo instabilities in the kinetic phase, analytical approaches cannot generally explore the complexity of real scenarios. Thus, computational experiments, i.e., numerical simulations, are currently the primary research tool for studying and reproducing observed physical phenomena. MHD, and more specifically dynamo theory, are topics where numerical simulations have enormously expanded our knowledge, although many open questions remain.

After a general introduction to MHD and dynamo theory in Chapter 2, the first topic of this thesis is developed in Chapter 3. It examines the growth of vorticity and the magnetic field using random irrotational subsonic expansion flows, a toy model for supernova explosions in the interstellar medium (ISM). The main driving forces supplying energy to the ISM are believed to be supernova explosions and stellar winds. Such localized sources are assimilable to curl-free velocity fields as a first approximation. They need to be combined with other physical processes to replicate real galactic environments, such as the presence of turbulence and a dynamo-sustained magnetic field in the ISM. We used the *Pencil code* to run resistive MHD numerical simulations in a periodic box with resolutions up to $512^3$ to model subsonic expansion flows. I explore the necessary ingredients required for spherical irrotational accelerations to create a turbulent environment that models the ISM. Specifically, I determine how the presence of rotation, baroclinicity, and shear, or a combination of these, affects subsonic expansion flows and whether it leads to the occurrence of a dynamo. I aim to identify the minimum ingredients required to trigger a dynamo instability, as well as the relationship between the dynamo and the growth of vorticity. I explore the parameter space set by several physical and numerical quantities of the model. Specifically, in comparison to previous works (Mee and Brandenburg, 2006; Del Sordo and Brandenburg, 2011), I test a wider parameter space in terms of Reynolds numbers (up to a few hundred), the magnetic Prandtl number, the forcing scale, and the cooling timescale in Newtonian cooling.

Additionally, I test whether imposing a curl-free acceleration rather than a curl-free velocity affects the system, and I evaluate the specific terms responsible for vorticity generation. I report the absence of a small-scale dynamo in all cases where only rotation is included, regardless of the given equation of state and rotation rate. When a shearing velocity field is introduced as a background sinusoidal profile, I find a hydrodynamic instability that produces an exponential vorticity growth at all scales, starting from the smallest ones. The onset of this instability occurs after a rather long temporal evolution of several thousand turbulent turnover times. The vorticity dynamo, in turn, drives an exponential growth of the magnetic field, which initially occurs at small scales, followed by large ones. The instability is then saturated, and the magnetic field approximately reaches equipartition with the turbulent kinetic energy. During the saturation phase, I can observe a magnetic field winding in the direction of the shearing flow. By varying the intensity of the shear, I see that the growth rates of this instability change. These results have led to two publications (Elias-López et al., 2023; Elias-López et al., 2024).

The second part of the thesis is devoted to the internal dynamos of gas giant planets, with a view toward exoplanets. Chapter 4 introduces planetary magnetism from an observational point of view and spherical shell convective numerical dynamos, which are commonly used to model convection and dynamo processes in planets and stars. The discovery of thousands of exoplanets has started a new era of planetary science, expanding our ability to characterize diverse planetary features. However, magnetic fields remain one of the least understood aspects of exoplanetary systems. A deeper understanding of planetary dynamos and the evolution of surface magnetic properties throughout a planet's lifetime is a key scientific purpose, with implications for planetary evolution, habitability, and atmospheric dynamics.

In this thesis, I explore specific aspects of magnetism and convection for gas giants. Chapter 5 is dedicated to the modeling of the evolution of magnetic fields generated by dynamo action in cold Jupiters. I examined the changes in the morphology and strength of the magnetic field at various evolutionary stages, offering a comprehensive view of the planetary life cycle. I solved the resistive MHD equations under the anelastic approximation with the 3D pseudo-spectral spherical shell MHD code *MagIC*. I employed 1D thermodynamical hydrostatic profiles taken from gas giant evolutionary models (using the evolutionary code *MESA*) as the background states of our MHD models. These calculations were performed using radial profiles corresponding to different planetary ages, allowing me to interpret them as different snapshots of the planetary dynamo during long-term planetary evolution. I obtained saturated dynamo solutions, which I used to characterize the magnetic field at various stages in the evolution of a cold gaseous planet. I find that a likely transition from a multipolar to a dipolar-dominated dynamo regime occurs throughout the life of a Jupiter-like planet. During the planetary evolution and the cooling-down phase, I infer a decrease in the average magnetic field strength near the dynamo surface as $\approx t^{-0.2} - t^{-0.3}$. This trend is consistent with previously proposed scaling laws (Christensen and Aubert, 2006; Reiners et al., 2009), which were obtained using completely independent methods. I also find that some of the dimensionless parameters characterizing the dynamo evolve differently for the multipolar branch compared to the dipolar branch, reflecting a change in the balance of forces acting on the fluid. I capture the long-term evolution of the internal dynamo phases of magnetic fields by considering snapshots at different ages. I find a slow decay and a transition in the dynamo behavior. This approach is a versatile method for predicting the magnetic properties of giant planets and for identifying promising candidates for exoplanetary low-frequency radio emission. The results of this project are published in Elias-López et al. (2025b).

Chapter 6 is dedicated to dynamos in hot Jupiters (HJs), gas giants orbiting very close to their host stars, therefore highly irradiated by them (Fortney et al., 2021). HJs are seen to be inflated, calling for an internal deposition of heat which scales with the irradiation (Thorngren, 2024), among which the Ohmic dissipation of atmospherically-induced currents has been arguably the most explored and promising scenario so far (Batygin and Stevenson, 2010; Batygin et al., 2011; Perna et al., 2010b; Wu and Lithwick, 2013; Ginzburg and Sari, 2016; Knierim et al., 2022; Viganò et al., 2025). HJs are commonly thought to host the strongest dynamo-generated magnetic fields among exoplanets, up to one order of magnitude larger than Jupiter. However, a firm observational or theoretical proof of this expectation is currently missing. Thus, they have often been regarded as the most promising exoplanets for displaying magnetic star-planet interaction signals and magnetically driven coherent radio emission, which, unfortunately, remains elusive despite numerous diversified observational campaigns. I investigate the evolution of internal convection and dynamo properties of HJs using one-dimensional models to rigorously explore the magnetic field expectations of HJs. I explore the dependency on orbital distance, planetary and stellar masses, and types of heat injection. I employ one-dimensional evolutionary models to obtain internal convective structures. Specifically, I get the Rossby number Ro as a function of planetary depth and orbital period, after showing that tidal synchronization is likely valid for all HJs. When the heat is applied uniformly, the convective layers of almost all HJs remain in the fast rotator regime, Ro $\lesssim 0.1$, with the only exception being possibly the most massive planets with large orbital distances (but still tidally locked). The concept of fast vs slow rotators root on the observed trend between magnetic activity and rotation (or Ro) in stars (Reiners et al., 2014), which saturates for Ro $\lesssim 0.1$, meaning that the intensity of the magnetic field scales with the convective heat flux rather than the rotation. This scaling law (Christensen et al., 2009) allows us to model the magnetic field strengths for HJs by applying well-known scaling laws for fast rotators. I also consider additional internal heating to reproduce the inflated radii, implementing them with practical functional forms as derived by Thorngren and Fortney (2018), but varying the depth of the layers at which it is deposited. When strong heat sources are applied primarily on the outer envelope and outside the dynamo region, as realistic Ohmic models predict (Batygin and Stevenson, 2010; Batygin et al., 2011; Wu and Lithwick, 2013; Viganò et al., 2025), convection in the dynamo region often breaks down. Consequently, the heat flux and the derived surface magnetic fields can be significantly reduced to or below Jovian values, contrary to what is commonly assumed, which negatively affects estimates for coherent radio emission and may explain the failure in detecting it so far. The results can be found in Elias-López et al. (2025a).

The short Chapter 7 encompasses two other applications for 3D planetary spherical shell dynamos oriented. In the first part, I discuss the expected values of magnetic field curvature within the active dynamo regions and compare them with Juno observations. These calculations are a part of the submitted work on evolutionary HJ models, including Ohmic dissipation (Viganò et al., 2025). Finally, I include the preliminary results for saturated dynamo solutions using HJ interior models as a function of orbital distance as described in Chapter 6, which are part of a work in preparation. In Chapter 8, I give the overall conclusions and future prospects.

List of related publications:

1. Elias-López, A., Del Sordo, F., Viganò, D. (2023).
   *Vorticity and magnetic dynamo from subsonic expansion waves*,
   A&A, 677, A46. doi:10.1051/0004-6361/202346696.

2. Elias-López, A., Del Sordo, F., Viganò, D. (2024).
   *Vorticity and magnetic dynamo from subsonic expansion waves II: Dependence on forcing scale, Prandtl number and cooling time*,
   A&A, 690, A77. doi:10.1051/0004-6361/202450398.

3. Elias-López, A., Del Sordo, F., Viganò, D., Soriano-Guerrero, C., Akgün, T., Reboul-Salze, A., Cantiello, M. (2025)
   *Planetary dynamos in evolving cold gas giants*
   A&A, 696, A161. doi: 10.1051/0004-6361/202453372.

4. Elias-López, A., Cantiello, M., Viganò, D., Del Sordo, F., Kaur, S., Soriano-Guerrero, C. (2025)
   *Rossby number regime, convection suppression, and dynamo-generated magnetism in inflated hot Jupiters*
   ApJ 990 38. doi: 10.3847/1538-4357/adf057

5. Viganò, D., Sengupta, S., Soriano-Guerrero, C., Perna, R., Elias-López, A., Kumar, S., Akgün, T. (2025)
   *Inflated hot Jupiters: inferring average atmospheric velocity via Ohmic models coupled with internal dynamo evolution*
   Accepted to A&A in July 2025. arXiv:2403.11501.

List of related computing resources:

1. BSC, RES, AECT-2023-2-0034. *Global High-Resolution simulations of stellar dynamos: the effect of tidal star-planet interaction.* Del Sordo, F., Elias-López, A., Akgün, T., 4728 Kh

2. BSC, RES, AECT-2024-2-0011. *3D MHD dynamo in Hot Jupiters at different evolutionary times.* Elias-López, A., Del Sordo, F., Viganò, D., Soriano-Guerrero, C., Reboul-Salze, A., 2322 Kh

3. BSC, RES, AECT-2024-2-0003. *Global High-Resolution simulations of stellar dynamos: the effect of tidal star-planet interaction. Continuation.* Del Sordo, F., Elias-López, A., Viganò, D., 5250 Kh

4. BSC, RES, AECT-2025-2-0017. *3D MHD dynamo in Hot Jupiters at different orbital distances.* Elias-López, A., Del Sordo, F., Viganò, D., Soriano-Guerrero, C., 3041 Kh

5. Hidra, ICE, CSIC: $\sim$ 1500 Kh

6. Drago, CSIC: $\sim$ 1500 Kh

7. Rusty, CCA, Flatiron Institute: $\sim$ 1950 Kh

# 2

# Concepts of magnetohydrodynamics and dynamo theory

This chapter can be considered a brief introduction to the theory of MHD and, more specifically, dynamo theory, without pretense of being exhaustive or complete. The goal is to introduce the basic concepts that will be used throughout the rest of the thesis. MHD is the study of the magnetic properties and behavior of electrically conducting fluids, providing a mathematical framework for describing the flow of, for example, plasmas, liquid metals, saline water, or electrolytes in the presence of magnetic fields. Initiated by the pioneering work of Hannes Alfvén (who was awarded the Nobel Prize in Physics in 1970), MHD has become fundamental in various domains of engineering, such as plasma confinement and liquid-metal cooling in nuclear reactors, as well as in geophysical and astrophysical applications.

The diverse phenomenology that arises from MHD comes from a complex physical mechanism: the motion of the conductive fluid through a magnetic field induces currents, which in turn affect the magnetic field itself, creating a dynamic feedback system that can often be challenging to model. This is mathematically expressed as a non-linear coupling between the equations of fluid dynamics and Maxwell's equations, specifically, the Navier–Stokes, magnetic induction, and energy equations. For a more comprehensive and deeper view, there are many standard extensive books dedicated to MHD, such as Davidson (2001) and Freidberg (2014).

The complexity and properties of MHD solutions strongly depend on the specific problem, including the initial and boundary conditions, the relative differences in the length scales involved, the force balances, and the fundamental assumptions used to neglect certain terms in the physical laws. Throughout this work, I adopt the non-relativistic, fully collisional fluid approximation. The first statement is appropriate when the characteristic fluid velocities are small compared to the speed of light, or equivalently, when the time it takes for light to travel a typical fluid length scale is much shorter than any relevant dynamical timescale, such as the eddy turnover time. In practice, relativistic corrections to the fluid equations are only necessary if the bulk of the fluid particles are extremely energetic (special relativity) or to take into account spacetime distortions near compact objects like black holes or neutron stars (general relativity).

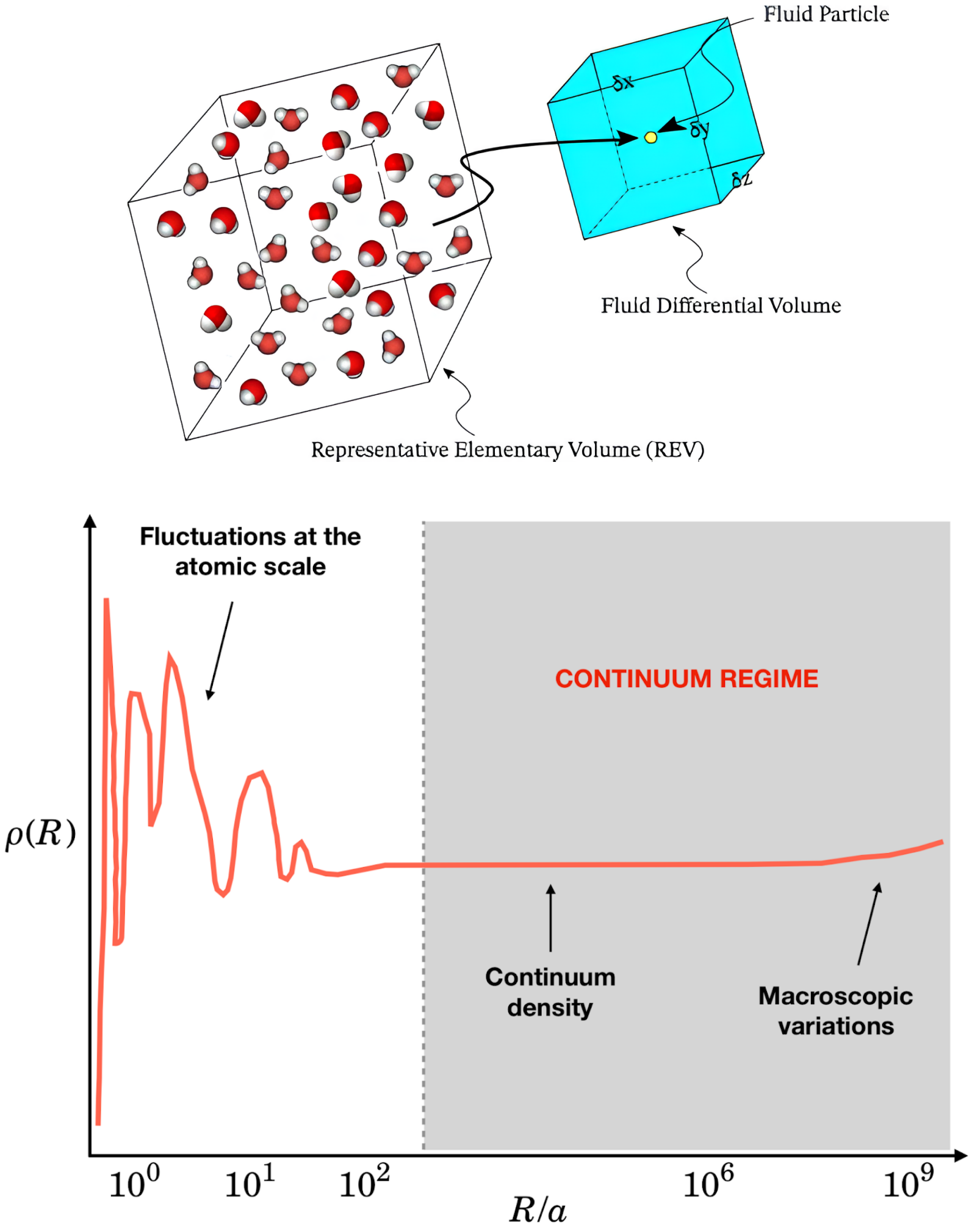

Figure 2.1: *Top:* Taken from Ramabathiran (2025). Illustration of a representative fluid element, depicting molecules and the fluid differential volume. *Bottom:* Taken from Katopodes (2019) Density fluctuations as a function of the ratio between the size of the volume over which averages are taken, $R$, and mean particle size, $a$.

Secondly, a fluid is considered collisional when both the particle mean free path and the gyroradius (or Larmor radius, i.e., the radius of the trajectory of a charge particle spiraling around a magnetic field line) are much smaller than the characteristic length scales of interest. Under such conditions, individual particle trajectories can be neglected, and the complex kinetic plasma equations can be reduced to a more tractable fluid description. This approximation is generally valid when the plasma density is sufficiently high and the non-collisional terms of the momentum equation can be neglected. These assumptions naturally lead to the continuum formulation of physical quantities. This approximation treats the fluid as a continuous medium, describable by smooth scalar and vectorial fields such as density $\rho(\mathbf{x}, t)$, pressure $p(\mathbf{x}, t)$, and velocity $\mathbf{u}(\mathbf{x}, t)$. The continuum description is valid when the representative volume of fluid contains enough particles to ensure that local averages of $\rho$ and $\mathbf{u}$ are well-defined and smooth. In other words, the fluid differential volume is much greater than the fluid particle itself. Figure 2.1 illustrates this principle, showing how density fluctuations depend on the chosen ratio between the length of the fluid parcel considered, $R$, and the particle size $a$. Throughout this text, any use of infinitesimal integration or differentiation implicitly assumes that we are within the continuum regime.

## 2.1 MHD equations

The governing equations for electrically conductive fluids are collectively known as the MHD equations, which comprise the fundamental conservation laws of mass, momentum, magnetic fields, and energy. The set of equations is closed by specifying two relations: one between thermodynamic quantities, the equation of state (EoS), and one between the electric, magnetic fields, and electrical currents, Ohm's law. MHD equations are then a set of coupled partial differential equations that describe the movement of an electrically conducting fluid in a magnetic field. Any conservation law of a quantity $q$ can be mathematically written as:

$$\frac{\partial q}{\partial t} + \boldsymbol{\nabla} \cdot \mathbf{F_q} = S_q \ ,$$

where $\mathbf{F_q}$ is the flux related to $q$ and $S_q$ represents the sources and sinks. Below, I present the full MHD equations; however, the reader can refer to Appendix A for a more detailed, albeit brief, derivation of these differential equations.

Applying mass conservation and the divergence theorem in a small fluid volume, one can obtain the continuity equation in its differential form:

$$\frac{\partial \rho}{\partial t} + \nabla \cdot (\rho \mathbf{u}) = 0, \tag{2.1}$$

where $\rho$ is the mass density and $\mathbf{u}$ is the velocity field of the fluid. This equation is equivalent to the continuity equation of electromagnetism, which is a statement of electric charge conservation. Instead of charge flux, i.e., the electrical currents, Eq. 2.1 uses the mass flux, $\rho\mathbf{u}$.

Similarly, momentum conservation leads to the equation of motion, which is known as the Navier-Stokes equation in fluid dynamics:

$$\rho\left(\frac{\partial}{\partial t} + \mathbf{u} \cdot \boldsymbol{\nabla}\right)\mathbf{u} = -\boldsymbol{\nabla} p - \frac{1}{\mu_0}\mathbf{B} \times (\boldsymbol{\nabla} \times \mathbf{B}) + \rho \mathbf{F}_\nu + \rho \mathbf{f} \ , \tag{2.2}$$

where $\mu_0$ is the vacuum magnetic permeability, $P$ is the pressure, $\mathbf{B}$ is the magnetic field, $\mathbf{f}$ is the net sum of any external force per unit mass (i.e., body acceleration), except the ones explicitly appearing, which are the fluid pressure gradient, the Lorentz force, and the viscous term, $\mathbf{F}_\nu$, defined as:

$$\mathbf{F}_\nu = \frac{1}{\rho}\boldsymbol{\nabla} \cdot (2\rho\nu\mathbf{S}) = 2\nu\mathbf{S} \cdot \boldsymbol{\nabla} ln\rho + 2\nu\boldsymbol{\nabla} \cdot \mathbf{S}$$

where $\nu$ is the kinematic viscosity in units of $L^2/t$, $\mathbf{S}$ is the rate-of-strain tensor, the components of which are given by:

$$S_{ij} = \left[e_{ij} - \frac{1}{3}\delta_{ij} \, \nabla \cdot \mathbf{u}\right], \qquad e_{ij} = \frac{1}{2}\left(\frac{\partial u_i}{\partial x_j} + \frac{\partial u_j}{\partial x_i}\right).$$

where $\delta_{ij}$ is Kronecker's delta.

To derive the induction equation, we can start from the complete Maxwell's laws, where the electric field $\mathbf{E}$ and magnetic field $\mathbf{B}$ are treated symmetrically. The fundamental limit employed in MHD is that the fluid is conductive enough so that the timescales of the $\mathbf{E}$ variation can be considered instantaneous compared to the other dynamic timescales. The displacement currents, $\varepsilon_0 \frac{\partial \mathbf{E}}{\partial t}$, become negligible. This is mathematically expressed as follows:

$$\begin{cases} \boldsymbol{\nabla}\cdot\mathbf{E} = \dfrac{\rho_{\rm q}}{\varepsilon_0} \\ \boldsymbol{\nabla}\cdot\mathbf{B} = 0 \\ \boldsymbol{\nabla}\times\mathbf{E} = -\dfrac{\partial\mathbf{B}}{\partial t} \\ \boldsymbol{\nabla}\times\mathbf{B} = \mu_0\left(\mathbf{J} + \varepsilon_0\dfrac{\partial\mathbf{E}}{\partial t}\right) \end{cases} \quad\Rightarrow\quad \begin{cases} \boldsymbol{\nabla}\cdot\mathbf{E} = 0 \\ \boldsymbol{\nabla}\cdot\mathbf{B} = 0 \\ \boldsymbol{\nabla}\times\mathbf{E} = -\dfrac{\partial\mathbf{B}}{\partial t} \\ \boldsymbol{\nabla}\times\mathbf{B} = \mu_0\mathbf{J} \end{cases}$$

where $\varepsilon_0$ is the vacuum permittivity and $\rho_q$ is the charge density, which is assumed to be zero based on the quasi-neutrality assumption, i.e., there is an equal number of positive and negative charges. Note that the electric current can now be defined with the magnetic field itself: $\mathbf{J} = \boldsymbol{\nabla}\times\mathbf{B}/\mu_0$. These equations can be closed using Ohm's law and rewritten to consider only the evolution equation for the magnetic field (Faraday's induction equation) and the solenoidal condition (i.e., magnetic flux conservation). In this thesis, Ohm's law contains the advective and resistive term, where $\lambda$ is the magnetic diffusivity[1] and $\sigma = 1/\mu_0\lambda$ is the electric conductivity:

$$\frac{\partial\mathbf{B}}{\partial t} = \nabla\times(\mathbf{u}\times\mathbf{B}) - \nabla\times(\lambda\nabla\times\mathbf{B}) \ , \qquad \nabla\cdot\mathbf{B} = 0 \ . \tag{2.3}$$

See App. A for the full derivation.

Finally, from energy conservation, one can deduce the internal energy equation:

$$\begin{aligned} \rho\left(\frac{\partial}{\partial t} + \mathbf{u}\cdot\nabla\right)\left(e + \frac{1}{2}u^2\right) + \frac{\partial}{\partial t}\left(\frac{B^2}{2\mu_0}\right) + \boldsymbol{\nabla}\cdot\left(\frac{1}{\mu_0}\mathbf{E}\times\mathbf{B}\right) = \\ = \Phi_\nu + \nabla\cdot(\kappa\nabla T) + \frac{\lambda}{\mu_0}(\nabla\times\mathbf{B})^2 - \nabla\cdot\left(p\mathbf{u} + \frac{\mathbf{B}\times(\mathbf{u}\times\mathbf{B})}{\mu_0}\right) , \end{aligned} \tag{2.4}$$

where $e$ is the internal energy density, $T$ is the temperature, $\kappa$ is the thermal conductivity which appears in the heat flux (approximated as $\kappa\boldsymbol{\nabla}T$, Fourier's law), and $\Phi_\nu$ is the viscous dissipation defined as:

$$\Phi_\nu = 2\rho\nu\mathbf{S}^2 = 2\rho\nu\left[e_{ij}e_{ji} - \frac{1}{3}(\boldsymbol{\nabla}\cdot\mathbf{u})^2\right]$$

where $\mathbf{S}^2 = S_{ij}S_{ji}$ in index summation notation, meaning that any repeated index is summed over.

In scenarios with convection, the equation for specific entropy, $s$, often replaces the energy equation:

$$\rho T\left(\frac{\partial}{\partial t} + \mathbf{u}\cdot\boldsymbol{\nabla}\right)s = \Phi_\nu + \boldsymbol{\nabla}\cdot(\kappa\boldsymbol{\nabla}T) + \frac{\lambda}{\mu_0}(\boldsymbol{\nabla}\times\mathbf{B})^2 \ , \tag{2.5}$$

Entropy is a more natural magnitude for convection processes[2], as the sign of its gradient determines how stable the system is for convection.

In the scenarios addressed in this thesis, the diffusivities can be considered isotropic, i.e., neglecting any anisotropy in conduction that may occur in other scenarios with low density and/or high magnetic fields (e.g., Koskinen et al. (2014) for the outermost atmospheric layers of magnetized planets). The set of equations (i.e., mass continuity, momentum conservation, energy conservation, the induction equation, and the solenoidal condition for the magnetic field) must be solved simultaneously to evolve the fluid or plasma flow together with the magnetic field.

[1]It is usual in MHD to define the magnetic diffusivity with $\eta$, but in this thesis, following the notation used by *MagIC*, $\lambda$ denotes the magnetic diffusivity and $\eta$ the dynamo region geometry factor (see Table 4.2). Chapter 3 is an exception: I use $\eta$ for magnetic diffusivity to keep notation consistent with the *Pencil Code*.

[2]The Schwarzschild criterion (Schwarzschild, 1906) states that to have a convection instability (i.e., buoyant fluid parcels rise while denser ones sink) one needs to have a negative entropy gradient $\partial s/\partial z < 0$.

## 2.2 Non-dimensional parameters

Table 2.1: Non-dimensional parameters usually in fluid dynamics and MHD. Other parameters used in stellar and planetary dynamo literature, such as the Rossby, Ekmann, or Lorentz numbers, are defined in Chapter 4.

| Parameter | Definition | Physical interpretation |
|---|---|---|
| Prandtl number | $\mathrm{Pr} = \dfrac{\nu}{\kappa}$ | $\dfrac{\text{Viscous diffusion}}{\text{Thermal diffusion}}$ |
| Magnetic Prandtl number | $\mathrm{Pm} = \dfrac{\nu}{\lambda}$ | $\dfrac{\text{Viscous diffusion}}{\text{Magnetic diffusion}}$ |
| Reynolds number | $\mathrm{Re} = \dfrac{UL}{\nu}$ | $\dfrac{\text{Inertial forces}}{\text{Viscous forces}}$ |
| Magnetic Reynolds number | $\mathrm{Rm} = \dfrac{UL}{\lambda}$ | $\dfrac{\text{Magnetic induction}}{\text{Magnetic diffusion}}$ |
| Péclet number | $\mathrm{Pe} = \mathrm{RePr} = \dfrac{UL}{\kappa}$ | $\dfrac{\text{Advection}}{\text{Thermal diffusion}}$ |

Fluid dynamics and MHD use non-dimensional parameters that describe the relative influence of the terms in the MHD equations by evaluating their ratios. Synthesizing the complexity of the system in one number requires assuming typical values, such as the length scale $L$, or a representative value of the velocity, $U$, or magnetic field strength $B$. Some of the most important parameters, relevant for the thesis, are defined in Table 2.1, with their corresponding physical interpretation.

These ratios are proxies for the relative importance of different physical processes within the usually complex MHD problem. For example, if $\mathrm{Pm} \gg 1$, viscous forces dominate over the resistive ones. Thus, fluctuations in velocity (i.e., small eddies) will dissipate faster than small-scale magnetic fields. Vice versa, if $\mathrm{Pm} \ll 1$, the fluid is much more resistive than viscous; thus, the small-scale magnetic fields will dissipate faster than the typical velocity fluctuation.

The Reynolds numbers, Re and Rm, are among the most important dimensionless numbers that discern between important flow regimes, respectively. When $\mathrm{Re} \ll 1$, viscosity dictates fluid motion, leading to essentially laminar flow and making dynamo action much less probable. On the contrary, when $\mathrm{Re} \gg 1$, the flow becomes inviscid and possibly turbulent, conditions much more favorable for magnetic field amplification. The last and most crucial dimensionless number for the occurrence of the dynamo process is the magnetic Reynolds number. A detailed discussion of both limits (i.e., $\mathrm{Rm} \gg 1$ or $\mathrm{Rm} \ll 1$) is included below.

### 2.2.1 Non-dimensional induction equation

The use of these non-dimensional numbers allows one to express MHD in a scale-invariant way, i.e., with the relevant dimensionless fields which can be given physical meaning with dimensional units by assigning values to a minimal set of physical magnitudes (e.g., velocity, lengthscale, and magnetic fields). Let's take the induction equation as an example, Eq. (2.3), and consider a constant magnetic diffusivity $\lambda$. Using $L$, $t_0$, $U \equiv L/t_0$, and $B_0$ as the characteristic values for the length-scale, time-scale, velocity, and magnetic field, then the transition from dimensional to dimensionless form is performed by substituting the following terms: $\mathbf{r} \to L\tilde{\mathbf{r}}$; $\mathbf{t} \to t_0\tilde{\mathbf{t}}$; $\mathbf{u} \to U\tilde{\mathbf{u}}$; $\mathbf{B} \to B_0\tilde{\mathbf{B}}$. The induction equation then becomes:

$$\begin{aligned}
\frac{\partial \mathbf{B}}{\partial t} &= \nabla \times (\mathbf{u} \times \mathbf{B}) + \lambda \nabla^2 \mathbf{B} \quad \to \\
\frac{B_0}{t_0}\frac{\partial \tilde{\mathbf{B}}}{\partial \tilde{t}} &= \frac{B_0 U}{L}\tilde{\nabla} \times (\tilde{\mathbf{u}} \times \tilde{\mathbf{B}}) + \frac{B_0}{L^2}\lambda \tilde{\nabla}^2 \tilde{\mathbf{B}} \quad \to \\
\frac{\partial \tilde{\mathbf{B}}}{\partial \tilde{t}} &= \tilde{\nabla} \times (\tilde{\mathbf{u}} \times \tilde{\mathbf{B}}) + \frac{\lambda}{UL}\tilde{\nabla}^2 \tilde{\mathbf{B}} \quad \to \\
\frac{\partial \tilde{\mathbf{B}}}{\partial \tilde{t}} &= \tilde{\nabla} \times (\tilde{\mathbf{u}} \times \tilde{\mathbf{B}}) + \frac{1}{\mathrm{Rm}}\tilde{\nabla}^2 \tilde{\mathbf{B}}
\end{aligned}$$

These dimensional transformations reflect the scaling nature of fluid dynamics. In other words, systems with vastly different scales might lead to very similar flow dynamics if their set of dimensionless numbers, i.e., the relative importance of each term in the MHD equations, is the same. This procedure is performed, but not explicitly repeated, in Chapter 4 to obtain the non-dimensional MHD equations for convection in a rotating spherical shell, as used in *MagIC*. With this form, one can easily see the two different limits for the induction equations depending on whether $\mathrm{Rm} \gg 1$ or $\mathrm{Rm} \ll 1$.

### 2.2.2 Induction equation limits

When $\mathrm{Rm} \gg 1$, the induction equation approaches the so-called ideal case, e.g., with an infinite conductivity (negligible resistivity):

$$\frac{\partial \mathbf{B}}{\partial t} \simeq \nabla \times (\mathbf{u} \times \mathbf{B}) \ . \tag{2.6}$$

In 1943, at the infancy of MHD theories, Hannes Alfvén put forward the *frozen-in flux theorem*, also known as *Alfvén's theorem*. It states that in ideal MHD, the electrically conducting fluid and embedded magnetic field are constrained to move together (see Sect. A.5 for a short proof of the theorem). This implies that the magnetic topology of a fluid in the limit of a large magnetic Reynolds number cannot change. In other words, even though the specific configuration of the field can change, that is, lines can stretch, bend, and twist, the connectedness or linkage of the lines remains constant.

The low magnetic Reynolds number limit ($\mathrm{Rm} \ll 1$) means that the advective term is negligible compared to the resistive one, which leaves the electric currents and magnetic field evolution determined by diffusion effects. Mathematically, this can be seen with the induction equation in the diffusive limit form, which, if $\lambda$ is constant, reads:

$$\frac{\partial \mathbf{B}}{\partial t} \simeq \lambda \nabla^2 \mathbf{B} \ . \tag{2.7}$$

This equation is very similar to the heat diffusion equation developed by Joseph Fourier in 1822. A diffusive equation tends to erase any gradient of the field, which in this case means dissipating the electrical currents and approaching a potential configuration ($\nabla \times \mathbf{B} = 0$, including the $\mathbf{B} = 0$ case), with a characteristic diffusive timescale $\tau_{\mathrm{diff}} = L^2/\lambda = \mu_0 \sigma L^2$.

In Table 2.2, I show the different $\tau_{\mathrm{diff}}$ for the Earth, Jupiter, Sun, and the galactic ISM. The evolutionary lifetimes of planets and stars are much longer than these magnetic diffusive timescale estimates. For the ISM, $\tau_{\mathrm{diff}}$ is of the same order of magnitude as stellar formation and collapse. It could be considered its dynamical timescale, but it is still much shorter than the galactic evolution. Thus, for all these objects, there is a need for dynamo action; otherwise, there could be no current magnetization of planets, stars, and galaxies.

Table 2.2: Estimates for astrophysical objects with magnetic field amplification or sustainment. Note that the usual ISM conductivity is $\sigma_{\mathrm{ISM}} = 1 \cdot 10^{-6}$ S/m; which leads to $\tau_{\mathrm{diff}} \sim 4 \cdot 10^{22}$ y. When turbulence effects are taken into account, the conductivity is estimated to be $\sigma_{\mathrm{ISM,turb}} = 3 \cdot 10^{-15}$ S/m.

| Object | $L$(m) | $\sigma$(S/m) | $\tau_{\mathrm{diff}}$ (y) |
|---|---|---|---|
| Earth | $2.2 \cdot 10^6$ | $1 \cdot 10^6$ | $\sim 2 \cdot 10^6$ |
| Jupiter | $3.6 \cdot 10^7$ | $1 \cdot 10^6$ | $\sim 5 \cdot 10^8$ |
| Sun | $2 \cdot 10^8$ | $1 \cdot 10^3$ | $\sim 1.6 \cdot 10^6$ |
| ISM | $3 \cdot 10^{17}$ | $3 \cdot 10^{-15}$ | $\sim 1 \cdot 10^7$ |

## 2.3 Dynamo theory

As we have seen, the origin and persistence of magnetic fields are observed to happen well beyond the resistive timescales, therefore calling for a continuous regeneration able to replace the dissipated electrical currents. Dynamo theory is the main framework that explains how magnetic fields can be spontaneously generated and sustained by the motion of electrically conducting fluids. While there is no absolute, universally accepted definition, dynamo theory is broadly regarded as the branch of physics that describes mechanisms by which a system containing a conductive fluid can spontaneously and continuously convert part of its kinetic energy into magnetic energy. Two well-established textbooks for introductory and advanced dynamo topics are those by Moffatt and Dormy (2019) and Rincon (2019), from which the main ideas of this introduction are taken.

Dynamo processes are invoked to explain the magnetic fields of a wide range of celestial bodies. For instance, Earth's geomagnetic field, the solar magnetic cycle, and the large-scale magnetic fields of galaxies such as the Milky Way are all believed to originate from some form of dynamo action. These systems differ widely in scale, structure, and plasma conditions. Yet, they share key underlying physics rooted in MHD turbulence and the non-linear feedback between fluid motion and magnetic fields.

The essential physical principle at the heart of dynamo theory is the so-called dynamo instability, i.e., the onset of the dynamo process. This process is characterized by the fast amplification of an existing weak magnetic field (a seed), thanks to its interaction with the flow of an electrically conducting fluid. This self-reinforcing process allows the fluid flow to rapidly amplify the seed magnetic field until it becomes dynamically relevant, and it is then self-sustained. The growth and sustainment of magnetic energy from kinetic energy is known as dynamo action.

Physical processes involving magnetic field growth can be well described by the induction equation (Eq. 2.3) if the corresponding classical MHD approximations are fulfilled. In other words, if the electrically conductive fluid in question can be modeled by a collisional, non-relativistic single fluid involving low-frequency and large-scale phenomena. Stellar and planetary convective interiors (as assumed in Chapters 5, 6 and 7), as well as laboratory plasma and molten metals, are examples of such scenarios. On the other hand, classical MHD breaks down for relativistic, quantum, or collisionless fluids. In the later cases, ion and electron dynamics decouple, leading to two-fluid effects with great importance at small scales. Dynamo processes around the solar corona or neutron stars require collisionless and relativistic/quantum effects, respectively. Even though the ISM is weakly collisional, quasi-hydrodynamic behavior is exhibited over large scales (tens of parsecs) and long times (Myr). This means that MHD can be effectively applied for modeling large-scale magnetic field evolution, as done in Chapter 3.

In summary, astrophysical dynamo theory provides a unifying framework for understanding how magnetic fields are generated and sustained in conducting fluids under astrophysical conditions. The interdisciplinary nature of this theory —drawing on fluid dynamics, electrodynamics, plasma physics, and turbulence theory—makes it both rich and complex, and its continued development remains central to our understanding of magnetic phenomena throughout the universe.

### 2.3.1 Anti-dynamo theorems and complex flows

Although there are almost no analytical solutions for the MHD equations, several important *anti-dynamo theorems* were proven in the past century. These place constraints on the kinds of fluid motions that can sustain magnetic fields or restrict the magnetic field morphology that can be created by dynamo action. These theorems prove that under certain symmetries or dimensional restrictions, dynamo action is impossible. Cowling's theorem (Cowling, 1934) is one of the earliest anti-dynamo results. It states:

> *An axisymmetric magnetic field cannot be maintained by dynamo action according to the MHD equations.*

This theorem implies that in a perfectly axisymmetric system (where all physical quantities are independent of the azimuthal angle $\phi$), any magnetic field will decay over time due to resistive dissipation. In practical terms, this means that 3D effects or asymmetries are necessary to sustain a magnetic field via dynamo action. The other proofs for the following theorems are very similar. Zel'dovich extended the idea further with his planar flow theorem (Zel'dovich, 1957):

> *A purely planar (two-dimensional) flow cannot sustain a magnetic field.*

Mathematically, if the velocity field $\mathbf{u}(x, y, t)$ lies entirely in a plane and does not vary in the third dimension (e.g., $z$), then dynamo action is not possible. This result demonstrates that three-dimensional flow structures are essential for maintaining magnetic fields. Additionally, theorems put further requirements on the properties of the fields, $\mathbf{B}$ and $\mathbf{u}$. For instance, Hide and Palmer (1982); Ivers and James (1984) expanded the proofs to include statements such as:

> *A magnetic field of the form $\mathbf{B}(x, y, t)$ (i.e., dependent only on two spatial coordinates and time) cannot be maintained by dynamo action.*

or;

> *A purely toroidal flow (where fluid moves in circular paths around an axis, such as in a doughnut-shaped torus) cannot sustain a magnetic field.*

One of the bottom-line conclusions of these theorems was that dynamo cannot be achieved with a simple flow, and a simple field cannot be generated by dynamo action, where the term *simple* means having specific spatial symmetries in the flow or the field. In a more scientific language, anti-dynamo theorems highlight the necessity of three-dimensional, non-axisymmetric, and often chaotic flows that twist the magnetic field for the successful operation of a dynamo. These insights are critical for understanding magnetic field generation in planets, stars, and galaxies, where complex fluid motions are present.

Several types of flow are known to lead to dynamo action. Turbulence randomly stretches and folds magnetic field lines, enhancing amplification at small scales in so-called fast dynamos.

Shear flows (i.e., velocity gradients) provide a more systematic, directional amplification due to the stretching and reorienting of magnetic field lines in a preferential direction. In Chapter 3, I show an example of a shear-driven dynamo triggered by an irrotational type of turbulent forcing. Differential rotation, a widespread feature in stars and disks, is a specific form of shear that systematically winds field lines due to varying angular velocity, triggering a dynamo by converting poloidal fields into toroidal ones. Convection, driven by thermal or compositional buoyancy, induces organized flows that play an essential role in large-scale planetary and stellar dynamos. Helical flows are a more refined type of flow, often resulting from the combination of rotation and convection, and are crucial for large-scale dynamo action, especially in mean-field theory via the $\alpha$-effect (see Sect. 2.3.4). The dynamo simulations seen in Chapters 5 and 7 are examples of dynamo numerical models driven by convection under the influence of rotation.

### 2.3.2 Kinematic and dynamic dynamo phases

In many dynamo studies, either theoretical or numerical, the initial magnetic field considered is a seed magnetic field, i.e., dynamically weak, meaning that the magnetic energy is negligible in comparison with the kinetic energy and that the Lorentz force on the fluid is initially negligible compared to the other forces and terms in the momentum equation. When a magnetic field is amplified, its evolution typically occurs in two stages: the kinematic phase and the dynamic phase.

The kinematic or linear phase is the initial stage of the dynamo process, when the weak initial magnetic field has no significant influence on the motion of the fluid. Mathematically, it means that the Lorentz force term (i.e., $\boldsymbol{B} \times (\boldsymbol{\nabla} \times \mathbf{B})/\mu_0$) has no effect in the momentum equation, Eq. 2.2. Therefore, for a given flow, the magnetic field is a passive quantity that is advected and possibly stretched and twisted. The main aim of several studies is to determine whether a particular fluid flow can amplify a small, seed magnetic field. The MHD equations can be linearized and solved, analytically or numerically, to find magnetic field eigenmodes and their growth. This is a linear regime, where the magnetic field either grows or decays, typically exponentially, based on the properties of the flow. Chapter 3 studies a linear dynamo phase with an expansive wave forcing. Even though we reach saturated states, we focus on characterizing how the kinematic phase depends on various MHD and forcing parameters.

As the magnetic field grows, it eventually affects the motion of the fluid through the Lorentz force, marking the beginning of the dynamic or saturated phase. The influence of the magnetic field on the fluid flow changes the structure of the flow, which in turn affects how the magnetic field evolves. In other words, the magnetic field and the fluid flow become fully coupled and evolve nonlinearly. This feedback loop continues until the system reaches a state of balance, where the magnetic energy stops growing and reaches a saturated state. See Vainshtein and Cattaneo (1992); Brandenburg and Subramanian (2005) for general reviews about nonlinear dynamo theory. In Chapters 5 and 7, I analyze the saturated phases of numerical dynamo solutions in the context of planetary magnetic fields. There, I am not interested in the kinematic phase, which happens due to the artificially imposed initial small field, but rather in the dependence of the saturated, self-sustained dynamo state on its characteristic dimensionless numbers.

### 2.3.3 Winding mechanism

In the presence of a differential velocity profile and once a sufficiently large magnetic field exists, there is a relevant effect known as winding (e.g., Fujisawa, 2015). For instance, given such a shearing profile of a velocity field directed along the $y$-direction and its gradient in the $z$-direction, $u_y^S(z)$, the advective term of the induction equation provides an increase of the $B_y$ component:

$$\frac{\partial \mathbf{B}}{\partial t} = \nabla \times (\mathbf{u} \times \mathbf{B}) \quad \rightarrow \quad \frac{\partial B_y}{\partial t} = -B_z \frac{\partial u_y^S}{\partial z} \ . \tag{2.8}$$

In contrast with the exponential growth in the kinetic phase of the dynamo, the winding mechanism provides a linear-in-time increase of the magnetic field along the shearing direction, which will stop when magnetic resistivity and/or dragging become important and smooth out the shear profile; namely, when the Lorentz force becomes relevant and the system becomes non-linear. This expression works for a general shearing profile $u_i^S(j)$ when $i \neq j$. This mechanism is seen in action in Chapter 3. The most common scenarios for winding are disks or spherical bodies with strong differential rotation, i.e., with azimuthal flows $u_\phi(r, \theta)$, for which the magnetic field lines are greatly stretched around the rotational axis (see $\Omega$-effect below).

### 2.3.4 Large-scale dynamos and the alpha-effect

A cornerstone of modern dynamo theory is mean-field electrodynamics, particularly as it applies to turbulent flows. Within this framework, the magnetic and velocity fields are decomposed into mean and fluctuating components (see App. A.7), allowing the derivation of an averaged induction equation. Large-scale dynamos (LSD) are dynamo processes that amplify coherent and ordered magnetic fields with correlation lengths larger than the underlying turbulence. LSDs are crucial for explaining coherent, global-scale magnetic structures observed in planets, stars, and galaxies. The theoretical backbone of LSD theory is *mean-field electrodynamics* (Moffatt, 1978; Brandenburg and Subramanian, 2005; Krause and Rädler, 1980), in which the magnetic and velocity fields are decomposed into mean and fluctuating parts:

$$\mathbf{B} = \overline{\mathbf{B}} + \mathbf{b}', \quad \mathbf{u} = \overline{\mathbf{U}} + \mathbf{u}' \ , \tag{2.9}$$

following the Reynolds rules (App. A.7).

A critical term arising in this approach is the so-called $\alpha$–effect, which encapsulates the contribution of small-scale turbulent motions to the evolution of the large-scale magnetic field. Physically, the $\alpha$–effect describes a mean electromotive force aligned (or anti-aligned) with the mean magnetic field, typically arising in rotating, stratified turbulence. This effect is especially important in astrophysical bodies where rotation and convective turbulence are both present, such as in stellar and planetary convective layers.

The development of mean-field dynamo theory, including the mathematical formalization of the $\alpha$–effect, represented a significant conceptual advance in understanding the maintenance of large-scale astrophysical magnetic fields. In particular, it offered a plausible explanation for the observed cyclic variability of solar and stellar magnetic activity, the dipolar symmetry of planetary magnetic fields, and the coherent spiral structure of galactic magnetism. While mean-field models have been successful in capturing many key features of astrophysical dynamos, they also raise subtle questions regarding nonlinear saturation, quenching of the $\alpha$–effect, and the role of magnetic helicity conservation—all of which are the subject of ongoing theoretical and numerical research.

The mathematical description of the $\alpha$–effect is obtained by substituting the decomposed velocity and magnetic fields into the induction equation, corrections and averaging over small scales gives the mean-field induction equation:

$$\frac{\partial \overline{\mathbf{B}}}{\partial t} = \nabla \times \left( \overline{\mathbf{U}} \times \overline{\mathbf{B}} + \boldsymbol{\mathcal{E}} - \lambda \nabla \times \overline{\mathbf{B}} \right) , \tag{2.10}$$

where the mean electromotive force (EMF) is defined as

$$\boldsymbol{\mathcal{E}} = \overline{\mathbf{u}' \times \mathbf{b}'} \,. \tag{2.11}$$

Let's assume that we have an initial, small-scale seed field, with no large-scale component. Dynamo action will mean the growth of $\overline{\mathbf{B}}$ from small-scale electromotive force, $\boldsymbol{\mathcal{E}}$, as it is the only remaining term in 2.10. The $\alpha$-effect is the most relevant concept of mean-field dynamo theory because it gives a theoretical mechanism by which small-scale helical turbulence induces a mean electromotive force aligned with the mean magnetic field. Physically, it results from the combined motion that twists and amplifies the magnetic field. If the turbulent velocity field has non-zero kinetic helicity (i.e., $\overline{\mathbf{u}' \cdot (\nabla \times \mathbf{u}')} \neq 0$), then one can express the EMF in terms of the large-scale:

$$\boldsymbol{\mathcal{E}} = \alpha \overline{\mathbf{B}} - \beta \nabla \times \overline{\mathbf{B}} + \cdots \,, \tag{2.12}$$

where $\alpha$ and $\beta$ are proportionality and they are related to the $\alpha$-effect and turbulent magnetic diffusivity, respectively. These terms can model systems with general flows like convection, helical turbulence, or shearing velocity profiles, such as differential rotation. Here, the $\alpha$ coefficient can be rewritten as:

$$\alpha \sim -\frac{\tau_c}{3} \overline{\mathbf{u}' \cdot (\nabla \times \mathbf{u}')}, \tag{2.13}$$

where $\tau_c$ is a characteristic correlation time of the turbulence.

In spherical and/or rotating astrophysical bodies such as disks, stars and planets, it is convenient to decompose the magnetic field in two components, poloidal and toroidal, hence having $\mathbf{B} = \mathbf{B}_{\text{pol}} + \mathbf{B}_{\text{tor}}$. The toroidal field wraps around the object and is defined by being orthogonal to the radial unit vector $\mathbf{e}_r$, i.e., $\mathbf{B}_{\text{tor}} \cdot \mathbf{e}_r = 0$. Similarly, the curl of the poloidal field is orthogonal to $\mathbf{e}_r$, i.e., $(\boldsymbol{\nabla} \times \mathbf{B}_{\text{pol}}) \cdot \mathbf{e}_r = 0$. In axial symmetry, this means that the toroidal field is purely azimuthal ($\phi$ direction), while the poloidal field is contained in the meridional plane ($r - \theta$ direction). The $\alpha$-effect plays a fundamental role in the operation of large-scale dynamos by enabling the regeneration of poloidal magnetic fields from toroidal ones (Brandenburg and Subramanian, 2005; Hughes, 2018). This mechanism is the consequence of the turbulence taking a preferred direction due to the Coriolis force. As a result, small-scale swirling motions twist and loop toroidal field lines, generating a poloidal component.

Contrary to the $\alpha$-effect, the $\Omega$-effect in mean field dynamo theory describes the generation of a toroidal magnetic field from a poloidal magnetic field due to differential rotation. Mathematically, the $\Omega$-effect can be represented by the induction equation, Eq. 2.3 in the presence of differential rotation (e.g., $\mathbf{u}_\phi = r \sin\theta \, \Omega(r, \theta)$). The shear in the angular velocity $\Omega(r, \theta)$ stretches the poloidal field lines into the azimuthal (toroidal) direction, amplifying the toroidal field component $\partial B_\phi / \partial t \approx (\mathbf{B}_p \cdot \nabla) \, \Omega$. This process efficiently converts the poloidal magnetic field $\mathbf{B}_p$ into the toroidal field $B_\phi$, and is essential for sustaining the solar and stellar magnetic cycles. Often, LSDs involve the combination of the $\alpha$-effect and the $\Omega$-effect, forming the basis of the so-called $\alpha\Omega$ dynamo models. These models are particularly accurate for stellar dynamos: they reproduce some features such as the solar magnetic cycle, magnetic polarity reversals, and the migration of sunspots. Planetary dynamos are instead thought to be of the type $\alpha^2$ (e.g. Moffatt, 1978), for which the $\alpha$-effect alone closes the loop between the toroidal and poloidal magnetic field components. However, for fast rotating gas giants like Jupiter and Saturn, which have stronger differential rotation in certain regions, leading to possible non-negligible $\Omega$-effects, leading to the hybrid dynamo model known as the $\alpha^2\Omega$ mechanism.

However, as the magnetic field grows for LSDs, the Lorentz force modifies the flow, reducing the kinetic helicity and suppressing the $\alpha$-effect, a phenomenon known as *quenching* or magnetic drag. Several models attempt to quantify these nonlinear feedback mechanisms, either via analytic expressions (e.g., $\alpha(B) = \alpha_0/(1 + (B/B_{\rm eq})^2)$, where $B_{\rm eq} = \sqrt{\mu_0 \rho u^2}$ acts as a normalization scale and represents the magnetic field strength at which the magnetic energy density equals the kinetic energy density of the turbulent motions) or through more sophisticated dynamical models that incorporate the conservation of magnetic helicity. They are known as *catastrophic quenching* models (Vainshtein and Cattaneo, 1992) as the magnetic field generation suppression is stronger when Re increases. Astrophysical systems tend to have both Re, Rm $\gg 1$, thus creating the conditions for it to happen. Understanding the saturation mechanisms, helicity transport, and scale separation remains an active area of research.

### 2.3.5 Small-Scale dynamo

Small-scale dynamos (SSDs) refer to magnetic field amplification at or below the length scale of the turbulent MHD motions. Contrary to large-scale dynamos, SSDs do not rely on large-scale flow structures, global rotation, or helicity. Instead, they exploit the random stretching, twisting, and folding of magnetic field lines by turbulent eddies. When Rm is sufficiently large (i.e., when the magnetic diffusivity, $\lambda$, is small enough), the magnetic field lines are stretched faster than they can diffuse, leading to an exponential growth of magnetic energy.

SSDs are particularly relevant in strong, turbulent environments lacking large-scale organization. A clear example is the ISM, or the primordial intercluster medium, during the formation of the first stars and galaxies. Another example is stellar and planetary convective envelopes at length scales much smaller than the object's radius. In these regimes, SSDs are capable of amplifying weak seed magnetic fields to dynamically significant strengths.

A theoretical framework for small-scale dynamos in the kinematic regime is provided by the Kazantsev model (Kazantsev, 1968), which predicts that the magnetic energy, initially concentrated at the small scales, grows exponentially. The nonlinear back-reaction of the magnetic field eventually saturates this growth. Usually, this leads to a complex non-linear steady state regime where magnetic and kinetic energies equilibrate, at least in some range of spatial scales. This is known as equipartition.

## 2.4 Some considerations about MHD codes

Usually, the equations that model astrophysical scenarios do not admit analytical solutions. Nonlinear dynamics, multi-scale interactions, and high-dimensional parameter spaces often produce analytically intractable scenarios. The MHD equations are applied to model vastly different problems, usually requiring 3D domains where both the magnetic and velocity fields are evolved. Due to computational reasons, chemical or nuclear processes are only included in basic forms.

In numerical simulations, the set of partial differential equations in MHD (2.1)-(2.3) is discretized. The vast range of relevant physical length and time scales usually implies high computational costs, and proper numerical recipes and schemes are needed to deal with such nonlinear and highly complex problems. Not surprisingly, projects that started from a single researcher have led to whole astrophysics departments spending decades-long efforts to develop efficient and accurate numerical codes. The two public MHD codes used in this thesis are the *Pencil Code*[3] (Pencil Code Collaboration et al., 2021), and *MagIC*[4] (Gastine and Wicht, 2012). All of them are international collaborations with benchmark examples and publications, ensuring their extensive verification and robustness with other code. All of them are regularly maintained with the help of software engineers to allow their compatibility with different operating systems and computer clusters.

[3]https://github.com/pencil-code

[4]https://github.com/magic-sph/magic

The discretization of the MHD equations can be performed by different numerical methods and schemes. Some common methods are: finite differences, finite volumes, spectral methods, and particle-in-cell. Each numerical approach has its strengths and limitations, and the choice often depends on the physical characteristics of the problem at hand. Other alternative methods arise with hybrid strategies, such as the finite elements method, i.e., a mixture of volume and difference methods, or pseudospectral approaches, i.e., a combination of a spectral and a finite difference method in orthogonal directions.

Finite difference methods approximate the simulation domain at discrete grid points. At each point, the discretized equations are solved at every timestep. Derivatives are approximated by evaluating differences at contiguous points. For instance, a first-order derivative $\partial f/\partial x$ can be obtained using a central difference scheme as:

$$\left.\frac{\partial f}{\partial x}\right|_i \approx \frac{f_{i+1} - f_{i-1}}{2\Delta x}.$$

Where $f$ can be, for example, the velocity or the magnetic field, depending on which are the fundamental quantities of the code. This method is simple to implement and well-suited to structured grids, but can struggle with conservation properties. The accuracy can be improved by increasing the number of cells used to approximate the derivative (accuracy order). The *Pencil Code* is a publicly available, high-order finite-difference code designed for simulating weakly compressible, turbulent flows. It employs sixth-order accuracy in space and uses a third-order Runge–Kutta explicit time integration scheme. Being a finite-difference code, it is non-conservative by design. The magnetic field is implemented with the magnetic vector potential to fulfill the solenoidal condition at every timestep. The code follows a modular architecture, enabling users to activate or deactivate specific physics modules (HD, MHD, particle dynamics, radiative transfer, chemistry, etc.), which makes it versatile for astrophysical and cosmological fluid dynamics simulations. Written primarily in `FORTRAN` and `C`, the *Pencil Code* supports MPI-based parallelization, allowing it to scale efficiently across high-performance computing environments. It is employed in Chapter 3 to solve the non-ideal compressible MHD equations.

Finite volume methods are conservative by construction and are thus particularly well-suited for solving the MHD equations in their conservative form (i.e., the integral form as shown in App. A). Instead of a grid, the domain is divided into volumes where fluxes across cell boundaries are computed:

$$\frac{d\mathbf{f}_i}{dt} = -\frac{1}{\Delta x}\left(\mathbf{F}_{i+1/2} - \mathbf{F}_{i-1/2}\right) + S_i,$$

where $\mathbf{f}_i$ is the cell-averaged conserved quantity, $\mathbf{F}$ is the numerical flux, and $S_i$ is the source or sink term in that cell volume. As they are conservative by construction, finite volume methods can implement high-resolution shock capturing methods; thus, they are the best option for handling shocks and discontinuities satisfactorily, making them very attractive to the astrophysics community. The results of Chapter 3 are compared with Achikanath Chirakkara et al. (2021), who use the finite-volume *Flash code*. As said above, we use a high-order finite difference code, which in principle leads to accurate solutions in the subsonic or weakly compressible regimes.

Spectral methods represent the fundamental MHD quantities as a sum of global basis functions, typically Fourier polynomials, Chebyshev polynomials, or spherical harmonics. The MHD equations time-evolve the set of constants, which can be used to reconstruct the real physical quantity at any point. This usually leads to exponential convergence with the size of the basis functions. As an example, for a Fourier decomposition:

$$f(x) = \sum_k \hat{f}_k e^{ikx}, \quad \frac{\partial f}{\partial x} = \sum_k ik\hat{f}_k e^{ikx},$$

the time-evolved quantity is $\hat{f}_k$ instead of $f$. Derivatives can usually be computed analytically in the spectral space, but other nonlinear terms are typically handled via hybrid/pseudo-spectral approaches, which require some computations in a physical grid. These methods usually offer the highest accuracies for smooth flows. They are less effective with strong discontinuities than the finite volume methods. Their intrinsic dissipation is much less than that of finite difference methods for a given number of points/multipoles. *MagIC*, used in this thesis, is a pseudo-spectral code that uses the spherical harmonic decomposition in the angular directions, that is, $\theta$ and $\phi$, and Chebyshev polynomials in the radial direction $r$. An illustration of the most relevant *MagIC* features and schemes can be found in App. D. It is mainly written in FORTRAN and provides many postprocessing Python functions, allowing rapid and useful data analysis. It is designed for super-computing clusters and relies on a hybrid parallelization scheme using both OpenMP in the horizontal and MPI in the radial direction. For scientific transparency, it is public, constantly maintained, and regularly updated with extensive user-friendly documentation. *MagIC* has been extensively used for both stellar and planetary models, including for Jupiter and Saturn dynamos, for instance, (Duarte et al., 2018; Wicht et al., 2019b; Gastine and Wicht, 2021; Yadav et al., 2022), and it was tested in convection and dynamo benchmarks (Christensen et al., 2001; Jones et al., 2011). In Chapter 5 and 7, we employ *MagIC* under the anelastic approximation.

A key property that an MHD code should have is to maintain the divergence-free condition for the magnetic field ($\nabla \cdot \mathbf{B} = 0$). If the magnetic field components are directly evolved via the induction equation, numerical artifacts could appear due to the violation of the solenoidal condition, leading to nonphysical fields, if there is no special care in enforcing this constraint. There are several strategies to avoid this. One is to use staggered grids, such as the Yee lattice. Another method is the use of divergence cleaning algorithms, where a new term in the induction equation is added to continuously remove deviations from $\boldsymbol{\nabla} \cdot \mathbf{B} = 0$. In this thesis, we use codes that employ another strategy. Instead of evolving $\mathbf{B}$ itself, they evolve magnetic potentials so that the constraint is fulfilled by construction. The *Pencil Code* treats the magnetic field evolution by evolving the magnetic vector potential $\mathbf{A}$. *MagIC* decomposes the magnetic field with poloidal and toroidal components, which, in turn, are expressed as two scalar potentials (see App. D). The magnetic field components are then derived from the potentials whenever the user requests a snapshot or some specific output.

# 3

# Vorticity and dynamo from expansion Gaussian waves

The occurrence of vortical flows is of high importance in astrophysics due to their close connection with turbulence. Vorticity is a method for characterizing turbulent flows, which are ubiquitous in astrophysical contexts. However, the connection between turbulence and the processes responsible for amplifying astrophysical magnetic fields has not been fully elucidated. As explained in Chapter 2, magnetic fields can be amplified at different scales in various astrophysical environments, from planets to the extragalactic medium. When a substantial fraction of the energy is spread over small scales, we can have turbulence, which, despite not having a rigorous definition, can be generally described as a state where chaotic movements of the fluid prevail over bulk, large-scale, coherent movements, implying also small-scale changes in pressure and, if present, in magnetic field direction and intensity. Since the solutions to MHD equations are highly model-dependent, turbulence can be associated with different underlying mechanisms. One of them is the natural energy cascade, for which the kinetic and magnetic energies of a system are spread over a large range of spatial scales. The prototype study in hydrodynamics comes from Kolmogorov model, which, based on basic considerations about the spatial correlations arising from isotropic, homogeneous hydrodynamic turbulence (no magnetic fields involved), predicts a slope of the magnetic energy spectral distribution $\hat{E} \sim k^{-5/3}$, where $k$ is the wavenumber of the spatial scale.

Numerical models are commonly used to simulate many of these astrophysical contexts, from planets to galaxies (see, e.g., Brandenburg and Ntormousi, 2023, for a recent review). In simulations, how the turbulent state is reached depends a lot on the setup of the numerical problem, such as initial and boundary conditions; physical ingredients that set the force balance and the relevant spatial and temporal scales at play; the existence of a preferred direction due to e.g., rotation, stratification, gravity, or background magnetic field. In many numerical simulations, turbulence is triggered through either a perturbation of an initial setup prone to instability or a forcing mechanism, i.e., a source in the momentum equation that injects energy at some given scales (Schekochihin et al., 2004a,b). Related to this, turbulence can trigger a dynamo action at small scales: the exponential growth of an initially dynamically weak magnetic field up to saturated values.

In this chapter, we focus on a simplified model that can be a simple representation of the perturbation and associated dynamo in the ISM. We focus on two open questions, namely (a) whether the occurrence of dynamo action depends on the forcing mechanism, and (b) the minimum ingredients needed to trigger a dynamo in an MHD turbulent medium.

The primary sources of energy in the ISM are of stellar origin, and, therefore, they are likely the drivers for the observed turbulence in the ISM. Supernova explosions (SN), stellar winds, protostellar outflows, and even ionizing radiation (in decreasing order of importance) are arguably the main contributors to ISM energy injections (Mac Low, 2002; Klessen and Glover, 2016). In particular, SN are believed to be the main contributor of energy and can solely provide the energy needed to produce the turbulence observed in the ISM (e.g., Gressel et al., 2008). They are also considered as one of the main ingredients in determining galactic dynamos (see Brandenburg and Ntormousi, 2023, for a recent review). Still, their effect on the amplification of galactic magnetic fields remains poorly understood. It is unclear how they contribute to the seeding and amplification of such fields to the observed values both for Milky Way-like galaxies (e.g., Ntormousi, 2018; Ntormousi et al., 2020) and high-redshift ones (Ntormousi et al., 2022). If they are seen as purely spherical or point sources, SN should not mathematically produce any vorticity in the surrounding medium.

However, many other aspects of the ISM should be taken into account. Stratification, differential rotation, shear, or shocks are physical mechanisms that might help generate vortical flows from the available energy. Three-dimensional (3D) simulations of such environments can be quite demanding, especially considering the supersonic feature of the flow. This field was pioneered by Korpi et al. (1999), who studied how SN produce turbulence in a multiphase ISM, and then by Mac Low et al. (2005), who analyzed the distribution of pressure in a magnetized ISM. However, in these simulations, the responsible mechanism for vorticity generation remains somewhat unclear. The question of whether vorticity can be amplified by a purely curl-free (compressible, or irrotational) forcing (e.g., Federrath et al., 2010; Federrath et al., 2011) or not (Mee and Brandenburg, 2006; Del Sordo and Brandenburg, 2011) is still a matter of debate.

Mee and Brandenburg (2006) and Del Sordo and Brandenburg (2011) used localized random expansion waves as forcing terms only in the momentum equation. This potential forcing induces spherically symmetric successive accelerations in random places during a given time interval, reaching locally trans-sonic regimes and leading to a somewhat similar SN driving. Such an irrotationally forced flow was found not to generate any vorticity up to $512^3$ mesh points (Mee and Brandenburg, 2006). However, when rotation, explicit shear, or baroclinicity were added, Del Sordo and Brandenburg (2011) found that vorticity is produced. Magnetic fields were added to the most straightforward setup (no rotation or shear, nor baroclinicity) by Mee and Brandenburg (2006), who found no evidence of magnetic field amplification. On the contrary, they found that this type of flow has a highly destructive effect on the magnetic field at the most minor scales. In general, the role of SN feedback on regulating star formation rate and the structure of the galactic disk is still not completely understood (e.g., Hennebelle and Iffrig, 2014; Iffrig and Hennebelle, 2017), and it may depend on how SN are implemented in simulations. The nature of the forcing may have an effect on several features of the ISM, besides the growth of the magnetic field. For instance, Mathew et al. (2023) showed how a purely compressible forcing in MHD simulations of star cluster formation influences the initial mass function.

In this chapter, we expand on the work by Mee and Brandenburg (2006) and Del Sordo and Brandenburg (2011). We aim to study how shear, rotation, and baroclinicity, in combination with the same irrotational forcing in the form of localized random expansion waves, contribute to the decay or growth of an initial magnetic field seed. We want to explore the possible dynamo action in a parameter space defined by the forcing parameters, the rotation and shear rates, the Reynolds and Prandtl numbers, and the initial magnetic seed. Our work concentrates on the subsonic and transonic regimes to establish the conditions under which a purely compressible forcing may amplify vorticity and magnetic fields.

Other works have also studied the appearance of small-scale dynamo (SSD), which is dynamo action at length scales equal to or smaller than the forcing scale, as a consequence of irrotational forcing. For a curl-free forcing based on early universe inflationary models (Kahniashvili et al., 2012), the Lorentz force was found to produce some vorticity in isothermal models. Still, the magnetic field eventually dissipates without undergoing any amplification. Additionally, Dosopoulou et al. (2012) found analogous results in the context of magnetized and rotating cosmological models. Porter et al. (2015) performed simulations of subsonic turbulence in an intra-galactic cluster medium with a purely curl-free (compressible) forcing, finding that dynamo can be excited when divergence-free (solenoidal) modes arise.

Nevertheless, other approaches found vorticity and magnetic fields to be exponentially amplified when a purely irrotational forcing is added in the form of a stochastic function in Fourier space (e.g., Federrath et al., 2010; Achikanath Chirakkara et al., 2021; Seta and Federrath, 2022). They use a turbulent driving forcing in Fourier space, which can be purely solenoidal, compressive, or a combination of both. Their isothermal dynamos can operate efficiently in both subsonic and supersonic regimes, both solenoidal and compressive cases, and even without large-scale contributions to the forcing, such as rotation or shear. Consequently, it remains unclear which are the minimum ingredients needed to excite a dynamo.

Apart from Hennebelle and Iffrig (2014) and Iffrig and Hennebelle (2017), other studies have achieved magnetic field growth with more complex forcings inspired by the ISM. Gent et al. (2013a,b) simulated a multi-phase ISM randomly heated and stirred by SN, in the energy equation, with a stratified, rotating, and shearing local domain with a galactic gravitational potential and shock diffusivities. The obtained flow dynamics are robust when three major phases are used (defined with temperature ranges) and when the parameters are adjusted to reproduce individual SN remnants. Käpylä et al. (2018) simulated such events with a combination of mass transfer during SN and stellar formation, energy deposition in the energy equation, and a stably stratified, rotating medium mimicking the galactic plane. They also obtained an SSD in such simulations and analyzed the specific vorticity source terms. This was also confirmed in subsequent studies by Gent et al. (2021, 2023) of a multi-phase ISM. Recently, Seta and Federrath (2022) studied turbulent dynamo in a multi-phase ISM as well, providing a detailed analysis of vorticity sources in their simulations.

These models are fairly complex as they want to reproduce what happens in a galactic environment. Our approach has instead been to work with a much simpler model to shed light on the minimum ingredients needed to amplify a magnetic field up to equipartition values in the presence of a purely irrotational forcing. This is in the spirit of studying general aspects of dynamo-generated magnetic fields and discussing their applications in planetary, stellar, or galactic environments. The analytical mean-field approach by Krause and Rädler (1980) and Rädler and Stepanov (2006) predicts the presence of an electromotive force and, hence, an amplified mean magnetic field, in the presence of a mean flow. While this is undoubtedly happening in the case of helical flows (see e.g., Brandenburg and Subramanian, 2005; Rincon, 2019), a mean electromotive force can (in principle) be observed also with non-helical turbulence in the presence of rotation Rädler (1968, 1969) or with a large-scale flow with associated vorticity (Rogachevskii and Kleeorin, 2003). LSDs were found in numerical experiments employing a non-helical forcing to drive small-scale turbulence embedded in a large-scale shear flow (e.g., Yousef et al., 2008; Brandenburg et al., 2008; Singh and Jingade, 2015). However, the forcing functions in these calculations did allow the injection of vorticity on small scales.

In this chapter, we test the possibility of driving a dynamo process with turbulence driven in a purely compressible way, either in the absence or in the presence of large-scale flows. We further investigate the effect of varying the explosion width and strength, the magnetic Prandtl number, and the usage of a Newtonian cooling function on the possible amplification of the vorticity and magnetic field. This chapter is based on two articles (Elias-López et al., 2023; Elias-López et al., 2024) and is organized as follows: in Sect. 3.1 we describe the MHD model and the organization of our numerical experiments. In Sect. 3.2 we show the result of our study. We observe that we do not obtain an HD or MHD instability unless a background shearing flow is imposed, despite our fairly wide exploration of the parameter space in terms of forcing scales (see Sects. 3.2.6 and 3.2.7), magnetic Prandtl number (Sect. 3.2.8), cooling times (Sect. 3.2.9), and Reynolds numbers. The models, including a shearing profile instead, develop an exponential increase in vorticity, followed by an exponential increase in the magnetic field, unless the scale of the forcing is too small, as explained in Sect. 3.2.7. In Sect. 3.1.5, we examine the contribution of every source term in the vorticity equation. Finally, in Sect. 3.3, we discuss the implications of these results in the framework of galactic dynamos. In App. B we show all tabulated runs and diagnostics described in this chapter.

## 3.1 Model and numerical methods

### 3.1.1 MHD equations with rotation and shear

We ran numerical simulations using the *Pencil Code*[1] (Pencil Code Collaboration et al., 2021), which is a non-conservative, high-order, finite-difference (sixth-order accurate in space and third-order Runge-Kutta in time) 3D MHD public code. We solved the non-ideal fully compressible MHD equations in a Cartesian box (coordinates $x, y, z$) following an approach similar to what was done by Del Sordo and Brandenburg (2011), that is, either in a rigidly rotating frame, with angular velocity $\mathbf{\Omega} = \Omega \mathbf{e}_z$, or with a differential velocity (shear) given, in their case, by $\mathbf{u}^S = u_y^S(z)\mathbf{e}_y$. We considered a homogeneous medium with no stratification. We solved the equations in the co-rotating reference frame, for which the mass and momentum conservation equations are expressed as:

$$\frac{D\ln\rho}{Dt} = -\nabla\cdot\mathbf{u}\ , \tag{3.1}$$

$$\frac{D\mathbf{u}'}{Dt} = -\frac{\nabla p}{\rho} + \frac{\mathbf{J}\times\mathbf{B}}{\rho} - 2\mathbf{\Omega}\times\mathbf{u} - u_z\frac{\partial u_y^S}{\partial z}\hat{\mathbf{y}} + \mathbf{F}_\nu + \mathbf{f} + \mathbf{f}_s\ , \tag{3.2}$$

where: $\mathbf{u}(t) = (\mathbf{u}^S + \mathbf{u}'(t))$ is the total velocity field combining the fixed shearing velocity with the residual (possibly turbulent) velocity $\mathbf{u}'$; $\rho$ is the mass density; $p$ is the pressure; $\mathbf{B}$ the magnetic field; and $\mathbf{J}$ the electrical current density. We define the turbulent components for the velocity, $\mathbf{u}'$, and vorticity, $\boldsymbol{\omega}'$, with the usual Reynolds decomposition rules (see App A.7) where the shearing velocity profile sets the large component of the flow.

Importantly, the advective derivative operator, $D/Dt := \partial/\partial t + \mathbf{u}\cdot\nabla$, applies to the total velocity, that is, to both the evolving residual velocity $\mathbf{u}'$ (kept in the left-hand side) and the fixed shearing velocity, $\mathbf{u}^S$. We keep the latter on the right-hand side, in the form of the only term $-\mathbf{u}\cdot\nabla\mathbf{u}^{\mathbf{s}} = -u_z\frac{\partial u_y^S}{\partial z}$. The other source terms are associated to the viscous force, $\mathbf{F}_\nu = \rho^{-1}\nabla\cdot(2\rho\nu\mathbf{S})$,

[1] https://github.com/pencil-code

where the traceless rate of strain tensor $\mathbf{S}$ is $S_{ij} = (1/2)(u_{i,j} + u_{j,i}) - (1/3)\delta_{ij}\nabla\cdot\mathbf{u}$, the external expansion wave forcing, $\mathbf{f}$, and a shearing forcing, $\mathbf{f}_s$ (see below for their definitions).

To close the system of equations, we consider two types of EoS: 1) a simple barotropic EoS (i.e., the pressure is $T$-independent and function of density only) $p(\rho) = c_s^2\rho$, where we fix the value of the sound speed $c_s = 1$ or 2) an ideal EoS (denoted as the baroclinic case), $p(\rho, T) = \rho R_g T$, with $R_g$ the specific gas constant and $T$ the temperature; in this case, the sound speed squared is $c_s^2 = (\gamma - 1)c_p T$, where we fix the adiabatic index $\gamma = c_p/c_v = 5/3$ (corresponding to a monatomic perfect gas), and $c_p$ and $c_v$ are the specific heats at constant pressure and constant volume, respectively.

In the latter case, the energy equation is expressed in terms of $\rho$, the specific entropy, $s$, and $T$:

$$T\frac{Ds}{Dt} = 2\nu\mathbf{S}^2 + \rho^{-1}\nabla\cdot(c_p\rho\chi\nabla T) + \rho^{-1}\frac{\eta}{\mu_0}(\boldsymbol{\nabla}\times\mathbf{B})^2 - \frac{1}{\tau_{cool}}(c_s^2 - c_{s0}^2) \;, \tag{3.3}$$

where $\chi$ is the thermal diffusivity[2], $\eta$ is the magnetic diffusivity, $c_{s0}$ is the initial, uniform sound speed (proportional to the initial temperature) and $\tau_{cool}$ regulates the timescale of the simple effective (also called Newtonian) cooling term, introduced to avoid indefinite heating by viscous and resistive dissipation.

The time evolution of the magnetic field is done by uncurling the usual induction equation shown in Eq. 2.3, using $\nabla\times\mathbf{A} = \mathbf{B}$, where $\mathbf{A}$ is the vector potential:

$$\frac{\partial\mathbf{A}}{\partial t} = \mathbf{u}\times(\nabla\times\mathbf{A}) - \eta\nabla^2\mathbf{A} \;. \tag{3.4}$$

Since the evolved field is the vector potential, and the divergence of the curl of any vector is zero, the solenoidal condition, $\nabla\cdot\mathbf{B} = 0$, is therefore fulfilled by construction.

### 3.1.2 Forcing

We impose an irrotational forcing, as a gradient of a Gaussian potential function, considering two slight variants:

$$\mathbf{f}_{\rm acc}(\mathbf{x}, t) = \nabla\phi(\mathbf{x}, t) = K\nabla e^{-(\mathbf{x}-\mathbf{x_f}(t))^2/R_f^2} \;, \tag{3.5}$$

$$\mathbf{f}_{\rm mom}(\mathbf{x}, t) = \frac{\rho_0}{\rho}\nabla\phi(\mathbf{x}, t) = K\frac{\rho_0}{\rho}\nabla e^{-(\mathbf{x}-\mathbf{x_f}(t))^2/R_f^2} \;, \tag{3.6}$$

where $\mathbf{x} = (x, y, z)$ is the position vector, $\mathbf{x}_f(t)$ is the random position corresponding to the center of the expansion wave, $R_f$ is the radius of the Gaussian, and $N$ is the normalization factor. We will make use of $k_f$ as the wavenumber corresponding to the scale of the adopted forcing:

$$k_f = \frac{2}{R_f} \;.$$

The time dependence of the forcing enters in the correlation time $\Delta t$ of the expansion waves (i.e., the time interval after which we consider a new random position, $\mathbf{x}_f$). We consider two different cases: in the first one $\mathbf{x}_f$ changes at every time-step (which is adaptive), $\Delta t = \delta t$, while in the second case we define an interval forcing time $\Delta t > \delta t$. We chose the normalization factor to be:

$$K = \phi_0\sqrt{c_{s0}R_f/\Delta t} \;,$$

where $\phi_0$ controls the forcing amplitude and both it and $K$ have dimensions of velocity squared.

Note that $f_{\rm acc}$, creates, by definition, an irrotational acceleration field (i.e., $\nabla\times\mathbf{f}_{\rm acc} = 0$), since the curl of any field gradient is zero. Mee and Brandenburg (2006) proved that the inclusion of

[2] The *Pencil Code* uses another definition from the one used in Chapter 2, i.e., $\chi = \kappa/c_p\rho$

$f_{\rm acc}$ alone indeed produced no measurable vorticity: the amount of vorticity produced showed a decreasing dependence on numerical resolution, hence they concluded its nature was numerical, essentially due to the discretization errors. The second forcing, $f_{\rm mom}$, corresponds instead to an irrotational force field which, due to the $\rho$ factor, will not necessarily lead to an irrotational acceleration. In other words, the interaction between the $f_{\rm mom}$ and fluctuations in $\rho$ might gradually create vorticity. The names have been chosen because $f_{\rm acc}$ directly changes the acceleration of the fluid, instead $f_{\rm mom}$ acts as a force and changes the momentum.

### 3.1.3 Non-dimensional parameters

After an initial transitory phase, the simulations reach a stationary state, throughout which the main average quantities maintain a saturated value. In particular, we will look at the root mean square of the velocity, $u_{\rm rms}$. In turn, this is used to define the fundamental timescale of our problem, which we will call turnover time, as

$$t_{\rm turn} = (k_f u_{\rm rms})^{-1} \ . \tag{3.7}$$

The turnover time can be understood as the average time for the fluid to cross the typical scale of the enforced expansion wave.

The root mean square values of velocity $u_{\rm rms}$ and vorticity $\omega_{\rm rms}$ (see Sect. 3.1.5 for its definition) are used to define the following dimensionless numbers:

$$\begin{aligned} {\rm Re} = \frac{u_{\rm rms}}{\nu k_f} \ , \qquad {\rm Rm} = \frac{u_{\rm rms}}{\eta k_f} \ , \\ {\rm Re}_\omega = \frac{\omega_{\rm rms}}{\nu k_f^2} \ , \qquad {\rm Ma} = u_{\rm rms}/c_s \ , \\ {\rm Pm} = \nu/\eta \ , \qquad {\rm Ro} := u_{\rm rms} k_f / 2\Omega \ , \end{aligned} \tag{3.8}$$

which are the Reynolds number, magnetic Reynolds number, vorticity Reynolds number, Mach number, magnetic Prandtl number, and Rossby number, respectively. Some of the dimensionless numbers were already shown in Table 2.1, where here we have used $U = u_{\rm rms}$ and $L = 1/k_f$. Analogously, we shall consider the maximum values of the velocity, $u_{\rm max}$, to define the maximum values of the parameters, denoted with $^M$, for instance:

$${\rm Re}_M = \frac{u_{\rm max}}{\nu k_f} \ , \qquad {\rm Ma}_M = u_{\rm max}/c_s \ . \tag{3.9}$$

Regarding the magnetic fields, we shall consider the root-mean-square of the intensity, $B_{\rm rms}$, which is closely related to the magnetic energy density (i.e., $E_{\rm mag} \propto B_{\rm rms}^2$).

### 3.1.4 Shear

To maintain a given shear velocity background along the $y$-direction $u_y^S$, we use a forcing term in the $y$-component of the velocity: $\mathbf{f}_s = (u_y^S - u_y)\hat{\mathbf{y}}/\tau_s$ , where $\tau_s$ is the timescale of the forcing (which we keep unity in all cases).

We choose to apply a sinusoidal background flow for the shearing term:

$$u_y^S(z) = A\cos(kz) \ , \tag{3.10}$$

with $k$ the shear wavenumber and $u_i$ indicate the unit vector in the direction $i$. As z ranges from $-\pi$ to $\pi$ (see Sect. 3.1.6), we have set $k = 1$, which allows simple periodic boundary conditions in the three directions. The same $k$ is used as the variable for the definition of spectrum, see

Sect. A.9. This profile is similar to what is used by Skoutnev et al. (2022) for studying dynamo in stellar interiors with a non-helical forcing, and was employed by Käpylä et al. (2009); Käpylä et al. (2010) in the context of a stratified convective medium.

We discarded the use of a more standard linear shear term $u_y^S = Sz$ since the implemented shearing boundary conditions were giving a spurious growth of vorticity at the boundaries (see Sect. 3.2.4 for a more detailed explanation).

### 3.1.5 Vorticity

The vorticity is defined as the curl of the velocity field ($\omega = \nabla \times \mathbf{u}$), and it is a way of quantifying the turbulent motions generated in fluid flows. Vortical motions are directly connected to the turbulent dynamo action, which is the mechanism by which our model may grow or maintain the magnetic field. The evolution of vorticity can be obtained by taking the curl of the total velocity evolution equation ($\partial \mathbf{u}/\partial t$):

$$\frac{\partial \boldsymbol{\omega}}{\partial t} = \boldsymbol{\nabla} \times (\mathbf{u} \times \boldsymbol{\omega}) + \boldsymbol{\nabla} \times \mathbf{F}_\nu + \frac{\nabla \rho \times \nabla p}{\rho^2} + \nabla \times \left(\frac{\mathbf{J} \times \mathbf{B}}{\rho}\right) \\ -2\nabla \times (\boldsymbol{\Omega} \times \mathbf{u}) + \nabla \times \mathbf{f} + \nabla \times \mathbf{f}_S \tag{3.11}$$

Here the first term on the right-hand side is analogous to the advective term $\nabla \times (\mathbf{u} \times \mathbf{B})$ in the induction equation, the second term represents the viscous forces acting on the system, the third is the baroclinic term, related to the EoS and calculated as $\boldsymbol{\nabla} T \times \boldsymbol{\nabla} s$ (see App. A.6), the fourth is the effect of the Lorentz force, the fifth appears if the system is rotating, the sixth is due to the impact of the implemented forcing and the seventh appears if some large-scale shear acts on the system.

Each right-hand side term in Eq. 3.11 can be considered a source or a dissipative term. The first two terms on the right-hand side are regarded as the amplification and the viscous forces terms, respectively. The vorticity equation notoriously resembles the induction equation. This equation is not evolved directly, but by analyzing its components, it is fundamental to understand the production mechanisms of vorticity, as each of the different terms can be either a source or a dissipative term to the total vorticity.

As we want to address what are the sources of vorticity in our numerical setup, we can take the dot product with $\boldsymbol{\omega}$, averaged over the volume (indicated by $\langle \cdot \rangle$), and using the vector identities $(\boldsymbol{\nabla} \times \mathbf{a}) \cdot \mathbf{B} = \boldsymbol{\nabla} \cdot (\mathbf{a} \times \mathbf{B}) + \mathbf{a} \cdot (\boldsymbol{\nabla} \times \mathbf{B})$ and $\nabla^2 \mathbf{a} = \boldsymbol{\nabla}(\boldsymbol{\nabla} \cdot \mathbf{a}) - \boldsymbol{\nabla} \times (\boldsymbol{\nabla} \times \mathbf{a})$, we obtain:

$$\frac{1}{2}\frac{\partial}{\partial t}\langle \boldsymbol{\omega}^2 \rangle = \langle (\mathbf{u} \times \boldsymbol{\omega}) \cdot \mathbf{q} \rangle - \nu \langle |\mathbf{q}|^2 \rangle + 2\nu \langle \mathbf{S} \boldsymbol{\nabla} \ln \rho \cdot \mathbf{q} \rangle - \\ -\langle (\boldsymbol{\nabla} T \times \boldsymbol{\nabla} s) \cdot \boldsymbol{\omega} \rangle + \left\langle \frac{\mathbf{J} \times \mathbf{B}}{\rho} \cdot \mathbf{q} \right\rangle - 2\Omega \langle (\boldsymbol{e_z} \times \mathbf{u}) \cdot \mathbf{q} \rangle + \\ + \langle \mathbf{f} \cdot \mathbf{q} \rangle + \langle \mathbf{f}_S \cdot \mathbf{q} \rangle \ , \tag{3.12}$$

where $\mathbf{q} = \boldsymbol{\nabla} \times \boldsymbol{\omega}$. We did not evolve $\boldsymbol{\omega}$ itself; thus, all the quantities are obtained from the evolution of $\mathbf{u}$. We use these diagnostic magnitudes to discriminate which are the most relevant vorticity amplification or destruction terms. Additionally, we can address the accuracy in numerically conserving the vorticity, i.e., assessing how small the relative difference between the sum of all right-hand side terms and the numerical time derivative of $\langle \boldsymbol{\omega}^2 \rangle$.

As mentioned above, the forcing $\mathbf{f}_{\rm acc}$ is irrotational by construction, while $\mathbf{f}_{\rm mom}$ is not. Therefore, the forcing $f$ in Eq. 3.11 and 3.12 is zero if $\mathbf{f} = \mathbf{f}_{\rm acc}$.

### 3.1.6 Boundary and initial conditions

The simulation domain consists of a uniform, cubic grid mesh $[-\pi, \pi]^3$, with triply periodic boundary conditions. We consider resolutions varying from $128^3$ up to $512^3$. Since MHD equations are dimensionally scalable, as commonly done in numerical codes, we adopt non-dimensional variables by measuring speed in units of the initial sound speed, $c_{s0}$, length in units of $1/k_1$ where $k_1$ is the smallest wave number in the periodic domain, implying that the non-dimensional size of the domain is $(2\pi)^3$. Regarding the initial conditions, pressure (and entropy and temperature in the baroclinic case), and density are set constant and with value 1 throughout the box. The flow is initially still, $\mathbf{u} = 0$.

Finally, we set a dynamically weak initial magnetic field. The flow is static at $t = 0$, but it is soon shaken by the expansion waves. However, the initial weak magnetic field does not influence the flow. For the initial magnetic topology, we consider either a uniform field in one given direction or a spatially random field. The latter is assigned by picking, at each point, uncorrelated random values to the three components of the initial magnetic vector potential, and it corresponds to an initial magnetic energy spectrum $E_{\rm mag}(k) \propto k^4$, as reported by Mee and Brandenburg (2006).

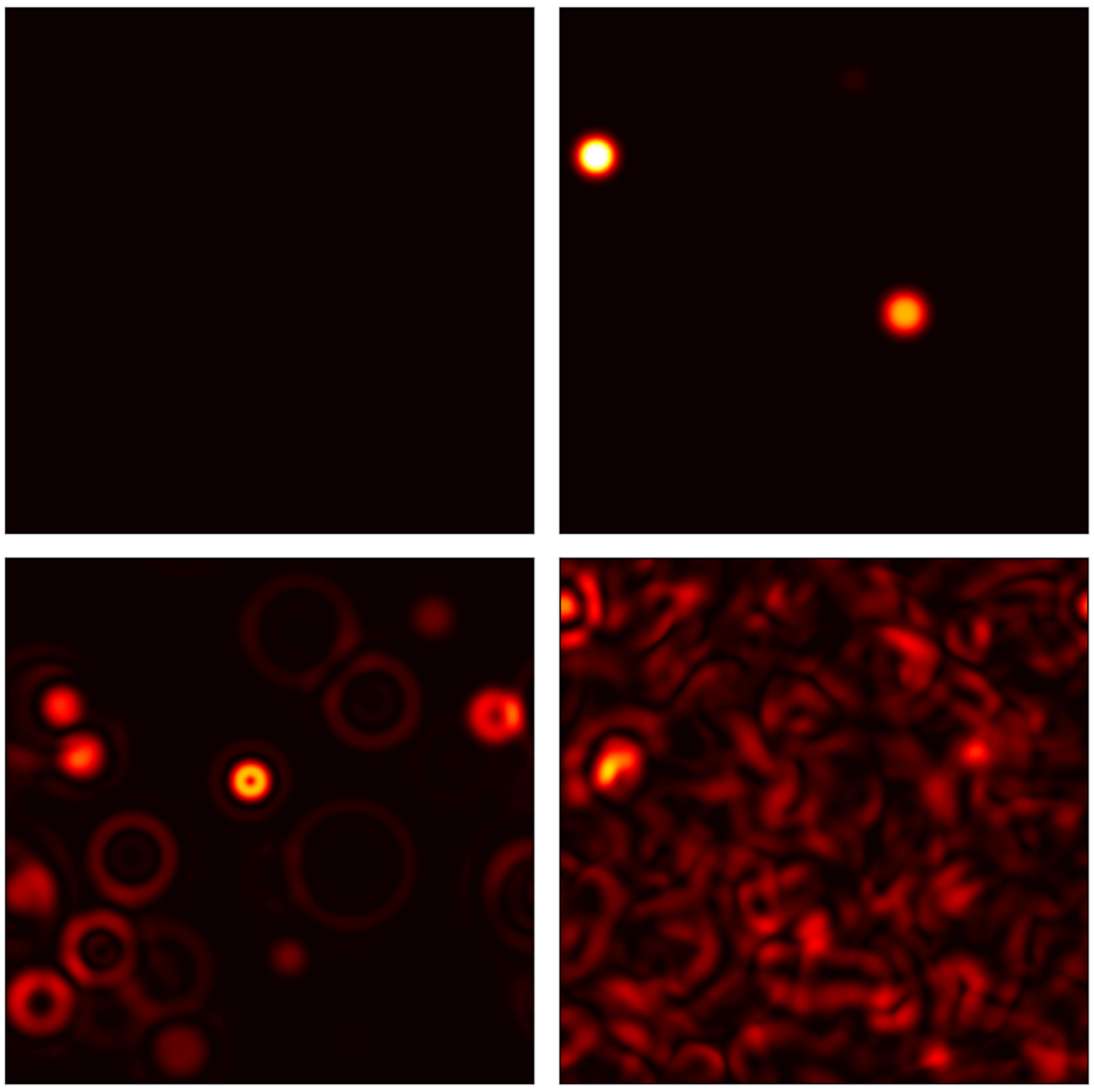

Figure 3.1: Slice in the $xy$-plane showing the value of $u^2$ at $t/t_{\rm turn} = 0, 0.01$ (top), 0.1, and 1 (bottom) for the non-rotating model H_0cW1 of Table B.1.

## 3.2 Results

### 3.2.1 Explored parameters

We ran a series of different numerical models to explore the role of each term in the right-hand side of Eq. (3.2). The forcing and viscous terms are always included in all models, while we separately consider the activation of rotation and/or shear. The forcing amplitudes are high enough to create transonic flows, but we do not investigate supersonic flow. We have restrained ourselves from the usage of shock viscosities to avoid the introduction of numerical artifacts and to concentrate on the effect of uniform viscosity. Most simulations have Pm = 1 (and therefore Rm = Re) up to values of a few hundred. In some runs, we went to Pm up to 100. We started from some basic hydrodynamic (HD) simulations, similar to Del Sordo and Brandenburg (2011). We then expanded them to magnetized cases, exploring the parameter space broadly. Unless otherwise specified, the expansion width has been chosen to be $R_f = 0.2$ (thus, $k_f$ is 10).

In the tables of App. B we list the input parameters of the performed simulations. The tagging names starting with "H" indicate HD runs, and those beginning with "M" indicate MHD runs. By default, we employed the barotropic EoS, while "B" identifies the baroclinic cases. The first number stands for the value of Ω. By default, we employ $\Delta t = 0.02$, $s$ at the end of the name identifies $\Delta t = 1$, while $c$ means $\Delta t = \delta t$. Unless stated otherwise (by the suffix 128, 512), the resolution is $256^3$. The width ($R_f$) and constant ($\phi_0$) of the Gaussian are marked with the number following W and F in the run names, and, if not stated, they are assumed to be 0.2 and 1, respectively. The cooling term of the baroclinic cases is used only in the runs listed in Table B.4. The initial magnetic field by default takes the form of random values of the potential vector component unless a background uniform field in a given $i$ direction is assigned (Bi).

We monitor the volume-integrated energies and the above-mentioned root mean square values of quantities. Additionally, we look at the spectral energy distribution of velocity (total or turbulent), vorticity, and magnetic field, as a function of the wavenumber, $k$. Spectra are also averaged over different times to filter out fluctuations, which are quite noticeable on the smallest scales. We also use the Helmholtz decomposition for the velocity field to evaluate the flow's rotationality, as defined in App. A.8. To do so, we have chosen only one single snapshot (once saturation is reached) to perform the decomposition, for computational practical reasons. For several specific runs, we verified that averaging among many snapshots does indeed lead to the percentage of rotational flow oscillating between values that vary by no more than $\sim 5\%$.

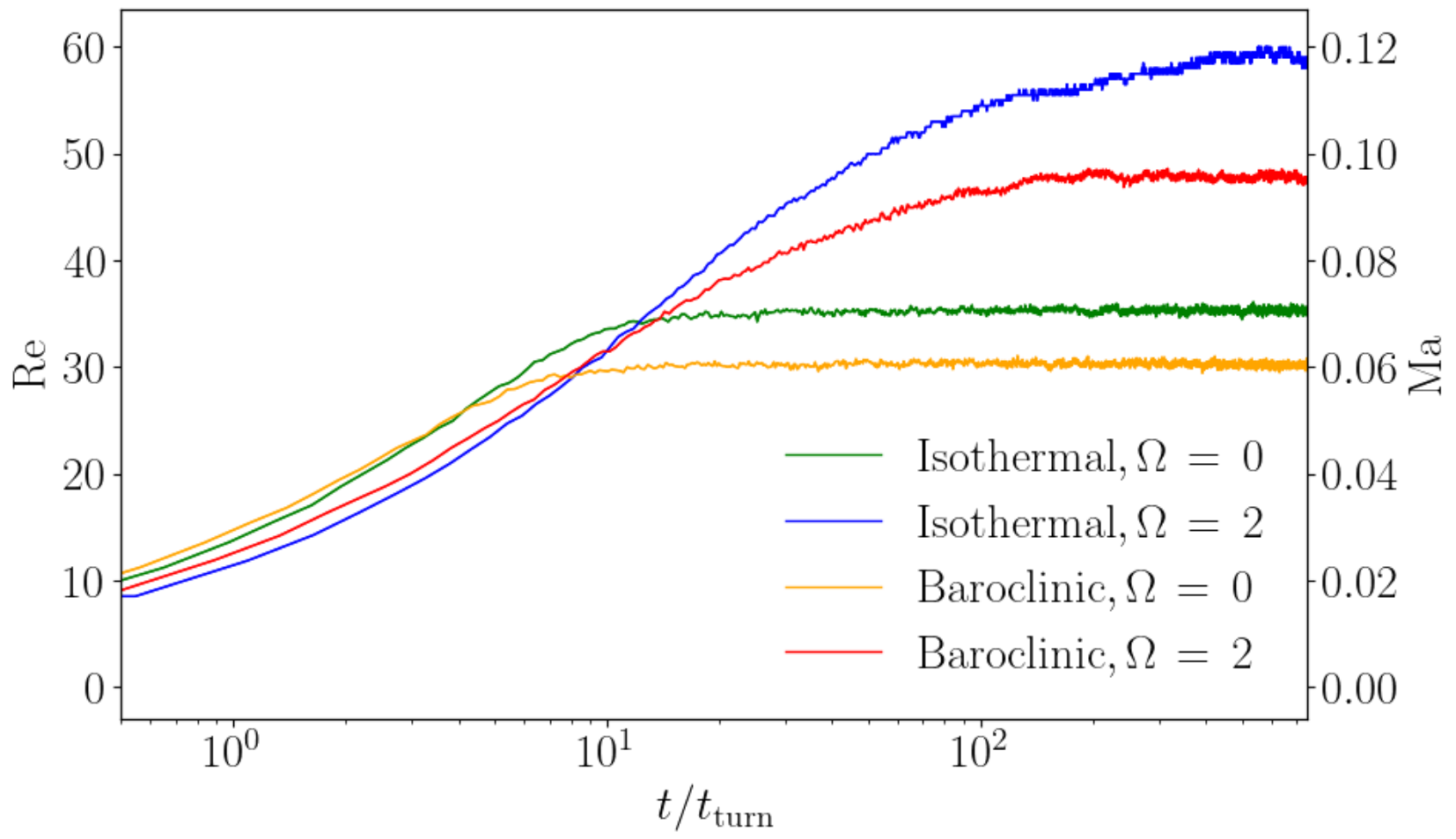

Figure 3.2: Comparison of Re and Ma evolution in the HD cases, considering or not the rigid rotation ($\Omega = 0$ or 2) and for isothermal and baroclinic scenarios, i.e., different EoS.

### 3.2.2 Effects of forcing and rigid rotation

Table B.1 lists some examples of HD runs, without magnetic fields. To visualize the general behavior at early times, Fig. 3.1 shows the values of $u^2$ in the $xy$-plane for one of them, the representative

model H_0cW1. After a turbulent state is reached, the imprints of the most recent expansion waves are visible on top. Such an evolution leads to a homogeneous turbulent flow, which becomes stationary when the dissipating forces counterbalance the forced energy injection.

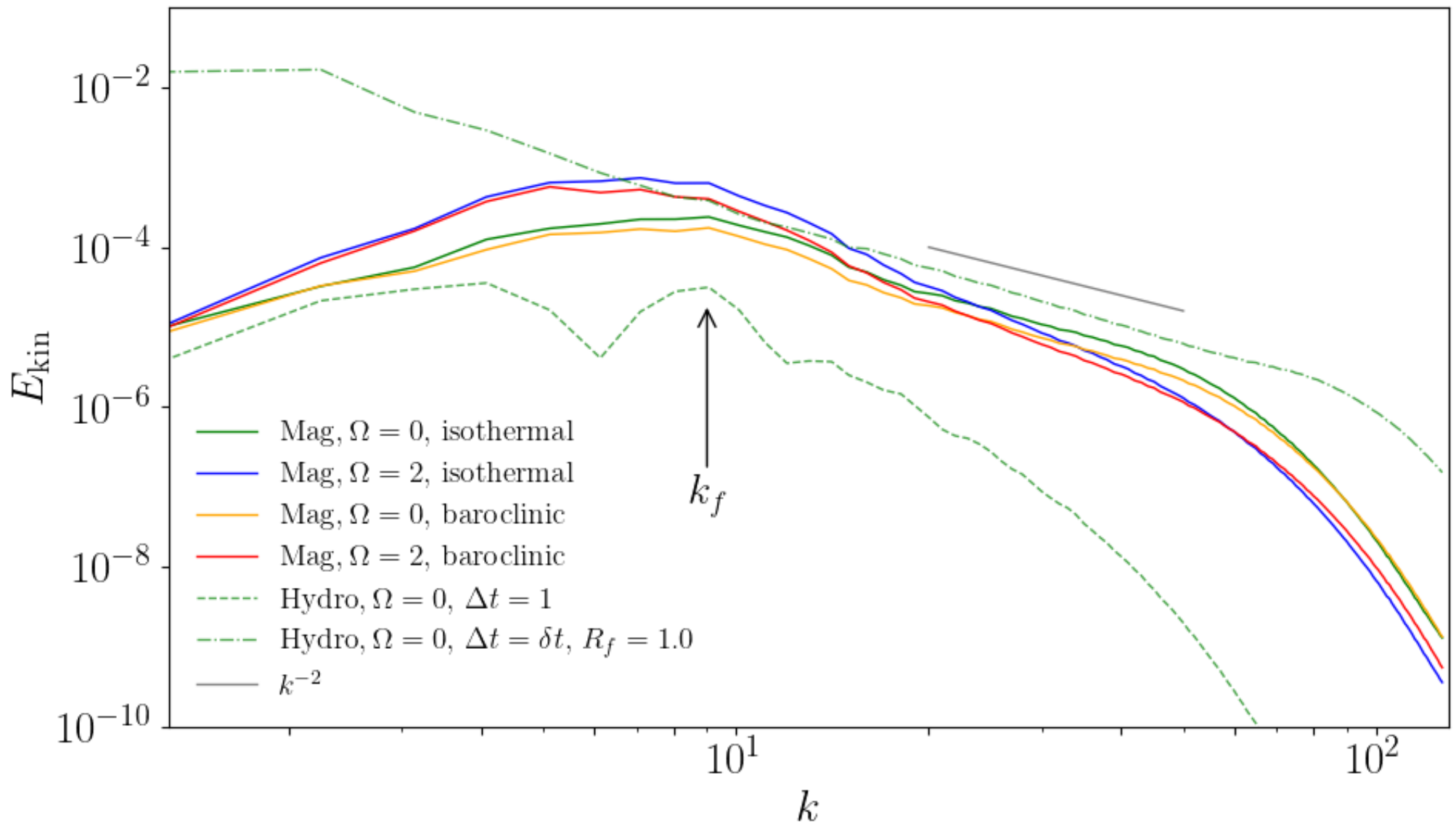

Figure 3.3: Time-averaged kinetic spectra at saturation of some representative simulations of Tables B.1, B.2, and B.3. We notice how the results on large scales (small k) are very similar for all the models, excluding H_0cW1 (dash-dotted green line), which has a forcing with $k_f = 2$. The spectrum with the lowest power is H_0s (dashed green line), which has $\Delta t = 1$. Models marked with other colors explore variations of $\Omega$ and EoS. The magnetic field decays rapidly in all these models, and kinetic spectra coincide with the corresponding HD runs of Fig. 3.2.

Such features are also seen in the MHD cases without shear, for which Tables B.2 and B.3 list the barotropic and baroclinic configurations, respectively, considering or not rigid rotation. In these same tables, we provide the diagnostic magnitudes. As a first comparison, in Fig. 3.2 we can see how different representative simulations (with or without rotation, and for the two EoS) saturate their Re and Ma values within $\sim 1000\ t_{\rm turn}$. A faster rotation increases the final saturation value for the velocities, but the growth takes place at a similar rate, so simulations with faster rotation take longer to saturate. The ideal EoS (i.e., baroclinic) cases show a lower saturation value than the barotropic ones. We interpret this result as a consequence of the presence of the extra dissipating terms in the energy equation. The maximum values for Re and Ma for these simulations, $\mathrm{Re}_M$ and $\mathrm{Ma}_M$, oscillate much more (but with values that are similar for each), namely, ranging from 150 to 300, and 0.3 to 0.6, respectively. These values slightly increase with the rotation rate, and again are smaller in the baroclinic case.

Figure 3.3 shows the comparison between kinetic spectra at saturation for different EoS and $\Omega$. In general, for all cases, kinetic spectra peak around the value $k_f$ (simulations M_0s and M_0 having $R_f = 0.2$, show the bump for $k \sim 5 - 10$, while H_0cW1 accumulates energy at the largest scales, since $R_f = 1$). This is in agreement with the characteristic forcing wavenumber $k_f$ of the adopted forcing, which in the Fourier space is $\propto \exp(-k^2/k_f^2)$. The inertial part of the spectra shows a $\sim k^{-2}$ slope. This irrotational forcing does not lead to any vorticity production, which might help explain the difference concerning the isotropic turbulent slope of $k^{-5/3}$. These general spectral features (peak and slope) are compatible with those found by Mee and Brandenburg (2006)

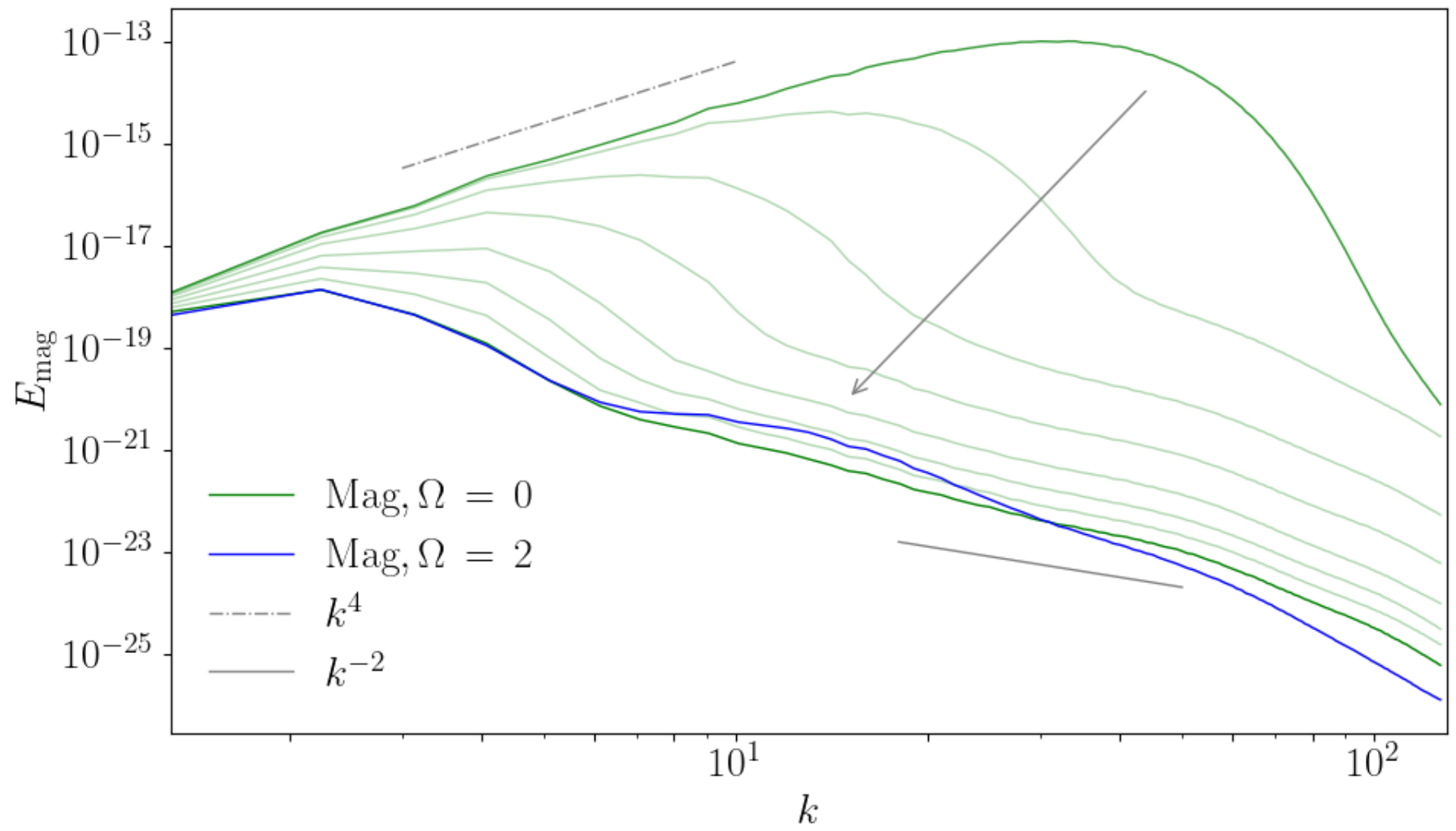

Figure 3.4: Magnetic spectra decay for the M_0 and M_2 models. The spectra evolution is only shown for M_0, and the first corresponds to less than a turnover time when it still shows the initial slope of $k^4$. The arrow indicates the direction of temporal evolution. The last spectrum is approximately at turnover time 20, and it is shown for both models.

and Del Sordo and Brandenburg (2011).

Rotation tends to inhibit smaller scales and promotes the accumulation of kinetic energy at the largest scales, also slightly displacing the spectra peak to the left. This results in a steeper slope in the inertial range. A similar (but less noticeable) effect is seen in the baroclinicity, as compared to the barotropic cases. This is in agreement with the lower total kinetic energy, due to the extra diffusive terms.

To quantitatively measure the dependence of vorticity on rotation, we compared the saturation values of $\mathrm{Re}_\omega$ (see Table B.2). The saturation value increases with $\Omega$, but reaches a maximum around $\Omega = 5$, from which rising rotation does not lead to more vorticity. This translates into a saturation of Ro of the order of $u_{\rm rms}$. The low amount of vorticity appearing with $\Omega = 0$ seems to be spurious since it decreases when resolution is increased, as already reported by Mee and Brandenburg (2006). Such a spurious contribution is much smaller than the physical one when the rotation is considered.

If rotation is the main contribution to vorticity, then the average vorticity should be proportional to $\Omega$:

$$\frac{\partial \boldsymbol{\omega}}{\partial t} = \ldots - 2\nabla \times \boldsymbol{\Omega} \times \mathbf{u}$$
$$\rightarrow \omega_{\rm rms} \approx 2\ \Omega\ t_{\rm turn}\ u_{\rm rms}\ k_f \rightarrow \frac{k_\omega}{k_f} \approx 2\ \Omega\ t_{\rm turn}\ , \tag{3.13}$$

where $k_\omega = \omega_{\rm rms}/u_{\rm rms}$. In Table B.2, some physical quantities of these runs are listed. We find a linear trend valid up to $\Omega \sim 3$, which we quantify by a trend between some simulations (marked in blue in Table B.2) as: $k_\omega/k_f = (0.180 \pm 0.004)\ \Omega + (0.010 \pm 0.009)$. The slope seems much smaller for $\Omega \gtrsim 4$ (but further simulations would be needed to quantify it).

Finally, the saturated values increase as we decrease $\Delta t$ from 1 (H_0s) to 0.02 (H_0). There is not much of a difference between $\Delta t$ being 0.02 or $\delta t$ since the latter is indeed on the order $\mathcal{O}(10^{-2})$. Overall, these results are compatible with the idea that this irrotational forcing alone does not produce any significant amount of vorticity.

### 3.2.3 Absence of dynamo with rigid rotation only

In all the models of Tables B.2 (barotropic) and B.3 (baroclinic), the initial magnetic field quickly decays, with or without rotation. In other words, despite the vorticity growth induced by rotation, the system does not experience SSD. Indeed, there is no significant difference in the flow between these MHD models and the purely HD ones that correspond to them.

An example of a decay of the magnetic spectrum is shown in Fig. 3.4, where it can be observed how the smallest scales decay faster and the $k^4$ slope rapidly changes. Washing away first the smaller scales is natural, as both numerical and physical diffusivities have more influence on the very small scales. After some turnover times, the inertial range of the magnetic spectra resembles (in some way) the kinetic ones. Rotation also tends to favor larger scales and increase the slope, although both slopes are steeper than in the kinetic spectra. There is a minimal interaction between the flow and the magnetic field through the first term of the induction, as expressed by Eq. (3.4).

The decaying evolution of $B_{\rm rms}$, for the initial random magnetic field case, was found to follow power laws roughly. The parameters fitting these power laws do not depend on $\Omega$ for our runs. Without forcing, we obtain very similar decays with the same exponent, hence, finding no substantial difference between M_0 and the corresponding simulation without forcing. This leads us to ascribe the magnetic decay to the small-scale dissipation controlled by the resistivity (physical and numerical) alone, with no relevant impact from the forcing, a specific aspect which seems to contradict what Mee and Brandenburg (2006) noted (an enhancement of the decay caused by the forcing). However, the difference might be due to differences in the setup and the numerical dissipation. In any case, the setup does not lead to any growth of the magnetic field either.

We have explored this for a large range of parameters, especially for the baroclinic cases, with rotation only. There was no SSD seen for any of the cases, despite thousands of runs at various turnover times. In particular, this conclusion is reached in each of these cases: (i) changing $R_f$, $\Delta t$ (Tables B.2 and B.3); (ii) using a simple uniform initial magnetic field, which experiences an early fast (exponential) decay, followed by a power law, with or without rotation; (iii) changing resolution; and (iv) changing $\tau_{cool}$ in the baroclinic cases (Table B.4).

These results are in contrast with Achikanath Chirakkara et al. (2021), who found SSD in similar turbulent setups. In particular, even with a completely irrotational driving (implemented in the Fourier space) and without any rotation, they see that after $\sim 10^3\ t_{\rm turn}$ there is an increase of magnetic energy, which saturates at about 1/1000 of the kinetic energy. To compare them, simulations MB_0_128 and MB_2_128 were run to approximately $10^4\ t_{\rm turn}$. We also ran simulations with high values for the forcing amplitude (up to $\phi_0$=500, see Table B.5 for non-rotating cases), which required higher values of viscosity and thermal diffusivities, leading to overall lower values of Re. However, we still could not observe any growth in the magnetic field.

We found that by not including the cooling term in the entropy equation, the sine of the baroclinic angle (defined as $\sin\theta = \langle \nabla T \times \nabla s\rangle / \langle \nabla T\rangle\langle \nabla s\rangle$) takes peak values that are higher than one. The mean angle takes similar values to those found in Korpi et al. (1999).

To investigate the parameter space as thoroughly as possible and search for possible dynamo action, we pushed the code's capability to its limits by increasing $\Omega$ values or decreasing diffusivities. In both cases, we observed that the growth of the field was only evident in cases that quickly became numerically unstable. In these cases, we were using either $\Omega = 10$ or diffusivities of the order of $\nu = 10^{-6}$. In cases like these, it was not possible to establish growth and saturation phases, so we did not consider them reliable and, therefore, excluded them from our analysis.

To summarize, the irrotational forcing in combination with solid body rotation can produce vorticity. The chosen ideal EoS favors it more than the barotropic one. However, SSD is never activated, regardless of the rotation, EoS, and correlation length of the seed magnetic field, at least in the explored range of Rm up to a few hundred units.

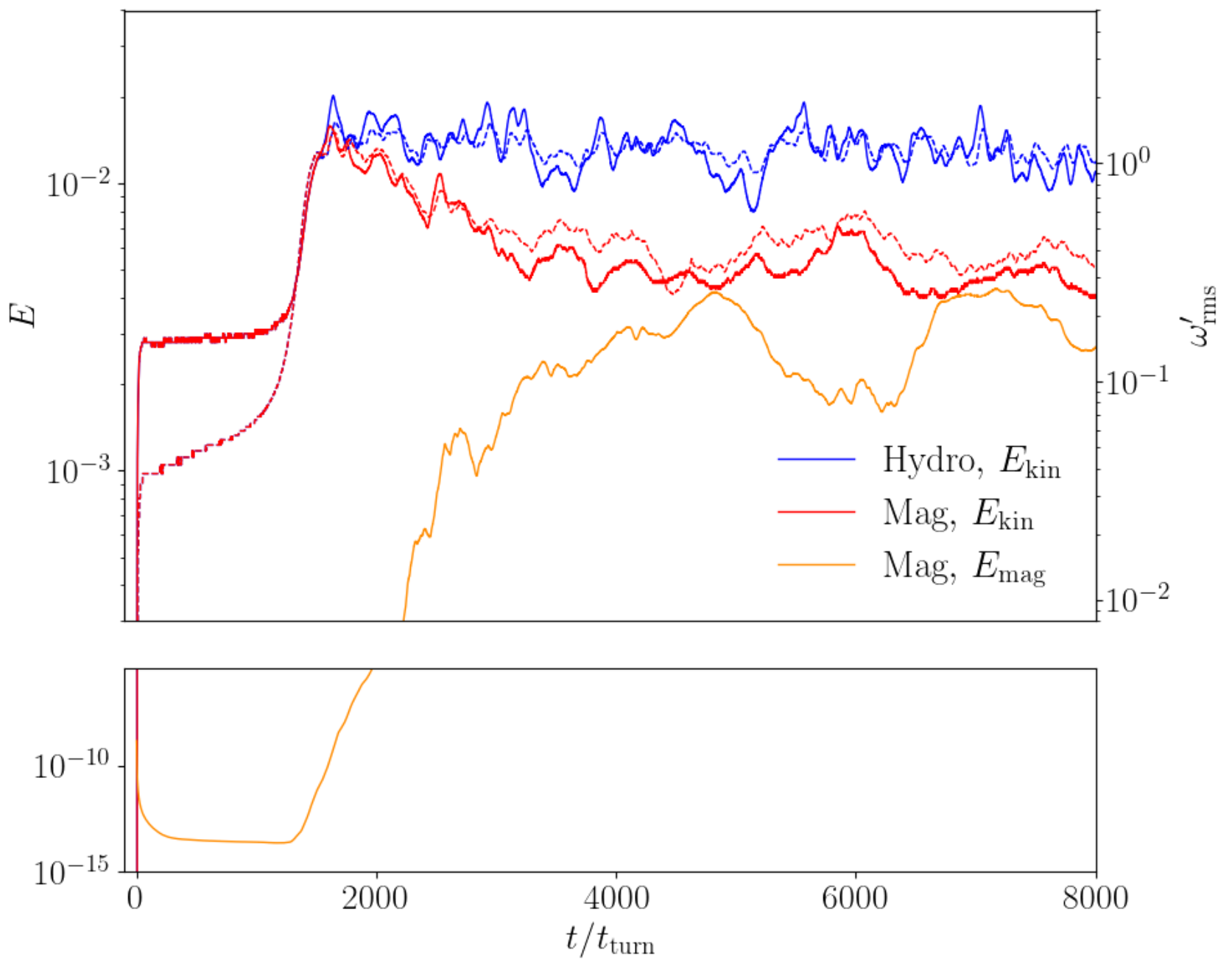

Figure 3.5: Time evolution of models including shear. We plot the average turbulent kinetic energy density (solid blue and red lines), the average magnetic energy density (yellow line), and the root mean square of the turbulent vorticity $\omega'_{\rm rms}$ (dashed blue and red lines) for both hydro and magnetic scenarios with similar parameters, i.e. $A = 0.2$, $N_{\rm mesh} = 128^3$.

### 3.2.4 Linear shear

We could opt for the use of a linear shear of the form $\mathbf{u}^S = (0, Sx, 0)$. With this choice, the $yz$-faces of the box cannot satisfy periodic boundary conditions; thus, shearing-periodic boundary conditions are needed: the $x$ direction is periodic concerning positions in $y$ that shift in time:

$$f(-\frac{1}{2}L_x, y, z, t) = f(\frac{1}{2}L_x, y + L_x St, z, t),$$

where $f$ is each of the evolving fields, this condition is routinely used in numerical studies of shear flows in Cartesian geometry since it was introduced by Wisdom and Tremaine (1988) and Hawley et al. (1995). These boundary conditions are known to potentially produce spurious vorticity at the boundary, which, also due to the irrotational nature of our forcing, we notice, as we discuss below.

We show the tested cases in Table B.7. For reasons of numerical stability, we set higher values of viscosity for the $128^3$ tests with the linear shear term. This leads us to a lower Re compared to most of the other shear-less simulations. Regarding the magnetic field evolution (last column of Table B.7), in most cases we find a decay of magnetic energy, which depends on the shear parameter $S$. Only in cases with a low magnetic diffusivity ($2 \cdot 10^{-5}$), we do observe some magnetic field exponential growth. However, we cannot draw any reliable conclusion about dynamo action for the following reasons.

Looking at the trend between the vorticity indicator (at saturation) and the shear amplitude, $S$, $k_\omega/k_f = (0.108{\pm}0.001) + (0.151{\pm}0.005)\ S$, we notice a nonphysical non-zero (and non-negligible) vorticity for vanishing shear, $S$. We checked that it is not cured by increasing the resolution. Del Sordo and Brandenburg (2011) already made this observation and attributed it to numerical artifacts due to the interaction between the expansion waves laying on both sides of the shearing boundary.

A visual inspection of the results reveals indeed that almost the entire vorticity and magnetic field production were near the shearing boundaries. As an attempt to mitigate this artifact, we spatially constrained the expansion waves in the middle half of the box (i.e., as far away as possible from the shearing boundaries). This indeed resulted in more spatially uniform and slower growth of vorticity and magnetic field. However, we still could not keep the system from being affected by the aforementioned spurious contributions at the shearing boundaries after a specific time. Higher values of $S$ or the addition of rotation seem to amplify some magnetic field in the middle of the box, but the boundary contribution was still dominant. This numerical issue led us to use the more complex sinusoidal shearing profile, which allows simple periodic boundary conditions, with no spurious effects.

### 3.2.5 Dynamo in the presence of a shear

A further effect to be investigated is the shearing flow in combination with this irrotational forcing. In Table B.6 we report the numerical experiment performed with the sinusoidal shearing flow $\mathbf{u}^S$ introduced above in Sec. 3.1.4. Dynamo growth is always found, unless the shearing profile is not steep enough, or the forcing acts on a relatively small length-scale (case M_0A020_W0.10_128), or if the forcing is not acting, meaning there is only the background flow (case M_0A020_F0_128). In terms of Reynolds numbers, we tentatively indicate a threshold value of $\mathrm{Rm}_{\mathrm{crit}} \sim 50$ for the chosen setups.

The system evolves as illustrated in Fig. 3.5. There is an initial growth of the vorticity that, after a few turnover times, seems to saturate, along with the value of $u_{\mathrm{rms}}$. The small-scale magnetic field initially decays. For about 1000 turnover times, the system slowly evolves by slightly increasing the vorticity and keeping both $u_{\mathrm{rms}}$ and $B_{\mathrm{rms}}$ almost constant. We then observe a sudden growth of vorticity, followed by that of the kinetic energy, and then also of the magnetic energy. This process occurs with a strong enough shear amplitude for a given set of diffusivities and forcing parameters. We understand it as a vorticity dynamo produced by a Kelvin-Helmholtz instability, which develops in the system after it is perturbed beyond a certain threshold.

In Fig. 3.5, we show the evolution of the average magnetic energy density, $E_{\mathrm{mag}} = \langle B^2 \rangle / 2\mu_0$, the average turbulent kinetic energy density, $E'_{\mathrm{kin}} = \langle \rho u'^2 \rangle / 2$, (i.e., neglecting the shearing contributions) and the root mean square of the turbulent vorticity $\omega' = \nabla \times \mathbf{u}'$, for one MHD and one HD representative runs. All averages are taken over the entire domain. We notice that $E'_{\mathrm{kin}}$ and $\omega'_{\mathrm{rms}}$ grow almost the same way in the HD and MHD cases during the initial instability. Then the MHD run shows SSD shortly after the vorticity dynamo. When the magnetic field increases to a dynamical regime, the MHD run sees a decrease in $E'_{\mathrm{kin}}$ and $\omega'_{\mathrm{rms}}$, until we almost reach equipartition between the magnetic and the turbulent kinetic energies. If, instead, the shear velocity is included, it dominates the total kinetic energy by a factor of a few, the value depending on the shear amplitude.

Such an exponential amplification of vorticity, or vorticity dynamo, was initially predicted by Blackman and Chou (1997) who, however, showed how a helical forcing was needed. The effect we see in our simulations was predicted by Elperin et al. (2003) for non-helical homogeneous turbulence with a mean velocity shear. It was then observed in the HD case by Käpylä et al. (2009) in a setup with linear shear and plane wave forcing. Since the Kelvin-Helmholtz is a kinetic instability, the MHD case leads to SSD just after the vortical structures are formed: the magnetic fields are twisted and stretched by advection, until the magnetic energy is strong enough to provide feedback on the fluid, slightly reducing the mean kinetic energy density. At a later stage, also a large-scale component of the magnetic field is amplified: this can be interpreted as the shear-current effect described by Rogachevskii and Kleeorin (2003), where an electromotive force proportional to vorticity is produced.

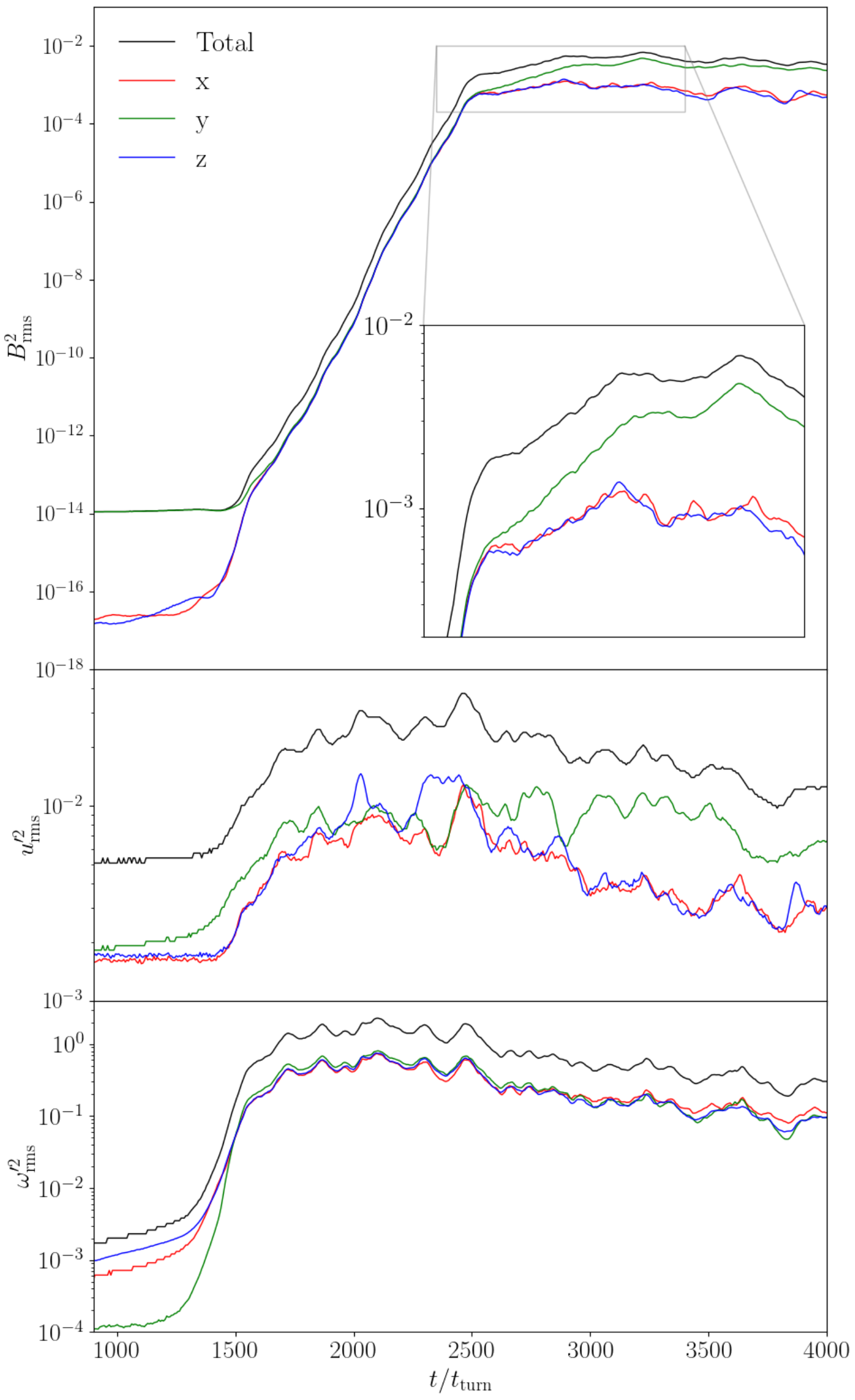

Figure 3.6: Vector component evolution for the magnetic field, turbulent velocity, and turbulent vorticity of the M_0A020 run. The $y$ direction of the magnetic field shows an evolution different from the other components before and after, but not during, the dynamo phase. This is only slightly observed in the turbulent velocity field, while vorticity shows no preferential direction at all.

As mentioned in Sect. 2.3.3, a shearing velocity profile in the presence of a sufficiently large magnetic field triggers the winding mechanism. This effect is relevant for our simulations, and can be seen for the model M_0A020 in the upper plot of Fig. 3.6, where during the time of dynamo growth, each component of the magnetic field is equally important until the non-linear problem sets in. Although our setup is more complex, as the potential forcing produces flow in all directions, this results in different growth of the field in the y direction during the latest stage of the field growth. In the middle and bottom panels of Fig. 3.6, we also show the root mean square of turbulent vorticity and turbulent velocity. They show a mild preferred contribution only in the $y$-component of the turbulent velocity (probably for the feedback of the strong magnetic field).

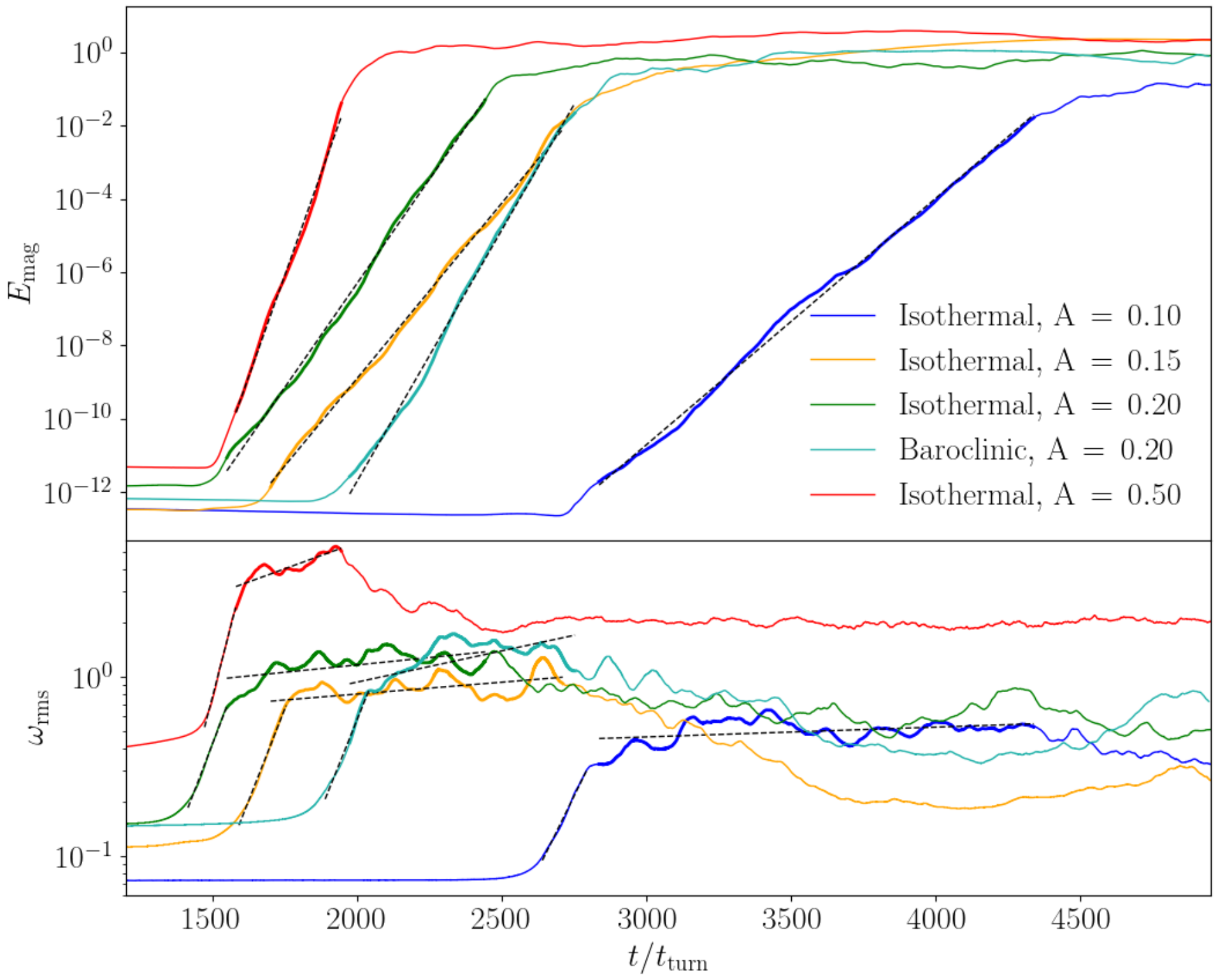

Figure 3.7: Evolution of the volume-integrated magnetic energy and the root mean square of the vorticity, $\omega_{\rm rms}$, for some dynamo runs (as in the legend). The dynamo and vorticity growth rates increase with the shearing amplitude. The baroclinicity slightly enhances the dynamo growth rate, even though it takes longer for the instability to start. The vorticity continues to grow after its initial exponential growth during the magnetic dynamo phase. It then decays to lower values. The straight lines indicate the fits we obtained for the growth rates of the dynamo, r, of the vorticity dynamo, $r_\omega$, and of the growth of the vorticity during the first part of its saturation phase, $r_{\omega,sat}$. Their values are reported in Table B.6.

In Table B.6, we show the different growth rates (in units $t_{\rm turn}^{-1}$) obtained for all dynamo runs. In Fig. 3.7 we also show the volume-integrated magnetic energy $E_{\rm mag}$ and the root mean square of the vorticity $\omega_{\rm rms}$ evolution for some of the $256^3$ runs. We have found that the growth rate of the magnetic field is roughly proportional to the shear rate, similar to what was seen by Käpylä et al. (2010). The dynamo growth seems to have saturated already in the $256^3$ run, but the vorticity growth seems to vary more. The $512^3$ run was stopped after the dynamo growth for computational reasons; thus, we do not show the last slope. Baroclinicity does not change much the vorticity instability growth, but allows for a more rapid dynamo growth. For the runs with constant amplitude ($A$=0.2) and increasing resolution, there is little difference between the $256^3$ run and the $512^3$ run, which indicates numerical convergence. Comparing the $128^3$ dynamo runs with different forcing, we may notice that the dynamo and vorticity growth rates increase with both $\phi_0$ and $R_f$, but when they exceed a certain value, both rates decrease. This is probably caused by the flow reaching a transonic regime and decreasing the growth efficiency Haugen et al. (2004); Federrath et al. (2011); Schleicher et al. (2013).

We did not reach dynamo for the cases with a shear amplitude $A \lesssim 0.1$, which can be translated as a critical Reynolds value of $\mathrm{Re}_{\mathrm{crit}} \sim 50$, and a minimal rotational component of the flow of about 50% which in our models corresponds to $k_\omega/k_{f,\mathrm{crit}}$ of 0.073. Although these specific values might depend on the resolution, this transition from non-dynamo to dynamo-generating flows was not observed in the rigidly rotating cases.

Finally, in Fig. 3.8 we represent the spectra and snapshots of M_0A020 to give a more visual and detailed description of the instability. This figure summarizes the evolution of the main quantities during the three main evolutionary stages of our simulations. We display three different phases of the evolution of the system, evolving from left to right. Slope lines in the kinetic energy spectra are proportional to $k^{-2}$ before the dynamo and get a bit steeper afterward. Magnetic energy spectra take a typical Kazantsev $k^{3/2}$ slope during the kinematic phase of the dynamo (Kazantsev, 1968). We note that the scale of the kinetic energy and vorticity spectra (or enstrophy spectra) is kept constant, but the magnetic ones are not, to follow the magnetic field growth.

### 3.2.6 Dependence on the forcing scale without shear

We explored the role played by the forcing scale $R_f$, which can be interpreted as the physical scale of the interaction between our toy-model SN explosion and the ISM. As a reference in this sense, the typical sizes of a SN remnant are typically 20-50 pc (Franchetti et al., 2012; Asvarov, 2014). The size of the box can be converted into physical scales if the shear is included: for the values that we employed here, the box corresponds to about 500 pc. Anyways, we stress that our box simulations aim to investigate basic aspects of the vorticity dynamo, rather than a detailed model of the effect of SN on the ISM.

We note that when we changed $R_f$, we changed $\phi_0$ accordingly to reach a similar $u_{\mathrm{rms}}$. This allowed us to compare simulations with different Reynolds numbers (which scale with $R_f$), different energy injection scales, and different filling factors within the box. This is especially relevant when the shear is included, since it represents another physical scale and the vorticity production and dynamo depend on both (see Sect. 3.2.7).

In the left panel of Fig. 3.9, we plot the temporal evolution of $\omega_{\mathrm{rms}}$ in terms of turnover time for different, representative runs without rotation, with isothermal conditions. Independent of the forcing scale $R_f$, vorticity reaches a steady state after less than 15 turnover times, except for the smallest value of $R_f = 0.1$, which takes less than 5 turnover times. In the model with a forcing scale of $R_f = 0.1$, we see the development of local transonic flows as a consequence of the attempt to reach values of the kinetic energy similar to the cases with larger forcing widths. This is accomplished by increasing the parameter $\phi_0$ in Eq. 3.5, and it leads to a different behavior of these models. The mean value of vorticity is observed to decrease with $R_f$, while its fluctuations do increase.

In the right panel of Fig. 3.9, we show the different kinetic spectra obtained from such runs. The corresponding forcing widths change the forcing wavelength $k_f$, thus changing the inertial range interval, as $\nu$ is kept constant. The slope of $\hat{E}_{\mathrm{kin}} \sim k^{-2}$ seems to be independent of $R_f$, as already seen by Mee and Brandenburg (2006). No dynamo was observed in these runs despite having reached Rm larger than 200. When rotation is added, we obtain steeper slopes but a similar inertial range behavior as the nonrotating cases plotted here.

This can be seen more clearly in Fig. 3.10, where we plot the average of various diagnostics calculated during the saturated stage for the isothermal and baroclinic models, with or without rotation, using the forcing $f_{\rm acc}$. We note that for $R_f = 0.1$ the non-isothermal runs are not shown because they reach locally supersonic flows and remain numerically stable for too few time steps. Despite attempting to select a $\phi_0$ that makes $u_{\rm rms}$ approximately independent of $R_f$, we succeeded in doing so only for the isothermal nonrotating case. In contrast, for the other non-isothermal or rotating cases, $u_{\rm rms}$ mildly depends on $R_f$. Therefore, the more crucial dimensionless quantity is $k_\omega/k_f$, as it is a measure of vorticity normalized with velocity and forcing wavelength. Figure 3.10 also illustrates many other features. First, we see how the Reynolds number increases with width for all cases, but no instability is found. We notice that the nonrotating cases seem to have a lower overall $\omega$, but it tends to grow with smaller $R_f$. In contrast, for the rotating runs, there is a maximum at $R_f = 0.4$ (i.e., when expansion waves are about a fifth of the simulation domain, $k_f = 5$). We also see how isothermal rotating case show the highest values for $\omega_{\rm rms}$, $u_{\rm rms}$, and Re, but in $k_\omega/k_f$ they are closely matched by the non-isothermal rotating with the addition of a the thermal cooling time. As expected, the cooling term creates an additional source of dissipation, which we had to compensate for by doubling the values of $\phi_0$ (for $\tau_{cool} = 0.1$).

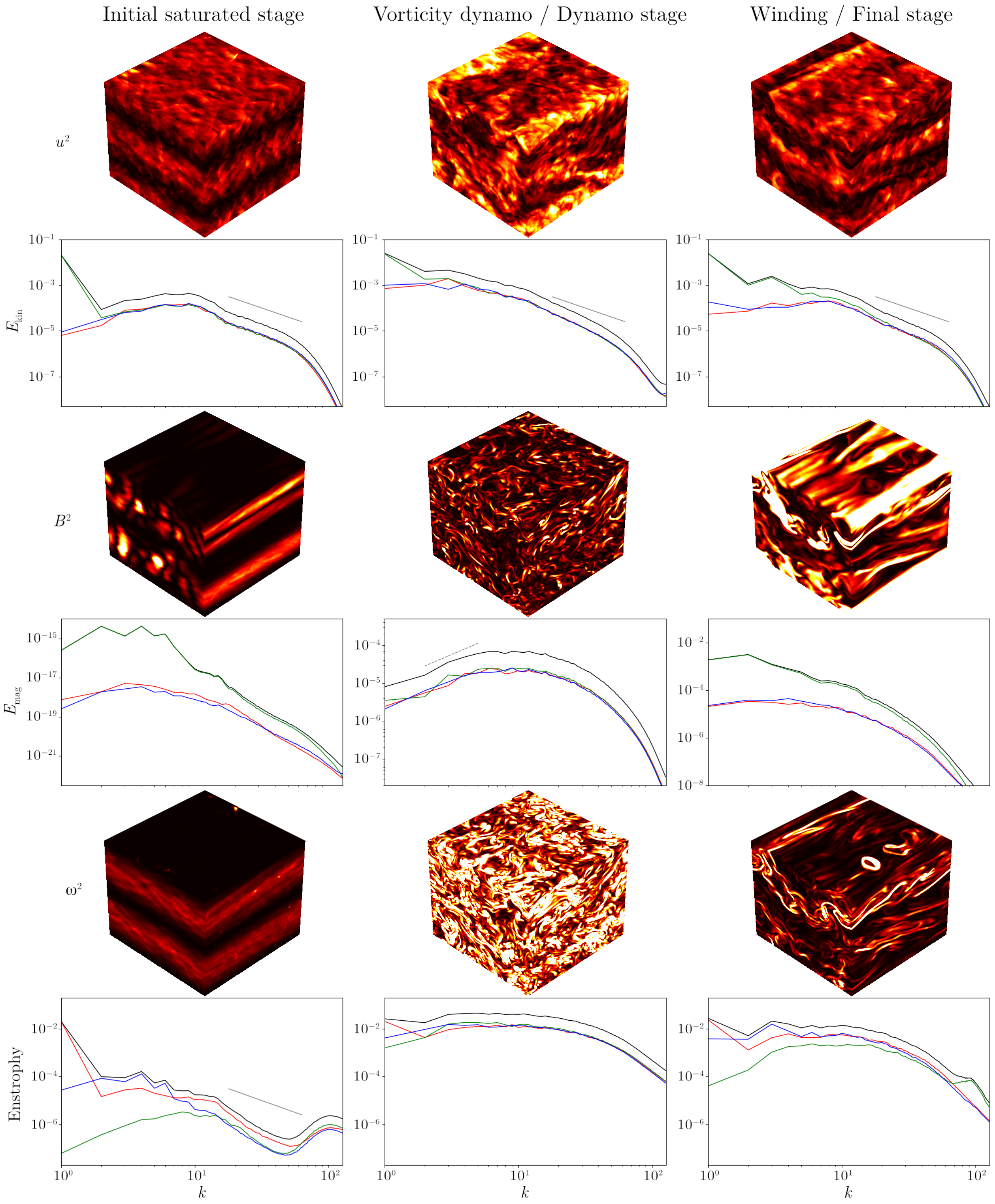

Figure 3.8: Spectra and snapshots over three faces of the domain, for the squares of velocity, $u^2$, magnetic field, $b^2$, and vorticity, $\omega^2$, of the run M_0A020. The $x$, $y$, and $z$ components of the spectra correspond to the red, green, and blue lines, respectively. Color bars have different ranges to allow a better visualization: for $u^2$ the range is (0,0.2); for $B^2$ the range is (0,1e-13) for the first snapshot and (0,0.015) for the others; for $\omega^2$ the range is (0,0.2) for the first snapshot and (0, 2.0) for the others, all in code units.

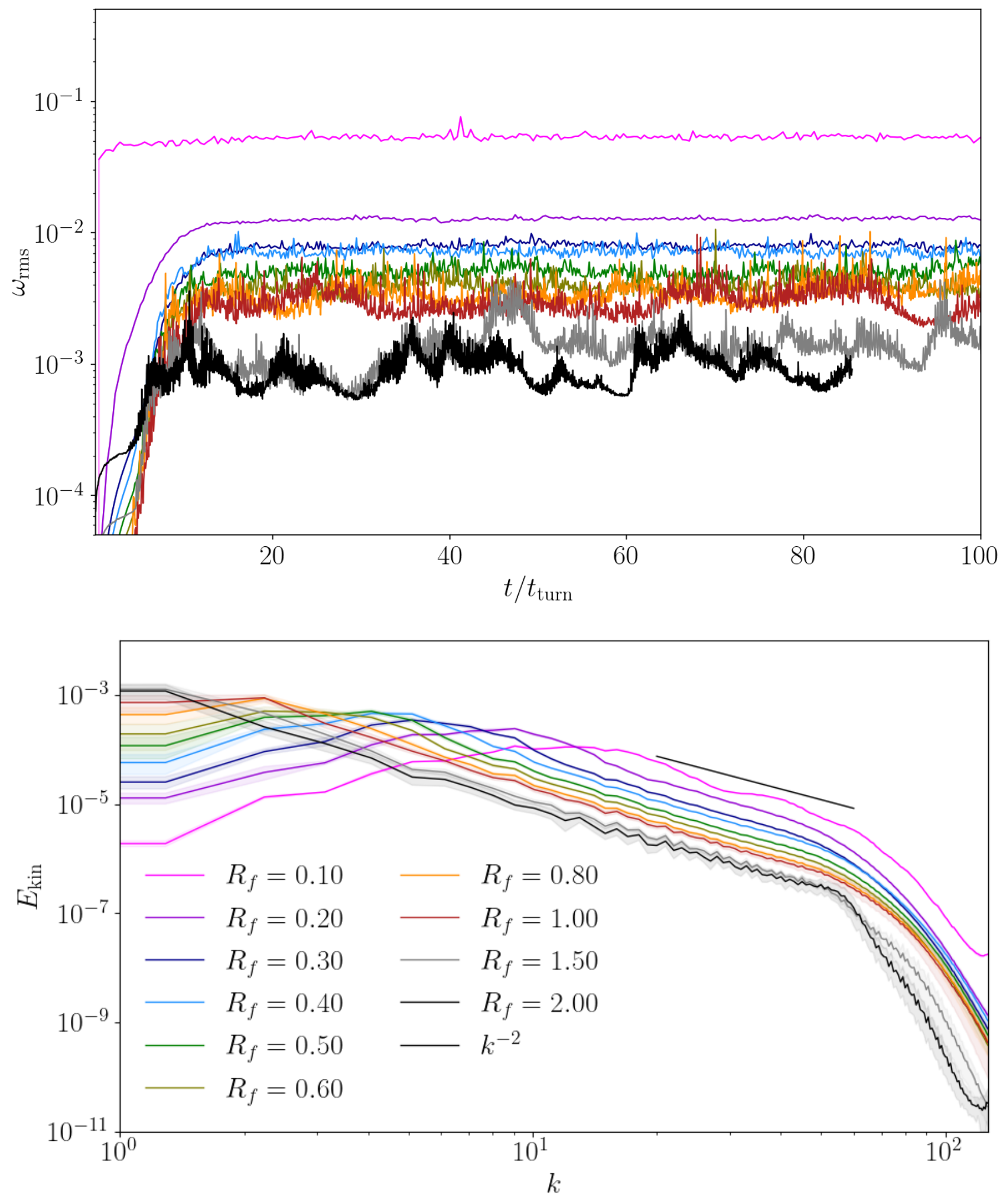

Figure 3.9: Time evolution for $\omega_{\rm rms}$ (left) and time-averaged kinetic spectra at saturation (right) for nonrotating isothermal runs with different explosion widths and similar total energy (see Table B.2). Notice that the peaks in $k$ are close to the corresponding forcing wavenumber $k_f = 2/R_f$.

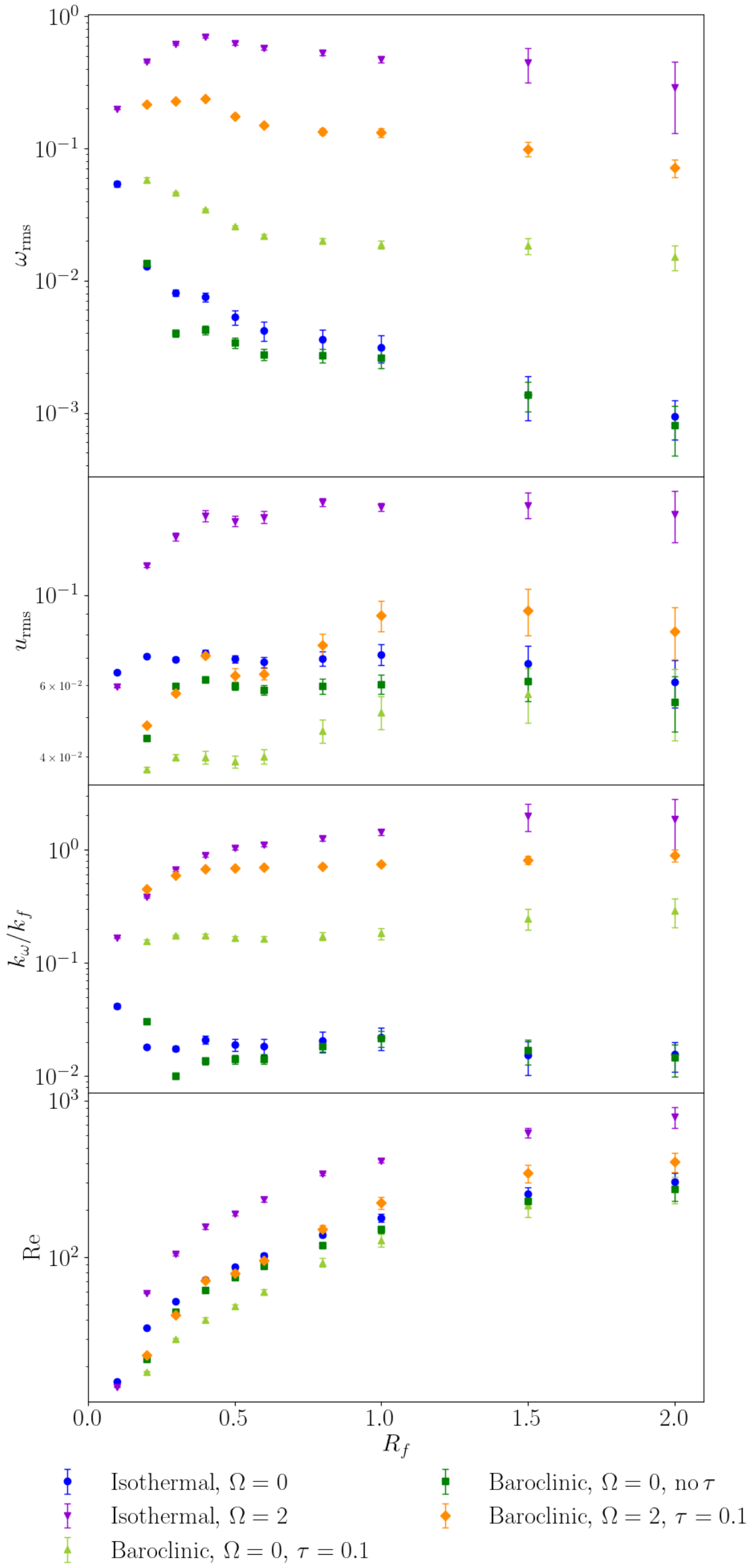

Figure 3.10: Different diagnostic quantities, $\omega_{\rm rms}, u_{\rm rms}, k_\omega/k_f$, and Re (from top to bottom), as a function of explosion width, $R_f$, for the forcing $f_{\rm acc}$. These runs have Pm set to 1, i.e., Rm = Re, with a grid size of $256^3$ and no dynamo present. The data are presented in Tables B.2 and B.3.

In Fig. 3.11 we plot the same quantities as in the left panel of Fig. 3.10, comparing runs with different resolutions and keeping $R_f$ constant with a value of 0.5. Most quantities seem to be resolution independent, even when we move to a resolution low enough to erase a good part of the inertial ranges. We observe that $u_{\rm rms}$ is more or less resolution independent for all except two cases: (i) the isothermal nonrotating case, and (ii) the isothermal rotating case. The first case is compatible with the idea of vorticity being created solely by numerical sources.

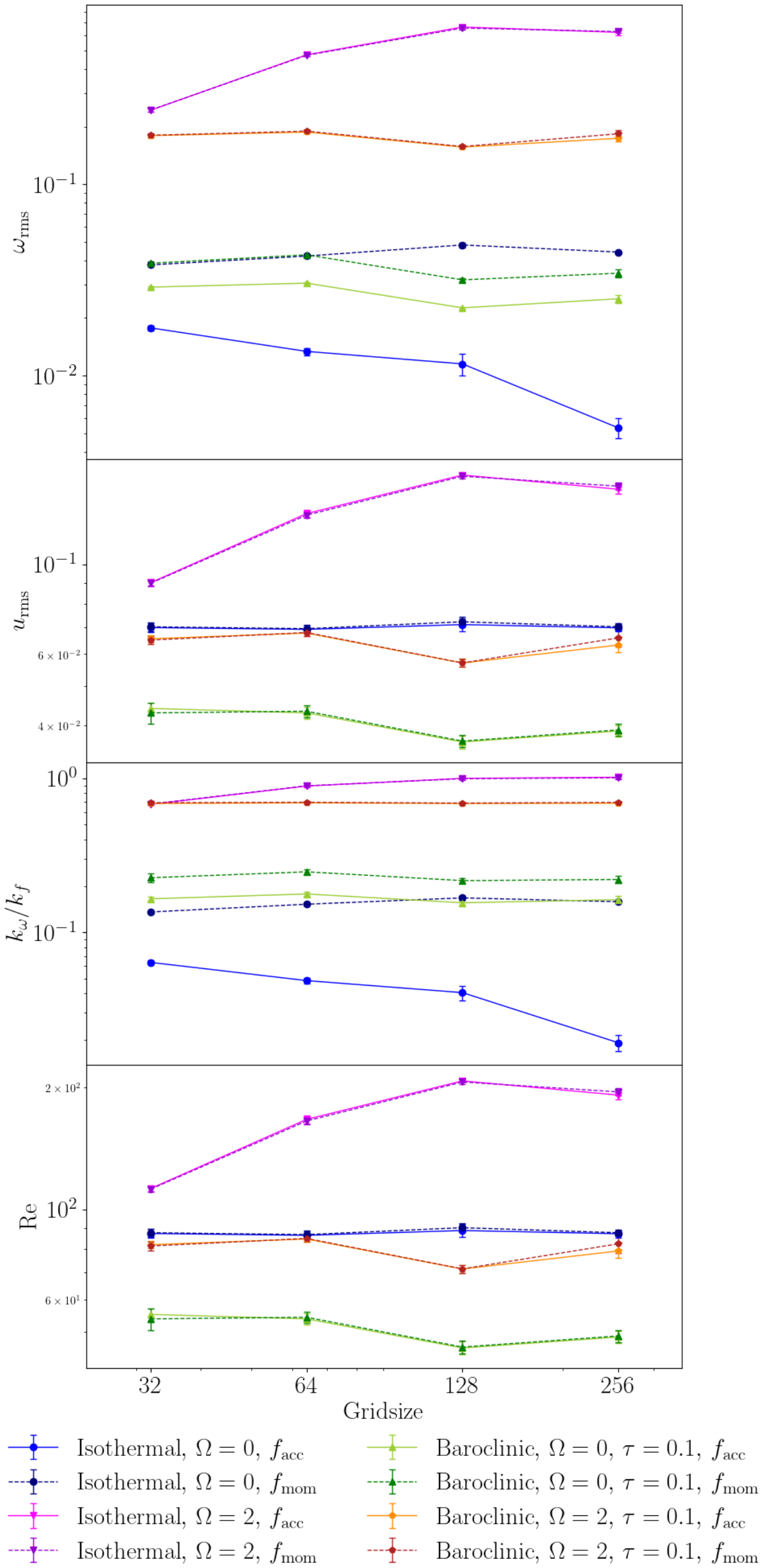

Figure 3.11: Same magnitudes as in Fig. 3.10 but now for runs with $R_f = 0.5$, i.e., $k_f = 4$, different resolutions, and varying the type of forcing.

As expected, the contribution of rotation to $\omega$ is much greater than the baroclinic one in non-isothermal models, but this difference diminishes at small values of the forcing scale $R_f$. No dynamo is found either for the cases with rotation and/or non-isothermality with an ideal gas law and different cooling times. The results still hold when $k_f$ grows nearly up to 1, and for Rm values of several hundred.

### 3.2.7 Dependence on the forcing scale in the presence of shear

The general behavior with the presence of shear is as described above: an HD instability develops first with an exponential growth of $\omega_{\rm rms}$, which is then closely followed by a magnetic instability leading to an exponential amplification of $B_{\rm rms}$. After the linear phase of the dynamo, a winding phenomenon is seen for all cases, independently of $R_f$, Prandtl number, and resolution. During this process, $B_y$ is further linearly amplified in the shearing direction by winding.

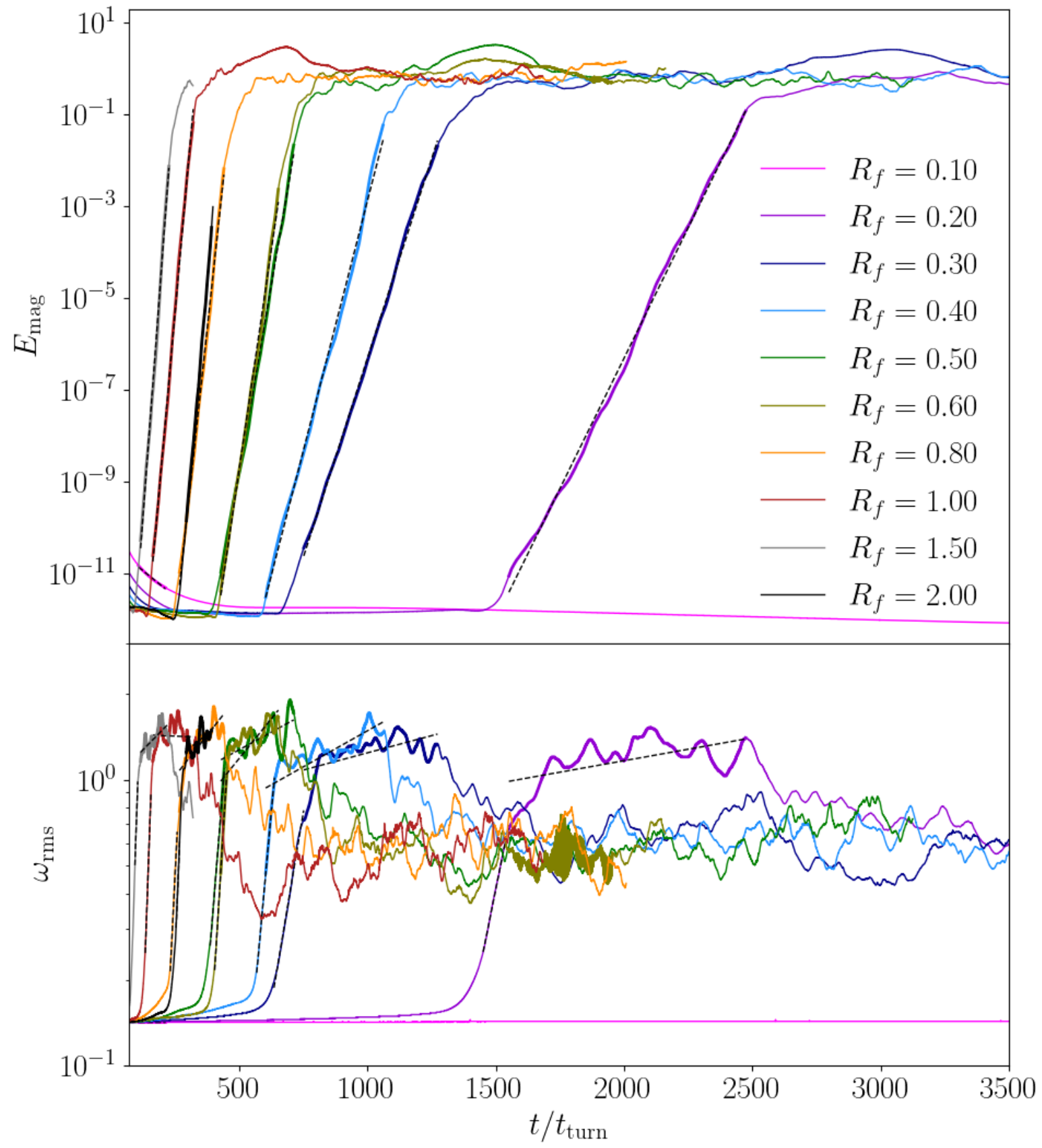

Figure 3.12: Time evolution (in units of $t_{\rm turn}$) of vorticity and magnetic energy of runs with the shearing profile and different $R_f$ (see Table B.6). If we use the shearing timescales, the magnetic instability is between 70 and 200 $t_{\rm shear}$ with more similar growth rates (see Fig. 3.13). In both cases, $R_f = 1.00$ and $R_f = 1.50$ take the least amount of time to reach the instability. Dashed lines represent the exponential fits for $E_{\rm mag}$ during dynamo growth (top panel), $\omega_{\rm rms}$ during vorticity growth, and $\omega_{\rm rms}$ during dynamo growth (bottom panel).

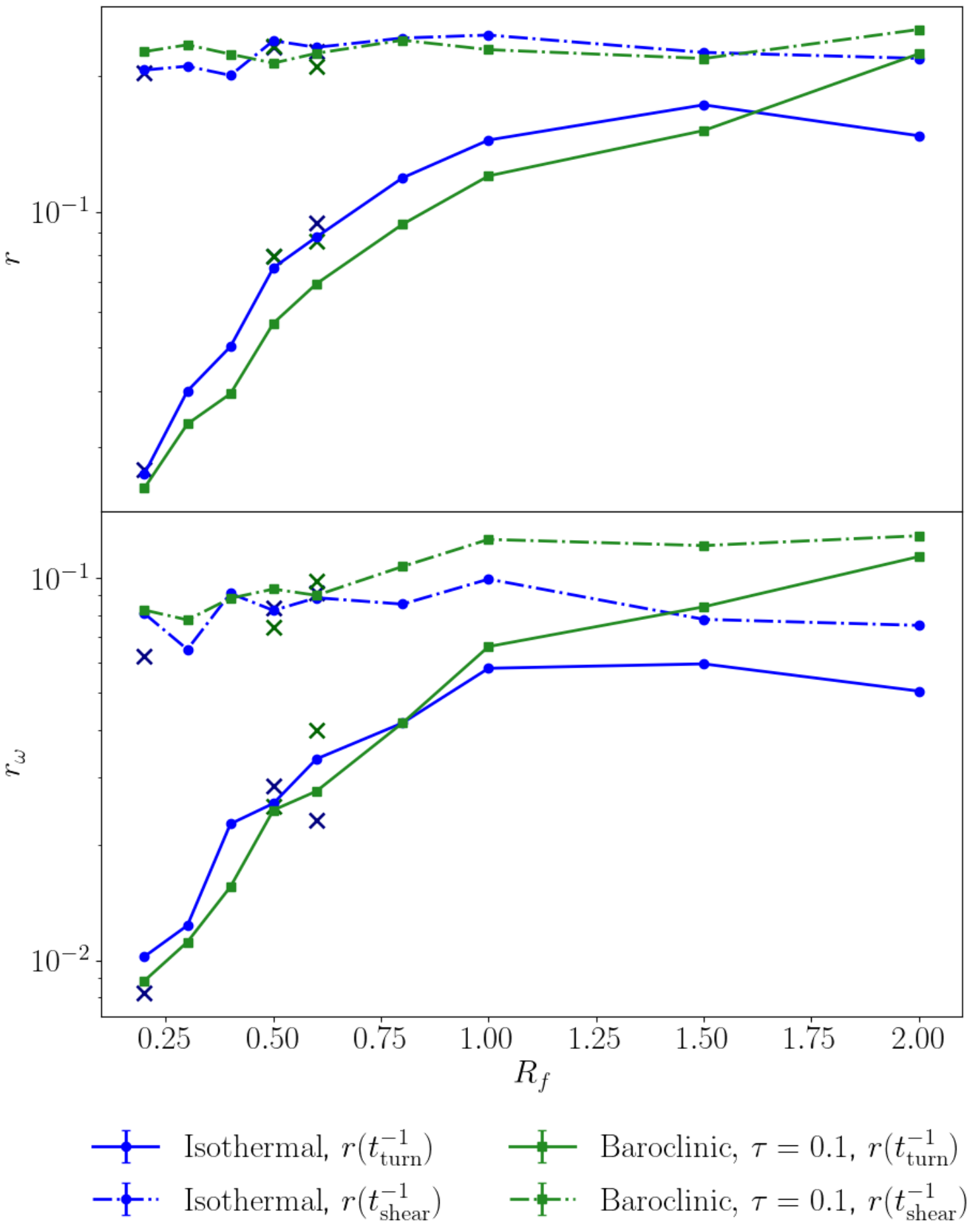

Figure 3.13: Vorticity and magnetic energy growth rates as a function of explosion width, $R_f$. The blue lines represent the isothermal models and the green ones the baroclinic cases. The dash-dotted lines correspond to growth rates in time units of $\mathrm{t_{shear}}$ and solid lines in forcing turnover times, $\mathrm{t_{turn}}$. Both time units are tabulated in App. B. The points marked with crosses are the corresponding dynamo runs with $\mathbf{f}_{\rm mom}$.

In Fig. 3.12 we plot the evolution of $E_{\rm mag}$ and $\omega_{\rm rms}$ for the isothermal runs with different $R_f$ and using the $f_{\rm acc}$ forcing. The only model that does not develop a dynamo is that of the smallest forcing scale (i.e., $R_f = 0.1$), and this is also true for the baroclinic case. Our interpretation is that, in this case, Re is subcritical, and this allowed us to estimate a critical value of Re $\sim 40$ for the vorticity instability to take place. We find instead a critical magnetic Reynolds number of slightly less than 20 (see the tabulated values) for the dynamo instability.

In Fig. 3.13 we show the growth rates $R_f$ as a function of $R_f$ for both isothermal and baroclinic cases. We show the values of $R_f$ calculated using different time units: either $t_{\rm shear}$ or $t_{\rm turn}$. When choosing $t_{\rm shear}$ as the time unit, $R_f$ approximately displays a constant behavior, while when choosing $t_{\rm turn}$, the growth rates vary considerably with $R_f$, as smaller expansion waves lead to much slower growth.

We observe that both magnetic and kinetic helicities grow in the dynamo cases, and start oscillating when $B_{\rm rms}$ and $\omega_{\rm rms}$ saturate. These oscillations resemble those seen in other instabilities such as the Tayler instability (e.g., Guerrero et al., 2019; Stefani et al., 2021; Monteiro et al., 2023).

### 3.2.8 Magnetic Prandtl number dependence

The magnetic Prandtl number (Pm) controls the scales at which the kinetic and magnetic energy cascades are truncated, respectively, by viscosity and resistivity. We varied Pm in our simulations to observe whether a difference in these truncation scales leads to different behavior for the amplification of vorticity and magnetic field. In the ISM, Pm is usually much larger than 1 (Ferrière, 2020), which is different from the range explored in our models. We also explored a range of Pm slightly smaller than unity, which is more typical of planetary environments.

We observe only a weak dependence on Pm for the models that do not develop a dynamo instability. In the isothermal case, we varied Pm from 0.25 up to 4 and see that $\mathrm{Re}_\omega$ either increases with Pm in the absence of rotation or slightly decreases with Pr when rotation is added (for more details see Table 3.1). In the baroclinic case, $\mathrm{Re}_\omega$ slightly decreases when Pm = 4 compared to the case of Pm = 0.25, independently of the presence of rotation. However, in the range of the explored values of Pm, we consistently report a decrease in the initial magnetic field.

Models with shear, conversely, develop a dynamo instability unless Pm = 4 or above (see the tabulated values). We interpret this as a consequence of the lack of vorticity instability that does not develop when the physical viscosity increases above a certain value. The growth rates for the magnetic field increase with Pm, even in the cases where $\eta$ is kept constant and $\nu$ is increased, of course, up until the point where viscosity does not allow the hydro instability. Conversely, the growth of vorticity is approximately constant in the explored range.

Table 3.1: Vorticity and magnetic energy growth rates for different values of Pm. We show the values for $\nu$ and $\eta$, to better illustrate the different growths. We have not allowed the physical diffusivities to reach numerical values. The other diagnostics can be seen in Table B.6.

| Pm | $\nu$ | $\eta$ | r ($t_{\mathrm{turn}}^{-1}$) | $\mathrm{r}_\omega$ ($t_{\mathrm{turn}}^{-1}$) |
|---|---|---|---|---|
| 0.1 | $2{\cdot}10^{-4}$ | $20{\cdot}10^{-4}$ | - | $10.52{\cdot}10^{-3}$ |
| 0.25 | $2{\cdot}10^{-4}$ | $8{\cdot}10^{-4}$ | $0.326{\cdot}10^{-2}$ | $8.57{\cdot}10^{-3}$ |
| 0.5 | $2{\cdot}10^{-4}$ | $4{\cdot}10^{-4}$ | $1.22{\cdot}10^{-2}$ | $9.33{\cdot}10^{-3}$ |
| 0.75 | $2{\cdot}10^{-4}$ | $2.667{\cdot}10^{-4}$ | $2.25{\cdot}10^{-2}$ | $10.96{\cdot}10^{-3}$ |
| 1 | $2{\cdot}10^{-4}$ | $2{\cdot}10^{-4}$ | $2.78{\cdot}10^{-2}$ | $10.22{\cdot}10^{-3}$ |
| 1.25 | $2.5{\cdot}10^{-4}$ | $2{\cdot}10^{-4}$ | $2.86{\cdot}10^{-2}$ | $9.18{\cdot}10^{-3}$ |
| 1.5 | $3{\cdot}10^{-4}$ | $2{\cdot}10^{-4}$ | $3.07{\cdot}10^{-2}$ | $10.01{\cdot}10^{-3}$ |
| 2 | $4{\cdot}10^{-4}$ | $2{\cdot}10^{-4}$ | $3.44{\cdot}10^{-2}$ | $9.38{\cdot}10^{-3}$ |
| 4 | $8{\cdot}10^{-4}$ | $2{\cdot}10^{-4}$ | - | - |
| 10 | $20{\cdot}10^{-4}$ | $2{\cdot}10^{-4}$ | - | - |

We find that for Pm = 0.1 the vorticity is amplified, but, differently from the other cases, this instability is not followed by an exponential amplification of the magnetic field. This can be seen as a consequence of having a Rm below the critical value, since, in general, it is also possible to excite a dynamo at a Pm value of less than 0.1 (e.g., Warnecke et al., 2023). Our results are therefore consistent with what is found in the literature.

### 3.2.9 Dependence on the cooling time

When the isothermal condition is relaxed, we let the temperature evolve according to Eq. 3.3, where we use a Newtonian cooling term, regulated by the timescale $\tau_{cool}$. The use of the cooling function leads to spectra cut at short wavelengths in models that do not develop instabilities. This is shown in Fig. 3.14 (compare the two purple lines). Differently, in the presence of shear and hence after a dynamo is excited, all the spectra recover their small-scale contribution and show again a wide dynamical range decreasing in a $k^{-2}$ fashion down to a dissipation scale, regardless of the cooling term (compare the orange curves). The use of the cooling term slightly diminishes the total amount of energy, but does not notably change the shape of the spectral distribution. However, we observe that in the presence of this cooling term, the instability kicks in at much earlier times, independently of the value of $\tau_{cool}$, at least within the explored range. This can be seen as a quicker injection of vorticity in the system due to the cooling function. We also observe that the average angle between $\nabla T$ and $\nabla s$ slightly increases when a cooling function is used, hence leading to a larger contribution of the baroclinic term in seeding the vorticity.

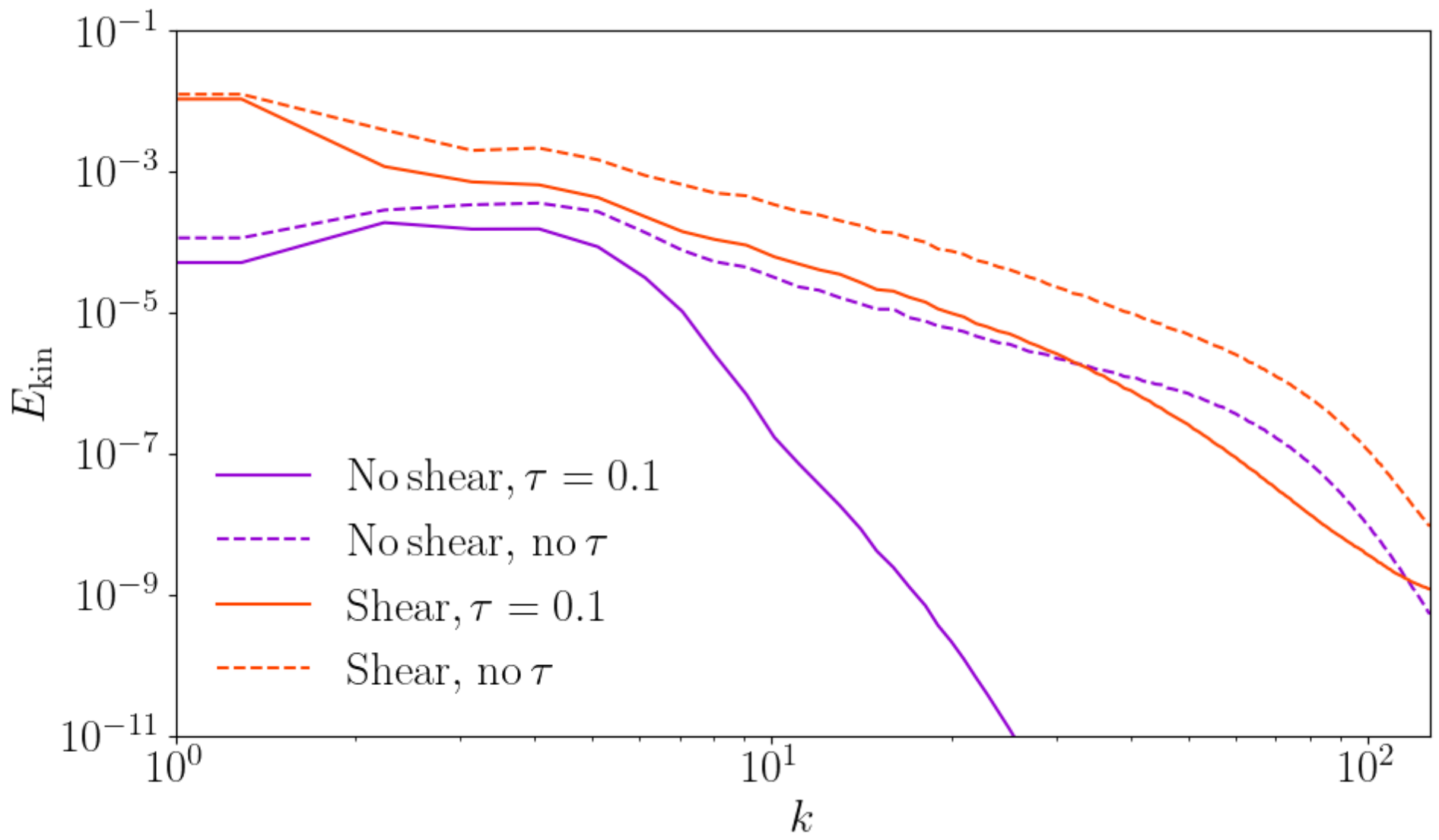

Figure 3.14: Kinetic spectra for models with $R_f = 0.5$ and an optional shear and cooling term. The models with shear are shown after dynamo growth.

Another possible interpretation is to invoke an effect similar to what was observed by Rädler et al. (2011). Irrotationally forced flows present peculiarities such as the possibility of having a negative magnetic diffusivity contribution from turbulent flows, especially at low Reynolds numbers, as shown analytically by Krause and Rädler (1980), Rädler and Rheinhardt (2007), and Rädler et al. (2011). The contributions to the diffusivity coming from the turbulence may affect the occurrence of dynamo instability even when one moves toward higher values of the magnetic Reynolds number. This regime is closer to that of astrophysical bodies. Rädler et al. (2011) found, with mean-field approaches based on the second-order correlation approximation (SOCA), that a negative contribution to the magnetic diffusivity can come from the presence of turbulence in irrotational flows in the case of small Péclet numbers $\mathrm{Pe} = uL/\kappa$, where $u$, $L$, and $\kappa$ are the typical velocity, length scale, and diffusivity of a system. In our case, the presence of a cooling time introduces an additional diffusion term for thermodynamic quantities, which results in Rm being smaller than it would be in the absence of cooling. However, the SOCA approximation is valid only for small magnetic Reynolds numbers, which is a condition that is not satisfied in our models.

Although our cooling function is not meant to model any specific astrophysical environment, we can attempt a comparison with typical values of the cooling in the ISM. Using the hydrogen cooling function $\Lambda$ (e.g., Sutherland and Dopita, 1993) and assuming a temperature of $10^4$ K,

$$\log \frac{\Lambda}{\text{erg cm}^3\text{s}^{-1}}(T = 10^4\text{K}) \approx -22 \qquad ; \qquad \frac{\partial e}{\partial t} = ... - \Lambda n$$
$$\rightarrow \quad \tau_{cool} = \frac{k_B T}{\Lambda n} \approx 10^{10} s \approx 0.5 \text{ kyr}.$$

If we consider a time unit of 8 Myr, then the cooling time $\tau_{cool}$ should be on the order of 0.001. In our models, we can reach down to $\tau \sim 0.01$ for numerical stability reasons. This is one order of magnitude lower than typical ISM values (i.e., lower cooling rates and higher temperature differences than a one-phase ISM).

### 3.2.10 Vorticity source terms

From Eq. 3.12 we can evaluate which are the most relevant terms in the vorticity equation for the different runs. We find that in the isothermal nonrotating runs, there are no positive terms comparable to the viscous ones. This fact, alongside the decrease in vorticity growth with resolution (see Fig. 3.11), makes us think that in such cases only numerical diffusive sources are in action, in agreement with what we already observed in Mee and Brandenburg (2006).

In Fig. 3.15 we can see the time series of the different source terms as a function of time, for a representative non-isothermal case. All terms are very small. The baroclinic term dominates as the most positive contribution. As vorticity grows, the turbulent contribution $\langle(\mathbf{u} \times \boldsymbol{\omega}) \cdot \mathbf{q}\rangle$ gains importance, and the sum of both is counteracted by viscous forces, so that $\omega_{\rm rms}$ saturates. From the viscous contributions, only $\nu\langle\mathbf{q}^2\rangle$ is relevant, while $2\nu\langle\mathbf{S}\boldsymbol{\nabla}\ln\rho \cdot \mathbf{q}\rangle$ is more than one order of magnitude lower. This last statement holds true for all the isothermal, non-isothermal, and rotating, nonrotating cases. Obviously, the Lorentz term is irrelevant in the cases without a dynamo, orders of magnitude below by comparison.

When the forcing is exactly irrotational, its corresponding term does not contribute to vorticity generation. But when it is applied in its second form (not exactly irrotational due to density fluctuations), there is indeed a small vorticity growth that leads to a similar behavior in strength and shape compared to the baroclinic source term. When applying this type of forcing in the non-isothermal runs, the forcing vorticity growth overtakes the baroclinic term to such an extent that this latter becomes negative.

When rotation is included, the generation of vorticity is more relevant. In this case, the Coriolis source creates an amount of vorticity that is later counteracted by viscous terms, so that the steady state is reached. In Fig. 3.16 we show the same plot as in Fig. 3.15 but having added rotation. We can see that in the beginning, the rotation term, $2\Omega\langle(\boldsymbol{e_z} \times \mathbf{u}) \cdot \mathbf{q}\rangle$, has a significant positive spike that leads to the initial growth of vorticity. We note that it is larger than the baroclinic term in Fig. 3.15 by more than one order of magnitude and oscillates substantially, even becoming negative at times, leading to an overall noisier $\omega_{\rm rms}$. At the beginning of the run, the rotation contribution is primarily positive, and, when $\omega_{\rm rms}$ saturates, the term is slightly positive on average, and of the same order as the viscous main contribution. Thus, in the presence of rotation, all other source terms become much less important.

For the cases with the shearing profile, both the HD and MHD instabilities make the contributions change substantially. In Fig. 3.17 we show all the relevant terms for an isothermal shearing run. As before, within the viscous forces, $\nu\langle\mathbf{q}^2\rangle$ still dominates over $2\nu\langle\mathbf{S}\boldsymbol{\nabla}\ln\rho \cdot \mathbf{q}\rangle$, but the latter becomes more relevant in comparison to the cases described above.

The background shearing profiles increase $\omega_{\rm rms}$ up to a certain value in a very few time steps. This value is kept approximately until the vorticity instability kicks in (for example, until $t \simeq$ $400\, t_{\rm turn}$ in Fig. 3.17). Afterward, when vorticity is amplified, the advective term (which includes both shear and turbulence), $\langle (\mathbf{u} \times \boldsymbol{\omega}) \cdot \mathbf{q} \rangle$, brings the main positive contribution, as expected. The viscous forces are not enough to counteract this completely, thus leading to growth in $\omega_{\rm rms}$.

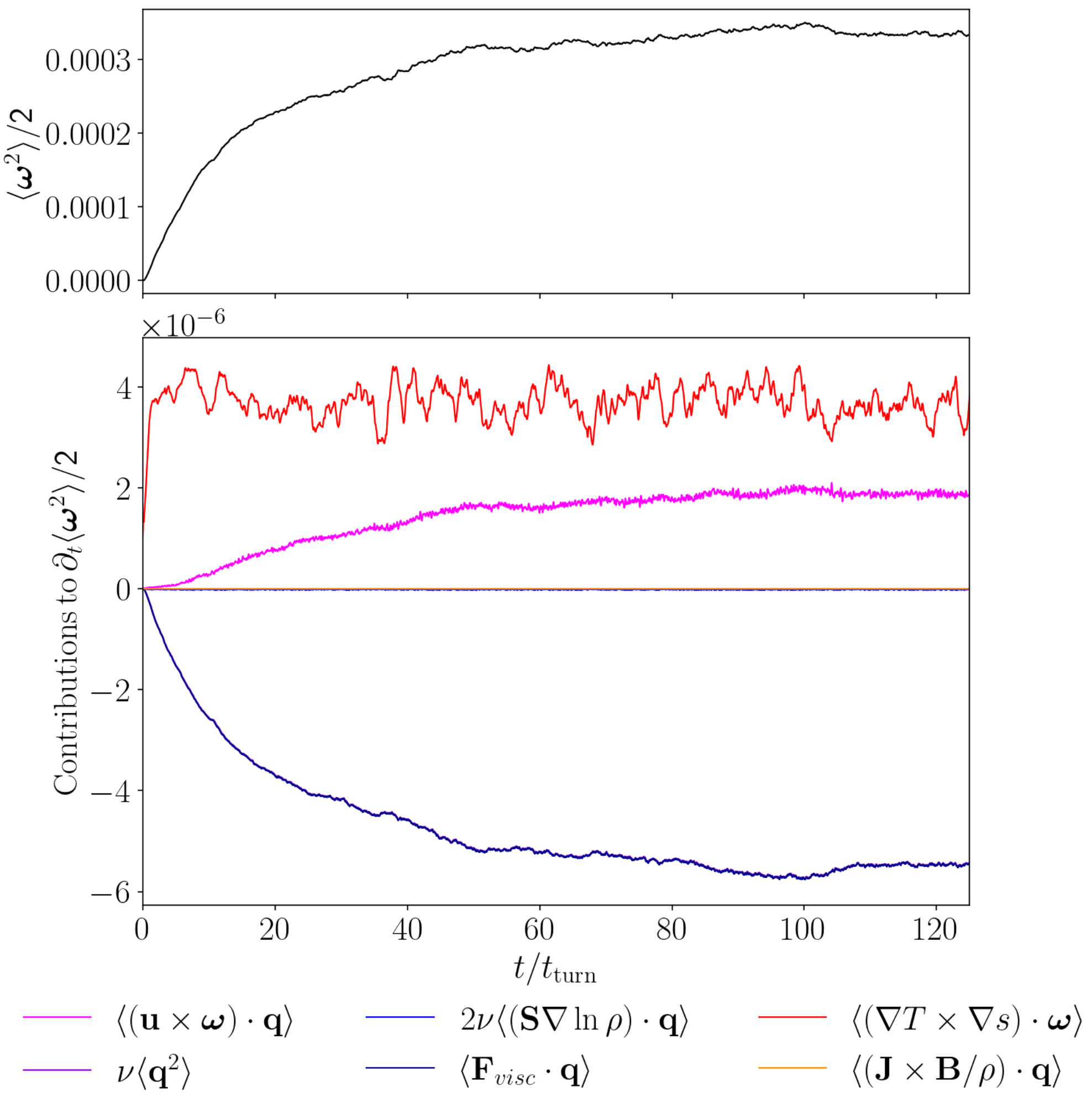

Figure 3.15: Time evolution for $\omega_{\rm rms}^2/2$ (left axis and gray line) and vorticity growth terms (right axis and various colored lines for each term) for a non-isothermal run with $R_f$=0.5 and forcing in acceleration form (and by construction its contribution to $\langle \partial_t \boldsymbol{\omega}^2/2 \rangle$ is zero). This plot zooms in on the beginning of the temporal evolution of all the terms until $t \simeq 130\, t_{\rm turn}$. However, the saturation regime does not present any relevant changes in time for more than 2000 $t_{\rm turn}$, and no instability is reached. Both the $2\nu \langle \mathbf{S} \boldsymbol{\nabla} \ln \rho \cdot \mathbf{q} \rangle$ and $\langle (\mathbf{J} \times \mathbf{B}/\rho) \cdot \mathbf{q} \rangle$ terms are negligible and very close to 0, so this makes the total contribution of viscous forces, i.e., $\langle \mathbf{F}_{visc} \cdot \mathbf{q} \rangle$, overlap with $\nu \langle \mathbf{q}^2 \rangle$.

The Lorentz term is negligible up to the point where the dynamo starts. Then there is a brief time when it becomes slightly negative, exactly when vorticity starts growing exponentially, but still the dynamo has not kicked in. This behavior was also observed by Seta and Federrath (2022). When the kinetic phase of the dynamo starts, the Lorentz term increases, but as the Lorentz forces act against the flow, $\langle (\mathbf{u} \times \boldsymbol{\omega}) \cdot \mathbf{q} \rangle$ decreases more than $\langle (\mathbf{J} \times \mathbf{B}/\rho) \cdot \mathbf{q} \rangle$ increases. This leads to a negative overall contribution and a decrease in vorticity, which later stabilizes at the end of the kinetic phase, and to the amplification of $B_y$ by winding. In all dynamo runs, the Lorentz term always ends up surpassing the advective term $\langle (\mathbf{u} \times \boldsymbol{\omega}) \cdot \mathbf{q} \rangle$.

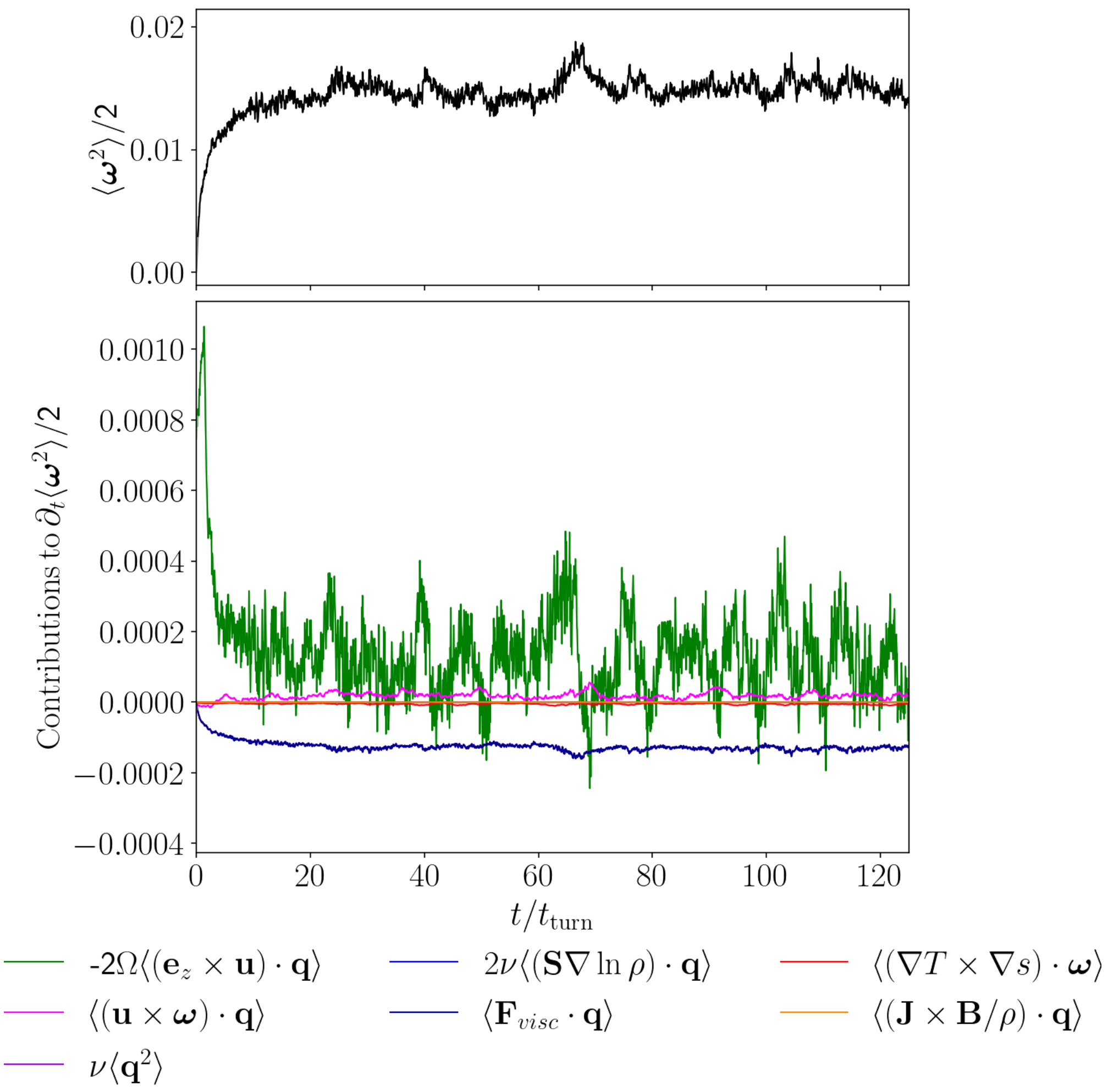

Figure 3.16: Time evolution for $\omega_{\rm rms}^2/2$ and vorticity growth terms for a non-isothermal rotating run with $R_f$=0.5 and forcing in acceleration form. The notation is the same as in Fig. 3.15, and similarly $2\nu\langle \mathbf{S}\boldsymbol{\nabla} \ln \rho \cdot \mathbf{q}\rangle$ and $\langle(\mathbf{J} \times \mathbf{B}/\rho) \cdot \mathbf{q}\rangle$ overlap near 0, also making $\langle \mathbf{F}_{visc} \cdot \mathbf{q}\rangle$ overlap with $\nu\langle \mathbf{q}^2\rangle$.

## 3.3 Discussion and possible astrophysical applications

The presented model is rather general and can therefore be applied to several astrophysical environments. In particular, the implemented forcing can be interpreted as the expansion waves coming from SN, the primary forcing in the ISM. Although SN tend to be highly supersonic and spherically asymmetric, the expansion waves under consideration here can be thought of as a rough approximation of a long-time expanding wave of the SN remnant.

We can interpret our box as a small cube with a side of 500 pc inside a galaxy (i.e., a resolution of about 2 pc). The forcing width of 1/10 of the box corresponds to 50 pc, slightly larger but of the same order of magnitude as some SN remnants. We can take the reference value for density of $10^{-23}$ g/cm$^3$, and the speed of sound of 10 km/s. Galactic rotation curves lead to a shearing amplitude of about 5 km/s for a 500 pc radial distance. This is of the same order of magnitude, $A = 0.2$, of the sinusoidal shear in our model.

The estimated SN rate of 2-3 SN per century per galaxy (Murphey et al., 2021) can then be translated to a 3-4 SN/Myr per (500 pc)$^3$-box, assuming a galactic volume of the order of $10^{12}$ pc$^3$. The value $\Delta t$ of 0.02 leads to a rate of $\sim$ 6 SN/My per (500 pc)$^3$-box. As an example, for the M_0A020 run, these units lead to a $u_{\rm rms}$ of the order of 2 km/s and a mean magnetic field of 10 $\mu$G, which is of the order of the estimated galactic one (e.g. Jansson and Farrar, 2012; Brandenburg and Ntormousi, 2023).

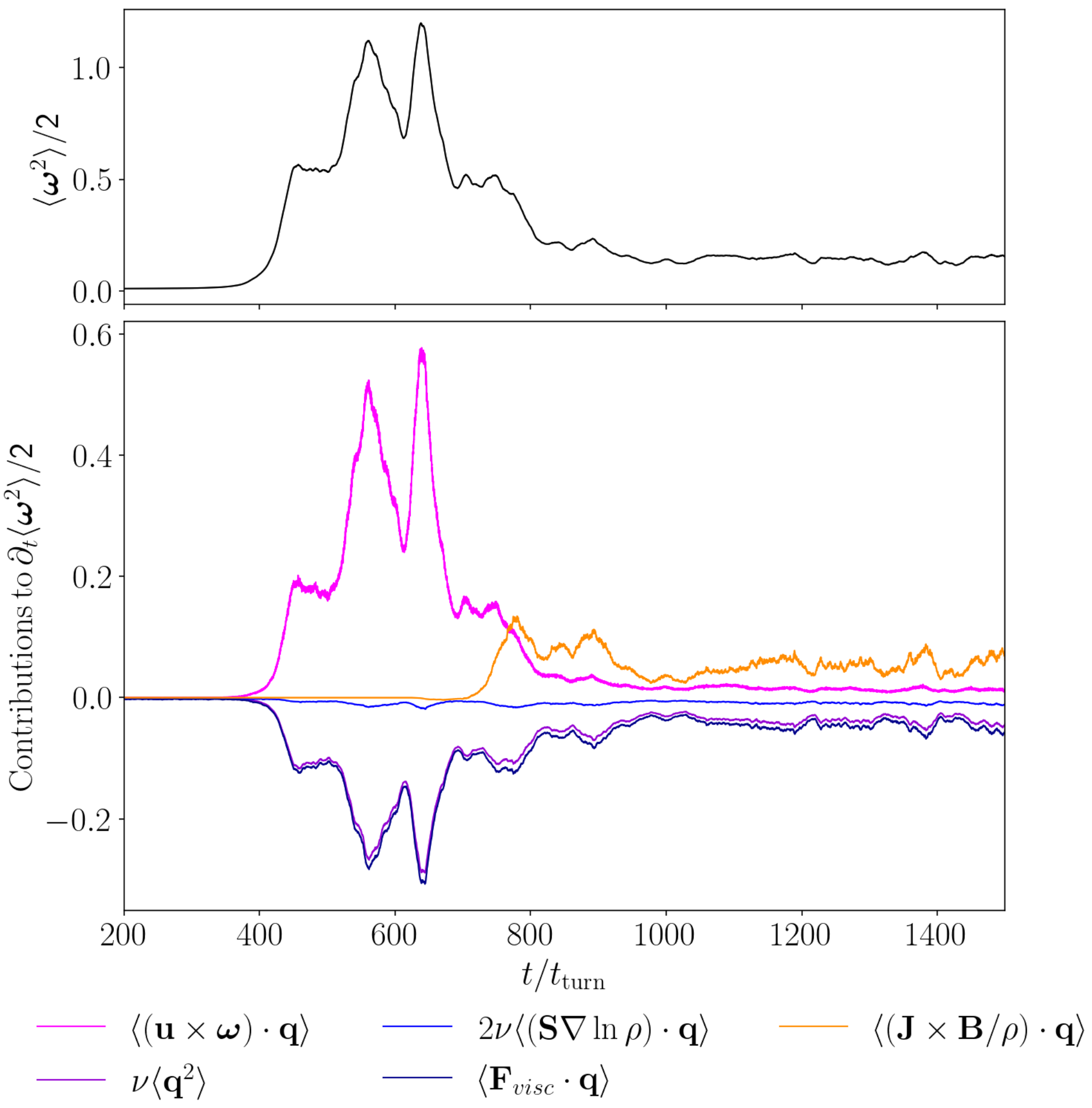

Figure 3.17: Time evolution for $\omega_{\rm rms}^2/2$ and vorticity growth terms for a isothermal run with the presence of shear, hence leading to instability, and with $R_f = 0.5$ and the exactly irrotational forcing. The notation is the same as in Fig. 3.15.

Therefore, the results can indeed have direct application for ISM. More realistic simulations could consider density variations and non-spherically symmetric expansion waves (which might lead to SSD). Finally, the shear here considered could be refined to replicate, for instance, the rotational curve of the galaxy or the shear corresponding to the spiral arms.

## 3.4 Summary

- We investigated the possibility of vorticity production and dynamo action under purely irrotational (curl-free) forcing, a toy model for SN explosions feeding the ISM.

- Irrotational forcing alone does not generate vorticity or dynamo action in either HD or MHD scenarios.

- Vorticity is produced when irrotational forcing interacts with solid body rotation and baroclinicity, but no dynamo action is observed across explored parameters, regardless of initial magnetic field topology.

- A background sinusoidal shear profile produces much greater initial vorticity in combination with the forcing. It later develops a hydrodynamical instability that induces an exponential vorticity growth.
- In the MHD scenario, the shear HD instability is followed by a dynamo phase, both in barotropic and baroclinic cases.
- Vorticity and magnetic field amplification start at small scales and cascade to larger scales during saturation.
- Final magnetic field structure dominated by large scales due to inverse cascade and winding effect.
- The forcing scale influences both the onset time of vorticity/magnetic amplification and the growth rate (when scaled by turnover time, but not by shear timescale).
- In the presence of baroclinic effects, the magnetic field grows slightly faster. When no cooling function is considered, vorticity growth is delayed in comparison with the isothermal case, but a Newtonian cooling function enables faster onset of instability.
- A cooling function and density-dependent forcing enhance vorticity, but no magnetic field growth. Shear is essential for dynamo instability with this subsonic expansion wave forced turbulence, regardless of the forcing scale.
- Magnetic Prandtl number increase leads to an order-of-magnitude increase in growth rate.
- In our simulations, we estimated critical magnetic Reynolds number, $\mathrm{Rm}_{\mathrm{crit}} \approx 20$, and a critical Reynolds number for shear-induced dynamo: $\mathrm{Re}_{\mathrm{crit}} \sim 40$.

# 4

# Planetary magnetism and spherical shell numerical dynamos

Naturally magnetized minerals were first documented in the ancient world by Thales of Miletus in Greece around 600 BC and independently in China around 300 BC. Although the existence of permanent rock magnetization, i.e., ferromagnetism, was known, the concept of geomagnetism emerged much later. This required the realization of a much larger, ordered magnetic field around the world that cannot be produced by the scattered distribution of rocks. The first accounts for the geomagnetic field coincide with the first records of the magnetic compass needle, dating back to the 11$^{\text{th}}$ century in China, and which, by the 12$^{\text{th}}$ century, had spread through Europe. These first records already document the directional properties, i.e., the tendency to point northward systematically. Specifically, they noted that compasses usually pointed towards a direction close, but not coincident, with the geographic North Pole. This difference is known as magnetic declination $D$. It was commonly believed that this force came from the celestial sphere.

The first scientific approach to geomagnetism did not come until the 17$^{\text{th}}$ century, when William Gilbert published his treatise *De Magnete* (Gilbert, 1600). He proposed that the Earth behaves like a giant solid magnet, based on experimental evidence with a spherical lodestone model he called the *terrella*, on which small compass needles aligned along its surface mimicking the behavior of compasses on Earth. This led Gilbert to conclude that the Earth's magnetic field originates internally, rather than from celestial influences as previously believed. Gilbert also made two other now accepted predictions: he made the distinction between the magnetic force of lodestones and the electric attraction produced by materials such as rubbed amber, coining the term *electric force*; and he attributed the movement of the celestial bodies to Earth's rotation around its axis, instead of the Ptolemaic model for concentric moving shells. On the other hand, other predictions from him were incorrect, such as his explanation for magnetic declination, which he attributed to variations in altitude, or the argument that magnetism maintained both Earth's rotation and the position of all heavenly bodies. Nonetheless, Gilbert is widely praised for his modern empirical methodology, emphasizing repeatable experimentation, careful observation, and conclusions based on evidence rather than adherence to ancient authorities or speculative philosophy.

Due to the existence of volcanoes, it was believed that the Earth's interior was completely molten and then surrounded by a thin crust. The first arguments against this theory suggested that there should be a solid, rigid mantle onto which oceanic tides can rise and fall. By the 19$^{\text{th}}$ century, Earth's density estimates predicted a denser core with a radius very close to the currently measured core radius, assuming a silicate mantle and ferrous interior. This structure was not confirmed until robust evidence came from seismic wave studies: P waves, which can travel through any medium, and S waves, which cannot transmit in a fluid. In 1906, Oldman proved the existence of a discontinuity in Earth's interior by showing different S and P waves shadows (Oldham, 1906), which were interpreted as a solid-liquid boundary between the rocky mantle and liquid metal core at a radius of 3470 km. This layer is now known as the core-mantle boundary (CMB). The solid inner core was discovered by Inge Lehmann in 1936, who showed that some P waves arrived in the P shadow zone after reflecting from another discontinuity layer at 1221 km (Lehmann, 1936). That article has a very creative and simple name: "P'", as it refers to a reflected P wave.

Until the end of the 19$^{\text{th}}$ century, it was accepted that geomagnetism arises from an internal permanent ferromagnetism of the Earth's interior. This view lost its appeal, however, when it became evident that the iron Curie temperature is exceeded in the Earth below a depth of about 30 km only. Thus, ferromagnetic materials in the Earth's interior cannot be the source of such a field. Moreover, magnetic diffusive processes in the liquid core have lifetimes much shorter than Gyr (see Sect. 2.2.2). An active dynamo in the Earth's fluid outer core became the most logical explanation for the geomagnetic field.

Permanent magnetic stations alongside geophysical evidence currently provide a much more complete description of the geomagnetic field, including its absolute strength, morphology, time variation, and, most importantly, origin. But due to a lack of observations, magnetism in other planets has remained an open question until the last decades. The first evidence came from the Jovian decametric coherent radio waves in 1955 (Franklin and Burke, 1958; Gardner and Shain, 1958), which are produced at the cyclotron frequency of the local magnetic field. They show that Jupiter has a $\mathcal{O}(10)$ G magnetic field intensity at the surface, ~one order of magnitude larger than the Earth and comparable to the average value at the solar surface. Since the 1970s, different space missions have measured *in situ* the magnetic field around the main planetary bodies of the solar system, including a few moons, for which the magnetic field is weaker. But the associated radio emission for other planets has too low frequencies to be detected from the ground, due to the shielding at $< 10$ MHz by the Earth's ionosphere. The interpretation of these results, which we briefly show in the following sections, leads to the fundamental evidence that all solar planets currently have internally generated magnetic fields, except Venus and Mars, and that there is a considerable variety in intensity and configuration, which brings information about the internal structure and dynamics. Ganymede is the only moon for which we could detect an internally generated non-negligible magnetic field. The Moon only shows signs of a weak (tens of $\mu$G) residual magnetization in its crust (similar to Mars), and for some other main moons and dwarf planets, we currently have only upper limits.

## 4.1 Structure and internal conducting fluids

As mentioned in Sect. 2.2.2, planetary magnetic fields must come from active dynamos in their interiors, i.e., the convective flow of an electrically conducting fluid under the effects of rotation must trigger the magnetic field growth and sustainment. The fluid's nature depends on the type of planet, as rocky planets, ice giants, or gas giants have vastly different interiors. The top panel of Fig. 4.1 illustrates the internal structure of the eight planets in the solar system, highlighting the layers that may serve as dynamo regions.

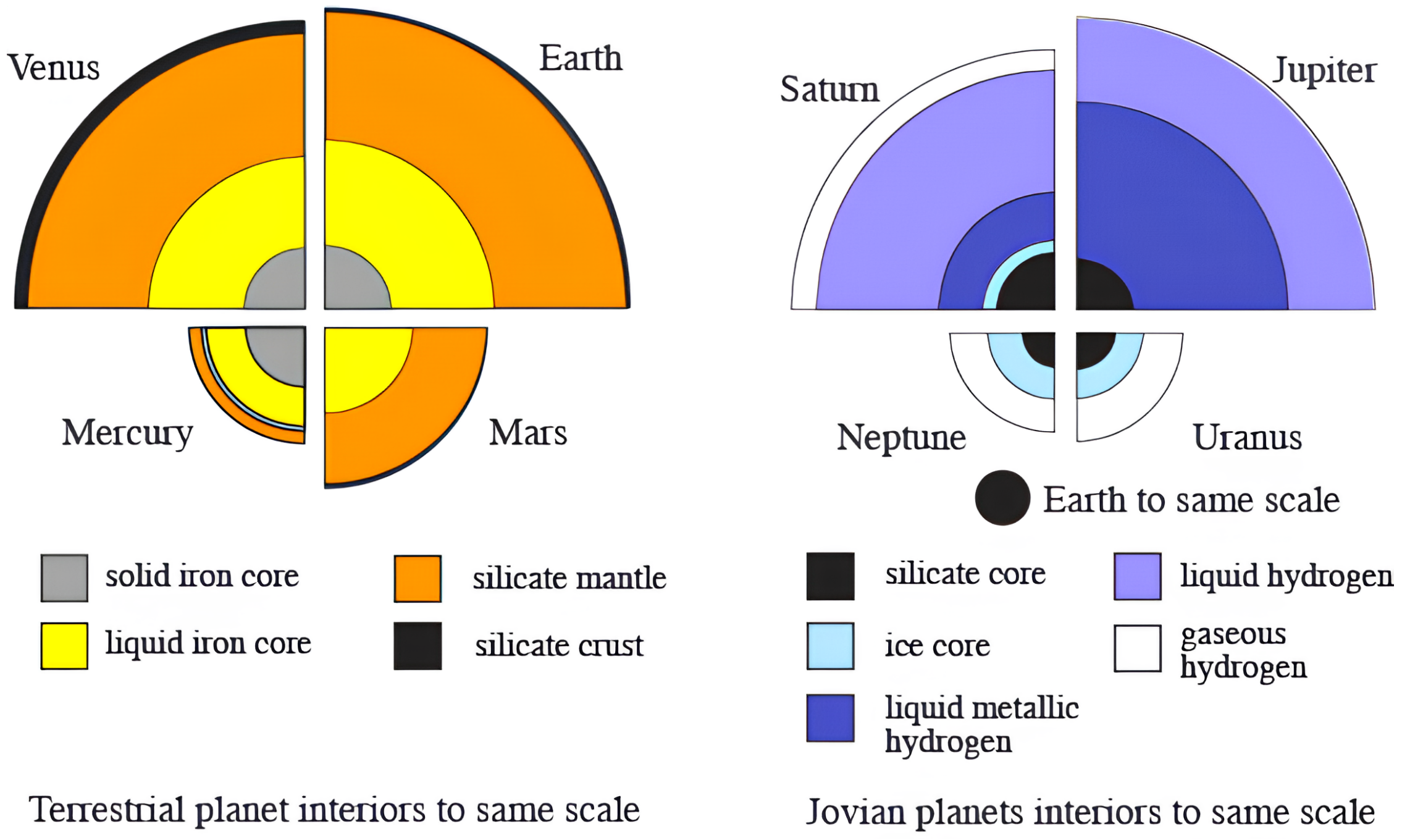

Figure 4.1: Internal structures of the eight planets at scale with each other. Image taken from Strobel (2020).

The primary conducting fluid for the terrestrial planets (i.e., Mercury, Venus, Earth, and Mars) is thought to be the molten iron-nickel-dominated mixture in their cores. As mentioned above, for the Earth, terrestrial planets have an outer thin rocky crust, an intermediate silicate mantel, and a central iron-alloy core. Earth possesses a well-developed dynamo driven by convection in its outer liquid iron core, surrounding a solid inner core. Mercury maintains a weak magnetic field, indicating a partially molten iron core. However, not all objects with fluid, electrically conducting layers generate such fields. The Moon, Mars, Venus, and Io are likely to have partially or completely molten cores, yet they do not possess dynamos or have ceased to function. Their lack of an active dynamo is thought to be indicative of insufficient core convection. In Venus, slow rotation and/or the lack of tectonic plates, which delays the cooling and avoids strong temperature gradients, could make the convection inefficient. On the other hand, direct measurements show that both Mars and the Moon have magnetized rocks on their surface. This reveals past active dynamos that may have cooled enough to stall convective motions and might have completely cooled. See Breuer and Moore (2007) for a review of the dynamics and thermal history of terrestrial bodies in the solar system.

In contrast, the outer giant planets' dynamos operate under very different conditions, related to the transport properties of their so-called warm dense matter. The gas giants have outer molecular envelopes of hydrogen and helium, fluid mantles composed of metallic hydrogen and helium, and central rocky cores, or diluted cores with compositional gradients. The thickness of these layers is not well known, except for the Earth. Therefore, an accurate description of their internal structures would help in understanding possible planetary dynamo morphologies. In Fig. 4.2, I show the phase diagram for hydrogen taken from Guillot and Gautier (2015). For atmospheric conditions, hydrogen takes its usual $H_2$ molecular form. For low pressures ($\lesssim 10^6$ bar) and temperatures below $\lesssim 10^4$

K, hydrogen can be solid, liquid, gas, or atomic. Still, these states have a very poor electrical conductivity due to the negligible ionization. For high temperatures, hydrogen dissociates and forms the plasma, characteristic of stellar interiors as seen by the Sun profile in Fig. 4.2. For higher pressures, $\gtrsim 10^6$ bar and not too high temperatures (warm dense matter), a new phase of hydrogen is reached.

Atoms become close enough that they start sharing electrons in the same way a metallic valence band does. The existence of a metallic hydrogen phase was already theoretically predicted by Eugene Wigner and Hillard Bell Huntington in 1935, and recent high-pressure shock wave laboratory experiments have claimed to have found it (Dias and Silvera, 2017; Loubeyre et al., 2020). Additionally, numerous numerical approaches, relying on density functional theory coupled with molecular dynamics, have assessed in more detail the hydrogen phase diagram and the transport properties. See Bonitz et al. (2024) for a review which includes simulations and diamond anvil and shock wave experimental results. Such *ab-initio* simulations predict that the interior adiabats (pressure-temperature profiles in the convective regions) of Jupiter and Saturn across the region where liquid metallic hydrogen is predicted (French et al., 2012; Preising et al., 2023). These works obtain hydrostatic thermodynamic radial profiles and transport coefficients for both gas giants. HD or MHD global planetary simulations commonly use these background states or at least compare them with similar functions, as I do in Chapter 5 and Chapter 7.

The ice giants Uranus and Neptune are believed to have outer molecular envelopes of hydrogen, helium, and ices, an intermediate ionic oceans of water, methane, and ammonia, and perhaps central rocky cores (Helled et al., 2020). The convective and conductive fluid layers are believed to be ionic oceans or high-pressure water-ammonia mixtures located between their rocky cores and outer envelopes. In the central part of ice giants, hydrogen metallization pressures are also likely reached. Such heterogeneous composition likely results in complex and highly tilted magnetic fields, but the lack of data from a dedicated mission (compared to Jupiter and Saturn) leaves many more questions open.

## 4.2 Planetary magnetic field properties

As seen in Fig. 4.3, the magnitude for the magnetic field at the Earth's surface ranges from 0.25 to 0.65 G. It can be approximated by a dipole positioned at the Earth's center with a tilt angle (with the rotation axis) of 11°. Interestingly, the geomagnetic south pole points approximately toward the geographic north pole, and vice versa. This is not by chance: the definition of north of a magnet derives from the needle side that points to the north star in a compass. Therefore, as one magnetic pole attracts the opposite polarity of another magnet, a magnetic south pole points to the geographic north, and vice versa.

The secular variation of Earth's magnetic field is the change of this internally generated dipole-dominated field. By the 18$^{th}$ century, it was already known that magnetic structures exhibit a westward drift. The axial dipole component changes more slowly, but measurably on human timescales: since 1840, the dipole moment has decreased by about 9%. Another clear time-dependent characteristic is that the north magnetic pole, i.e., the position where the field points perpendicular to the Earth's surface, shows a movement of tens of kilometers per year, currently going from Canada towards Siberia. Current measurements are precise enough to infer the main feature of the flow of liquid iron just below the CMB. It is dominated by a westward flow, more pronounced in the Atlantic hemisphere of the globe, with a typical velocity of $\sim$ 0.5 mm/s. This leads to a core equipartition level of $E_{mag}/E_{kin} \sim 10^3$ and a core turnover time is $\sim$ 100-500 years. Therefore, the

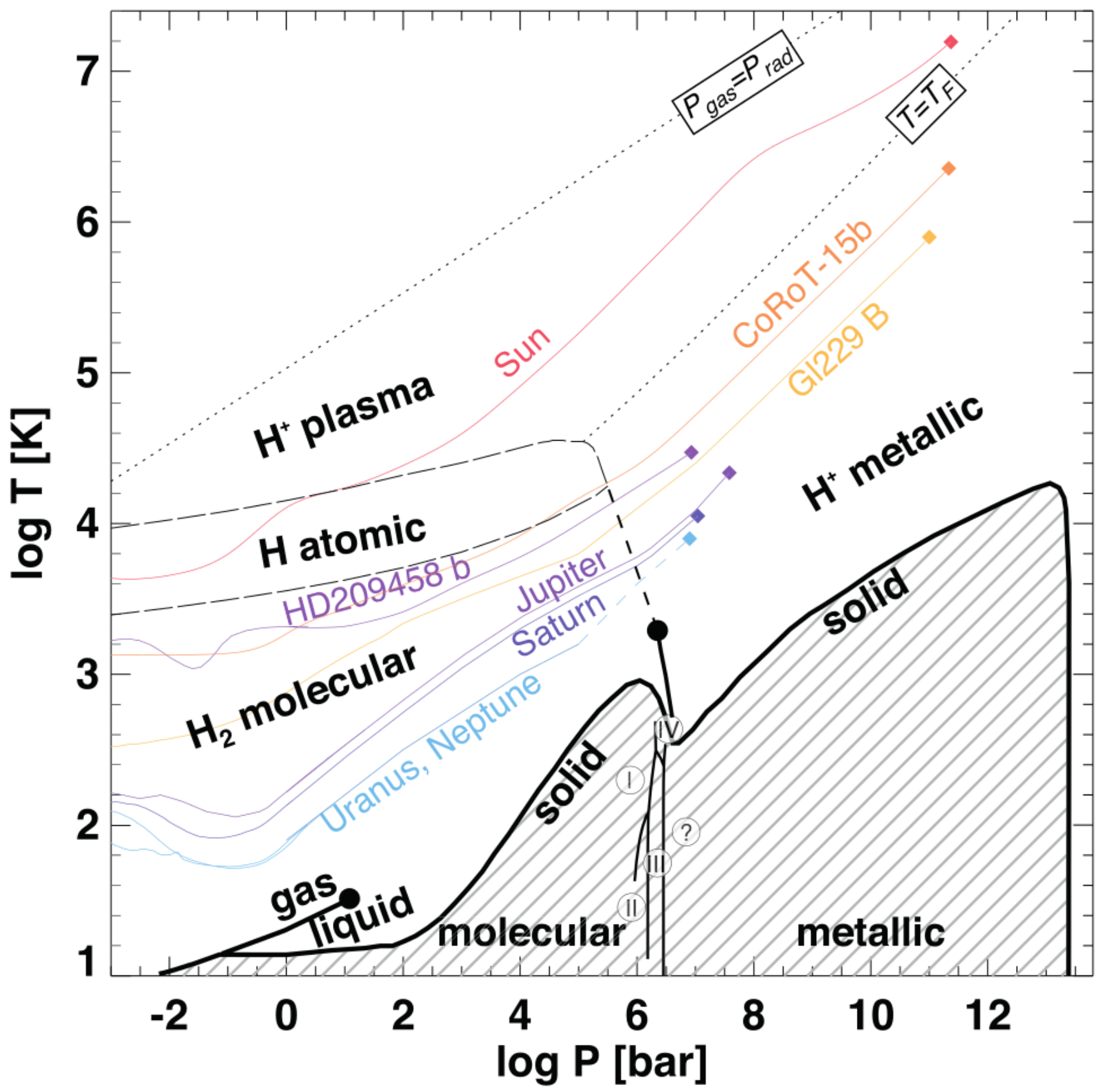

Figure 4.2: Pressure-temperature hydrogen phase diagram. Colored lines show the internal profiles for the Sun and a selection of noteworthy substellar objects: brown dwarfs, gas giants, and ice giants. Taken from Guillot and Gautier (2015), where all the phase transition lines are referenced.

400 years humans have taken magnetic measurements is only equivalent to ∼ 1 turnover time in core dynamics. See Nimmo (2007) for a review of Earth's core dynamics.

The evidence for long-period variability of the geomagnetic field comes from remnant magnetized ferromagnetic rocks. A hot magmatic rock aligns its magnetic domains in the same direction as the ambient magnetic field. When the rock cools down below its Curie temperature, it becomes a magnetic mineral insensitive to later changes in field direction. In other words, when liquid magma freezes, the resulting rock will carry information about the ambient field of that specific instant. This process, known as thermoremanence, has allowed us to see the orientation and sometimes the magnetic field strength of dated rock samples. The ocean floor has provided the most information about the geodynamo. Around the ocean ridges, where plate tectonics separate, magma is expelled, cools down, and records the magnetic field strength and geometry. This method allows one to track the large-scale magnetic field over the last ∼ 100 Myr, the typical lifetime of ocean floors before being pushed below a neighboring plate. The evidence shows that geomagnetism is prone to frequent polarity reversals. This is a stochastic process with a mean frequency ∼ 0.5 Myr, and between each reversal, the polarity is stable. The reversal timescale estimates are relatively rapid, ranging from a few thousand up to 28,000 years, with a mean of 7,000 years (Clement, 2004). During reversals, the dipole becomes much weaker while the multipole components stay approximately constant, leading to a multipolar field during reversal. There is also evidence for two other geomagnetic processes: excursions, short events when the dipole axis becomes strongly tilted but goes back to the original state; and superchrons, which are extended periods of tens of millions of years during which no reversals occurred. Further details on paleomagnetism and dipole reversals can be found, e.g., in

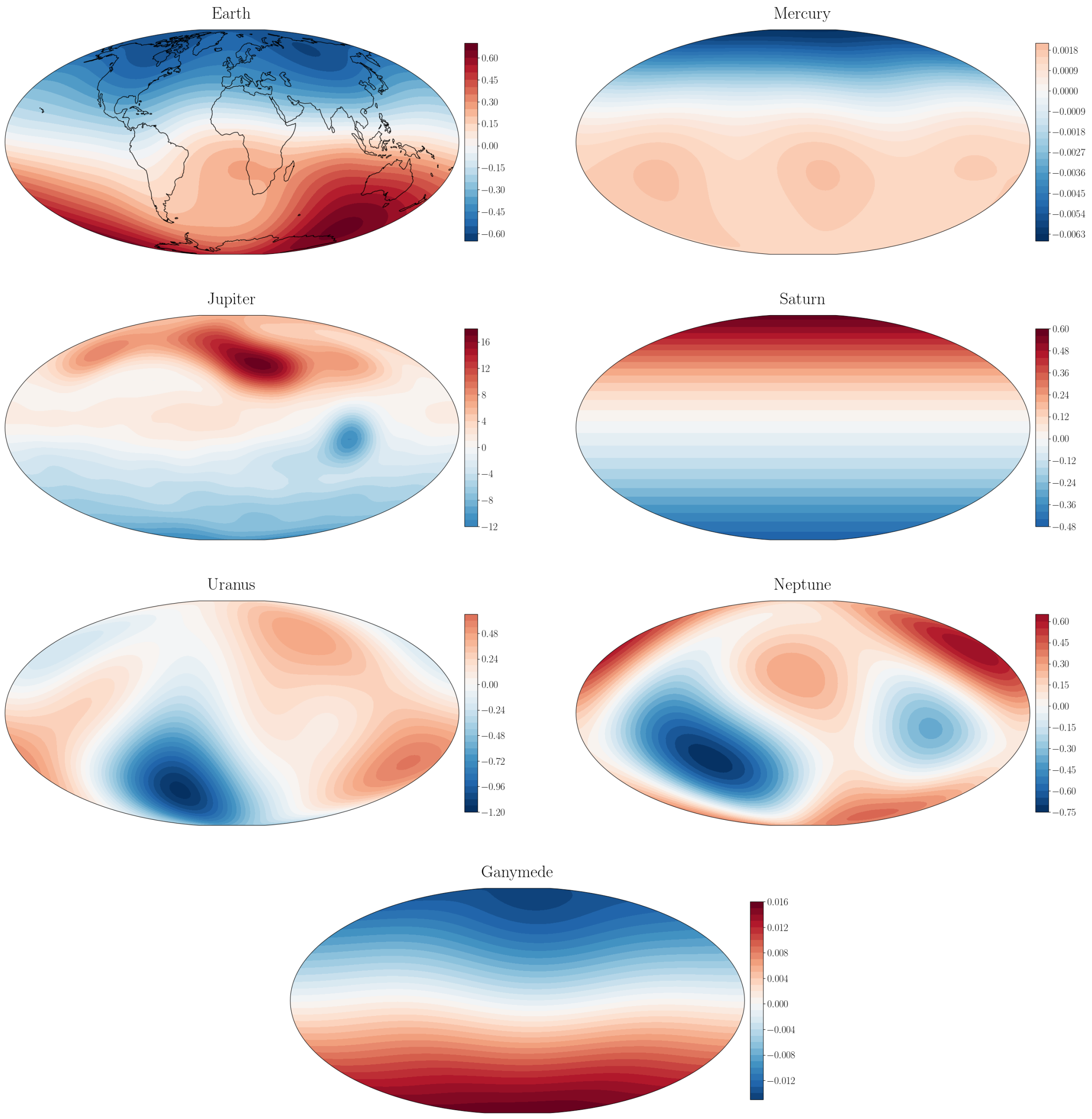

Figure 4.3: Radial components at the planetary surfaces of the magnetic field (in G) for the planetary bodies in the solar system with internal active dynamos. All are reconstructed from the publications shown in Table 4.1.

Merrill and McFadden (1999) or Glatzmaier and Coe (2007).

Other paleogeomagnetic studies have allowed us to assess rock magnetization processes within the last 3 billion years. The derived magnetic field strength fluctuates stochastically within a factor of two or three of the current strength, and the amplitude between the dipole component and higher multipoles has been similar to current values. Detailed features about the geometry of the field are more difficult, but the evidence supports a dipole dominance for at least the last 3 billion years.

In contrast with geomagnetic field knowledge, for other planetary bodies we only have crude measurements for the current configuration, and no information about their secular variation (besides the mentioned residual magnetization of Martian and Moon rocks). The current data show an impressive variety of magnetic field geometries and strengths, with the Jovian field being the most similar to the Earth's, but with a factor of $\sim 20$ stronger. Surface rock magnetization has been detected on Mars and the Moon, proving past dynamos which have since switched off. There-fore, the only solar planet with no evidence for current or past dynamo is Venus. This diversity in strength and morphology is still not fully understood, but several numerical simulations show possible systematic dependence on parameters such as convective energy flux, rotation rate, and internal structural geometries.

Mercury's magnetic field was unexpectedly detected by the Mariner 10 flyby in 1975 and confirmed by *MESSENGER* in 2008. It was thought that such a small planet would not have any internal activity left. It consists primarily of a dipole with a very small tilt angle and a significant offset of the magnetic equator northward. With an average surface field strength of only 0.003 G, it is much smaller in comparison to the other planetary fields, which still lacks a satisfactory explanation.

As mentioned above, the Jovian magnetic field was first detected by decameter, coherent radio emission in 1955. These are generated by energetic electrons that gyrate around magnetic field lines close to Jupiter's surface (Barrow and Carr, 1992). The physical mechanism is known as electron cyclotron maser instability (ECMI), and it is observed in several solar system planets, the Sun, M dwarfs, and brown dwarfs. It has now been proposed to be present in exoplanetary systems (see Sect. 6.4 for hot Jupiter estimates). Jupiter's magnetic field is about ten times stronger than the geomagnetic field, but shares a similar morphology with a $\sim 10^\circ$ dipole tilt and similar low multipole ($\ell = 2, \ell = 3$) weight ratios. An interesting feature is the equatorial magnetic south pole, which, inspired by the Great Red Spot atmospheric vortex, is known as the Great Blue Spot. Note that the blue color only comes from the arbitrary choice for magnetic field maps (i.e., red for north and blue for south) and has nothing to do with the cyclonic 400-year-old storm visible as a large red spot. The two features lie at different latitudes, and while the Great Red Spot shows a westward drift ($\sim 0.3^\circ$ to $0.6^\circ$ per day) primarily staying on a specific longitude, the Great Blue Spot does not show any changes in atmospheric timescales. In fact, among different *Juno* magnetic field models from different orbits, the Great Blue Spot has been found to drift eastward ($\sim 2^\circ$ in 1.75 years), which accounts for the only secular variation estimate in planetary magnetic fields, other than the geomagnetic field (Connerney et al., 2022).

Saturn's magnetic field is characterized by an almost zero dipole tilt, within the uncertainty error. Its strength is slightly weaker than Earth's at the surface. Measurements by passing spacecraft and the *Cassini* orbiter are fitted with models composed of only the dipole and zonal quadrupole and octupole multipoles. This fact seems in contradiction with Cowling's theorem, which states that an axisymmetric magnetic field cannot be generated by dynamo action. Still, more precise and better-mapped future measurements could better reveal the weak deviation from axial symmetry.

The most poorly measured planetary magnetic fields in the solar magnetized planets are, by far, Uranus and Neptune. Both magnetic field strengths are similar to each other yet distinct from all other planets: the surface field strength is comparable to that of Earth's, yet the geometric properties can be defined as the only multipolar dynamos in the Solar System. They are dominated by a strongly inclined dipole axis and quadrupole and octupole contributions comparable to the dipole to the dipole at the surface (which means that at the dynamo radius the quadrupole and octupole fields are stronger than the dipole, see Sect. 4.3.2). Their magnetic and gravity fields were characterized by a single *Voyager 2* flyby in the 1980s, which has concerning uncertainties. The main reason for the limited data is that spacecraft missions to the solar system's outer edge are expensive and lengthy, taking more than a decade to reach their targets. *The Uranus Orbiter and Probe* is a NASA orbiter mission concept that aims to study Uranus and its major moons with several flybys. It has the highest priority Flagship-class mission with an estimated launch in the mid to late 2030s. Similarly, *Neptune Odyssey* is a proposed mission but remains currently unscheduled due to logistical and cost reasons.

Ganymede is the only known satellite in the solar system with an internally generated magnetic field. Flyby measurements from the *Galileo* and *Juno* spacecraft have detected a dipole-dominated field with a surface strength of approximately 0.01 Gauss. Note that all Jovian moons experience weak induced magnetic fields as a result of their motion through Jupiter's powerful magnetosphere. These induced fields arise from electric currents generated within the moons' electrically conductive interiors. The strength of an induced field is typically comparable to that of Jupiter's magnetic field at the satellite's orbital distance. In Ganymede's case, this is about 0.0012 Gauss, significantly weaker than its intrinsic field. Therefore, although Ganymede's magnetic field includes an induced component, it represents only a small fraction of its overall magnetic strength.

The Moon and Mars (Acuña et al., 1992; Connerney et al., 1999; Mittelholz and Johnson, 2022) currently do not possess intrinsic magnetic fields. However, localized magnetic fields of crustal origin are found throughout the surface of both objects. For the Moon, the strengths reach $10^{-4}$ G at the lunar surface, while the martian values reach the order of $10^{-2}$ G, significantly stronger than Earth's crustal field. The most plausible explanations for this magnetization are ancient early lunar and martian dynamos. Estimates for the cessation of internal dynamos are based on the presence or absence of magnetization around impact basins. For the Moon, it is estimated to be about 3.3 Gyr ago, while for Mars, it is about 4.1 Gyr, coinciding with the late heavy bombardment period.

Similarly, Venus lacks a global magnetic field (Nimmo, 2002). However, unlike the Moon and Mars, it shows no evidence of small-scale remanent magnetization in its crustal rocks. This absence may be due to Venus's extremely high surface temperature at around 735 K. This value approaches or exceeds the Curie point of most ferromagnetic minerals, preventing them from retaining magnetic information. This agrees with some geological studies that suggest a planet-wide resurfacing event about 500 million years ago, erasing any possible magnetic crustal signatures. Consequently, the hypothesis of an ancient internal dynamo for Venus remains an open and unresolved question.

## 4.3 Planetary magnetic field formalism

Experimental magnetic field measurements for direction and strength are taken in situ from space missions, along a certain number of orbits, which unavoidably represent a tiny fraction of the magnetospheric volume. Therefore, the collected data is used to reconstruct planetary magnetic field models, and the accuracy depends on the spatial coverage of the measurements. These models assume that the field is observed in a region with no relevant currents (i.e., $\nabla \times \mathbf{B} = \mathbf{J} = 0$), thus the field can be expressed as the gradient of a scalar potential function, $V_{int}$, which by definition assumes an internal generation of magnetic field. Therefore, the magnetic field is the gradient of the full scalar potential:

$$\boldsymbol{B} = -\boldsymbol{\nabla} V, \tag{4.1}$$

For almost-spherical objects, it is natural to give these potentials as expansions of spherical harmonic functions that solve Laplace's equation $\nabla^2 V = 0$. In planetary sciences, these potentials are expanded in the Schmidt quasi-normalization (see App. C for a full review):

$$V_{\text{int}} = a \sum_{n=1}^{n_{\max}} \left(\frac{a}{r}\right)^{n+1} \sum_{m=0}^{n} P_n^m(\cos\theta) \left[g_n^m \cos(m\phi) + h_n^m \sin(m\phi)\right] , \tag{4.2}$$

where a is the planetary equatorial radius, $r$ is the radial distance to the planet's center, and the angles $\theta$ and $\phi$ are colatitude and longitude, respectively. The $P_m^n(\cos\phi)$ are Schmidt quasi-normalized associated Legendre functions of degree n and order m, and $g_n^m$, $h_n^m$ are the Schmidt coefficients that parameterize the internal and external magnetic field components, respectively[1]. Some models, such as the ones for Mercury and Jupiter, take into account the magnetospheric currents' contributions, which are related to the tenuous plasma disks around the planets and are usually second-order compared to the dynamo contribution. This is done including an additional scalar potential $V_{ext}$ with their own set of coefficients $G_n^m$ and $H_n^m$. Therefore the total magnetic field is reconstructed with $\boldsymbol{B} = -\boldsymbol{\nabla}(V_{\rm int} + V_{\rm ext})$. With the analytical expressions shown in App. C one can obtain all components of $\boldsymbol{B}$ in spherical coordinates at any given radius.

The latest publications of magnetic field models for each planet are shown in Table 4.1. The public data is always given in terms of the Schmidt coefficients, i.e., the set of constants $g_n^m$, $h_n^m$, $G_n^m$, and $H_n^m$ used in the spherical harmonic expansions. The specific multipole degree that is reached depends on the data available for each planet: the better the coverage, the higher the number of multipole contributions that can be constrained. Together with Eq. 4.1, this data has been used to produce Figs. 4.3 and 4.4. The repository used to produce these figures has been made public[2].

Table 4.1: Missions and publications dedicated to each planetary magnetic field, with the number of reconstructed multipole contributions. For Mercury and Jupiter, the external multipole contribution of $n_{\rm max} = 1$ is marked in parentheses.

| Planet | Multipoles | Mission | Publication |
|---|---|---|---|
| Mercury | 3 (1) | *MESSENGER* | Toepfer et al. (2021b)<br>Toepfer et al. (2021a) |
| Earth | 13 | *IGRF* | New model every 5 years (Alken et al. (2021)) |
| Jupiter | 18 (1) | *Juno* | Connerney et al. (2018)<br>Connerney et al. (2022) |
| Ganymede | 2 | *Juno*, *Galileo* | Weber et al. (2022) |
| Saturn | 6 | *Cassini* | Cao et al. (2023) |
| Uranus | 3 | *Voyager 2* | Connerney et al. (1987)<br>Ness et al. (1989) |
| Neptune | 3 | *Voyager 2* | Connerney et al. (1991)<br>Selesnick (1992) |

The geomagnetic field is the most measured and well-monitored planetary magnetic field. Direct data from the past few hundred years is available, and there are currently 150 geomagnetic observatories worldwide and low Earth orbit satellite systems that complete the coverage. This leads to a significant improvement in time and spatial resolution. The accepted empirical model for Earth's magnetic field is the International Geomagnetic Reference Field (*IGRF*), released by the International Association of Geomagnetism and Aeronomy (IAGA). The *IGRF* is essentially this set of coefficients, $g_n^m$ and $h_n^m$, up to coefficient 13, as well as their secular variations, which track the change of these coefficients per year. This model is updated every 5 years, the latest of which is currently the 14$^{\rm th}$ generation, published in December 2024 and is still waiting for scientific approval. The latest peer-reviewed model is the 13$^{\rm th}$ generation, published in Alken et al. (2021)

[1]This expansion is similar but not equal to the spherical harmonic expansion used by *MagIC* (see Appendix D). Moreover, the notation used in this chapter is consistent with the planetary literature; thus, I use $n$ and $m$ as the degree and order, instead of the $\ell$ and $m$ commonly used in spherical harmonics, and employed in the Chapters related to *MagIC*.

[2]https://github.com/csic-ice-imagine/magnetic_field_planets

and shown in Fig. 4.3. Magnetic field data is precise enough to derive a magnetic field model to more than 700 multipoles Maus (2008), but at multipoles higher than 13, white spectral noise due to crustal rock magnetization dominates. In other words, at small scales, the magnetic field intensity variations are dominated by the scattered distribution of ferromagnetic sources.

The Jovian magnetic field has now been accurately measured by the *Juno* spacecraft and published in 2018 (Connerney et al., 2018) from the first nine orbits and refined in 2021 after completion of the mission at 33 orbits (Connerney et al., 2022). The 2018 model, JRM09, has a maximum degree $n_{\max}$ of 10 for the internal component. On the other hand, the 2021 model, JRM33, includes data for the first 33 orbits and has reasonably well resolved coefficients up to $n_{\max} = 13$ (with some useful information up to $n_{\max} = 18$) for the internal component, and $n_{\max} = 2$ for the external component. Thus, the JRM33 model reached a remarkable milestone: Jupiter's dynamo is better resolved than Earth's dynamo, due to the absence of the magnetized rocks on Jupiter's surface, arguably making it the only property better known for a gas giant planet than Earth.

A point to stress further is that the 3D data measured by orbiters is limited. The obtained models can only reliably reconstruct the large-scale structures and have to omit the smaller ones. In other words, the number of multipoles for the magnetic potential expansions is chosen depending on the quantity and quality of data. For example, after 9 *Juno* orbits, the model JRM09 could only include multipoles up to degree 10 before overfitting. And after 33 orbits, the degree of multipoles for the JRM33 model could not be increased further than 18, as the coefficients for higher multipoles have high deviations and no further fitting is obtained. Therefore, the reconstructed configuration can look much simpler than in reality if the data are limited.

### 4.3.1 Magnetic spectrum

Similar to a fluid simulation spectrum (see Sec. A.9), one can calculate the power spectrum from the spherical harmonic model of the internal magnetic field of a given planet. This power spectrum was first defined by Mauersberger (1956) and Lowes (1974) in the context of the geomagnetic field, but is now applied to all planetary bodies. When a planet's magnetic field potential has been expressed as above, the power spectrum at its surface is defined as:

$$R_n = (n+1) \sum_{m=0}^{n} \left[ (g_n^m)^2 + (h_n^m)^2 \right]$$

For any other radii, each degree $n$ has a different contribution as they decay differently with distance $r$:

$$R_n(r) = \left(\frac{a}{r}\right)^{2n+4} R_n = (n+1) \left(\frac{a}{r}\right)^{2n+4} \sum_{m=0}^{n} \left[ (g_n^m)^2 + (h_n^m)^2 \right]$$

This is known as the Lowes-Mauersberger magnetic power spectrum, or simply the Lowes spectrum. One can prove that the total magnetic energy can be recovered:

$$2\mu_0 E_B(r) = \sum_{n=0}^{\infty} R_n(r)$$

### 4.3.2 Downward extrapolation

The scalar potential magnetic expansion shown in Eq. 4.2 is valid only in the current-free regions. Taking advantage of this fact, one can use this expansion as well as the Lowes spectra not only at the planetary surface $a$, but at other radii $r$. In other words, one can extrapolate below the planetary surface as long as there are no relevant currents, which means as low as the layers are insulating. For Earth and Jupiter, the usual distribution of $R_n$ as a function of $n$ is a falling exponential with a slope that depends on the radius, as seen in the top two panels of Fig. 4.4.

By construction of the potential, the higher the multipole, the higher its coefficient becomes when you downward extrapolate. At some radius, the downward extrapolation leads to an almost flat spectrum except for the multipole (for Earth) or other low-degree multipoles (for Jupiter). A common assumption is that the magnetic spectra will be flat at the dynamo surface except for the dipole contribution (Lowes, 1974). Based on the stochastic nature of MHD processes, the magnetic spectra inside the dynamo are expected to be flat within a specific range. This is backed up by numerical simulations, where equipartition between multipoles usually holds between low multipoles ($n \gtrsim 5$) and the resistive scale, where magnetic energy is dissipated. The lowest multipoles (the dipole $n = 1$ especially) do not usually follow the slope of the spectrum because they are influenced by large-scale flows caused by rotation. The dipole component of the magnetic field is enhanced in the rotational axis direction, breaking the equipartition.

In Fig. 4.4, I show the downward extrapolation procedure for Earth and Jupiter, starting at radii well above the planetary radius and finishing at the planetary depth where the spectra is approximately constant. From the Lowes spectra, we can infer that the geodynamo starts at about $\sim$ 55% of Earth's radius, roughly where the liquid outer core of the Earth begins. For Jupiter, it corresponds to $\sim$ 80% of the Jovian radius, which is also roughly where interior models predicted the appearance of metallic hydrogen (French et al., 2012; Bonitz et al., 2024). Note that the bottom panels in Fig. 4.4 describe the field at $r = 0.75\,R_J$, which is possibly a region where the dynamo is already active, so that currents circulate, and the potential field assumption doesn't hold. Therefore, it is shown only as an exercise in extrapolation, to show how small-scale structures dominate more and more, rather than a realistic description of the field at those depths.

Additionally, note that, even in the current-free region, since high multipoles acquire more and more importance inwards, the uncertainties in the reconstruction, which affect the small scales more, increase. The mean magnetic field strength and the flow components at the top of the dynamo region are usually determined using this procedure for downward continuation. As there might be unresolved toroidal field components, the derived magnetic field strength is only a lower limit.

This procedure was also used in gas-giant dynamo simulations by Tsang and Jones (2020). They used a hyperbolic conductivity profile $\sigma(r)$ that qualitatively reproduces the huge conductivity radial gradient in correspondence to the quite sharp transition to metallic hydrogen, at some specific planetary depth. Another common analytical function is a polynomial followed by an exponential drop (e.g. Gómez-Pérez et al., 2010) which is employed in Chapters 5 and 7. In their work, they used the most external spectra distribution of their simulations and performed a downward extrapolation. They recover the surface at which the actual magnetic field spectra differ from the extrapolated one, corresponding to the depth at which the fluid is conductive. They justify that this is a better dynamo surface definition, as they note that the usual definition of a flat spectrum usually underestimates the dynamo radius by a few percent. In Sec 5.2.5, I repeat this procedure for my steady state dynamo solutions to argue our definition of dynamo surface.

## 4.4 Exoplanetary magnetism

There are more than 5,700 confirmed exoplanets. Fig. 4.5 is the current exoplanet population, a readily made plot by the NASA exoplanet archive with the latest data. This extensive population has shed light on various aspects of exoplanetary science, such as their formation, structure, at-mosphere, and population. The planetary classification between rocky, ice-giants, and gas giants has become more complex as the full spectrum of planetary mass, radii, and orbital distances has been populated. Despite having thousands of confirmed exoplanets, their magnetism is still largely unknown. Knowledge about the presence or lack of internal dynamos has many consequences in planetary formation, internal structure, evolution, and habitability. Due to the lack of observational constraints, dynamo models for exoplanets are much less developed compared to those for solar system magnetic fields. The most direct method to infer exoplanetary magnetic fields would be the detection of Jovian-like, coherent, circularly polarized radio emission at sub-GHz. There are no confirmed cases yet of radio detections; see, for instance, the tentative claim by Turner et al. (2021), which was not confirmed by follow-up observations by Turner et al. (2024). Other approaches use the detection of modulations of activity indicators, CaII line in particular, with the orbital period of hot Jupiters, to infer planetary magnetic field values, via magnetic star-planet interaction models (Cauley et al., 2019). However, the inferred values are model-dependent (for instance, due to the unknown efficiency in converting the star-planet interaction energy burst into the power of the observed line). Therefore, the magnetic fields in exoplanets are substantially unconstrained, and one is forced to use the solar planets on one side, and the brown dwarfs and low-mass stars on the other, to derive phenomenological scaling laws Christensen and Aubert (2006); Christensen et al. (2009).

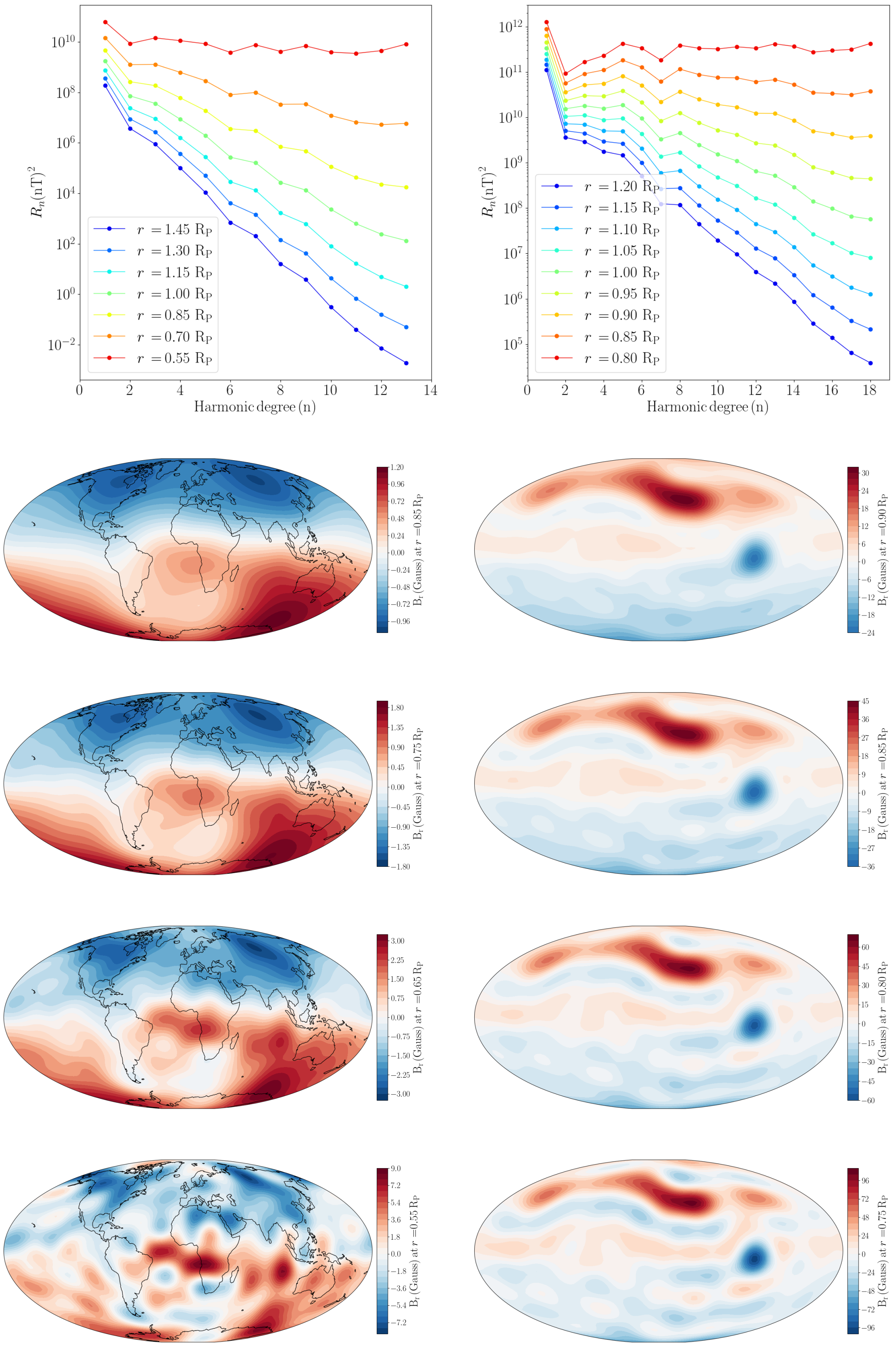

Figure 4.4: Top: Lowes-Mauesberg magnetic spectra for Earth (left) and Jupiter (right) at different distances from the center. Bottom: Radial magnetic field obtained from downward continuation for Earth (right) and Jupiter (right panels).

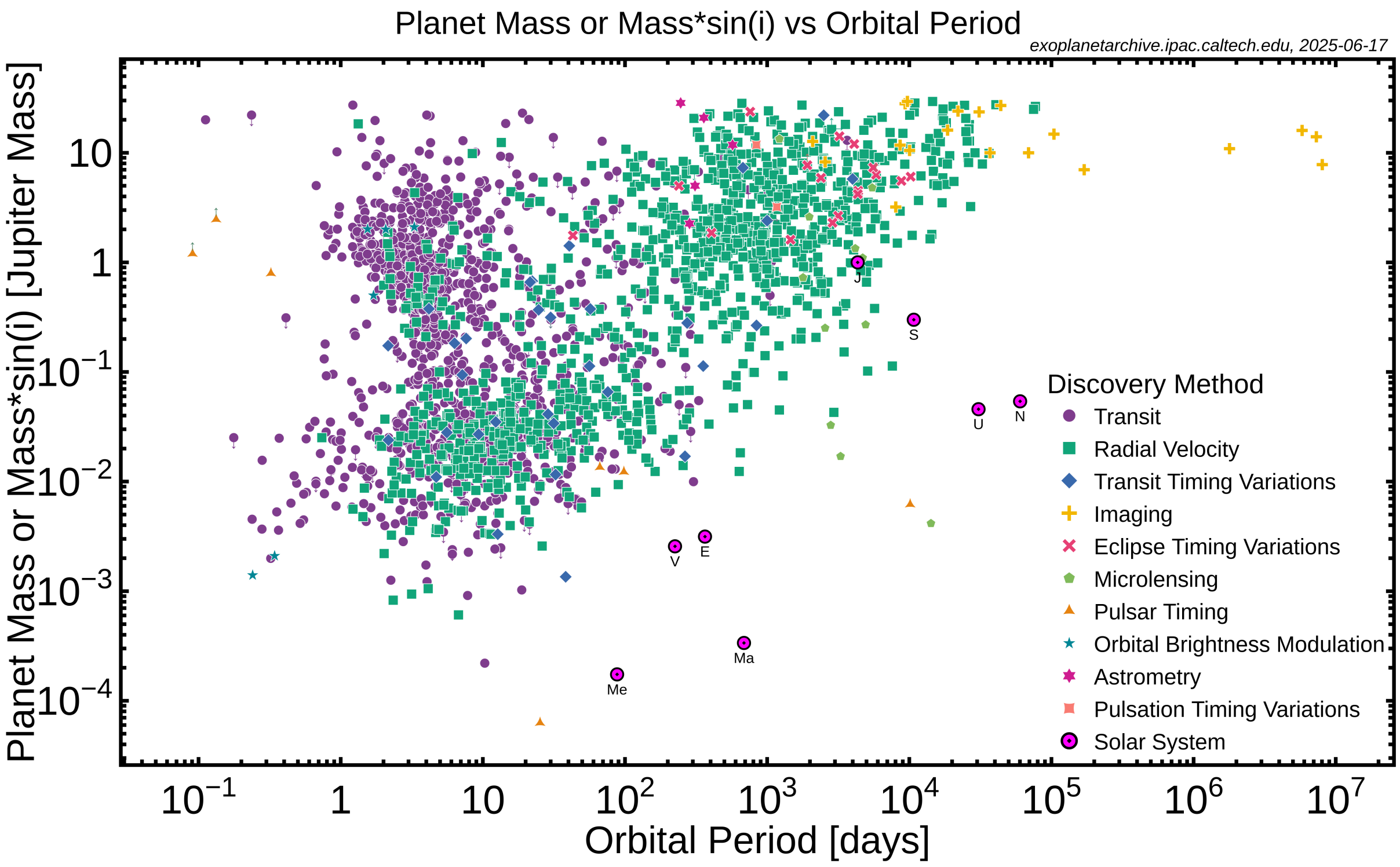

Figure 4.5: Taken from pre-generated plots from NASA exoplanet archive https://exoplanetarchive.ipac.caltech.edu/exoplanetplots/. Confirmed exoplanet population (as of June 2025) distribution in planet mass versus orbital period, color-coded by discovery method.

In analogy with the solar system and with basic energetic arguments, gas giants are the exoplanetary category with the highest magnetic field strength prediction, and therefore with the strongest observational cases. In Chapter 5, I use 3D MHD spherical shell simulations to model the Gyr evolution of the internal dynamos hosted by the cool down of a gas giant, which can be applied to cold to mildly irradiated Jupiter-like analogs. I will also focus on hot Jupiters, short-orbit giants highly irradiated by their host stars to which they are tidally locked, with masses ranging from 0.3 to 13 $M_J$. An introduction about hot Jupiters and their properties can be found in Chapter 6, where I model their interiors and infer their internal convective regime and magnetic properties.

## 4.5 MHD for convection in rotating spherical shell

Dynamo action is the most accepted mechanism for the current existence of planetary magnetism. To model such processes, the MHD equations shown in Chapter 2 need to be adapted to the convective and rotating interior of planets. The most important changes are the force terms in the momentum equations:

$$\rho\left(\frac{\partial \mathbf{u}}{\partial t}+\mathbf{u}\cdot\boldsymbol{\nabla}\;\mathbf{u}\right)=-\boldsymbol{\nabla}p+\frac{1}{\mu_0}(\boldsymbol{\nabla}\times\mathbf{B})\times\mathbf{B}+\rho\mathbf{g}-2\rho\boldsymbol{\Omega}\times\mathbf{u}+\rho\boldsymbol{\nabla}\cdot(2\rho\nu\mathbf{S}), \tag{4.3}$$

where $\boldsymbol{\Omega}$ is the angular velocity (already introduced in Chapter 3, and $\mathbf{g}$ is the gravity. The only differences with Eq. 2.2 are the rotation and the buoyant terms; all other quantities, including $\mathsf{S}$, are the same. Additionally, for geodynamo simulations, it is very common for chemical composition, $\xi$, to be considered. Bouyancy due to chemical gradients occurs in the Earth's molten outer core when the metallic mixture freezes only the heavy elements (Fe, Ni) onto the solid inner core and the release of the light constituent into the fluid outer core, possibly oxygen, silicon, or magnesium. The evolution of $\xi$ is then:

$$\rho\left(\frac{\partial \xi}{\partial t}+\mathbf{u}\cdot\boldsymbol{\nabla}\xi\right)=\boldsymbol{\nabla}\cdot(k_\xi\boldsymbol{\nabla}\xi)+\epsilon_\xi, \tag{4.4}$$

When considering chemical composition, there is an equivalent Pr, known as the Schmidt number, $\mathrm{Sc}=\nu/\kappa_{xi}$, where instead of the thermal diffusivity $\kappa$, the chemical or mass diffusivity $\kappa_{xi}$ is used. The spherical shell dynamos in this thesis model gas giant convection, where chemical buoyancy is believed to play a minimal role. Thus, Eq. 4.4 is not included in this thesis. The continuity, entropy, and induction equations are not shown to avoid repetition, as they are the same as Eq. 2.1 and 2.3. Given the initial conditions, this set of equations also makes a complete set of partial differential equations that, together with an equation of state, describe the evolution of the system.

Another peculiarity of planetary or stellar convective numerical dynamos is the geometry. The slightly modified MHD equations are usually solved in a spherical shell or, sometimes, if required, in a full sphere domain. In Fig. 4.6, I show the spherical shell and notation used by *MagIC*, which is consistently used in Chapters 5 and 7.

### 4.5.1 Non-dimensional numbers in convective rotating spherical shells

As mentioned in section 2.2, there are several nondimensional parameters in fluid dynamics. Here we focus on some of them that are relevant for stellar and planetary dynamos. Table 4.2 is an extension of Table 2.1, with definitions for the specific case of fluid convection in rotating spherical shells. They represent the ratio of different terms in the MHD equations, including buoyant and Coriolis terms. Therefore, each non-dimensional number defines different fluid regimes depending on how far they are from unity.

The Ekman number, E, is the quotient between the viscous and Coriolis terms. For planetary interior convection is usually $\mathrm{E} \ll 1$ (see Table 4.3), which implies that Coriolis forces dominate over viscous forces. This leads to strong rotational constraints, i.e., convective patterns tend to align with the rotation axis and form columnar structures (see Sect. 4.6.1).

The Rayleigh number, Ra, is the ratio between buoyant and diffusive forces. For a given planetary shell, there is always a critical value for the onset of convection, $\mathrm{Ra}_{\mathrm{crit}}$ (see also below).

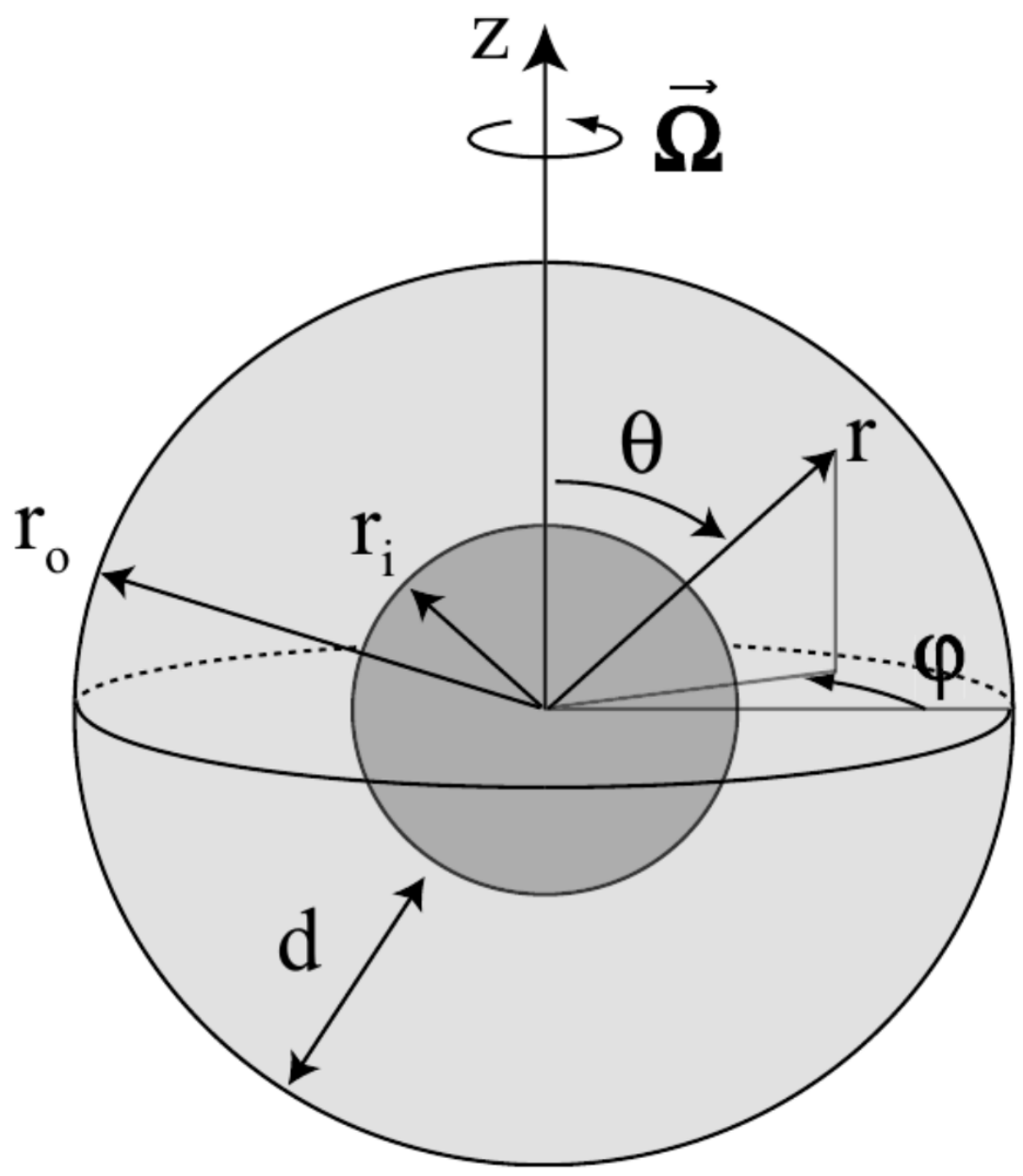

Figure 4.6: Taken from https://magic-sph.github.io/. Sketch of the spherical shell and its system of coordinates used by *MagIC*.

For planets and stars, one can estimate Ra $\sim 10^{20} - 10^{30}$, which is some orders of magnitude above their critical values. This means that thermal convection is extremely vigorous, leading to highly nonlinear and turbulent flows. There is a secondary critical value, above which dynamo action is triggered, given that a compatible set of E, Pr, and Pm (see Sect. 4.6).

Another crucial parameter is the Rossby number, Ro. It is the quotient of inertial and Coriolis forces. It thus can be understood as the ratio of the rotation period and the convective timescale, that is, the turnover time in the dynamo region. Similarly as E $\ll 1$, when Ro $\ll 1$, fluid movements are constrained by rotation and lead to columnar structures in the same direction as the rotation axis, and contrarily, when Ro $\gg 1$, convection is barely influenced by rotation.

The Elsasser number is the ratio between magnetic and Coriolis forces: when it is very small ($\Lambda \ll 1$), $\boldsymbol{B}$ is dynamically weak compared to the rotation, which makes its configuration follow the columnar flows. Instead, when $\Lambda \gg 1$, the magnetic field dominates fluid motion over rotation, breaking the axial symmetry of the flow.

In Table 4.3, some of the dimensionless parameters for Earth, Jupiter, and the Sun (or at least their order of magnitude) are shown. Earth's values have been taken from Schaeffer et al. (2017), Jupiter's from Gastine and Wicht (2021), and the Sun from Miesch (2005); Hanasoge et al. (2016); Pandey et al. (2021). Some assumptions are made to estimate most of these non-dimensional parameters. The reason is that many fluid properties, such as the flow speeds, the exact location of the convective shell, or the magnetic field strength, are poorly constrained. In spherical dynamo literature, it is common to use $r_o$ and $r_i$ for the outer and inner radius of the convective shell, respectively. Then, the usual parameter $d = r_o$ - $r_i$ is the thickness of the dynamo region, which becomes the reference length-scale ($L$ in Chapter 2), employed to evaluate the quantities in Table 4.2. Moreover, Re and Rm for planetary and stellar interiors are also usually defined with $d$. Therefore, the following dimensional transformations are performed:

$$\begin{aligned} &r \rightarrow r\, d, \quad t \rightarrow t\, (d^2/\nu), \quad s \rightarrow s\, \Delta s, \quad \xi \rightarrow \xi\, \Delta\xi, \\ &B \rightarrow B\, (\mu_0 f \lambda_i \tilde{\rho}_o \Omega)^{1/2}, \quad \rho \rightarrow \rho\, \rho_o, \quad p \rightarrow p\, (\rho_o \nu_o \Omega) \end{aligned} \tag{4.5}$$

where the sub-indexes $_o$ and $_i$ refer to the outer or inner boundaries of the shell. Repeating the

Table 4.2: Additional non-dimensional parameters usually used in stellar and planetary dynamo literature, besides the ones already shown in Table 2.1.

| Parameter | Definition | Physical interpretation |
|---|---|---|
| Ekman number | $\mathrm{E} = \frac{\nu}{\Omega d^2}$ | $\frac{\text{Viscous force}}{\text{Coriolis force}}$ |
| Rayleigh number | $\mathrm{Ra} = \frac{\alpha g_0 \Delta T d^3}{\nu \kappa}$ | $\frac{\text{Buoyancy}}{\text{Diffusion}}$ |
| Rossby number | $\mathrm{Ro} = \frac{U}{\Omega d}$ | $\frac{\text{Advection}}{\text{Coriolis forces}}$ |
| Elsasser number | $\Lambda = \frac{B^2}{\rho \mu \lambda \Omega}$ | $\frac{\text{Lorentz}}{\text{Coriolis forces}}$ |
| Dynamo region depth | $\frac{r_o}{r_P}$ | |
| Dynamo region geometry | $\eta = \frac{r_i}{r_o}$ | |

Table 4.3: Non-dimensional parameters estimates for the most characterized dynamos in the Solar System, as well as the typical values obtained with simulations.

| Parameter | Earth | Jupiter | Sun | Models |
|---|---|---|---|---|
| Ra | $\sim 10^6\, Ra_{crit}$ | $10^{31}$ | $10^{22}$-$10^{23}$ | 1-50 $\mathrm{Ra}_{crit}$ |
| E | $3 \cdot 10^{-15}$ | $10^{-18}$ | $10^{-12}$-$10^{-15}$ | $10^{-4}$-$10^{-10}$ |
| Pr | 0.1 - 10 | $10^{-2}$-1 | $\lesssim 10^{-6}$ | $10^{-2}$-$10^{3}$ |
| Pm | $2 \cdot 10^{-6}$ | $10^{-6}$ | $10^{-6}$ | $10^{-1}$-$10^{3}$ |
| Re | $10^9$ | $10^{12}$ | $10^9$ | $<10^4$ |
| Rm | 2000 | $10^6$ | $10^6$ | 50-$10^4$ |
| Ro | $3 \cdot 10^{-6}$ | $10^{-6}$ | 0.1 - 1 | $10^{-1}$-$10^{-4}$ |
| $\Lambda$ | $\lesssim 10$ | 10 - 100 | $\lesssim 1$ | 0.1 - 100 |

same procedure done in Sect. 2.2.1 for the induction equation, the set of MHD planetary equations becomes:

$$\boldsymbol{\nabla} \cdot (\rho \mathbf{u}) = 0, \tag{4.6}$$

$$\rho \left( \frac{\partial \mathbf{u}}{\partial t} + \mathbf{u} \cdot \boldsymbol{\nabla}\, \mathbf{u} \right) = -\boldsymbol{\nabla} p + \frac{1}{\mathrm{PmE}} (\boldsymbol{\nabla} \times \mathbf{B}) \times \mathbf{B} + \frac{\mathrm{Ra}}{\mathrm{Pr}} \rho g s \mathbf{e_r} + \frac{\mathrm{Ra}_\xi}{\mathrm{Sc}} \rho g \xi \mathbf{e_r} - \frac{2}{\mathrm{E}} \rho\, \mathbf{e_z} \times \mathbf{u} + \boldsymbol{\nabla} \cdot \mathsf{S}, \tag{4.7}$$

$$\rho T \left( \frac{\partial s}{\partial t} + \mathbf{u} \cdot \boldsymbol{\nabla} s \right) = \frac{1}{\mathrm{Pr}} \boldsymbol{\nabla} \cdot (\kappa_{\mathrm{norm}} \boldsymbol{\nabla} T) + \frac{\mathrm{Pr\,Di}}{\mathrm{Ra}} \Phi_\nu + \frac{\mathrm{Pr\,Di}}{\mathrm{Pm}^2\, \mathrm{E\,Ra}} \frac{\lambda_{\mathrm{norm}}}{\mu_0} (\boldsymbol{\nabla} \times \mathbf{B})^2 \tag{4.8}$$

$$\rho \left( \frac{\partial \xi}{\partial t} + \mathbf{u} \cdot \boldsymbol{\nabla} \xi \right) = \frac{1}{\mathrm{Sc}} \boldsymbol{\nabla} \cdot (\kappa_{\xi,\mathrm{norm}} \rho \boldsymbol{\nabla} \xi) \tag{4.9}$$

$$\frac{\partial \mathbf{B}}{\partial t} = \boldsymbol{\nabla} \times (\mathbf{u} \times \mathbf{B}) - \frac{1}{\mathrm{Pm}} \boldsymbol{\nabla} \times (\lambda_{\mathrm{norm}} \boldsymbol{\nabla} \times \mathbf{B}), \qquad \boldsymbol{\nabla} \cdot \mathbf{B} = 0, \tag{4.10}$$

where $\lambda_{\mathrm{norm}}$, $\kappa_{\mathrm{norm}}$, $\kappa_{\xi,\mathrm{norm}}$ are the magnetic, thermal, and chemical diffusivity normalized at their value at one of the radial boundaries ($r_i$ for $\lambda$ and $r_o$ for $\kappa$ and $\kappa_\xi$). All the tildes denoting the new dimensionless variables were already dropped[3]. The dimensionless parameters used in *MagIC* and the next chapters are:

$$\mathrm{E} = \frac{\nu_o}{\Omega d^2}, \quad \mathrm{Pr} = \frac{\nu_o}{\kappa_o}, \quad \mathrm{Pm} = \frac{\nu_o}{\lambda_i}, \quad \mathrm{Ra} = \frac{\alpha_o g_o T_o d^3 \Delta s}{c_p \kappa_o \nu_o}, \\ \mathrm{Sc} = \frac{\nu_o}{\kappa_\xi}, \quad \eta = \frac{r_i}{r_o}, \quad N_\rho = \ln\frac{\rho_i}{\rho_o} \tag{4.11}$$

### 4.5.2 The anelastic and Boussinesq approximations

If one were to evolve the dimensionless MHD equations numerically, most of the computing power would go into resolving different perturbations of the magnetic and velocity fields. In MHD, these wave modes are known as Alfvén waves and fast and slow magnetoacoustic waves, which are a whole field of research on their own. These phenomena have timescales much shorter than convection timescales, and due to their perturbative nature, they have little or no effect on the convection-driven planetary dynamos. In the context of buoyancy-driven flows, two different approximations restrict or filter the waves: the Boussinesq and the anelastic approximation.

The less restrictive one is the anelastic approximation. It assumes a radially dependent static thermodynamic background where small perturbations can develop. Following the first derivation for the anelastic approximation for convective shells (Gilman and Glatzmaier, 1981), we assume that a reference static state and perturbations can separate all thermodynamic variables: $p = \tilde{p} + p'$, $\rho = \tilde{\rho} + \rho'$, $T = \tilde{T} + T'$, $s = \tilde{s} + s'$ and $g = \tilde{g} + g'$. Their relative departures are very small compared to the background state:

$$\epsilon = \frac{|p'|}{\tilde{p}} = \frac{|\rho'|}{\tilde{\rho}} = \frac{|T'|}{\tilde{T}} = \frac{|s'|}{\tilde{s}} = \frac{|g'|}{\tilde{g}} \ll 1 \ .$$

The overall contributions for $\mathbf{u}$ and $\mathbf{B}$ are neglected at $0^{\text{th}}$ order (i.e., $\tilde{\mathbf{u}} = \tilde{\mathbf{B}} = 0$), thus they only influence the $1^{\text{st}}$ order terms, where the order refers to how many factors of $\epsilon$ each term has. When one performs these substitutions into Eqs. (4.6)-(4.10) one recovers both the hydrostatic equilibrium from the $0^{\text{th}}$ order terms contributions ($\boldsymbol{\nabla}\tilde{p} \propto \tilde{\rho}\tilde{g}$), and the complete set of the anelastic equations from the $1^{\text{th}}$ order terms:

$$\boldsymbol{\nabla} \cdot (\tilde{\rho}\mathbf{u}) = 0, \tag{4.12}$$

$$\left(\frac{\partial \mathbf{u}}{\partial t} + \mathbf{u} \cdot \boldsymbol{\nabla}\mathbf{u}\right) = -\frac{1}{\mathrm{E}}\boldsymbol{\nabla}\left(\frac{p'}{\tilde{\rho}}\right) - \frac{2}{\mathrm{E}}\mathbf{e_z} \times \mathbf{u} + \\ + \frac{\mathrm{Ra}}{\mathrm{Pr}}\tilde{g}\, s'\, \mathbf{e_r} + \frac{1}{\mathrm{Pm}\,\mathrm{E}\,\tilde{\rho}}(\boldsymbol{\nabla} \times \mathbf{B}) \times \mathbf{B} + \frac{1}{\tilde{\rho}}\boldsymbol{\nabla} \cdot \mathsf{S}, \tag{4.13}$$

$$\tilde{\rho}\tilde{T}\left(\frac{\partial s'}{\partial t} + \mathbf{u} \cdot \boldsymbol{\nabla}s'\right) = \frac{1}{\mathrm{Pr}}\boldsymbol{\nabla} \cdot \left(\kappa_{\text{norm}}\tilde{\rho}\tilde{T}\boldsymbol{\nabla}s'\right) + \frac{\mathrm{Pr}\,\mathrm{Di}}{\mathrm{Ra}}\Phi_\nu + \frac{\mathrm{Pr}\,\mathrm{Di}}{\mathrm{Pm}^2\,\mathrm{E}\,\mathrm{Ra}}\lambda_{\text{norm}}\left(\boldsymbol{\nabla} \times \mathbf{B}\right)^2 \ , \tag{4.14}$$

$$\frac{\partial \mathbf{B}}{\partial t} = \boldsymbol{\nabla} \times (\mathbf{u} \times \mathbf{B}) - \frac{1}{\mathrm{Pm}}\boldsymbol{\nabla} \times (\lambda_{\text{norm}}\boldsymbol{\nabla} \times \mathbf{B}), \qquad \boldsymbol{\nabla} \cdot \mathbf{B} = 0 \ . \tag{4.15}$$

The anelastic approximation is typically used for modeling the density-stratified low-Mach-number convection flows in gas giants and stars (Braginsky and Roberts, 1995; Lantz and Fan, 1999), where there are strong $\rho$, $T$, $P$ radial dependencies. I use the anelastic approximation in Chapters 5 and 7 to model dynamo processes in gas giants. Note that, for simplicity, I have already suppressed the chemical buoyancy term and the chemical composition equation in Eqs. (4.12)-(4.15), as they are not expected to play a relevant role in gas giants' interior dynamics.

[3]Tildes in this Chapter and the ones below refer to the background state for the anelastic approximation, do not confuse with the dimensionless procedure shown in Chapter 2.

For the sake of completeness, I show the Boussinesq approximation below, although it's not used in this thesis. This approximation is more well-suited for rocky planets, where radial density variations in their molten metallic cores are irrelevant (in the first order). A further restriction is added: spatial variations in fluid properties are ignored except for the density, the fluctuation of which produces the buoyant term $\rho = \rho_0 - \alpha\rho_0 T'$. Then Eqs. (4.6)-(4.10) become:

$$\nabla \cdot \mathbf{u} = 0.$$

$$\left(\frac{\partial \mathbf{u}}{\partial t} + \mathbf{u} \cdot \boldsymbol{\nabla}\ \mathbf{u}\right) = -\boldsymbol{\nabla} p' - \frac{2}{\mathrm{E}}\mathbf{e_z} \times \mathbf{u} + \frac{\mathrm{Ra}}{\mathrm{Pr}} T' \frac{\mathbf{r}}{r_o} + \frac{1}{\mathrm{EPm}}(\boldsymbol{\nabla} \times \mathbf{B}) \times \mathbf{B} + \nabla^2 \mathbf{u},$$

$$\tilde{\rho}\left(\frac{\partial T'}{\partial t} + \mathbf{u} \cdot \boldsymbol{\nabla} T'\right) = \kappa \nabla^2 T' + \epsilon,$$

$$\frac{\partial \mathbf{B}}{\partial t} = \boldsymbol{\nabla} \times (\mathbf{u} \times \mathbf{B}) - \frac{1}{\mathrm{Pm}} \boldsymbol{\nabla} \times (\lambda(r)\, \boldsymbol{\nabla} \times \mathbf{B}), \qquad \boldsymbol{\nabla} \cdot \mathbf{B} = 0,$$

$$\tilde{\mathbf{g}} = \tilde{g}_o \frac{\mathbf{r}}{r_o}, \partial s \approx \frac{\tilde{\rho} c_p}{\tilde{T}} \partial T, \kappa = \frac{k}{\tilde{\rho} c_p}.$$

As briefly introduced in Sect. 2.4, *MagIC* is one of the most currently used HD/MHD codes in a spherical shell. It solves the equations above either using the anelastic or the Boussinesq approximations. Chebyshev polynomials or finite differences are used in the radial direction, and spherical harmonic decomposition in the azimuthal and latitudinal directions. *MagIC* supports several Implicit-Explicit time schemes where the nonlinear terms and the Coriolis force are treated explicitly, while the remaining linear terms are treated implicitly. In Chapter 5 and 7 I employ *MagIC* under the anelastic approximation, which is typically used for modeling the density-stratified low-Mach number convection flows in gas giants and stars (Braginsky and Roberts, 1995; Lantz and Fan, 1999). See App. D for a more detailed description of the numerical technique that *MagIC* employs, and how to treat the dimensionless units.

To perform a numerical MHD simulation with rotating spherical shell codes, such as *MagIC*, a standard set of dimensionless control parameters is Ra, E, Pr, and Pm. When combined with the appropriate boundary conditions, these parameters fully define a simulation setup. In cases where the anelastic approximation is employed, it is also necessary to specify the background radial profiles. In contrast, quantities such as Re, Rm, Ro, and $\Lambda$), depend on output variables such as $u_{\mathrm{rms}}$ and$B_{\mathrm{rms}}$, and are therefore considered diagnostic parameters.

As mentioned above, Ra is one of the most fundamental dimensionless parameters, which has a specific value for the onset of convection, $\mathrm{Ra}_c$. For rotating convection, the onset of convective instability is delayed compared to the non-rotating case; thus, there is a dependence on E:

$$\mathrm{Ra_c} \propto \mathrm{E}^{-4/3}.$$

This scaling arises because stronger rotation (smaller E) suppresses convective motions, requiring a greater buoyant force (i.e., higher Ra) to initiate flow. To keep supercriticality constant when playing around with numerical dynamos, this relation needs to be fulfilled when increasing (or decreasing) E at the same time that Ra is decreased (or increased).

In terms of spatial resolution, the smallest viscous length scales in rapidly rotating convection scale as:

$$\ell_\nu \propto E^{1/3}.$$

This means that reducing E by a factor of 10 requires approximately a factor of $10^{1/3} \approx 2$ increase in grid resolution along each spatial direction to adequately resolve the flow. Moreover, the timestep $\Delta t$ must also decrease, typically by an order of magnitude, to satisfy the Courant stability condition, given the smaller resolved scales and faster flow variations. This increase in computational resources explains why, in numerical simulation, we often cannot explore the physical range of dimensionless numbers.

## 4.6 Planetary dynamo simulations

At the end of the last century, Glatzmaier and Roberts (1995, 1996) presented the first time-dependent 3D MHD under the anelastic solution of a self-sustained magnetic field triggered by thermal convection in a rapidly rotating spherical fluid shell surrounded by a solid conducting inner core. With a low degree of stratification to model Earth's outer core, the simulation ran for about 40000 years (about three times the magnetic diffusive timescale), reaching a saturated state with some geodynamo similarities. They recovered several important geodynamo features: a strongly dipole-dominated magnetic field, with a structure at the CMB and its secular variation closely mimicked Earth's patterns and amplitude, and a single successful magnetic field reversal. When assigning compositional buoyancy in addition to thermal buoyancy, they could also reproduce a westward drift in velocity with the solid inner core seismic studies.

After decades of countless other numerical solutions to the MHD equations in convective spherical shells, many studies have been obtained under various assumptions and parameters (see Christensen and Wicht (2007); Wicht and Tilgner (2010); Jones (2011) for planetary numerical dynamo reviews). Specifically, in the context of gas giants, amplification and self-sustaining of magnetic fields have been recovered, such as the magnetic field morphology or the latitudinal variations in the atmospheric jets (e.g., Jones, 2011; Schubert and Soderlund, 2011). High-resolution models are getting close to reproducing the internal dynamos of Earth and Jupiter (Schaeffer et al., 2017; Gastine and Wicht, 2021, respectively), including stochastic dipole reversals that resemble the geomagnetic field (Glatzmaier and Coe, 2007; Meduri et al., 2021).

Numerically, the above-mentioned dimensionless dynamo numbers represent a vast parameter space which is ideally worth exploring, to consider different relative weights of the various fluid forces. However, there is an unavoidable computational caveat: the parameter space accessible through numerical simulations diverges significantly from the physical reality, as mentioned above. Typically, the feasible range of some parameters, Rayleigh and Ekman in particular, differs by many orders of magnitude from the expected realistic ones (see Table 4.3). The reason is that the range of relevant spatial scales to be considered is too wide, spanning from microscopical diffusion to planetary scales for global rotation and convection patterns.

To partially overcome this intrinsic drawback, some studies have used many numerical models to find scaling laws between the different dimensionless numbers characterizing the dynamo, to interpret the results in real scenarios, via extrapolation to the numerically inaccessible ranges of dimensionless numbers, assuming that the same scaling laws apply there. For example, for rapidly rotating dipole-dominated dynamo solutions under the Boussinesq approximation, Christensen and Aubert (2006) derived scaling laws that connect the dynamo parameters spanning at least two orders of magnitude. First, they studied how the dipole solutions depended on Ra, Pm and E. The trends can be seen in Fig. 4.7. As usual, $\mathrm{Ra}_{\mathrm{crit}}$ is defined as the minimum Ra to reach the onset of convection. For a given E and Pm, there is a secondary critical value for Ra that makes a dipole-dominated dynamo solution. If Ra is further increased, solutions have multipole-dominated magnetic fields. As the E is decreased and approaches a more realistic physical regime, the parameter space with active dynamo increases for the same parameter space. Also, regions with dipolar-dominated solutions grow with decreasing E. It can also be appreciated how, for a given

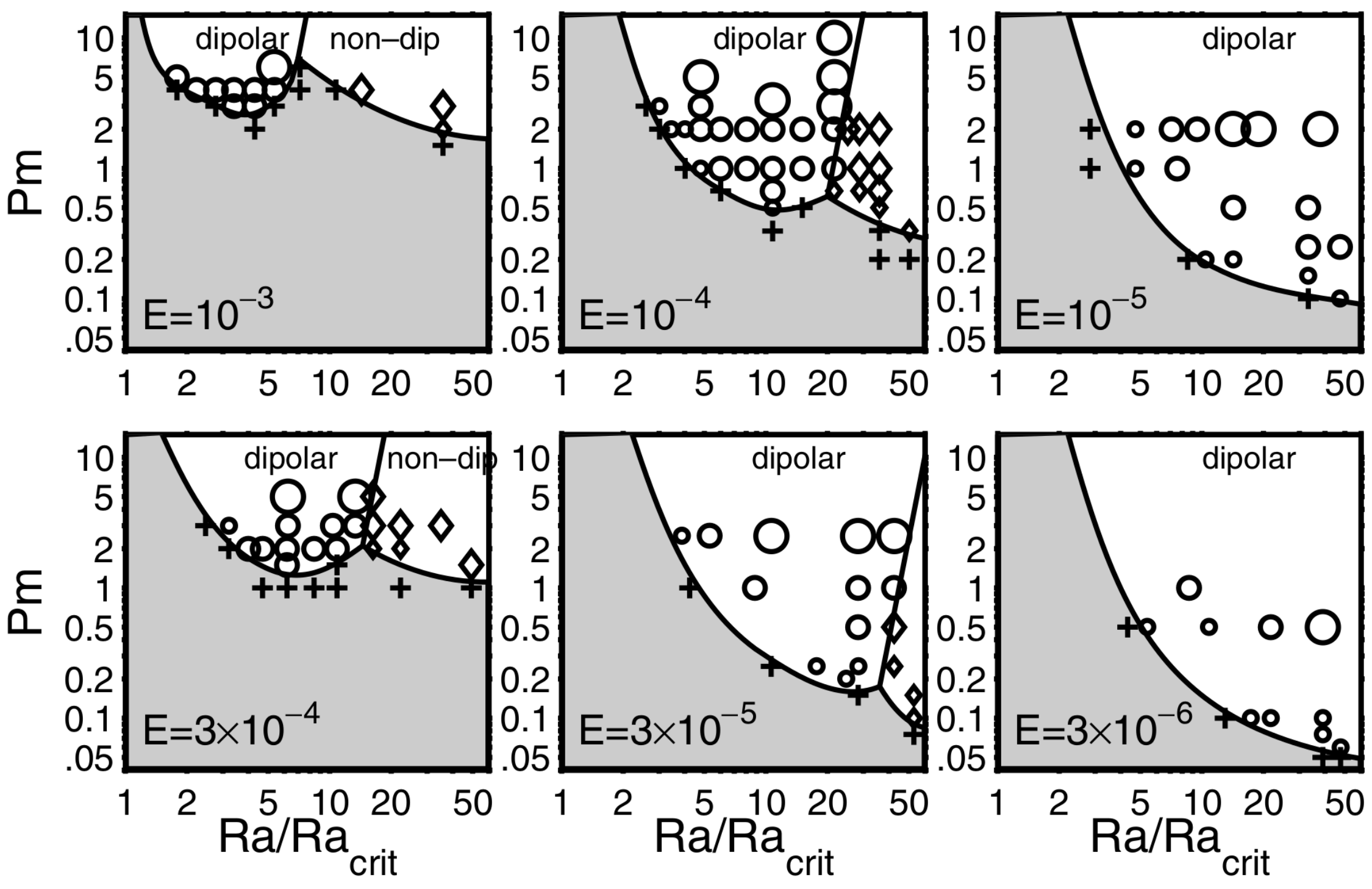

Figure 4.7: Taken from Christensen and Aubert (2006). Regime diagram for spherical shell dynamos under the Boussinesq approximation with Pr = 1, with different values for E. Circles, diamonds, and crosses denote dipolar, non-dipolar dynamos, and failed dynamos, respectively. The marker size shows the approximate value for $\Lambda$ for comparison, i.e., the relative magnetic field strength. Regime boundaries are tentative.

E and Ra, there is a minimum value Pm for which dynamos are excited. This behavior is better depicted in the left panel of Fig. 4.8, where the lowest Pm allowing dynamo is shown for different values of E. The right side of Fig. 4.8 is the minimum Rm for which a self-sustained dynamo is produced. Christensen and Aubert (2006) confirmed that Rm $>$ 50 is a necessary condition for a dipolar planet-like dynamo solution, a rule-of-thumb which is recurrently used in the planetary dynamo community. These trends are general for dynamo solutions and were used for the initial tests and parameter sweeps for Chapter 5.

More specifically, Christensen and Aubert (2006) obtained kinetic scaling laws by taking the dipole-dominated solutions and performing fits between Ro and the input variables Ra, E, Pm, Pr. They found the following relation: $\mathrm{Ro} = 1.07\mathrm{Ra_Q}^{0.43}\mathrm{Pm}^{-0.13}$, where E, and Pr differ only very marginally from zero. The same is done for the magnetic counterpart with Lo, which leads to the following scaling law: $\mathrm{Lo}/f_{\mathrm{ohm}^{1/2}} = 0.76 Ra_Q^{*0.32}\mathrm{Pm}^{0.11}$. In Fig. 4.8, both scaling laws are shown together with the scatter plot from their simulations. These relations are commonly assumed to work in the real planetary regime, meaning that the relative importance of each term in the Navier-Stokes equation (the force balance) is expected to be similar to numerical models (Davidson, 2013; Yadav et al., 2016).

For the modeling of gas giants, the anelastic approximation (e.g., Braginsky and Roberts, 1995; Gilman and Glatzmaier, 1981; Glatzmaier, 1984, 1985a,b; Lantz and Fan, 1999) is more appropriate than the Boussinesq approximation, as explained above, since it allows for density variations but still effectively filters out the sound and magnetosonic waves. It relies on using a static, adiabatic, and spherically symmetric background reference state that is specified by density, gravity, temperature, and other thermodynamic variables. In addition, the velocity and magnetic fields are evolved together with the deviations from the background. These equations have been extensively used to model the magnetic field of gas giants and stars. Usually, the no-slip boundary conditions employed by geodynamo models are replaced by stress-free conditions because they reflect the nature of the outer atmosphere of the gas giants better. The simulations are more computationally demanding, and the numerical solutions have several features that are not seen in the Boussinesq approximation. For example, solutions usually develop highly geostrophic zonal winds (see Sect. 4.6.1 for the notion of geostrophic balance), which are always highly geostrophic and thus extend through the whole gaseous envelope. These Jupiter-like zonal flows promote weaker multipolar fields, while strong dipole fields suppress the zonal flows via Lorentz forces (e.g., Grote et al., 2000; Simitev and Busse, 2003, 2009; Sasaki et al., 2011; Schrinner et al., 2012; Duarte et al., 2018).

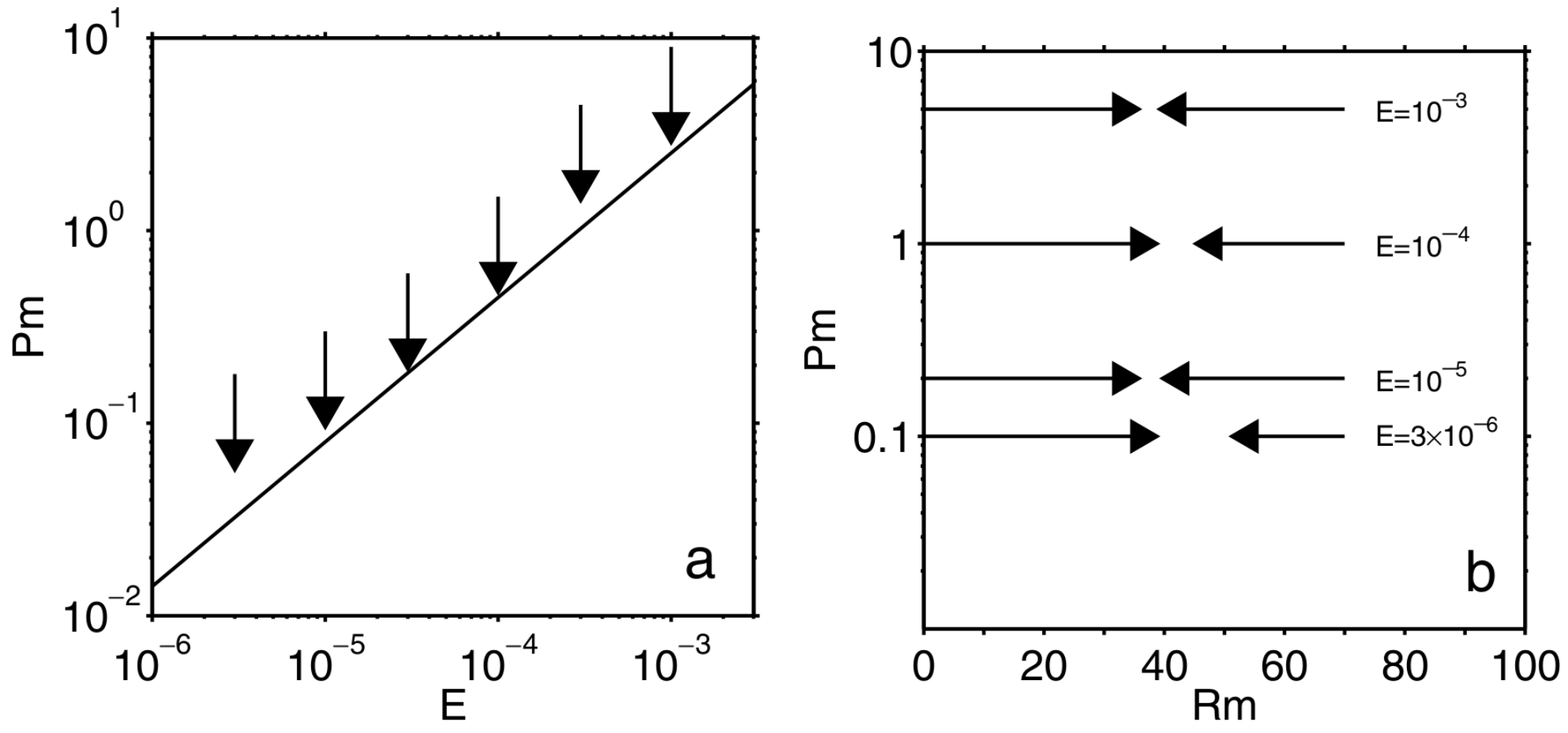

Figure 4.8: Taken from Christensen and Aubert (2006). (Left) The tip of the arrows indicates the lowest Pm with a non-decaying dipolar dynamo. (Right) The tip of the right arrow indicates the lowest Rm for self-sustained dipolar dynamos, and the left arrows indicate the highest Rm with a decaying field.

Another example is the bistability found for not too large Rayleigh numbers: dipolar-dominated and multipolar solutions coexist with identical parameters (Gastine et al., 2012), where different solutions can be reached by setting different initial conditions (Schrinner et al., 2012). In this context, Yadav et al. (2013) provided scaling laws for dynamo models under the anelastic approximation similar to those of Christensen and Aubert (2006). In this case, dipolar and multipolar solutions, as well as a different range of density stratification and radial-dependent diffusivities, were used. They produce very similar scaling laws and show equivalent plots to Fig. 4.9. In Chapter 5, I will analyze my dynamo simulations, showing that they also lie along the same scaling laws of Yadav et al. (2013).

For a more realistic modeling of gas giants, one of the main challenges is to incorporate the outer steep gradients of various thermodynamic profiles, since they imply very different timescales farther outward. In particular, there is a steep outward decrease in the electrical conductivity because of the transition of hydrogen from metallic to non-metallic state. This behavior was quantified along the Jovian adiabat by French et al. (2012) and was used during the past decade in different dynamo simulations (Gastine and Wicht, 2012; Jones, 2014; Gastine et al., 2014; Wicht et al., 2019b). These studies provided significant results in terms of a comparison with the data provided by the ongoing *Juno* mission and earlier Jovian missions, and they highlighted the importance of incorporating a realistic background, although with the caveat on the nondimensional dynamo numbers mentioned above. In recent years, some works (Gastine and Wicht, 2021; Yadav et al., 2022; Moore et al., 2022) have added a stably stratified layer just below the region in which metallic hydrogen starts to mimic a helium-rain region due to hydrogen-helium demixing (Klepeis et al., 1991; Nettelmann, 2015; Nettelmann et al., 2015). This layer helps to naturally obtain alternating east-west zonal winds centered around the equator for the Jovian dynamo, as well as a highly axisymmetric magnetic field for the Saturn model.

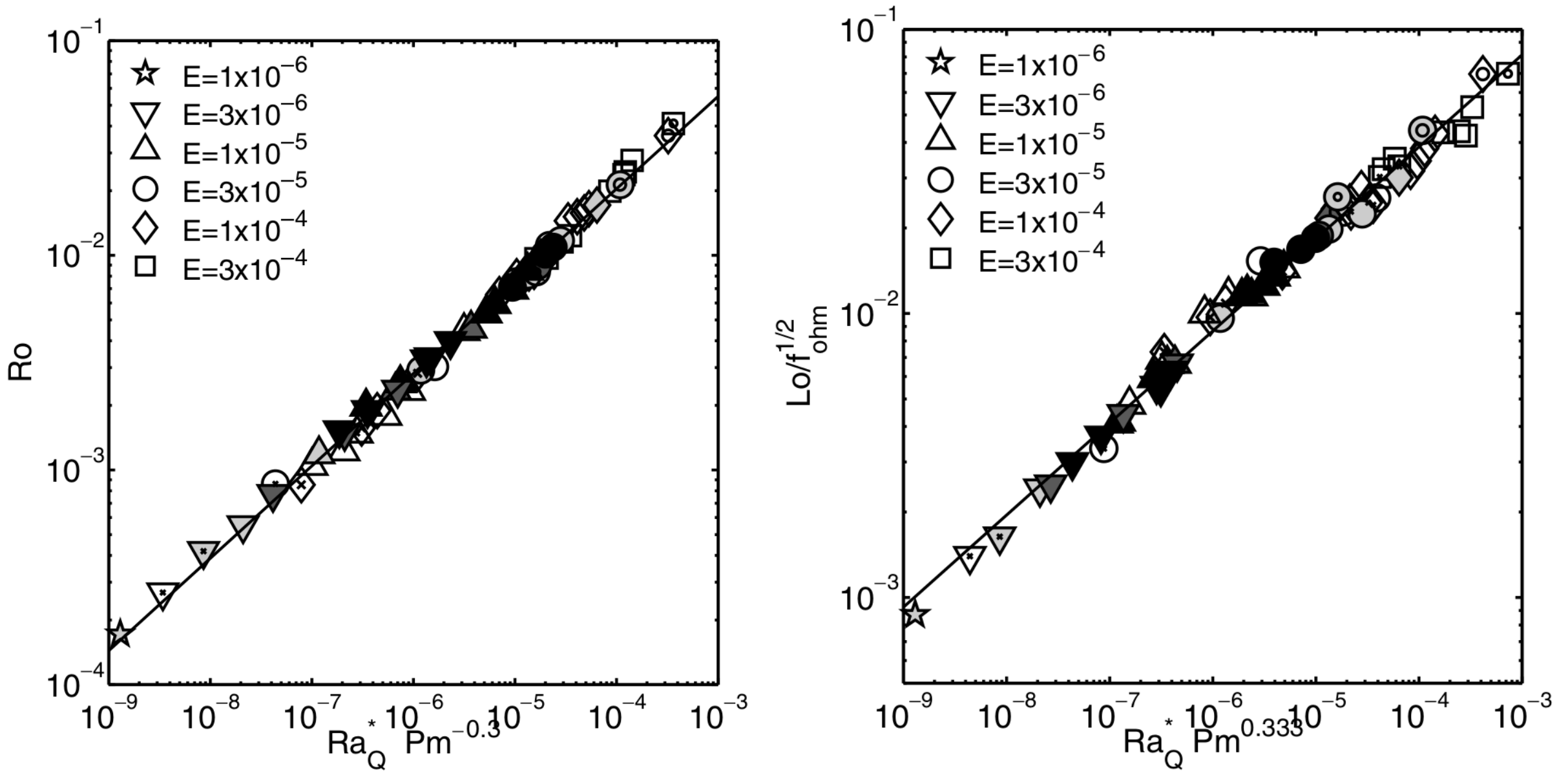

Figure 4.9: Taken from Christensen and Aubert (2006). (Left) Ro versus a combination of modified of Ra∗ and Pm. (Right) Lo corrected by $f_{ohm}$ versus a combination of modified of Ra∗ and Pm. The marker shape indicates E and the shade Pm (Pm < 0.3: black, 0.3 < Pm < 1: dark grey, Pm = 1: light grey, Pm < 1: white). Pr = 3 is indicated by a small cross, Pr = 10 by a larger cross, Pr = 0.3 by a small circle, and Pr = 0.1 by a larger circle (Pr = 0.1).

By using the aforementioned scaling laws as well as magnetic field measurements, Christensen et al. (2009) determined that for the Earth, Jupiter, and fast-rotating, fully convective, low-mass stars, the convective heat flux determines the magnetic field strength. The whole scaling law is given as the radial integral of the convective flux in the electrically conductive shell. To match observational constraints and give a simpler approach, Reiners et al. (2009) gave a simpler expression for this law in terms of the mass $M$, luminosity $L$, and radius $R$ in a more simplified form. Using this scaling law and the analytical evolutionary tracks for substellar objects in (Burrows and Liebert, 1993; Burrows et al., 2001), Reiners and Christensen (2010) provided a scenario for the evolution of the magnetic field, with which they obtained a steadily weakening (a factor of ≈10 over around 10 Gyr) of the magnetic field at the dynamo surface. These scaling laws are used in Chapter 5 to compare the results we obtain with a novel methodology, and in Chapter 6 to provide magnetic field estimates for hot Jupiter dynamos.

An additional note to be made is whether small-scale dynamo (SSD) action is excited in numerical planetary dynamos with Ro $\ll$ 1. Much of the explored space in the literature lies has Rm > 30 and Pm > 1, where SSD is known to be triggered. Planetary dynamo studies commonly involve detailed analysis of the saturated solutions with integrated averaged quantities. These allow addressing topics such as the level of equipartition, the magnetic field dipolarity, radial and spectral energy distributions, force balances, and so on. It is thus common to start simulations with already saturated dynamo solutions, where SSD is assumed to play little role, saving the computational time required for the kinematic dynamo phase. Similar to Warnecke et al. (2025) in the context of the solar dynamo, it would be interesting to address for planetary dynamos the interplay between SSD and LSD in the kinematic phase, as well as whether the fluctuations generated by SSD (possibly to the non-axisymmetric part) influence the overall large scale flow and magnetic field, this latter being maintained by LSD.

### 4.6.1 Planetary flow regimes and force balances

To study the force balance inside planetary simulations, one can follow the procedure done by e.g. Aubert et al. (2017); Schwaiger et al. (2019). Each term in the momentum equation (i.e., Eq. 4.13) is decomposed using the radial vector $\boldsymbol{r}$ in the following way:

$$\boldsymbol{f}(r,\theta,\phi,t) = \mathcal{Q}\boldsymbol{e_r} + r\boldsymbol{\nabla}\mathcal{S} + \boldsymbol{r}\times\boldsymbol{\nabla}\mathcal{T} \tag{4.16}$$

which are the radial, spheroidal, and toroidal components, respectively. By using spherical harmonics $Y_\ell^m$, as defined in Appendix D, each potential is separated from its radial components (i.e. $\mathcal{R} = \mathcal{R}_\ell^m(r)Y_\ell^m(\theta,\phi)$):

$$\boldsymbol{f} = \sum_{\ell=0}^{\ell_{\max}} \mathcal{Q}_\ell^m Y_\ell^m \boldsymbol{e_r} + S_\ell^m r\boldsymbol{\nabla}Y_\ell^m + \mathcal{T}_\ell^m \boldsymbol{r}\times\boldsymbol{\nabla}Y_\ell^m \tag{4.17}$$

Then the total energy associated with a specific force is obtained with the following identity:

$$F^2 = \int_V \boldsymbol{f}^2 dV = 2\int_{r_i}^{r_o} \sum_{\ell=0}^{\ell_{\max}}\sum_{m=0}^{\ell} |Q_\ell^m|^2 + \ell(\ell+1)\left(|S_\ell^m|^2 + |\mathcal{T}_\ell^m|^2\right) r^2 dr \tag{4.18}$$

This expression can be rearranged without summing the $\ell$ index:

$$F^2 = \sum_{\ell=0}^{\ell_{\max}} \mathcal{F}_\ell^2 \quad \text{where} \quad \mathcal{F}_\ell^2 = 2\int_{r_b}^{r_t} \sum_{m=0}^{\ell} |Q_\ell^m|^2 + \ell(\ell+1)\left(|S_\ell^m|^2 + |\mathcal{T}_\ell^m|^2\right) r^2 dr \tag{4.19}$$

For a steady state dynamo solution, $\mathcal{F}_\ell$ gives the relative importance between forces at a specific harmonic degree and thus is regarded as the spectral force balance.

All known planetary interiors fall under the rapid rotator regime, meaning the timescales for convection and diffusion are much longer than the planetary rotation period. In dynamical terms, this implies that both the convective and viscous terms in the Navier–Stokes equation are much smaller than the Coriolis term. This situation is characterized by very small values of E and Ro, i.e., E $\ll 1$, Ro $\ll 1$.

At large scales, the dominant force balance is between the Coriolis and pressure gradient forces, a regime known as geostrophic balance. Under this condition, the Coriolis acceleration nearly cancels the pressure gradient force:

$$2\rho\boldsymbol{\Omega}\times\mathbf{u} \approx -\nabla p\,. \tag{4.20}$$

This balance leads to invariant flow structures along any line parallel to the rotation, known as the Taylor-Proudman theorem (taking the curl of Eq. 4.20 leads to $\partial\mathbf{u}/\partial z = 0$). As a result, convection in this regime organizes into columnar vortices aligned with the rotation axis, commonly referred to as Busse columns. These structures are a hallmark of rotating convection and are essential for understanding planetary and stellar interior dynamics.

Geostrophic balance usually occurs on the largest scales (i.e., low-degree multiples). The ageostrophic contribution (the residual part of the balance, i.e., Coriolis minus pressure) can be balanced at 1$^{\text{st}}$ order by the sum of the buoyant and Lorentz forces. This is called the MAC balance, which stands for magnetic, Archimedean, and Coriolis balance. This can be seen in Fig. 4.10, where I show the force spectra for a Boussinesq saturated dynamo solution from Schwaiger et al. (2019). There is a geostrophic balance throughout all multiples, followed by a MAC balance. Note that at some specific scale $\ell_{\mathrm{MA}}$, the buoyancy and the Lorentz force are equally important and balance with the ageostrophic contribution. This situation is known as quasi-geostrophic Magneto-Archimedean-Coriolis (QG-MAC) balance (Aubert et al., 2017; Schwaiger et al., 2019). For some other simulations with strong equipartitions (i.e., high magnetic energies), the Lorentz force overruns the Coriolis force above some specific multipole $\ell_{\mathrm{MS}}$, with typically $\ell_{\mathrm{MS}} > \ell_{\mathrm{MA}}$. For multiples above $\ell_{\mathrm{MS}}$, geostrophy is broken and transitions to magnetostrophy, where pressure balances with magnetic forces towards smaller scales. Inside the dynamo regions, inertial and viscous forces remain negligible in comparison to geostrophy/magnetostrophy or the 1st order balance.

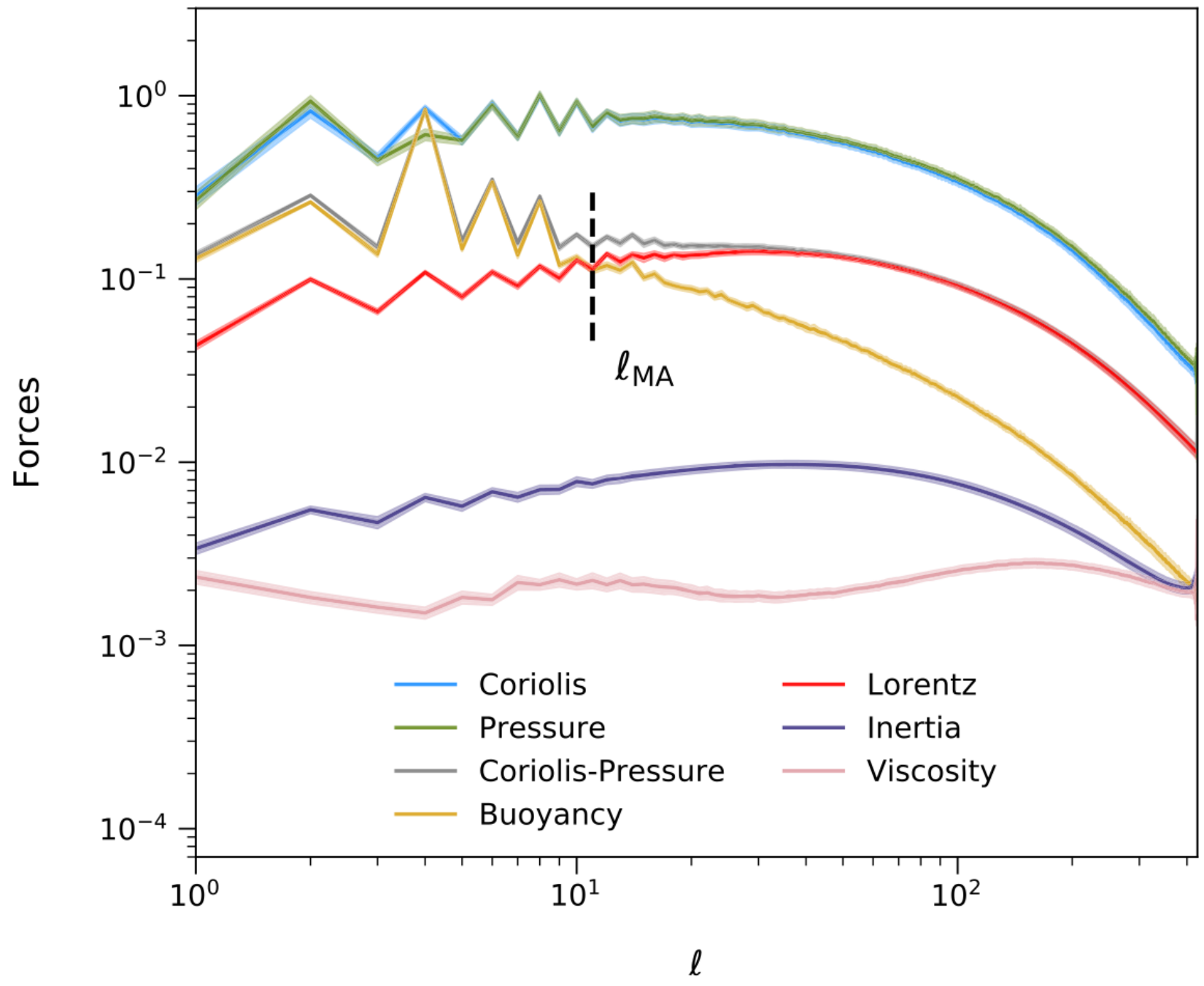

Figure 4.10: Taken from Schwaiger et al. (2019). Time-averaged force balance spectra for a saturated dynamo ($\mathrm{E} = 10^{-6}$, $\mathrm{Ra} = 5.5 \cdot 10^9$, $\mathrm{Pm} = 0.4$, $\mathrm{Pr} = 1$), normalized relative to the peak of the Coriolis force. Shaded regions represent the standard deviation in time.

When this analysis is done for gas giant models, i.e., anelastic simulations with external regions of low conductivity, the force balance outside dynamo regions or at the boundaries changes and transforms from MAC balance to IAC balance, inertial, Archimedean, and Coriolis balance (Gastine and Wicht, 2021). Similarly, there is a multipole, $\ell_{\mathrm{IA}}$, above which inertial forces overrun buoyant forces. In the dynamo region, inertia remains negligible, while in the outer regions, where the conductivity is much lower with zonal winds and no electric currents, magnetic forces are overrun by inertia forces. Viscous effects remain minimal both inside and outside the dynamo.

When considering the relative influence of buoyancy and Coriolis forces, one can define a dimensionless parameter known as the modified Rayleigh number, given by:

$$\mathrm{Ra}^* = \frac{\mathrm{Ra}\,\mathrm{E}}{\mathrm{Pr}},$$

This parameter determines the dominant force regime: if $\mathrm{Ra}^* \ll 1$, rotation dominates and constrains convection to form quasi-geostrophic structures like Busse columns; vice versa, if $\mathrm{Ra}^* \gg 1$, buoyancy dominates and the flow becomes more isotropic and turbulent.

# 5

# Planetary dynamos in evolving cold gas giants

In this chapter, we address the long-term evolution of the dynamo action in Jupiter-like planets through an alternative approach. We perform 3D anelastic dynamo simulations with a background corresponding to different ages of the long-term planetary evolution. By comparing how the solutions change from one age to the next and recalling the intrinsic caveats related to the accessible ranges of nondimensional numbers, we evaluate the topology and intensity changes that an internal dynamo in cold gas giants can undergo during its evolution over billions of years.

This work is organized as follows: The overall method and the internal thermodynamic profiles from the evolutionary code *MESA* are described in Sect. 5.1, where we also summarize our 3D dynamo models that were performed with the *MagIC* code. In Sect. 5.2 we show the main results of our parameter exploration, interpret simulations representative of different evolutionary stages, and compare them to results from other works.

## 5.1 Methodology

It is currently not possible to simulate the realistic time evolution of the radial dependence of the thermodynamic quantities of a 3D planetary dynamo. The main reason is that the timescale associated with the convection inside gas giants tends to be on the order of years or decades. In contrast, the secular planetary cooling and contraction are appreciable at timescales on the order of gigayears. Due to this timescale separation of at least six orders of magnitude, a set of fixed backgrounds can be considered, and the dynamo models can be evolved for each of them. In other words, we aim to obtain different snapshots, that is, 3D dynamo solutions, each with fixed radial thermodynamic profiles that correspond to a given age of the 1D long-term evolutionary models. To schematically summarize, we employed a method that followed these steps:

1. Evolve a standard evolutionary 1D model of a contracting, non-irradiated gas giant over 10 Gyr.

2. For a given age of the evolutionary model, implement the radially dependent thermodynamical profiles as the background state of an anelastic spherical shell MHD model, with a given choice of the nondimensional numbers.

3. Evolve the 3D MHD equations in a spherical shell domain under the anelastic approximation and reach a self-sustained dynamo solution. Then let it evolve long enough (typically up to some ten thousand years of physical timescales) to average out the typical fluctuations and ensure statistical significance.

4. Repeat the process from step 2, where some of the nondimensional dynamo numbers rescale (compared to the first simulation) according to the relative variation of the involved thermodynamic quantities. Note that all the nondimensional numbers can be rescaled in this way, so that further assumptions about the rotation rate and viscosity, for instance, are required, as discussed below.

The process can then be repeated for another planetary model or a different choice of the reference values of the nondimensional numbers. In this approach, the cooling information is included in the trend of the dynamo numbers in the simulation sequence.

### 5.1.1 Internal structure

#### Long-term evolution

To model the evolutionary change of radially dependent thermodynamic quantities of gas giants, we used the public code *MESA*[1] (Paxton et al., 2011, 2013, 2015, 2018, 2019; Jermyn et al., 2023). Modules for Experiments in Stellar Astrophysics (*MESA*) is an advanced, open-source 1D stellar evolution modular code written primarily in Fortran. It consists of several open-source libraries related to computational stellar astrophysics. The most widely used model is the one-dimensional stellar evolution physics module, which solves the fully coupled structure and composition equations and obtains the corresponding stellar evolutionary tracks. It uses adaptive mesh refinement and sophisticated timestep controls, and supports shared memory parallelism based on OpenMP. State-of-the-art modules provide equation of state, opacity, nuclear reaction rates, element diffusion data, and atmosphere boundary conditions. It can also be used for gas giant evolutionary models (Paxton et al., 2013), which are used here as well as Chapters 6 and 7. The equations we solved are the conservation of mass, hydrostatic equilibrium, energy conservation, and the energy transport equation, respectively:

$$\frac{dm}{dr} = 4\pi r^2 \rho \ , \tag{5.1}$$

$$\frac{dP}{dm} = -\frac{Gm}{4\pi r^4} \ , \tag{5.2}$$

$$\frac{dL}{dm} = -T\frac{ds}{dt} \ , \tag{5.3}$$

$$\frac{dT}{dm} = -\frac{GmT}{4\pi r^4 P}\nabla \ , \tag{5.4}$$

where $m$ is the mass enclosed within a radius $r$, $\rho$ is the density, $P$ is the pressure, $G$ is the gravitational constant, $s$ is the specific entropy, $T$ is the temperature, $L$ is the internal luminosity, and $\nabla \equiv d\ln T/d\ln P$ is the logarithmic temperature gradient, which was set to the smallest between the adiabatic gradient and the radiative gradient. In the energy Eq. 5.3, the only source term we considered is the gravitational contraction. We neglected additional sources such as stellar irradiation (Guillot et al., 1996) or internal heat deposition (Komacek and Youdin, 2017; Thorngren and Fortney, 2018) from tidal (Bodenheimer et al., 2001) or Ohmic dissipation (e.g., Batygin and Stevenson, 2010; Perna et al., 2010a), or chemical processes such as hydrogen dissociation and recombination (Tan and Komacek, 2019). These additional terms are fundamental for hot Jupiters (see Fortney et al. 2021 for a review), but negligible for cold, weakly irradiated planets. The set of equations is closed using the *MESA* equation of state (Paxton et al., 2019), which, for the gas giant ranges of interest, is substantially the interpolation of the Saumon-Chabrier-van Horn equation of state for H-He mixtures (Saumon et al., 1995).

[1] https://github.com/MESAHub/mesa

These 1D evolutionary models do not incorporate either diluted cores or hydrogen-helium demixing layers, that is, possible stratified layers in the convection interior. For all our models, we assumed an interior inert rocky core of 10 $M_\oplus$ with a homogeneous density $\rho_c = 10$ g cm$^{-3}$, and a fixed solar composition for the envelope. To illustrate the evolutionary changes, we show in Fig. 5.1 the different profiles for two different planets with masses 1 and 4 $\mathrm{M_J}$ at ages 0.5, 1, and 10 Gyr. For comparison, we also show the profiles from the widely employed results of French et al. (2012) for the interior of Jupiter. As reported by (Paxton et al., 2013), during the evolution of the planet, the radius shrinks slowly, and after a few million years of evolution, this is independent of the chosen initial planetary radius. The planet slowly shrinks, and at early ages, the total radius of the $1\,M_J$ model is therefore larger than that of the current Jupiter. The internal structure is always characterized by a very thin radiative layer with a thick, fully convective isentropic shell that encloses the inert core. The higher planetary mass ($4\,M_J$) shows a larger radius, higher gravity and temperature, and lower thermal expansion coefficients $\alpha$. However, the trends of the profile with age are similar to the $1\,M_J$ case. Moreover, all models exhibit a nontrivial oscillating behavior of the Grüneisen parameter $\Gamma$ (bottom panel; see below for the definition).

### Background state implementation

To run an anelastic MHD model, a series of thermodynamic quantities needs to be expressed as a function of radius. We implemented the *MESA* $\rho(r)$, $T(r)$, $g(r)$, thermal expansion coefficient $\alpha(r)$, and the Grüneisen parameter $\Gamma(r)$, all shown in Fig. 5.1, into a 3D model. We obtained $\alpha$ and $\Gamma$ in terms of the readily available *MESA* thermodynamic profiles

$$\Gamma = \left(\frac{\partial lnT}{\partial ln\rho}\right)_s - 1 \equiv \Gamma_3 - 1 \;, \quad \alpha = -\frac{1}{\rho}\left(\frac{\partial \rho}{\partial T}\right)_P = -\frac{\chi_T}{T\chi_\rho} \;, \tag{5.5}$$

where

$$\chi_\rho \equiv \left(\frac{\partial lnP}{\partial ln\rho}\right)_T \;, \quad \chi_T \equiv \left(\frac{\partial lnP}{\partial lnT}\right)_\rho \;.$$

Transport coefficients, which we discuss and prescribe below, are the other possible thermodynamic quantities that can show a radial dependence.

To define the radial domain for the 3D models, we cut the *MESA* profiles both at the inner and outer radial domains. The hydrostatic background is approximately isentropic due to the efficient convection. Close to the solid inert core, the profiles show a decrease in entropy, leading to a shallow stratified layer that is due to the boundary conditions. Since a more realistic modeling of the core is beyond the purpose of this study, we considered the inner boundary of the convective shell for each model to be the region in which the profile is isentropic and cut out the innermost $\approx$ 2-3% (in radius) of the shell, case by case. We do not expect these inner cuts to lead to significant differences because the results of Moore et al. (2022) showed that replacing a diluted core with a solid compact core of the same radius has little effect on the strength and morphology of the dynamo.

The density profiles usually spanned more than six orders of magnitude (the typical outermost layer in *MESA* is at a fraction of a bar), with the most significant drop in the 1% outermost part of the planet, where no dynamo is expected. Spherical shell dynamo models cannot handle too large density contrasts. Therefore, as commonly done in anelastic dynamo models, we cut the external layers to reduce the density ratio. We ensured that we always kept the pressure at which hydrogen metallization starts ($\approx$1 Mbar) inside the domain. The maximum ratio we considered is $\rho_i/\rho_o \approx 100$, and most of our models took values of $\rho_i/\rho_o \approx 20$, which corresponds to approximate values for the number of density scale heights, $N_\rho := ln(\rho_i/\rho_o)$, of 4.6 and 3.0, respectively. The thick profiles shown in Fig. 5.1 have $N_\rho \approx 3$, and the thin endings represent an extension up to $N_\rho \approx 4.6$. Throughout this work, we use the subindex notation $o$ ($i$) for the value at the outer (inner) shell. These cuts define $r_o$, $\rho_o$, $T_o$, and $P_o$ as well as their inner values, which are listed in Table E.1. The thin radiative outer layer usually ends at about $\approx$ 1 bar, well above the external cut. Finally, all profiles except for $\Gamma(r)$ were normalized to make the outer values equal to unity. This is a common practice for many 3D hydrodynamic codes, as the fundamental units are not the physical units. *MagIC*, for example, works with units of the shell thickness as the length scale and with viscous timescales for units of time (see Sect. 5.1.2 for the exact details).

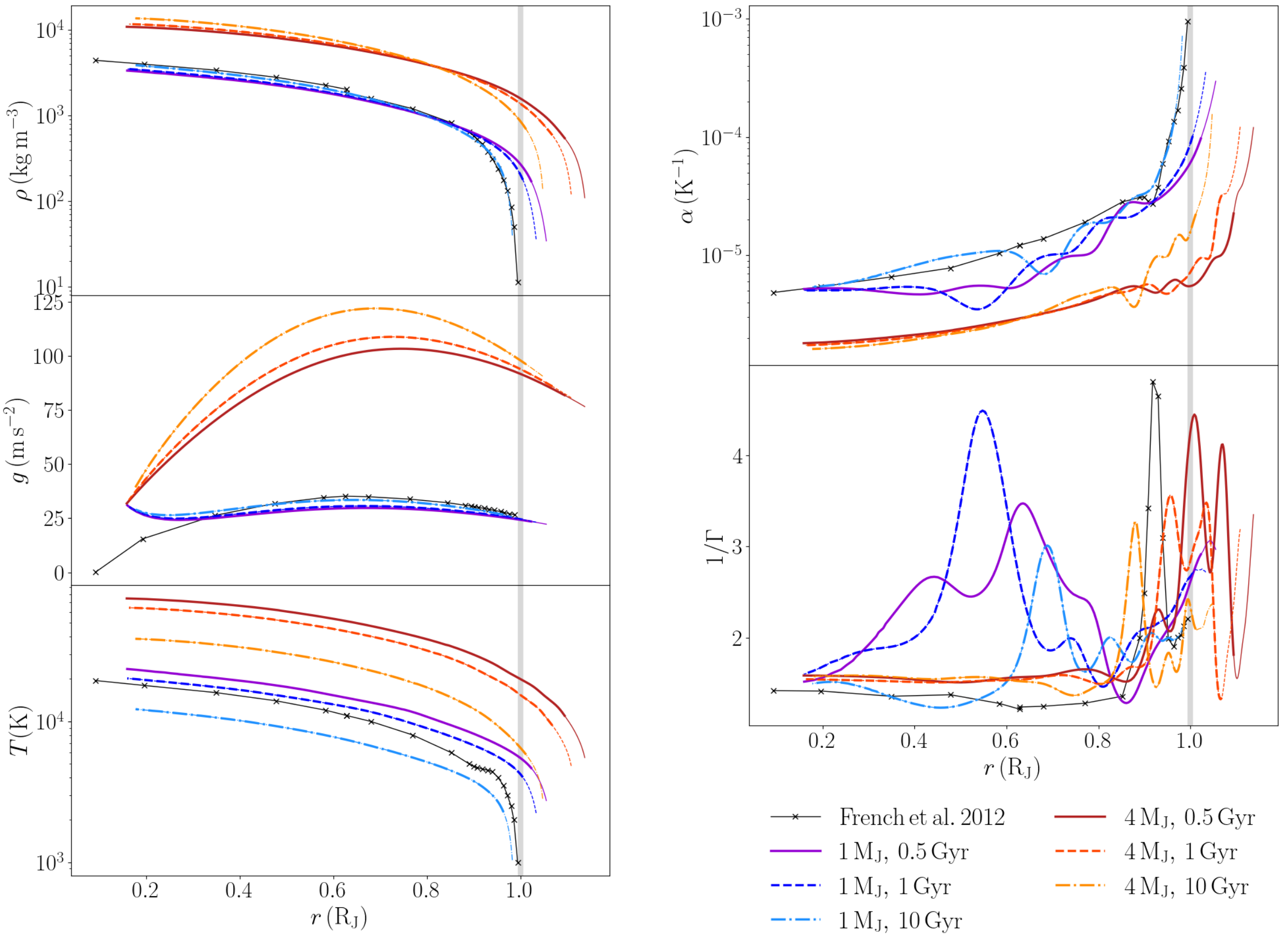

Figure 5.1: *MESA* hydrostatic profiles of 1 and 4 $M_J$ at different evolutionary times, cut at an outer density $\approx$ 100 times (thin extended lines) or $\approx$ 20 times (thick lines only) that is lower than the innermost radius of the isentropic shell, just outside the core-envelope boundary. The gray lines show the Jovian values according to the popular French et al. (2012) model, and the vertical gray band reflects the current Jovian radius as a reference. From top to bottom: density $\rho(r)$, temperature $T(r)$, gravity $g(r)$, thermal expansion coefficient $\alpha(r)$, and the inverse of the Grüneisen parameter $\Gamma(r)$.

After the profiles were cut and normalized to the inner or outer boundary (as required by the code implementation; see below), we used high-degree polynomials to fit any shape the profiles can take generally. For $\rho(r)$, $T(r)$, and $g(r)$, we employed a 20-degree Taylor series, which was enough to fit all cases tested here smoothly. $\alpha(r)$ and $\Gamma(r)$ have some peaks and valleys that complicate the fitting procedure, however. With polynomials with degrees below $\approx$ 100-120, we found that the wiggles were poorly fit. We adopted a 150-degree polynomial for both,

$$\begin{aligned} \Big(\rho(r), T(r), g(r)\Big) &= \sum_{n=0}^{20} \Big(\rho_n, T_n, g_n\Big) r^n \,, \\ \Big(\alpha(r), \Gamma(r)\Big) &= \sum_{n=0}^{150} \Big(\alpha_n, \Gamma_n\Big) r^n \,, \end{aligned} \tag{5.6}$$

where $r$ ranges from $r_i$ to $r_o$. A change in the reference point of the expansion did not quantitatively improve the fits (we also tried with powers of $(r - r_o)$, $(r - r_i)$ and $(r - r_o/2)$). Therefore, for simplicity, we opted for powers of $r$, that is, a MacLaurin series. We also verified that the radial resolution we employed in the MHD code, that is, the number of Chebyshev polynomials (see Sect. 5.1.2), was more than enough to satisfactorily reproduce even the most complex radial profiles given by the implemented MacLaurin expansion.

Unlike the fit model used by Jones (2014), the background profiles taken from *MESA* are almost but not exactly isentropic. To quantify this deviation, we considered the quantity $|ds/dr|\cdot r/s$, which usually takes values of $10^{-4} - 10^{-5}$ with maxima near the outer cut regions of $10^{-3}$. This might lead to a slight energy imbalance resulting from the radial background profile itself. To ensure that this did not influence the overall dynamics, we analyzed the energy balances, that is, we compared the buoyancy power with viscous and Ohmic dissipation (see Sec 5.1.2 for more details).

### Chebyshev polynomials fits

There is a potential concern over the high-degree MacLaurin radial expansion for thermodynamic variables. The initial background profiles are always well resolved by the expansion, but *MagIC* internally expresses them as a combination of Chebyshev polynomials. To ensure that this process does not change the radial dependency, we checked that the degree used in the Chebyshev polynomials (i.e., the radial resolution) used in our runs is enough to fit the most oscillating or complicated radial profiles. In Fig. 5.2, we show different order Chebyshev polynomial fits of the radial profiles for two representative models shown in Fig. 5.1, both with $\rho_i/\rho_o \sim 100$ and $t = 1$ Gyr: 1 $\mathrm{M_J}$ and 4 $\mathrm{M_J}$, the latter having very sharp peaks. As suspected, Chebyshev polynomials are very effective in reproducing the profiles and require fewer polynomials than the MacLaurin expansion. A Chebyshev expansion of degree $n_{\rm cheb} = 50$ already works very well for the most challenging cases (with minor deviations of $< 5\%$ in the sharpest peaks of i.e. $1/\Gamma$ for the 4 $M_J$ case), while lower values of $n_{\rm cheb} = 12, 25$ are not. Higher values (i.e., $n_{\rm cheb} = 150$, the same number of MacLaurin polynomials used, or $n_{\rm cheb} = 288$, the resolution we used in the simulations, reminding $N_r = n_{\rm cheb} + 1$) give reconstructed profiles which are indistinguishable from each other and the implemented one.

### Transport coefficients

The profiles of transport coefficients did not come directly from *MESA*. Although some important ingredients have evolved, such as the particle density, realistic profiles for diffusivities require proper ab initio calculations. In particular, the electrical and thermal conductivities have to take into account the degenerate state of the electron population. The pressure ionization (rather than thermal) is non-negligible in the dense but relatively cold (compared to stars) convective interior.

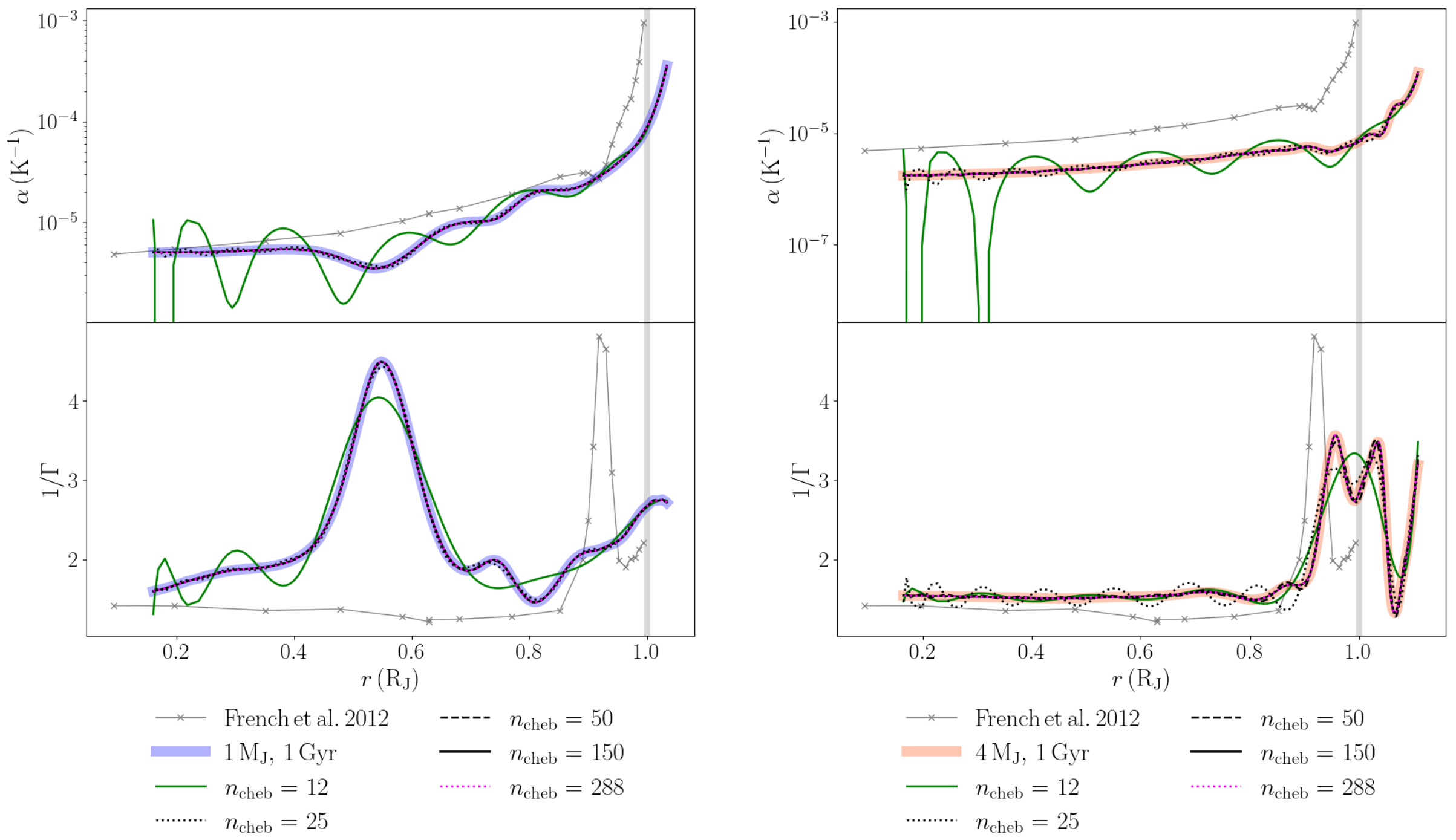

Figure 5.2: Chebyshev polynomial fits (performed with the `numpy.polynomial.chebyshev` library) for two of the most complex radial profiles coming from *MESA* (left: 1 $\mathrm{M_J}$ in blue, right: 4 $\mathrm{M_J}$ in pink, both at $t$ = 1 Gyr). Different lines mark different numbers of Chebyshev polynomials (see legends): $n_\mathrm{cheb} = 12, 25, 50, 150, 288$. In all cases, the $n_\mathrm{cheb} = 150$ and $n_\mathrm{cheb} = 288$ (used in the runs) are indistinguishable.

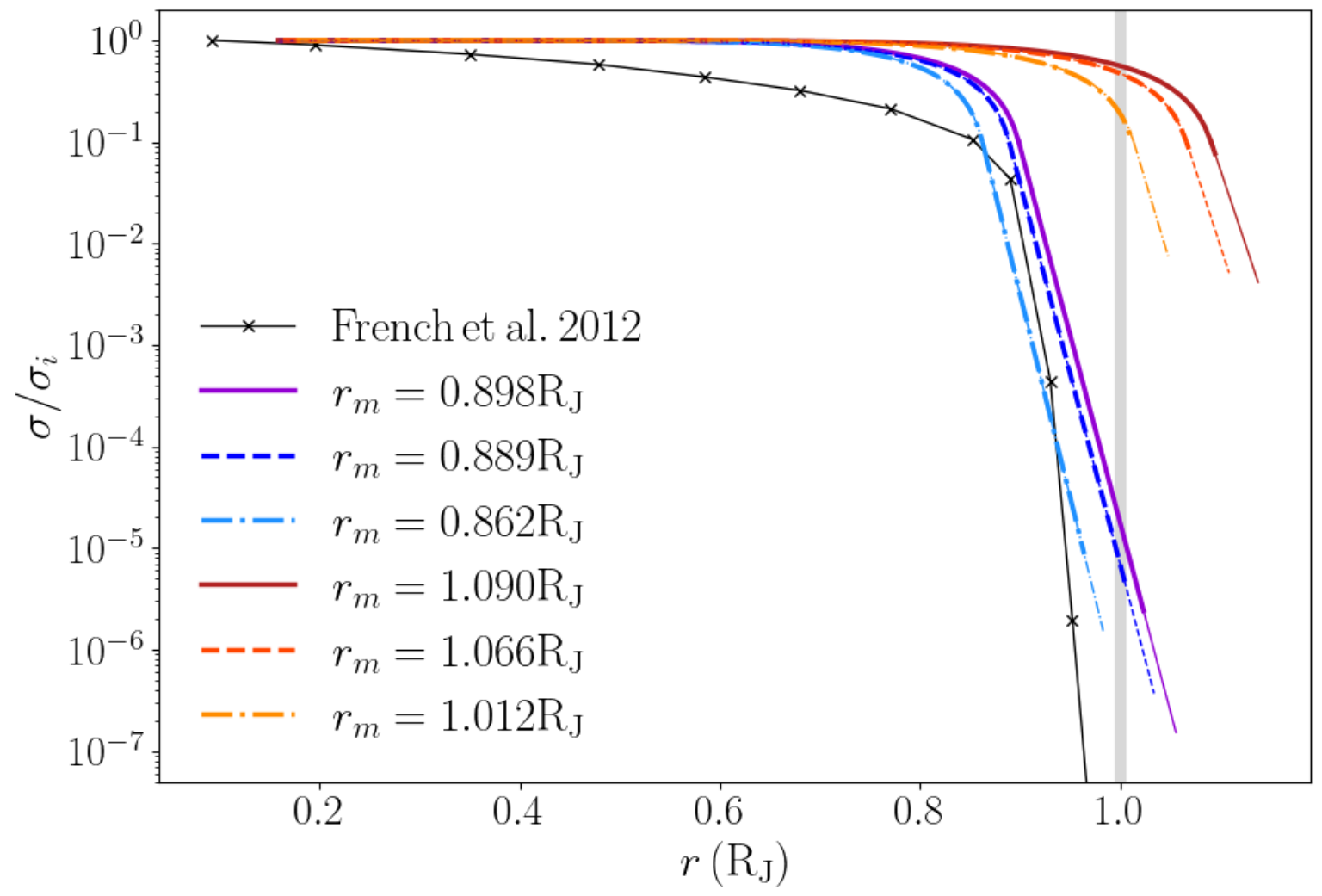

Figure 5.3: Electrical conductivities normalized to their inner values ($\sigma/\sigma_i$) obtained from Eq. (5.7) for the same models as shown in Fig. 5.1. The values of $r_\mathrm{m}$ in the legend correspond to the start of the exponential decay. The French et al. (2012) profile was also normalized.

Moreover, the dynamo region arguably corresponds to the transition to the metallic phase for hydrogen. In this sense, French et al. (2012) calculated the electric conductivity $\sigma$ for a set of $(T, \rho)$ pairs along the modeled Jupiter adiabat. Further models about transport coefficients in pure H-He mixtures have been developed (usually neglecting the contribution of thermally ionized alkali metals, which becomes relevant where hydrogen is molecular; Kumar et al. 2021), with relative differences in the values of $\sigma$ by factors of a few (see Bonitz et al. 2024 for a recent review). In any case, the trend is that there is a continuous and steep increase in the conductivity for increasing pressure up to around 1 Mbar, after which the dependence on both temperature and pressure (or density) is much milder.

To capture these fundamental properties, we adopted the electrical conductivity profile first defined in Gómez-Pérez et al. (2010), which consists of an approximately constant conductivity in the innermost hydrogen metallic region, with a polynomial plus exponential decay toward the outer molecular region,[2]

$$\frac{1}{\tilde{\lambda}(r)} := \tilde{\sigma}(r) = \begin{cases} 1 + (\sigma_m - 1)\left(\dfrac{r - r_i}{r_m - r_i}\right)^{-a} & r < r_m \ , \\ \sigma_m e^{-a\left(\frac{r - r_m}{r_m - r_i}\right)\frac{\sigma_m - 1}{\sigma_m}} & r \geq r_m \ , \end{cases} \tag{5.7}$$

where $\tilde{\sigma}$ and $\tilde{\lambda}$ are the normalized conductivity and magnetic diffusivity, respectively. The actual physical relation, $\lambda = (\mu_0\sigma)^{-1}$, with $\mu_0$ being the vacuum magnetic permeability, is simplified when the quantities are normalized to their innermost values: $\tilde{\sigma} = \sigma/\sigma(r_i)$ and $\tilde{\lambda} = \lambda/\lambda(r_i)$. This expression ensures that both $\tilde{\lambda}$ and $d\tilde{\lambda}/dr$ are continuous at $r_\mathrm{m}$, and that they qualitatively reproduce the main features. Moreover, it allowed us to compare results with several previous works that have employed this profile for gas giant convection and dynamo modeling (Duarte et al., 2013, 2018; Wicht et al., 2019b,a; Gastine and Wicht, 2021). They used values of $\sigma_m$ and $a$ that ranged from approximately 0.9 to 0.01 and from 1 to 25. In our models, we fixed $\sigma_m = 0.1$ and $a = 7$. At the same time, $r_\mathrm{m}$ was chosen as the radius at which each *MESA* planetary profile reached 1 Mbar, that is, the approximate pressure above which hydrogen is thought to undergo metallization. The profiles of $\tilde{\sigma}$ are shown in Fig. 5.3 for the same representative models as in Fig. 5.1.

On the other hand, for simplicity, we kept the kinematic viscosity $\nu$ and thermal diffusivity $\kappa$ constant within the same model. In Sect. 5.2.9, we return to the impact and caveats of this choice and the related assumption about the Prandtl numbers (Sect. 5.1.2).

### 5.1.2 3D numerical dynamo model

The next step is fundamental: We performed 3D MHD spherical shell simulations using the public code *MagIC*[3] with the anelastic approximation (Gastine and Wicht, 2012). In *MagIC* all the quantities marked with a tilde (Sect. 5.1.1) are static in time, radially dependent, and normalized to their outer values, except for $1/\lambda$, which, because of its fast decay in the outer radial regions, is normalized to its innermost value for practical reasons.

The shell is filled with a finitely conducting fluid rotating along the vertical axis $\hat{\boldsymbol{z}}$ with a constant angular velocity $\Omega$, and the background profiles, which we set up as explained in Sect. 5.1.1. The geometry is set by the aspect ratio $\eta := r_i/r_o$. We work in dimensionless units using the shell thickness $d := (r_o - r_i)$ as the length unit and the viscous diffusion timescale $d^2/\nu$ as the time unit. The magnetic fields are in units of $(\rho_o\mu_o\lambda_i\Omega)^{1/2}$, where $\lambda_i = 1/(\mu_0\sigma(r_i))$ is the magnetic diffusivity at the inner boundary. The convection is set by a fixed entropy gradient $\Delta s$, and this difference serves as the nondimensional units for $s$.

[2] A spurious contribution to Ohmic dissipation is produced very close to the outer boundary. We checked how relevant it is in terms of affecting the saturated solution by comparing representative runs with what is obtained by setting a thin perfectly insulating fluid layer in the outermost ≈1-2% (in radius) of the domain, by using the parameter *r_LCR* in *MESA*. In all cases, we found negligible differences in all relevant time-averaged diagnostics (Sect. 5.1.2) within the stochastic oscillations of the solutions.

[3] https://github.com/magic-sph/magic. We used the version 6.3: https://zenodo.org/records/3564626

The equations we solved were the mass continuity equation, the momentum equation, the entropy equation, the induction equation, and the solenoidal constraint for the magnetic field. These equations are already shown in Eqs. 4.12 - 4.15, alongside the dimensionless parameters in Eq. 4.11.

### Boundary conditions

We assumed stress-free and impenetrable boundary conditions for the velocity at the inner and outer radii, $r = r_i, r_o$,

$$u_r = \frac{\partial}{\partial r}\left(\frac{u_\theta}{r}\right) = \frac{\partial}{\partial r}\left(\frac{u_\phi}{r}\right) = 0 \ .$$

When stress-free boundary conditions are used at both boundaries, *MagIC* gives the option to ensure angular momentum conservation, which we have used. We employed constant entropy at both boundary conditions,

$$s'(r = r_0) = 0 \ , \quad s'(r = r_i) = 1 \ .$$

The material outside the outer radius was electrically insulating, that is, the magnetic field matched a potential field. At the inner boundaries, we imposed a perfectly conducting core.

### Numerical technique

*MagIC* solves the set of equations (4.12 - 4.15) with the above boundary conditions by expanding the mass flux and the magnetic fields into poloidal and toroidal potentials,

$$\tilde{\rho}\mathbf{u} = \boldsymbol{\nabla} \times (\boldsymbol{\nabla} \times W\,\mathbf{e_r}) + \boldsymbol{\nabla} \times Z\,\mathbf{e_r} \ ,$$
$$\mathbf{B} = \boldsymbol{\nabla} \times (\boldsymbol{\nabla} \times g\,\mathbf{e_r}) + \boldsymbol{\nabla} \times h\,\mathbf{e_r} \ .$$

The quantities $W$, $Z$, $g$, $h$, $s'$ and $p'$ are expanded up to $l_{\max}$ in spherical harmonic degree and $N_C$ in Chebyshev polynomials, where we always set $N_C = N_r - 1$. The equations are time-stepped by advancing nonlinear and Coriolis terms using an explicit second-order Adams-Bashforth scheme, and the remaining terms are time-advanced using the implicit Crank-Nicolson algorithm (for more details, see Glatzmaier (1984); Christensen and Wicht (2007)).

### Diagnostic parameters

To characterize the numerical dynamo solutions, we made use of several diagnostic quantities. We typically took averages in time or in space, either over the whole volume $V$ or over spherical surfaces, to show the radial dependence,

$$||a||\,(r,t) = \int a(r,\theta,\phi,t)\, sin\theta\, d\theta\, d\phi \ ,$$
$$\langle a\rangle(t) = \frac{1}{V}\int a(r,\theta,\phi,t)\, dV \ , \quad \bar{a} = \frac{1}{\Delta t}\int_{t'}^{t'+\Delta t} a(t)\, dt \ .$$

We performed the time averages after a stationary state had been reached, and we typically monitored the dimensionless hydrodynamic and magnetic Reynolds numbers, the Rossby number, and the Elsasser number:

$$\mathrm{Re} = \overline{\sqrt{\langle u^2\rangle}} \ , \quad \mathrm{Rm} = \overline{\frac{1}{V}\int_{r_i}^{r_o} \frac{\sqrt{||\mathbf{u}^2||}}{\tilde{\lambda}} r^2 dr} \ ,$$
$$\mathrm{Ro} = \frac{\mathrm{Rm}\ \mathrm{E}}{\mathrm{Pm}} \ , \quad \Lambda = \overline{\left\langle \frac{\mathbf{B}^2}{\tilde{\rho}\tilde{\lambda}} \right\rangle} \ .$$

The total kinetic and magnetic energy, which are in units of $\rho_0 d^5 \mathrm{E}^2 \Omega^2$, were

$$E_{\rm kin} = \frac{1}{2}\overline{\langle \tilde{\rho} \mathbf{u}^2 \rangle}\,, \qquad E_{\rm mag} = \frac{1}{2}\overline{\frac{1}{\mathrm{EPm}}\langle \mathbf{B}^2 \rangle}\,.$$

We defined the dipole fraction, $f_{\rm dip} = E_{\rm mag,l=1}/E_{\rm mag}$, as the ratio of the magnetic energy stored in dipolar components (axisymmetric and nonaxisymmetric), divided by the total magnetic energy.[4]

To study the energy dissipation, we used the buoyancy power,

$$P_\nu(t) \equiv \frac{\mathrm{RaE}}{\mathrm{Pr}}\langle \tilde{\alpha}\tilde{T}\tilde{g}s' u_r \rangle \ . \tag{5.8}$$

We used the subindex $\nu$ to emphasize that the quantity was calculated in viscous timescales. For comparison with other works, we also used the rotation timescale (see Sect. 5.2.10). For a well-resolved numerical run, the buoyancy power must be equal to the sum of viscous and Ohmic dissipation rates after a steady-state solution is reached. These are defined as

$$D_{\rm visc}(t) \equiv \langle S^2 \rangle\,, \qquad D_{\rm ohm}(t) \equiv \frac{1}{\mathrm{E\,Pm}^2}\langle \tilde{\lambda}\,(\boldsymbol{\nabla}\times\mathbf{B})^2 \rangle\,. \tag{5.9}$$

Another quantity to monitor is the fraction of energy dissipated by Joule heating alone, that is, the Ohmic fraction $f_{\rm ohm} = D_{\rm ohm}/P_\nu$. After a statistically steady state has been reached, the input buoyant power must balance the viscous and Ohmic diffusion. To evaluate whether the numerical solution has good time and spatial invariance and also if the background state affects the energy balance strongly, we assessed the power imbalance by its proxy $f_P = |\overline{P_\nu} - \overline{D_{\rm visc}} - \overline{D_{\rm ohm}}|/\overline{P_\nu}$.

Finally, we studied the time-averaged kinetic and magnetic spectra, that is, the distribution of the energy over different multipoles of order $\ell$, which *MagIC* already implemented as a user-friendly output. We inspected the spectra for each model, in particular, to ensure that the resolution was high enough to resolve the maximum dissipation, that is, $\ell(\ell+1)E(l)$, for the volume-integrated spectra and 2D spectra taken at relevant radii, for instance, $r_i$, $r_o$, $r_{\rm m}$.

### Parameter evolution and model descriptions

As explained in 5.1.1, when the *MESA* profiles were cut close to the desired $N_\rho$, we extracted $\Delta T$, $r_o$, $r_i$, and $r_{\rm m}$, from which we deduced $\eta$ and $\chi_m := r_m/r_o$. The corresponding physical values can be recovered by knowing the units in which each quantity is expressed and the values from the *MESA* profile, for example, using the thickness of the physical shell thickness $d_{phys} = r_{o,phys}(1-\eta)$. The real quantities of the planet, which reflect the evolutionary changes, were used to evolve the dynamo parameters. Since, as mentioned above, the real physical values E and Ra are computationally inaccessible, we can still use their dependence on the physical values that change during the long-term evolution. In particular, the shell thickness $d = r_o - r_i$ and the temperature difference $\Delta T$ enter the definition of the Ekman, $\mathrm{E}(t) \approx d(t)^{-2}$, and Rayleigh numbers, $\mathrm{Ra}(t) \approx d(t)^3 \Delta T(t)$. Therefore, we considered a series of ages, for which, after having found a suitable pair of $\mathrm{E}_0$ and $\mathrm{Ra}_0$ that produces convection and dynamo for the setup with $d_0$ and $\Delta T_0$ corresponding to a given age, the values $\mathrm{E}'$ and $\mathrm{Ra}'$ of the remaining models in that series were set up by scaling with $d(t)$ and $\Delta T(t)$,

$$\mathrm{E}' = \mathrm{E}_0 \frac{d_0^2}{d'^2}\ , \quad \mathrm{Ra}' = \mathrm{Ra}_0 \frac{d'^3 \Delta T'}{d_0^3 \Delta T_0}\ .$$

[4] Our definition differs from another widely used definition, that is, the ratio of the axisymmetric dipole component to the magnetic energy in the spherical harmonic degrees $l \leq 12$ at $r_o$, and the total, for instance, Christensen and Aubert (2006).

We made two assumptions: (i) The diffusivities at a given radius remain constant in time, which implies that we considered the same values of Pr and Pm along a sequence and that the change in E only comes from the contraction of the planet; and (ii) planetary rotation is constant in time. The latter assumption is well justified when we consider the possible relevant torque acting on a gas giant. Batygin (2018) studied the evolution of rotation, considering the magnetic coupling between the planetary interior and the quasi-Keplerian motion of the disk in the planetary formation stages. This resulted in efficient braking of the planetary spin that ceased to evolve after ≈1 Myr, when it reached a terminal rotation rate (probably similar to that of Jupiter), which can hardly change later. Our earliest model is at 100 Myr, for which the rotation can be safely considered constant. In this sense, cold giants are expected to be fast rotators (Ro $< 0.12$), and they might host planetary dynamos similar to those found in dipole-dominated numerical solutions with very low E (Davidson, 2013; Yadav et al., 2016; Schwaiger et al., 2019). In this scenario, there is a quasi-geostrophic balance (Coriolis and pressure forces) at the largest scales, followed by an ageostrophic magneto-Archimedean-Coriolis balance (Coriolis, buoyancy, and Lorentz forces).

With these assumptions, we considered five sets of dynamo models: (*i*) A long series with a total of 12 evolutionary stages, ranging from 0.1 to 10 Gyr for a 1 $M_J$ planet; (*ii*) different density ratios, that is, different cutoff radii, for the same 1 $M_J$ model at 1 Gy; (*iii*) different planetary masses with $N_\rho \approx 3.0$; (*iv*) several models with Pm and Pr different from 1; and (*v*) a 4 $M_J$ mass series with $N_\rho \approx 4.6$. When a different mass was chosen, the radial profiles changed (see Fig. 5.3), so that using Eq. (5.7) with the above-mentioned values of the free parameters $a$ and $\tilde{\sigma}_m$, we had to adapt the density contrast $N_\rho$ to include the drop in the conductivity in the outer layers of our shell, without at the same time, having too low values of $\sigma$. For this reason, the series of 4 $M_J$ has a higher $N_\rho$ (the exponential drop of $\sigma$ would have been cut out with $N_\rho \approx 3$). For the same reason, the 0.3 $M_J$ model has a lower contrast, $N_\rho \approx 1.1$. The alternative would have been to consider an equally arbitrary change of the free parameters in Eq. (5.7). In Table E.1 we show the input values for the 3D simulations together with the parameters of the background profiles from the *MESA* 1D long-term evolution.

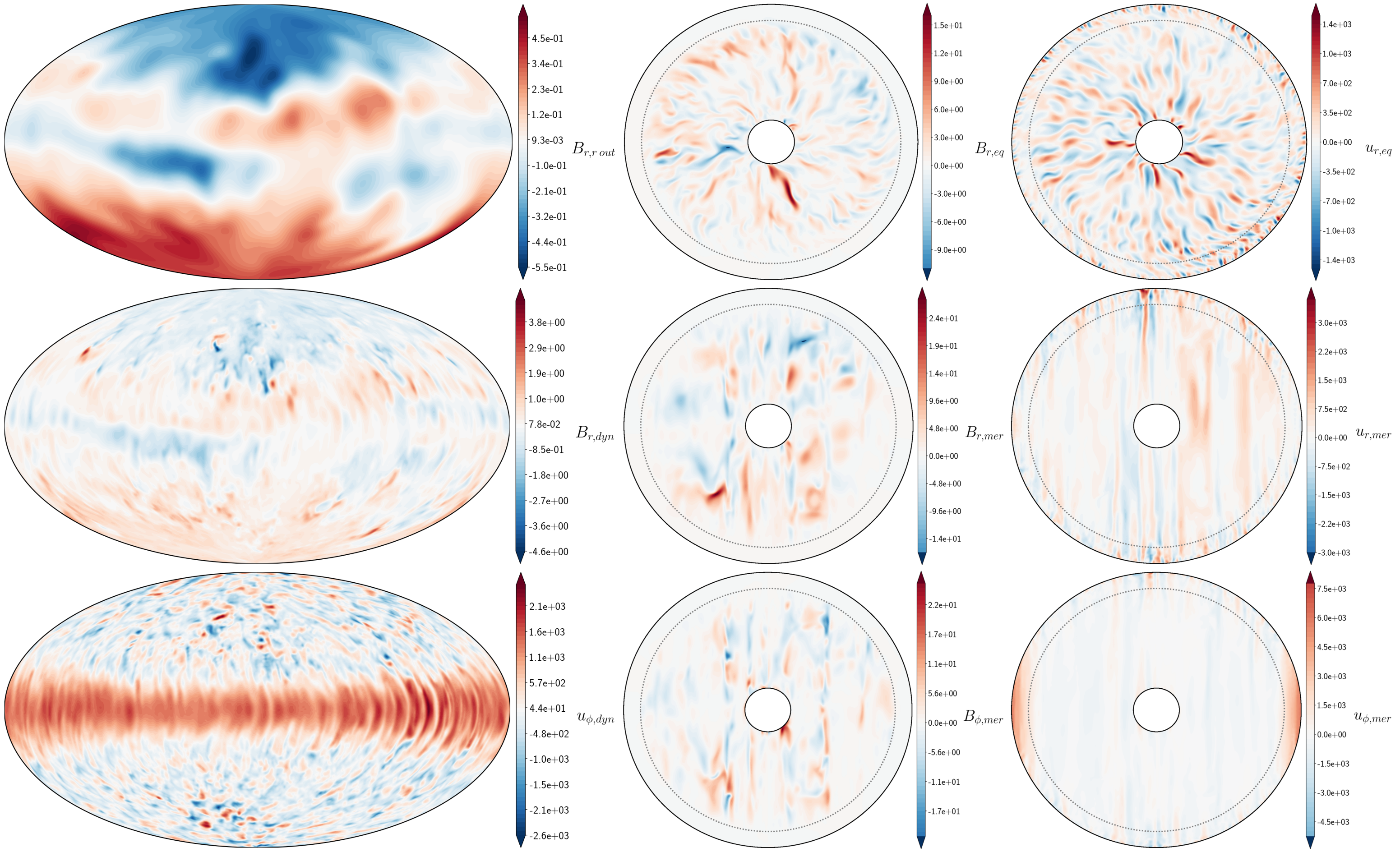

Figure 5.4: Snapshots of the saturated solution for the representative 1 $M_J$ 1 Gyr model. *Left column:* Maps of $B_r$ at the outermost layer of our domain, $B_r$ at $r = r_m$, $u_\varphi$ at $r = r_m$, from top to bottom. *Center:* Equatorial slice of $B_r$, meridional slice of $B_r$, and meridional slice of $B_\varphi$. *Right:* Equatorial slice of $u_r$, meridional slice of $u_r$, and meridional slice of $u_\varphi$. In the central and right panels, the location of $r_{\rm m}$ is marked with dotted gray lines in the equatorial and meridional cuts. The color bars indicate the values in code units.

## 5.2 Results

### 5.2.1 Preliminary exploration of parameters

Our first goal was to determine the dependence of the dynamo solutions on mass and age for a given evolutionary sequence of background thermodynamic setups (Table E.1). Therefore, the first requirement was to determine a range of parameters for which both convection and dynamo operated for all models. This practically meant that we needed to find a feasible range of Ra and E (which, within a sequence of models, have relative variations set by $\Delta T$, $T_o$ and $d$, Table E.1), for which convection and dynamo action are present in the entire sequence, keeping in mind that the chosen values were orders of magnitude away from the realistic values, as mentioned above. To determine these feasible ranges, we needed to perform a preliminary parameter exploration. At the same time, we assessed the sensitivity of results on other parameters. We summarize in this subsection the four main steps we took for this overall assessment.

First, to locate the region with viable dynamo solutions, we performed low- and medium-resolution runs for one specific model, the model with 1 $\mathrm{M_J}$ 10 Gyr. This is the oldest and coldest model of the 1 $\mathrm{M_J}$ sequence, that is, with the lowest value of $\Delta T$, which also implies the lowest value Ra of the series, that is, it is least favorable to convection. We spanned the ranges $10^{-5} < \mathrm{E} < 10^{-3}$, $10^{6} < \mathrm{Ra} < 10^{10}$ with Pm=Pr=1 and a relatively low resolution, $(N_r,N_\theta,N_\phi)$ = (193,192,384). For $\mathrm{E} \leq 10^{-4}$, we obtained convection for $\mathrm{Ra} \gtrsim 10^{7}$, and, additionally, magnetic field growth for $\mathrm{Ra} \gtrsim 5{\cdot}10^{7}$.

Second, for the 1 $M_J$ 10 Gyr case with $\mathrm{E}{=}10^{-5}$, $\mathrm{Ra}{=}5{\cdot}10^{8}$ (the reference model in the rest of this subsection), we explored the Prandtl numbers in a relatively easily accessible range 0.25 $< \mathrm{Pm}, \mathrm{Pr} <$ 4. This was done to discuss the impact of our assumptions of constant-in-time diffusivities, Sect. 5.1.1). The results of this exploration are shown in Sect. 5.2.9 for high-resolution models and different evolutionary ages.

Third, for the same reference model ($\mathrm{E}{=}10^{-5}$, $\mathrm{Ra}{=}5{\cdot}10^{8}$, Pm=Pr=1), we explored the sensitivity to the parameters $a$ and $\sigma_m$, which define the slope of $\tilde{\lambda}(r)$ in the outermost layers. For a wide range of values (i.e., $0.07 < \sigma_m < 0.9$, $5 < a < 15$), we indeed recovered the results of Duarte et al. (2013), that is, the strong equatorial jet remains confined to the weaker conducting outer region and does not interfere with the deeper dynamo action. As previously mentioned, we finally opted for rather low values for both $\sigma_m$ of 0.1 and $a$ of 7. These values are similar to those used by Gastine and Wicht (2021), whose model approximately reproduces the French et al. (2012) profiles. For numerical stability reasons, we did not use higher values of $a$, that is, steeper exponential drops, because the hydrogen metallic region of some of our models is already quite deep ($\chi_m >$0.85), which leads to too high values of $\tilde{\lambda}(r)$ near the outer surface. Similarly, we studied the effect of $N_\rho$ on several diagnostic quantities (see Sect. 5.2.3 for more details). Given the non-negligible effects of choosing different values of $N_\rho$, we compare below models with the same $N_\rho$, to avoid additional biases.

Fourth, we tested different boundary conditions again for the same reference model. Integrated quantities such as Rm or $E_{\mathrm{kin}}$ and convection patterns did not change appreciably when rigid boundary conditions were applied at the inner core. The radial distributions showed a drop in velocity in a very thin region (< 1%) of the radius. For the magnetic field, we tested for a perfect conductor and insulator for the inner core and an insulating, perfect conductor and pseudo-vacuum at the outer radii. We found no relevant differences in the internal dynamo.

Considering the explored values of Ra and E for these low-resolution test, we set $\mathrm{Ra} = 1.3{\cdot}10^{-5}$

and E $\approx 5 \cdot 10^8$ for the 1 $M_J$ 10 Gyr model with a $N_\rho \approx 3$. This choice set the rescaled values for the remaining sequence in a range that allowed dynamo action. Using the definition of E, Ra, we scaled their values in each run with the corresponding values of $T$, $\Delta T$, $d$, as described in Sect. 5.1.2. Most models use a resolution of $(N_r,N_\theta,N_\phi) = (289,256,512)$. As an exception, the 4 $M_J$ $N_\rho \approx 4.6$ runs use a grid of (385,320,640).

Finally, we employed a strategy to save computational time that was inspired by other spherical shell dynamo works (Christensen and Aubert, 2006). The idea is that the final solution does not depend on whether the initial conditions are taken from another saturated dynamo model or if they are the usual $\mathbf{u} = 0$ with a weak perturbation of both $\mathbf{B}$ and $T'/s'$. This suits our case in particular because the relative changes in the dynamo parameters from one setup to the next are small, and the solution of the new setup is reached much faster than starting from a $\mathbf{u} = 0$ state. In Sect. 5.2.4 we show the results of some specific numerical experiments that support this strategy.

We now proceed to the main results. We refer to App. E for a table with detailed values of the time-averaged output nondimensional numbers and the other quantities we used as diagnostics.

### 5.2.2 Dynamo solutions: general behavior

In Fig. 5.4 we show snapshots (maps and slices of some velocity and magnetic field components) of the 1 $M_J$ 1 Gyr saturated dynamo solution, which is a representative case. The other models we obtained are qualitatively similar to each other in terms of the morphology of the velocity and magnetic fields. The largest differences are the relative average strengths of $\mathbf{u}$ and $\mathbf{B}$ and the magnetic field dipolarity, which we discuss below. All the models have a strong equatorial flow that reaches deep down to the dynamo region. The velocity and magnetic fields show a westward drift in the inner parts. The magnetic field is mostly constrained under $r < r_m$, where convection is also stronger, as seen in $u_r$. As expected from rotation-dominated convection, we found columnar structures in the direction of the rotation axis (Zhang and Busse, 1987; Ardes et al., 1997; Simitev and Busse, 2003), as shown in the meridional slices. Generally speaking, our numerical solutions are similar to those reported by other works for Jovian-like dynamos with a nonconstant electrical conductivity (Jones, 2014; Duarte et al., 2018).[5]

As mentioned above, we can show in Fig. 5.6 both kinetic and magnetic spectra for at $r_i$, $r_o$, and $r_{\rm dyn}$ for one of the specific backgrounds, which is representative of our simulations. We do indeed see that the magnetic spectra are always resolved by two orders of magnitude anywhere in the domain, the rule-of-thumb indication of a good resolution. Kinetic spectra are also resolved in the convective region and dynamo boundary but are close to the limit of the rule-of-thumb at the outer boundary $r_o$, i.e. there are $\sim 2$ orders of magnitude between the lowest ($10^7$, $l = 1$) and the highest ($10^5$, $l = 200$) multipoles. In that sense, one can think that the boundary layer is (slightly) underresolved at the outer surface, but this should have a negligible impact on the paper's main results, as the region where the magnetic field is generated is well resolved.

The dynamo solutions we found are generally less dipole-dominated than their incompressible (Boussinesq) counterparts with similar dynamo parameters (i.e., Ra, E, Pr, and Pm). However, we note that we did not explore values for Pr lower than 0.1 with Pm $> 1$, where dipole-dominated solutions have been found (Jones, 2014; Tsang and Jones, 2020). We restricted ourselves to a less demanding parameter space with a wider liberty of parameter exploration, but with the caveat that we might obtain less dipole-dominated models. Similarly to Yadav et al. (2013), for all runs shown here, we also obtained Nu $> 2$ at the top and bottom surfaces, which ensures a fully developed convection. The Nusselt number Nu is the ratio of the total transported heat flux to the conducted heat flux.

[5] They use a different code with a strictly isentropic background profile that fits $T(r)$, $\rho(r)$ and $\sigma(r)$ from French et al. (2012).

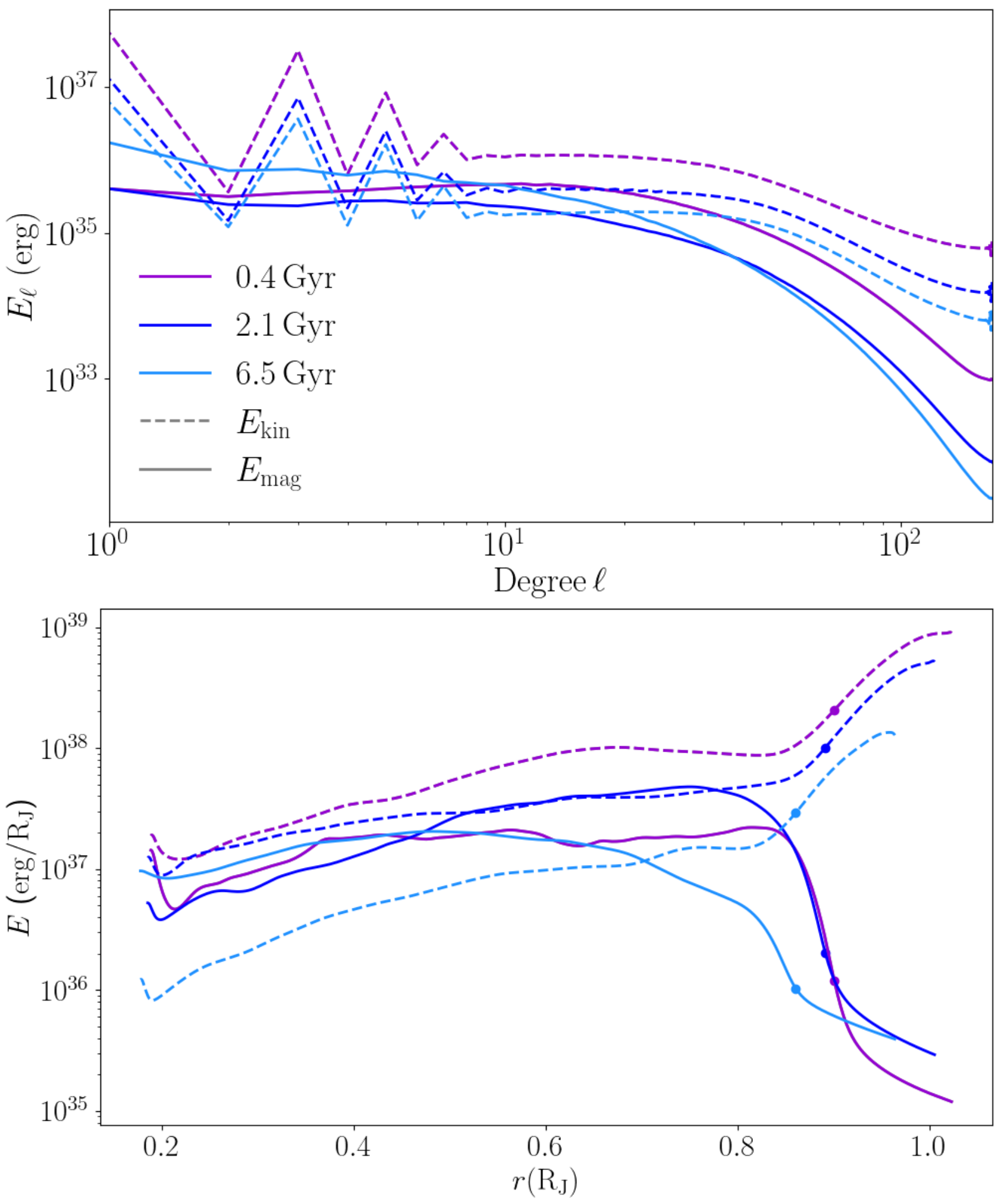

Figure 5.5: Magnetic (solid) and kinetic (dashes) energy distribution over the multipole degrees $l$ (up), and the radius (down), for three models representing the same 1 $M_J$ planet at different evolutionary stages (0.4, 2.1, and 6.5 Gyr). Spectra have been averaged in time over the saturated state. The location of $r_m$ is marked with a dot in the radial plots. The physical units are obtained by multiplying by the factor $\rho_0 d^5 E^2 \Omega^2$, where $\rho_0$, $d$ and E depend on each model and $\Omega = 1.76{\cdot}10^{-4}$, that is the Jovian value.

Moreover, for gas giants, the definition of the dynamo surface is not absolute. As hydrogen gradually transitions outward from metallic to molecular, the electrical conductivity and electrical currents are quickly (but not abruptly) damped over a finite region. For our models, the most obvious choice for the dynamo surface is the radius at which the exponential decay for $\sigma$ starts, that is, $r_m$. To obtain a more physically justified definition for the dynamo surface, we followed Tsang and Jones (2020) by computing the magnetic energy spectra at different radii, $F_l(r)$. They defined the dynamo surface, $r_{dyn}$, as the radius within which the slope of the Lowes spectrum (i.e., a potential solution extrapolated back from the outermost layer to the interior) diverges from the slope of the simulated $F_l(r)$. A similar analysis for some of our models is shown in Sect. 5.2.5, where we find that this definition of the dynamo surface always gives values very close to $r_m$.

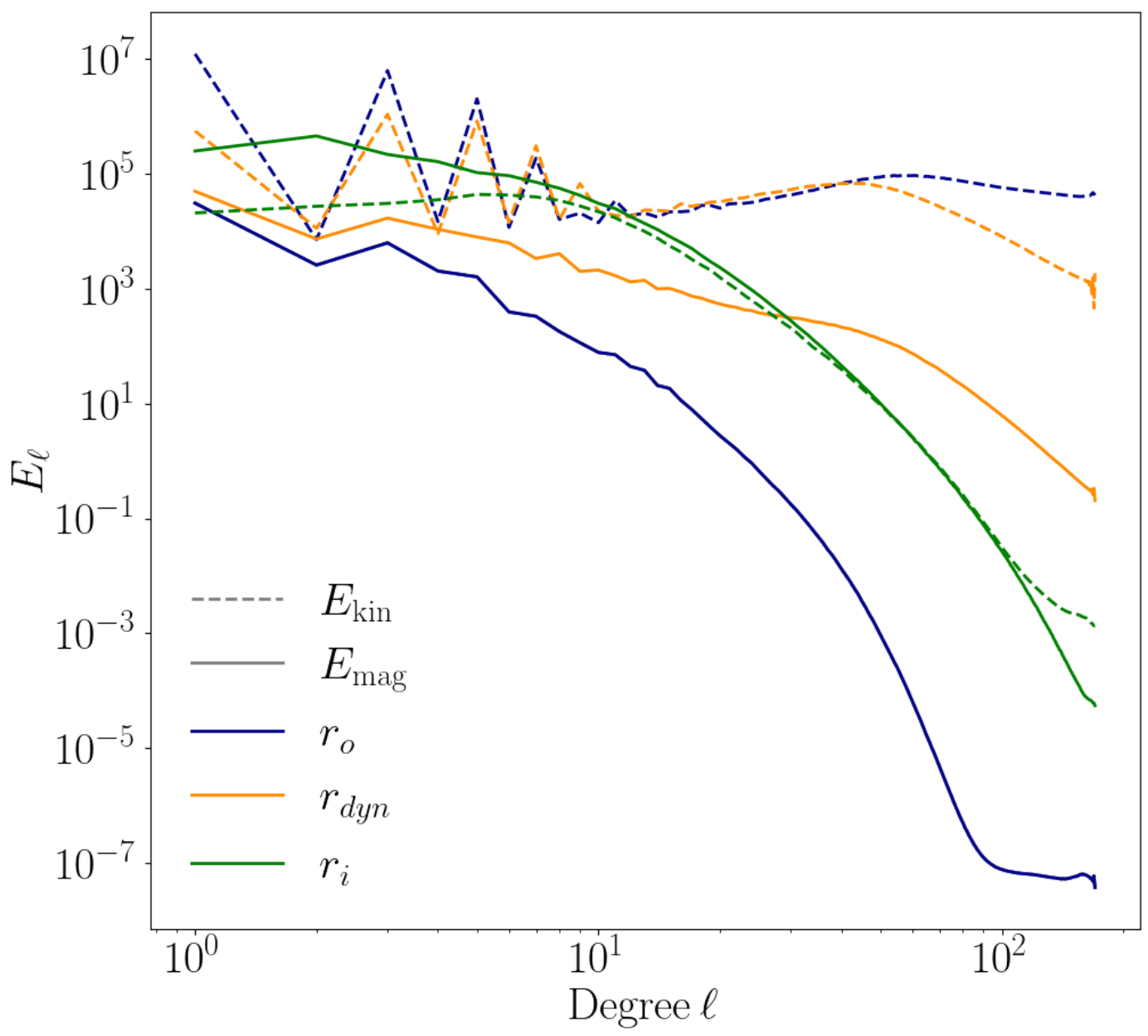

Figure 5.6: magnetic (solid) and kinetic (dashes) 2D energy distribution over the multipole degrees $l$, at relevant radii $r_o$, $r_{\rm dyn}$, $r_i$ for the representative 1 $\rm M_J$ 6.5 Gyr model. Energy is given in code units.

### 5.2.3 Sensitivity on the density ratio

To evaluate how the external cut applied to the 1D *MESA* profile influences the overall dynamo behavior, we compare the dynamo corresponding to $N_\rho \approx 1.1, 3.0, 3.7, 4.6$ (i.e., $\rho_o/\rho_i \approx 10, 20, 40, 100$) for the $1\rm M_J$ 1 Gyr model keeping the same 3D resolution of $(N_r, N_\theta, N_\phi) = (289,256,512)$. By looking at the spectra, the model with $N_\rho \approx 4.6$ seems slightly under-resolved (there is an overall drop of only 1 order of magnitude, less than what indicates a large enough grid), but the overall quantities seem to follow the same trend as with the other distinct $N_\rho$ models. We could not explore properly higher values of $N_\rho \gtrsim 5.3$, due to the excessive resolution required. In Table 5.1 we show the different diagnostics.

Table 5.1: General outputs for the 1 $\rm M_J$ 1 Gyr models with different densities.

| $N_\rho$ | 2.30 | 2.99 | 3.68 | 4.58 |
|---|---|---|---|---|
| Rm | 1097 | 732 | 461 | 301 |
| Ro | $1.35{\cdot}10^{-2}$ | $8.20{\cdot}10^{-3}$ | $4.98{\cdot}10^{-3}$ | $3.16{\cdot}10^{-3}$ |
| $\Lambda$ | 1.87 | 0.67 | 0.172 | 0.072 |
| $P$ | $8.74{\cdot}10^{10}$ | $7.16{\cdot}10^{10}$ | $5.10{\cdot}10^{10}$ | $3.83{\cdot}10^{10}$ |
| $E_{\rm mag}/E_{\rm kin}$ | 0.207 | 0.146 | 0.090 | 0.087 |
| $f_{\rm ohm}$ | 0.283 | 0.190 | 0.101 | 0.066 |
| $E_{\rm kin}$ (code) | $1.48{\cdot}10^{7}$ | $1.62{\cdot}10^{7}$ | $1.54{\cdot}10^{7}$ | $1.58{\cdot}10^{7}$ |
| $E_{\rm kin}$ (erg) | $1.39{\cdot}10^{38}$ | $9.3{\cdot}10^{37}$ | $4.80{\cdot}10^{37}$ | $1.82{\cdot}10^{37}$ |
| $E_{\rm mag}$ (code) | $3.07{\cdot}10^{6}$ | $2.37{\cdot}10^{6}$ | $1.38{\cdot}10^{6}$ | $1.37{\cdot}10^{5}$ |
| $E_{\rm mag}$ (erg) | $2.90{\cdot}10^{37}$ | $1.37{\cdot}10^{37}$ | $4.32{\cdot}10^{36}$ | $1.58{\cdot}10^{36}$ |
| $f_P$ (%) | 0.18 | 0.0033 | 0.31 | 0.43 |

Generally, many of the dimensionless quantities are affected by the value of $N_\rho$. Except for the $N_\rho \approx 100$ model, the overall kinetic energy seems to plateau (in code units). However, since a dominant fraction of the kinetic energy is located in the nonconductive outer layers (see Fig. 5.5), all magnitudes containing $u_{rms}$ may differ substantially. As we capture a higher density ratio, the aforementioned zonal flows gain importance and compete against the magnetic field in the interior, affecting the overall dynamics. For example, even though Ra increases, the total buoyant power, $P$, decreases because it depends only on $u_r$. Similarly, Rm and Ro also decrease even though they depend on $u_{rms}$. This is due to the decreasing dimensionless conductivity $1/\tilde{\lambda}$, which erases the zonal flow contribution and only captures the suppression of the internal convection with higher $N_\rho$. Consequently, the Elsasser number $\Lambda$ and $E_{\rm mag}$, also $N_\rho$, also decrease for the same suppression reasons. Correcting factors for the dimensionless mass $M$ and volume $V$ do not mitigate the differences among the $N_\rho$ series.

To overcome this systematic effect, we tend to compare models with a similar $N_\rho$, independently of the mass and age of the model. For most models, we choose $N_\rho \approx 3$, as it is more computationally feasible but still has a relevant nonconductive outer layer where the zonal jet develops. This restriction allows us to analyze the different saturated models for an evolutionary sequence.

### 5.2.4 Initial conditions from previous models

To reduce computational resources and avoid starting each model with random initial conditions, we have usually employed already saturated solutions as initial conditions for other models. As we are interpreting the different models as stages in planetary evolution, an obvious choice would be to use, for example, the saturated state of the 1 $M_J$ 0.5 Gyr model as initial conditions for the $M_J$ 0.7 Gy, and so on.

As a proof of concept, we made two tests: we used the final saturated state of a 0.5 Gyr model as initial conditions for a 1 Gyr model, and the same for a 1 Gyr and 10 Gyr models. In Fig. 5.7 we show the kinetic and magnetic energy time series for these transitions. The steady states reached are indistinguishable, that is, one cannot discern from the spectra, radial distribution, or final diagnostics in which initial conditions were used. The only quantities that showed a noticeable difference are the dipolarity indicators ($f_{\rm dip}$), but they are within one standard deviation of each other. To obtain satisfactorily similar means, much longer computing times would probably be required. Overall, there is convergence, starting from different initial conditions.

The activation of convection, followed by the dynamo kinematic phase, and finally the saturated phase, where Lorentz forces become relevant, is a lengthy computation. Starting from an already saturated solution, not too far from the expected one, highly reduces by more than a factor of 5 the computing time needed to reach the new steady state. Thus, in almost all models, we have used a high Ra model as initial conditions, specifically the saturated 1 $M_J$ 0.5 Gyr model with Pm = Pr = 1.

### 5.2.5 Spectral-radial distributions and dynamo surface definition

To compare our results with (Tsang and Jones, 2020), we obtained the spectral-radial energy distribution. In the left and middle panels of Fig. 5.8, we show these spectra for the saturated 1 $M_J$ 2.1 Gyr model. The radial integration of both spectra is already shown in Fig. 5.5. It can be observed that the radially dependent magnetic energy spectra, $F_l(r)$, decay rapidly for $r > r_m = 0.88\ R_J$. At the same region, the kinetic spectra start showing the equatorial jet pattern.

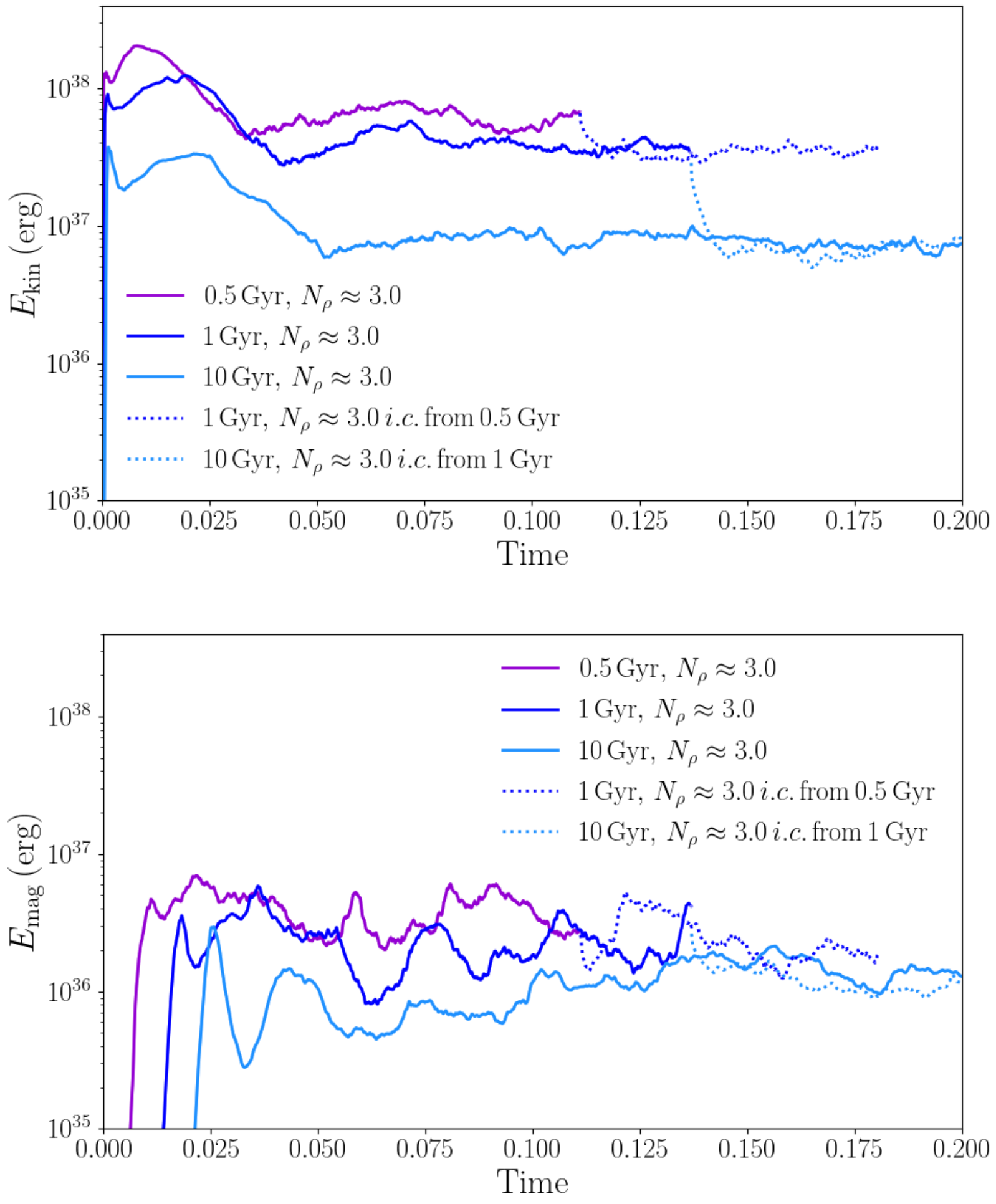

Figure 5.7: Kinetic (top) and magnetic (bottom) energy evolution time series for the 1 MJ models at 3 different ages (different colors). Solid lines are simulations starting from a $\mathbf{u} = 0$ initial condition, while the dotted lines take as initial condition a snapshot of the saturated solution of another model. Time is in viscous units.

In the region where $\mathbf{J} = 0$, we define the scalar potential $V$, for which $\mathbf{B} = -\nabla V$, with its usual spherical expansion shown in Eq. 4.1, which the spherical expansion coefficients $g_l^m$ and $h_l^m$ defined at the planetary surface $a$. As mentioned in Sect. 4.3, one can obtain the so-called Lowes-Mauersberger magnetic spectrum (Mauersberger, 1956; Lowes, 1974) at any distance $r$ from the planetary center:

$$R_l(r) = \left(\frac{a}{r}\right)^{2l+4} R_l(a) \ , \quad \text{where} \quad R_l(a) \equiv (l+1) \sum_{m=0}^{l} \left[ (g_l^m)^2 + (h_l^m)^2 \right] \ .$$

This expression gives the Lowes spectrum in the interior of the planet, that is, the downward extrapolation which would be valid in the absence of electrical current (potential field). Therefore, we expect a true Lowes spectrum in the domain outside the dynamo surface, since inside it $\mathbf{B}$ stops being potential. In other words, in the molecular insulating region as well as the outside vacuum, the potential field spectrum coincides with the poloidal field spectrum, which means that there is no toroidal field. Moreover, from equipartition reasons, it is usually assumed that the magnetic field in the dynamo region is equally distributed among different scales until the diffusive scale ('white source hypothesis', Backus et al. 1996). A flat spectrum at the dynamo surface would imply that at the planetary surface, there will be a linear relation $log_{10} R_l(r) \approx -\beta(r)$. For our saturated dynamo solutions, we define the Lowes spectrum as the magnetic spectra at the outermost radius, where we impose potential boundary conditions:

$$R_l(r_o) = F_l(r_o) \ .$$

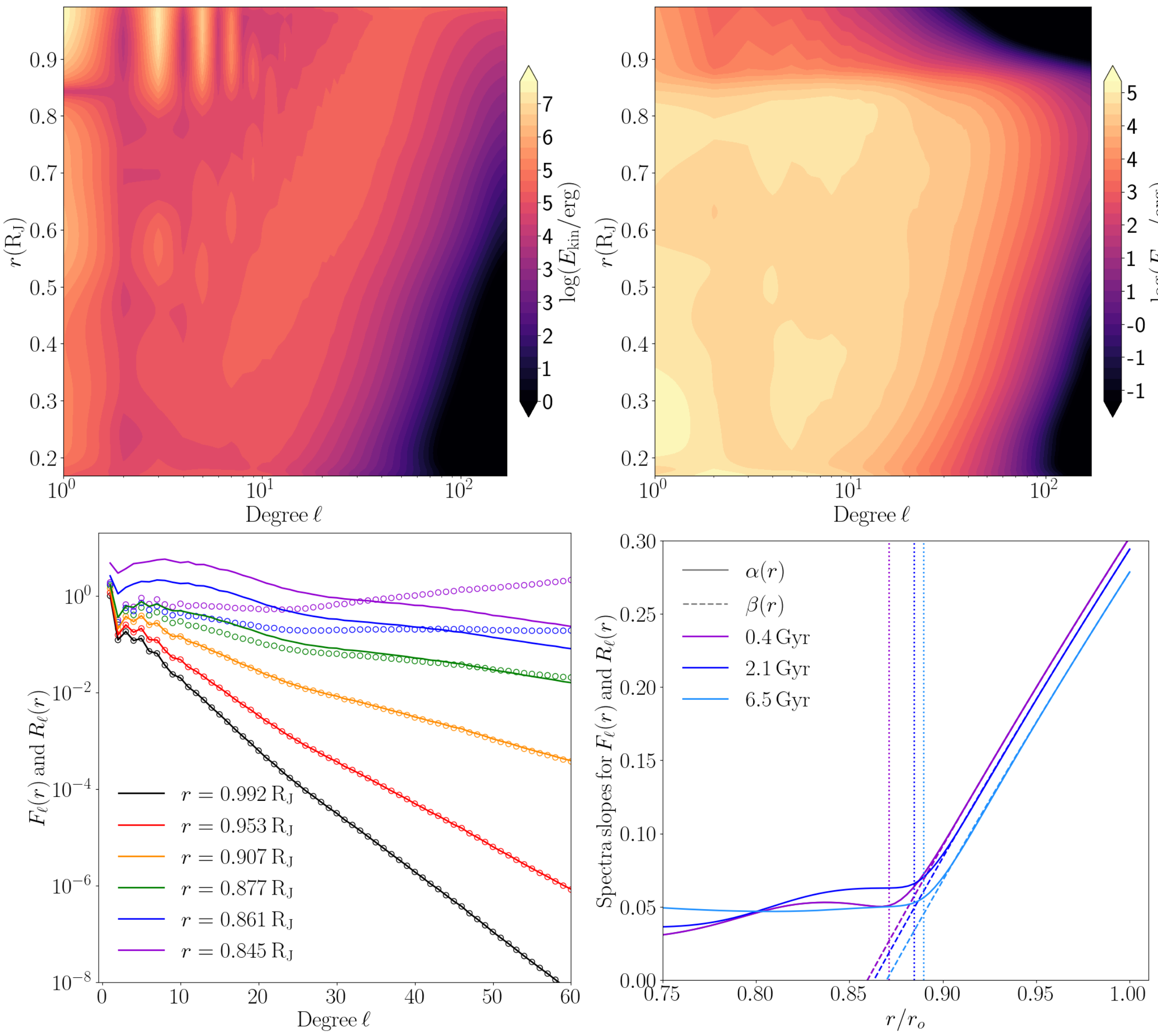

Figure 5.8: *Top*: Spectral-radial kinetic (left) and magnetic (central) energy distributions of the 1 $M_J$ 2.1 Gyr model. *Bottom left:* Lowes spectra $R_l(r)$ (circles) superimposed with $F_l(r)$ (solid lines) at different depths for the saturated dynamo solution of the same model. *Bottom right:* Spectral slopes at different dimensionless radial depths for the same runs as Fig. 5.5. The vertical lines are their respective $r_{\mathrm{m}}$.

On the bottom left panel of Fig. 5.8, we show both $R_l(r)$ and $F_l(r)$ at different radii for one specific saturated solution. The black line corresponds to $r_o$, and the others are at different depths. At some specific depth, $R_l(r)$ stops being similar to $F_l(r)$ and even has an unphysical negative slope. The Lowes radius is defined where $\beta(r) = 0$, and it is usually taken as the depth where the dynamo starts. As (Tsang and Jones, 2020) noted, $R_l(r)$ is already quite different from $F_l(r)$ at such radius. To surpass this discrepancy, they similarly defined the spectra slope for $F_l(r)$, i.e., $log_{10} F_l(r) \sim -\alpha(r)$. The slope $\alpha(r)$ is almost the same as $\beta(r)$ outside the dynamo region, but becomes more or less flat at positive values inside. They argue that the radius where $\alpha(r)$ and $\beta(r)$ show a discrepancy is where the effective dynamo surface is.

We similarly obtained the spectra slopes $\alpha(r)$ and $\beta(r)$ by a least square minimization between multipoles 10 and 50, which is the region where spectra are exponential and have not reached the dissipation scales set by our resolution. In the bottom right panel of Fig. 5.8, we show the resulting spectra slopes for the same models shown in Fig. 5.5. For some of our models, we do not obtain a flat spectrum inside. Still, if we take the average value of $\alpha(r)$ in the dynamo region and interpolate it for the decaying outer part, we obtain values very similar to the ones of $r_{\rm m}$, shown as vertical lines. Therefore, we can safely assume that the radial position where the conductivity starts the exponential decay, $r_{\rm m}$, is a good definition as the dynamo surface for our models.

### 5.2.6 Evolutionary changes

We now focus on the variation in the solutions along the longest sequence of models, the 1 $M_J$ planet with $N_\rho \approx 3$ and Pm = Pr = 1. The shell-averaged spectral distribution and radial distribution of the magnetic (solid lines) and kinetic (dashed lines) energy are shown in Fig. 5.5. We focus on three representative ages: 0.4, 2.1, and 6.5 Gyr.

In the left panel, the kinetic spectra show a drop of about 1.5 orders of magnitude or more from the integral to viscous scales, while the magnetic spectra decrease by at least 3 orders. The sawtooth shape on the lowest multipole side is associated with the external jet that we see in all of our runs. This behavior is not seen for the $m$ spectrum, which is dominated by the zonal flows, $m = 0$. For higher multipoles, the spectrum plateaus before it reaches the viscous scale and drops off. A comparison of the three different models shows that the overall shape of the kinetic spectra does not significantly change, other than a constant decrease in time throughout all harmonic degrees. The magnetic spectra show a similar diffusive scale, which is approximately located at the same $l$ as the viscous diffusive scale (compatible with Pr = Pm = 1), although the knee is less pronounced. With the usual measure of dipolarity, it ranges from $0.3 > f_{\rm dip,l<12}^{\rm axi,surf} = E_{\rm mag}(r_0)_{l=1,m=0}/E_{\rm mag}(r_0)_{l\leq 12} > 0.8$. Depending on the work, this could be considered multipolar or dipolar (Christensen and Aubert, 2006; Yadav et al., 2013; Zaire et al., 2022). The magnetic spectra show a clear evolution with age: The relative weights of high multipoles tend to decrease, while the strength of the large scales slightly increases. This inversion leads to an increase in the total dipolarity (see the discussion below).

The radial energy distributions, shown in the right panel of Fig. 5.5, shows that $E_{\rm kin}(r)$ increases almost monotonically outward, but the steepest changes occur in the outermost layers, $r \gtrsim r_m$, due to the appearance of the equatorial zonal wind, where the density is lower and the magnetic drag is weaker than in the interior. Similarly to the spectra, the kinetic radial distribution does not show a clear variation with age. The radial profile $E_{\rm mag}(r)$ and its change with age shown also in Fig. 5.5 is instead more complex. The radial profile peaks at radii slightly smaller than $r_{\rm m}$, after which it significantly drops, following the $\sigma(r)$ profiles (Fig. 5.3). A comparison of different evolutionary ages shows that the innermost region of the radial distribution does not show a clear trend, with a slight increase in the deepest regions for late-age models. On the other hand, the layers $\approx$ 10-20 % below $r_{\rm m}$ show a steady decrease with age (see Sect. 5.2.8).

The overall changes between the runs are gradual, and we found that a few models can already predict the general behavior. Several time and volume-averaged diagnostic quantities are shown as a function of evolutionary models in Fig. 5.10. As expected during the planetary cool-down of a gas giant, Rm decays approximately like a power law in time. Ro and $P_\nu$ also behave similarly, and they are therefore not shown to avoid repetition. All of these quantities are dependent on $u_{rms}$ or at least on one of its components. This reflects the fact that the mean velocity is dictated by the buoyancy input parameter Ra, which is proportional to the temperature difference in the convective shell. Fig. 5.10 also shows that the dipolarity $f_{\rm dip}$ does not seem to change significantly in time, except for a mild increase between 1.5 and 3.8 Gyr. This appears to indicate a transition between a multipolar or weakly dipolar-dominated regime to a strong dipolar-dominated regime. This increasing trend is obscure, with $f^{\mathrm{axi,surf}}_{\mathrm{dip,l<12}}$ (for the 1 $\mathrm{M_J}$ series, they take values from 0.3 to 0.8). As an alternative, we obtained the average $f_{\mathrm{dip}}$ over the volume 5% near the dynamo surface, $f_{\mathrm{dip,dyn}}$. Fig. 5.10 shows that $f_{\mathrm{dip,dyn}}$ grows gradually, with a similar jump seen in $f_{\mathrm{dip}}$.

This transition from multipolar to dipolar dynamo was also observed by Zaire et al. (2022) for dynamos in stratified stellar interiors, which they modeled with shallower shells ($r_i/r_o = 0.6$) and different $N_\rho$ and Ra. They obtained a threshold $\mathcal{F}_I/\mathcal{F}_L$ (i.e., the relative importance of the inertial over Lorentz forces) below which multipolar dynamos collapse into dipolar dynamos. They also reported that $E_{\mathrm{kin}}/E_{\mathrm{mag}}$ can equally well capture this magnetic morphology transition and found this transition at about $E_{\mathrm{kin}}/E_{\mathrm{mag}}$=0.7. In Fig. 5.9 we show both $f^{\mathrm{axi,surf}}_{\mathrm{dip,l<12}}$ and $f_{\mathrm{dip,dyn}}$ as a function of $E_{\mathrm{kin}}/E_{\mathrm{mag}}$ for the 1 $\mathrm{M_J}$, Pr = Pm = 1 series. We also report two distinctive populated areas, low dipolarity with high $E_{\mathrm{kin}}/E_{\mathrm{mag}}$ and high dipolarity with lower $E_{\mathrm{kin}}/E_{\mathrm{mag}}$. This abrupt change in the magnetic field morphology seems to be better reflected with $f_{\mathrm{dip,dyn}}$. A good definition for a dipole-dominated dynamo could be $f_{\mathrm{dip,dyn}} > 0.1$, which might be compatible with the definitions of Yadav et al. (2013) or Zaire et al. (2022) of $f^{\mathrm{axi,surf}}_{\mathrm{dip,l<12}} > 0.3$ and 0.5, respectively. These dipole-related quantities are usually the most fluctuating integrated quantities in saturated dynamo solutions because they are susceptible to the specific magnetic field configuration.

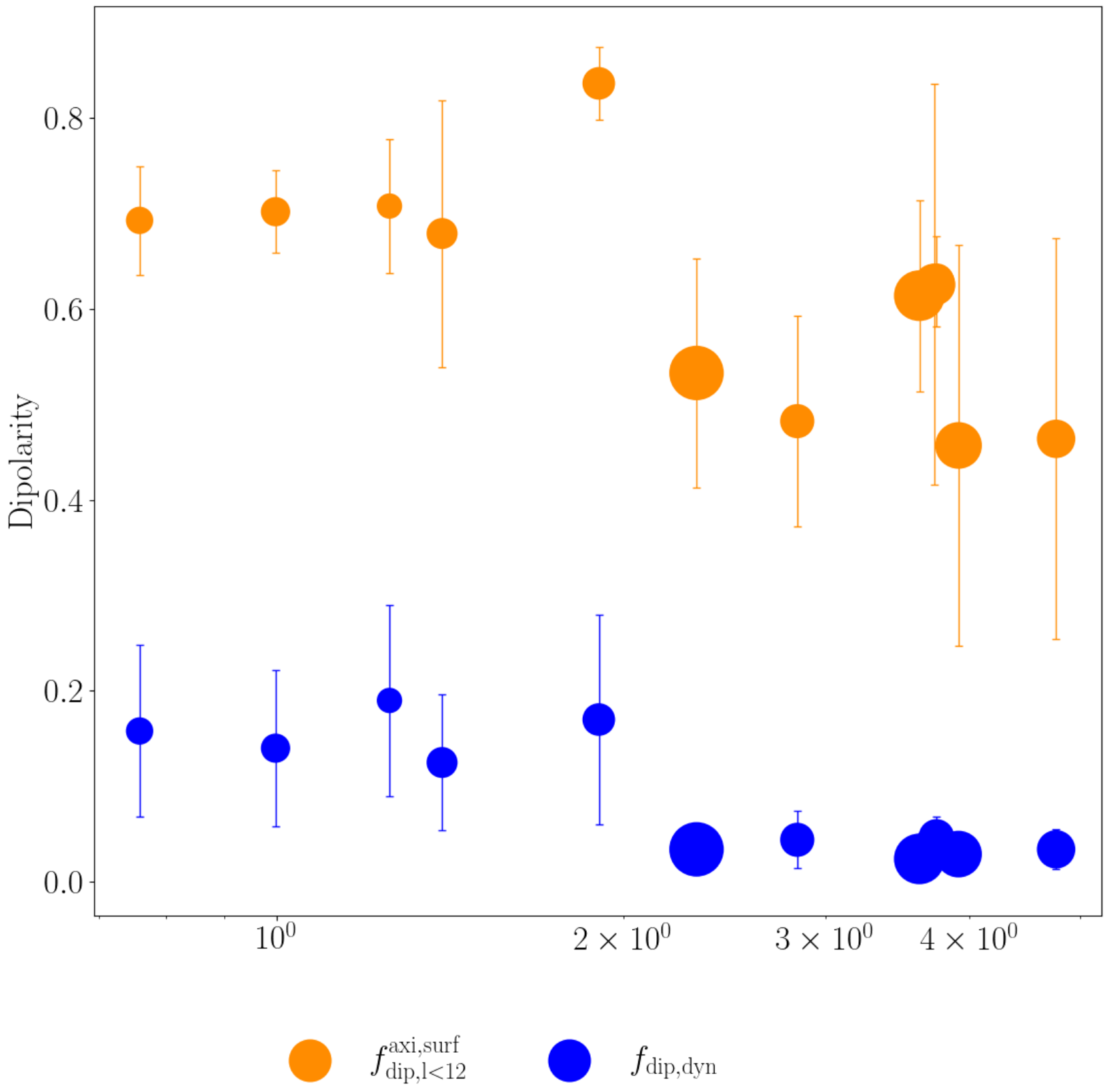

Figure 5.9: Dipolarity measurements as a function of the inverse of equipartition for the 1 $\mathrm{M_J}$, Pr = Pm = 1 models.

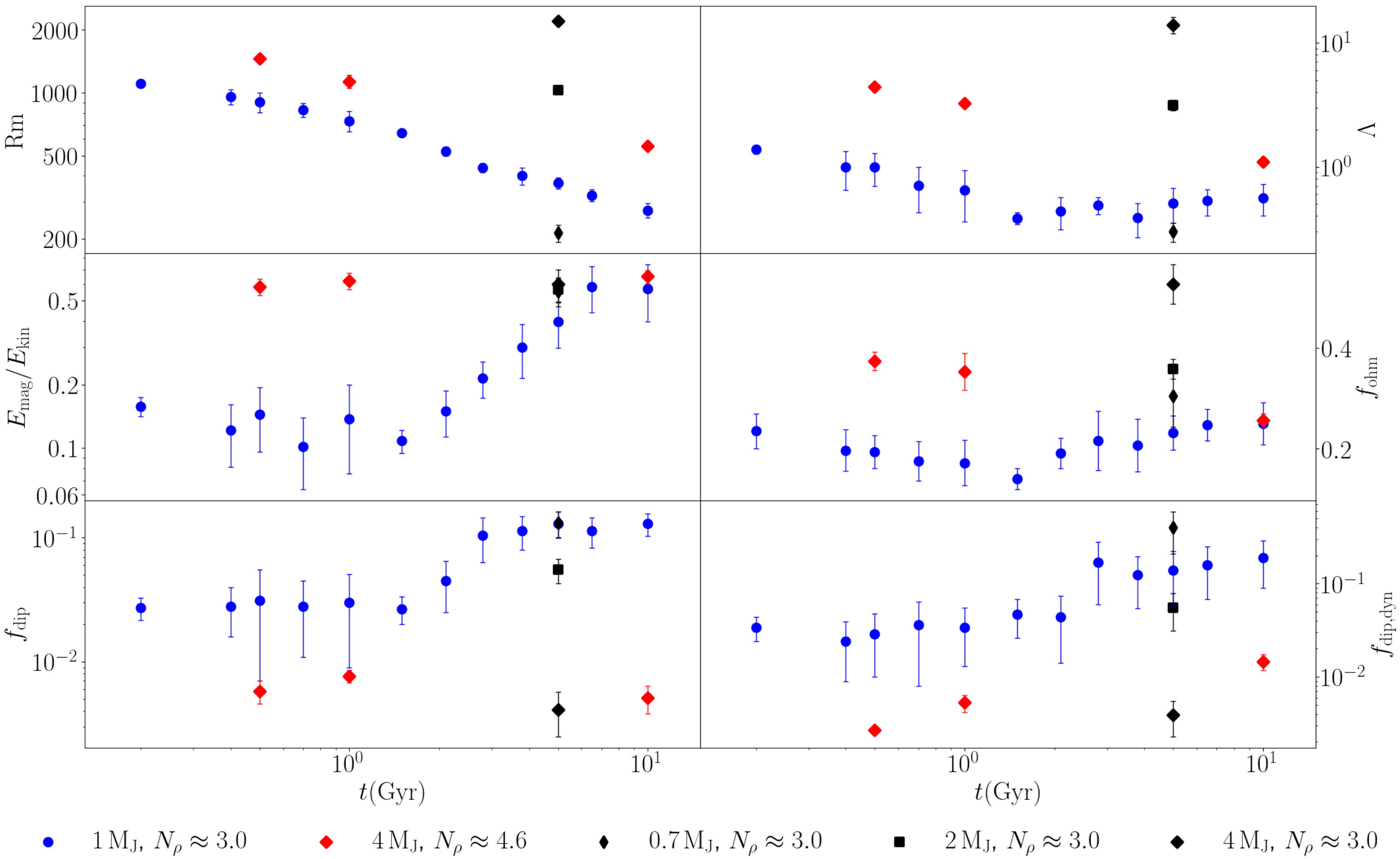

Figure 5.10: Diagnostics as a function of the age for different masses, all with Pm = Pr = 1. From left to right and top to bottom, we show the magnetic Reynolds number, the Elsasser number, the magnetic-to-kinetic energy ratio, the Ohmic fraction, total dipolarity, and dipolarity at the dynamo surface. The 4 $M_J$ runs at 0.5, 1, and 10 Gyr have a higher $N_\rho$ (see text).

In any case, our strongest evolutionary trend shows a transition from a multipolar to a dipolar regime in the middle of our series. Within the more multipolar part of the series, $\Lambda$ and $f_{\rm ohm}$ show a minor decay in time, and $E_{\rm mag}/E_{\rm kin}$ also seem to decrease, but more subtly. In the dipolar regime, these trends reverse: $\Lambda$ and $f_{\rm ohm}$ plateau and show a slight increase; the ratio $E_{\rm mag}/E_{\rm kin}$ grows noticeably. Following the magnetic fields of Jupiter and Saturn (Connerney et al., 2022; Cao et al., 2023), gas-giant dynamos are expected to live in a parameter space region with dipole-dominated solutions, and thus, the expected evolution would be of the latter part of our series. Moreover, the Jovian value for Pm is expected to be $\approx 10^{-6}$, meaning that $D_{\rm ohm} >> D_{\rm visc}$, and thus, $f_{\rm ohm}$ is expected to be close to one. This would mean that our predicted $f_{\rm ohm}$ trend is not physically noticeable. An increase in $E_{\rm mag}/E_{\rm kin}$ and $\Lambda$, the integrated nondimensional magnetic energy, might contradict the aforementioned scaling laws, but we show their compatibility in Sect. 5.2.8.

The values of $E_{\rm mag}/E_{\rm kin}$ that we show are below equipartition (i.e., $0.1 < E_{\rm mag}/E_{\rm kin} < 0.6$ for the long $1\rm M_J$ sequence) because we analyzed a volume including the nonconducting outer layer. When we restrict the energy integration within the metallic region $r < r_m$, then $0.25 < E_{\rm mag}/E_{\rm kin}|_{r<r_m} < 1.4$. This tendency can be sensed from the radial distributions in Fig. 5.5 and 5.11.

### 5.2.7 Dependence on planetary mass

We have obtained saturated models for five different masses at 5 Gyr. We directly compared only four of them because the conductivity of the 0.3 $\rm M_J$ model was not numerically feasible with $N_\rho \approx 3$ (see above). The model of 4 $\rm M_J$ is slightly under-resolved, possibly because of its higher Ra and a very little conductivity drop, which does not help to stabilize the stress-free boundary conditions. We did not use higher-mass 1D models (specifically, the 8 and 12 $\rm M_J$) because of the resolution constraints required by the E, Ra combination.

The main dimensionless diagnostics for these runs were already shown in Fig. 5.10. The overall trends are dictated by the increase of Ra with a decrease in E that comes with the 1D profiles themselves. Therefore, as the mass increases, both $f_{\rm dip}$ and $f_{\rm dip,dyn}$ decrease, while $f_{\rm ohm}$, $\Lambda$, and Rm increase. The energy ratio is approximately maintained.

In Fig. 5.11, we plot the energy spectra and radial distribution for these models. The key features are very similar to the feature described above (Fig. 5.5). A noticeable difference is that the magnetic spectra seem to become flatter for higher masses. The sawtooth shape in the kinetic spectra diminishes with mass because the depth of the outer nonconductive layer decreases as the hydrogen metallization pressures are reached faster. To overcome this difference and to observe whether the evolutionary trends shown in Fig. 5.10 changed for other masses, we obtained saturated dynamos for the 4 $\rm M_J$ 0.5, 1, and 10 Gyr $N_\rho \approx 4.6$ models. For these three, we obtained similar dynamos with larger equatorial jets; in other words, we recovered the sawtooth shape seen in the kinetic spectrum. The evolutionary trends match the multipolar side of the 1 $\rm M_J$ long series.

### 5.2.8 Evolution of the magnetic field strength at the dynamo surface

Relying on the original work Christensen et al. (2009) who calibrated dynamo scaling laws with observed magnetic fields in solar planets and low-mass fast-rotating stars (see Sect. 6.4.1 for more details), Reiners et al. (2009) gives the mean strength of the magnetic field at the dynamo surface, $B_{\rm dyn}$, in terms of the mass $M$, luminosity $L$, and radius $R$ of the substellar object,

$$B_{\rm dyn} = 4.8^{+3.2}_{-2.8} \left(\frac{M}{M_\odot}\right)^{1/6} \left(\frac{L}{L_\odot}\right)^{1/3} \left(\frac{R_\odot}{R}\right)^{7/6} \text{ kG} . \tag{5.10}$$

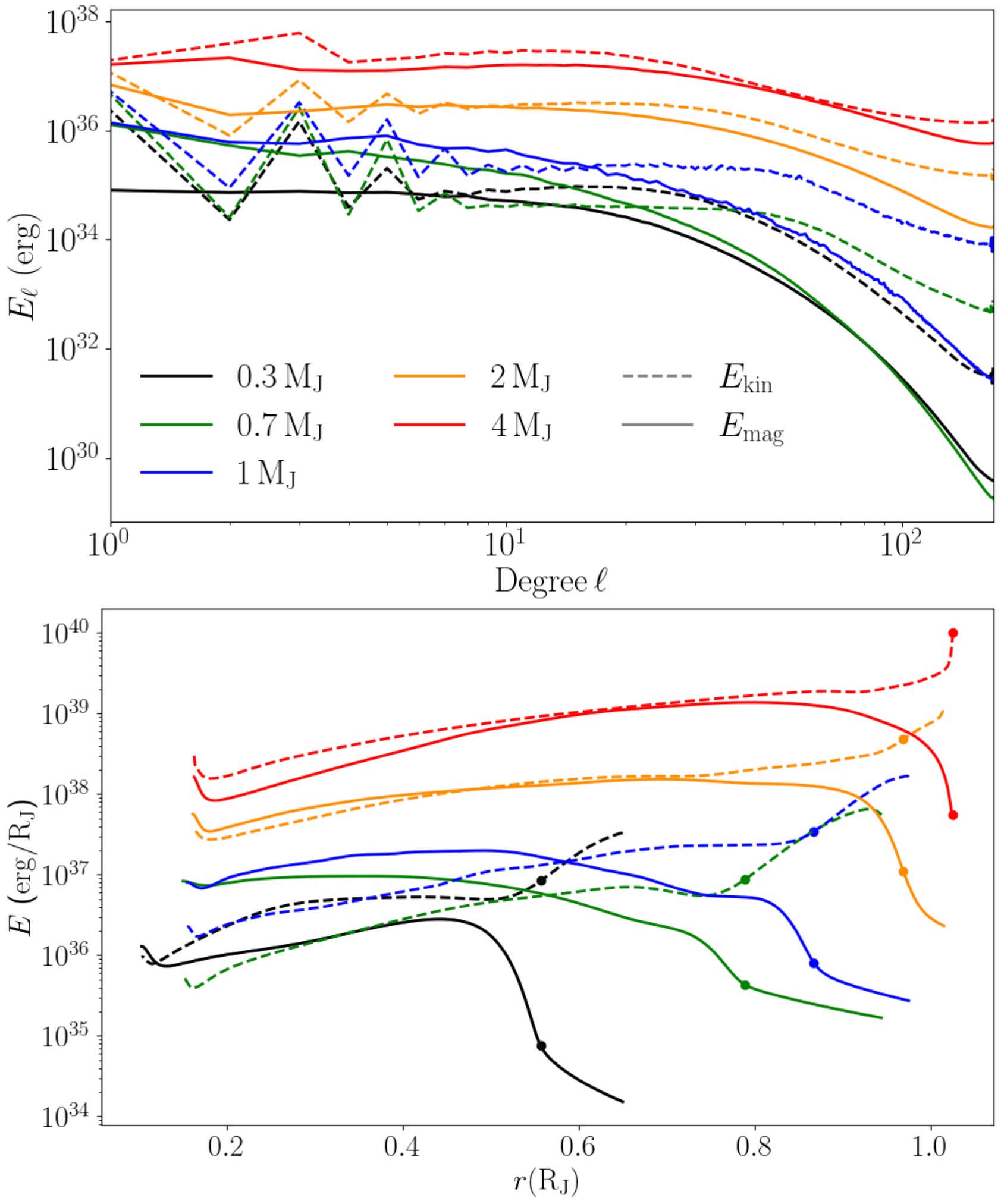

Figure 5.11: magnetic (solid) and kinetic (dashed) energy distribution over the multipole degrees $l$ (left), and over the radius (right) for planets with different masses at 5 Gyr. The location of $r_m$ is marked with a dot in the radial plots.

Or for a more planetary-centered view, this scaling law can be translated into Jovian units by using $M_J = 9.54 \cdot 10^{-3}\, M_\odot$, $L_J = 8.67 \cdot 10^{-10} L_\odot$ and $R_\odot = 9.731 R_J$:

$$B_{\rm dyn} = 20.4^{+13.6}_{-11.9} \left(\frac{M}{M_J}\right)^{1/6} \left(\frac{L}{L_J}\right)^{1/3} \left(\frac{R_J}{R}\right)^{7/6} \text{ G} \tag{5.11}$$

We evaluated $B_{\rm dyn}$ by using the $L(t)$, $R(t)$ output from our *MESA* simulations (pink lines in Fig. 5.12).

Another slightly different estimate comes from inserting the analytical expressions for $L(t)$ and $R(t)$ of Burrows and Liebert (1993); Burrows et al. (2001) in Eq. (5.10) as given for a substellar-mass solar-metallicity object,

$$L \approx 4 \cdot 10^5 L_\odot \left(\frac{1Gy}{t}\right)^{1.3} \left(\frac{M}{0.05 M_\odot}\right)^{2.64}, \tag{5.12}$$

$$R \approx 6.7 \cdot 10^4 km \left(\frac{10^5\,\text{cm}\,\text{s}^{-2}}{g}\right)^{0.18} \left(\frac{T_{eff}}{1000K}\right)^{0.11}. \tag{5.13}$$

Using these estimates, Reiners and Christensen (2010) obtained a slow decay of the dynamo magnetic field of about one order of magnitude in about 10 Gyr. Fig. 1 in their paper shows that the approximate power-law relation is $B_{\rm dyn} \approx t^{-0.3}$ (marked with a gray line in Fig. 5.12).

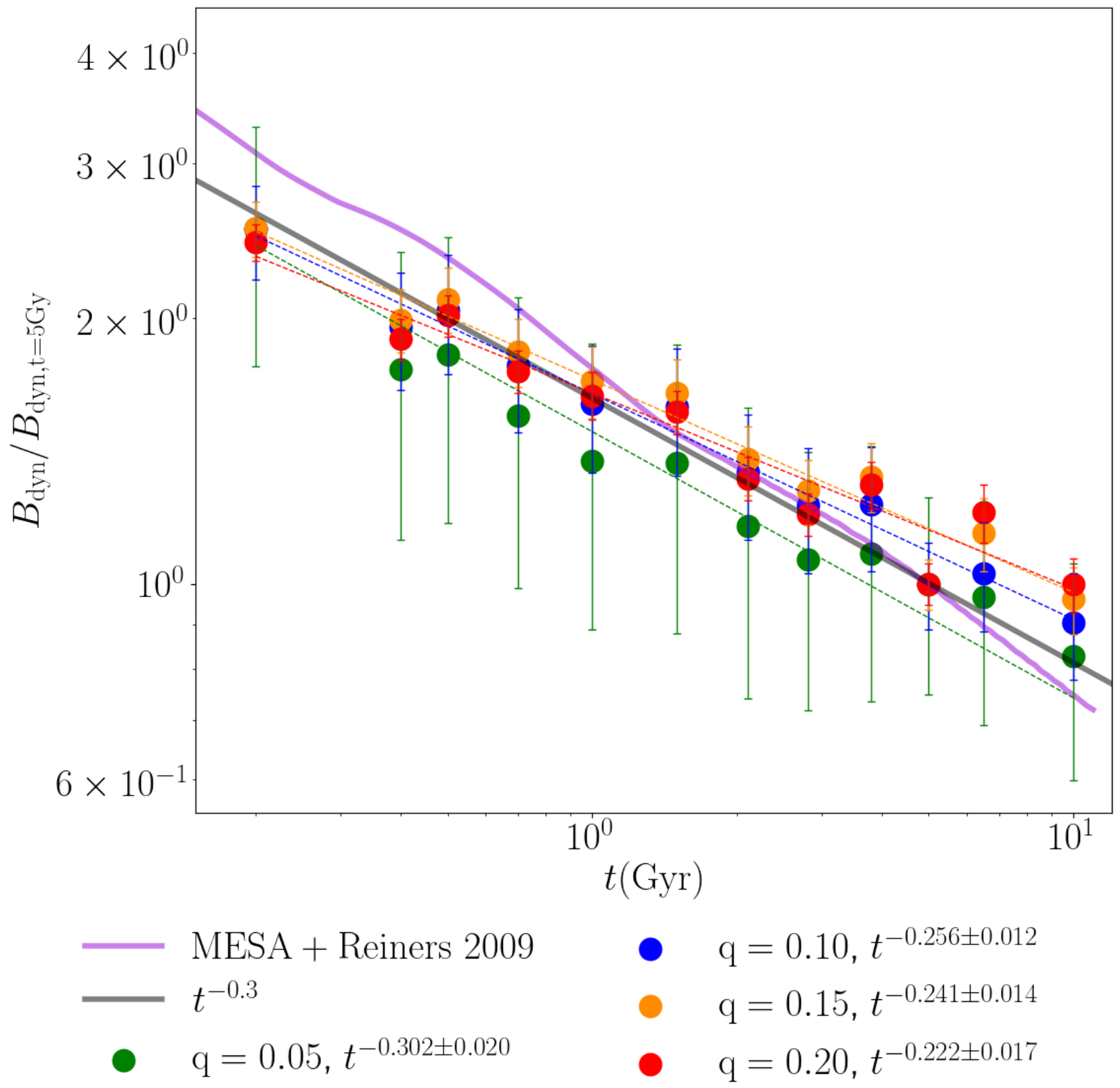

Figure 5.12: Evolution of the magnetic field strength at the dynamo surface averaged in time, using the scaling laws and calculating the average value of the magnetic field over different relative thicknesses $q$, around $r_{\rm m}$ (green, blue, orange, and red). The error bars are associated with the radial variation and are larger for thinner integration shells over which we evaluated Eq. 5.14. The dotted lines show the corresponding best-fit power laws. The solid lines indicate Eq. (5.10) applied to our *MESA* output (pink), and the prediction by Reiners and Christensen (2010) (gray).

We then compared these methods with ours. To do this, we evaluated the values of $B_{\rm dyn}$ for our models. As mentioned above, we made use of the fact that the effective dynamo surface is located at $r_{\rm m}$. We decided to obtain the volume average of $E_{\rm mag}$ over a spherical shell from $r_{\rm m}$ to some not-too-deep layer,

$$E_{\rm mag,dyn}(q) = \frac{1}{r_m - r'_m} \int_{r'_m}^{r_m} E_{\rm mag}(r) dr \ , \tag{5.14}$$

where $r'_m(q) = (\chi_m - q) r_o$, and $E_{\rm mag}(r)$ was obtained from the previously shown radial distributions. The shell thickness is therefore controlled by the parameter $q$, which in turn allowed us to evaluate the surface dynamo field as $B_{\rm dyn}(q) = \sqrt{2\mu_0 E_{\rm mag,dyn}(q)/MV}$. In Fig. 5.12 we show the estimated $B_{\rm dyn}$ for $q = 0.05$, 0.1, 0.15, and 0.2 (colored points in Fig. 5.12, with the related statistical error). The best-fitting slope (dotted lines) decreases with thicker integrating regions, that is, larger $q$. Within the standard deviations, we recover the previously mentioned slope of $\approx t^{-0.3}$ for the most uncertain slope (thinnest averaging region). The others are slightly shallower than the trends obtained by Reiners and Christensen (2010) and Reiners et al. (2009).

### 5.2.9 Dependence on the Prandtl numbers

To assess the impact of the assumption of constant Pr and Pm, we obtained several saturated dynamo states with $\mathrm{Pr}, \mathrm{Pm} \neq 1$ for some 1 $\mathrm{M_J}$ models. We performed runs with 0.5, 1, and 10 Gyr, which we deemed enough for assessing general properties of the trends. We investigated within the following range: $0.5 < \mathrm{Pm} < 4$ and $0.5 < \mathrm{Pr} < 2$. The results are shown in Fig. 5.13.

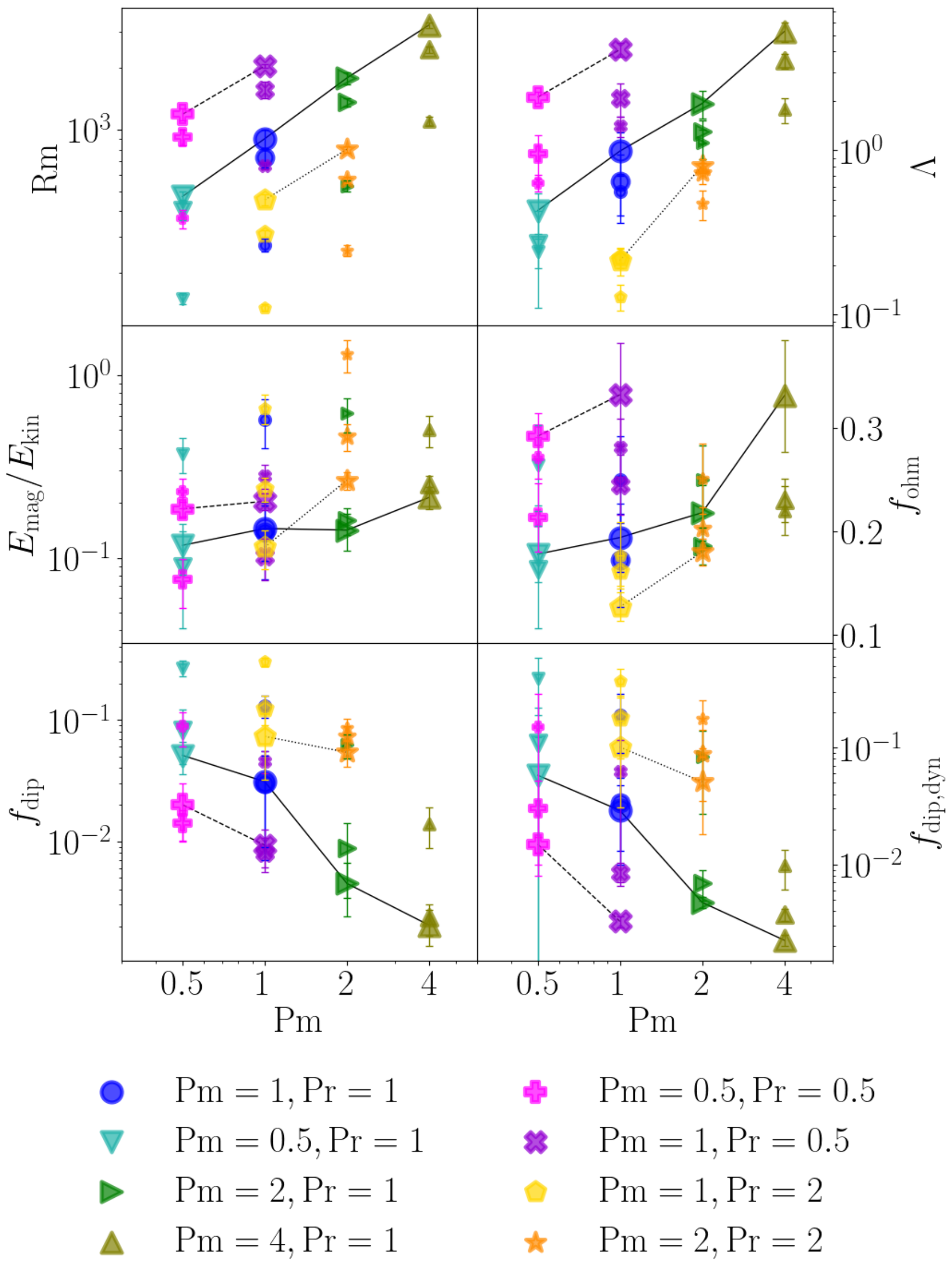

Figure 5.13: Same diagnostics as in Fig. 5.10, shown as a function of Pm, with different values of Pr. The decreasing size of the mark indicates the increase in age (0.5, 1, and 10 Gyr). The colors and shapes help to distinguish the evolutionary changes and are the same as in Fig. 5.14 and 5.15. The lines of constant Pr were added for the 0.5 Gyr models.

The increase in Pm can be understood as lowering $\lambda$ while keeping $\nu$ constant, or similarly, increasing $\nu$ with constant $\lambda$. Both effects lead to an increase in the magnetic energy concerning the available buoyant power. Therefore, Fig. 5.13 shows that Rm and $\Lambda$ tend to increase with Pm, which is a more efficient dynamo mechanism (Elias-López et al., 2024). The same applies for $E_{mag}/E_{kin}$ and $f_{ohm}$, as the decreasing $\lambda$ increases the magnetic energy percentage and the Ohmic dissipation contribution. The surface and volumetric dipolarities both decrease with increasing Pm, which is compatible with what was found by Tsang and Jones (2020): A higher Pm means a less steep magnetic spectrum, or in other words, a more weakly dipole-dominated magnetic spectrum.

In contrast, Rm, $\Lambda$, and $f_{ohm}$ decrease for increasing Pr. This can easily be understood by considering a higher Pr as increasing values of $\nu$ in comparison to $\kappa$. A higher viscosity will lead to lower kinetic as well as magnetic energy, and therefore, to lower Rm and $\Lambda$. The decrease in $f_{\rm ohm}$ means that Ohmic dissipation becomes less important than viscous dissipation. The magnetic energy ratio, $E_{\rm mag}/E_{\rm kin}$, and the two dipolarities, $f_{\rm dip}$ and $f_{\rm dip,dyn}$, increase with Pr because it leads to a more efficient dynamo mechanism.

With these trends in mind, we describe the effect of the constant Pr and Pm assumptions on the obtained trends in Figs. 5.10 and 5.12. A slight increase in Pr is expected to occur during the long-term evolution of the planets because, while it cools down, the ratio of the thermal and electrical conductivities (inversely proportional to Pr) decreases according to the Widemann-Franz law, which is valid in the metallic region (French et al., 2012). In contrast, the viscosity and conductivity themselves are not thought to vary appreciably with temperature (i.e., in time) (French et al., 2012; Bonitz et al., 2024). Therefore, by using the trend of $\Lambda$ with Pr, we expect that for an evolutionary change in Pr, the trend $B_{\rm dyn}(t)$ would be slightly steeper, which might agree even better with the Reiners and Christensen (2010) scaling law trend. However, a firm conclusion about this based on a modified setup for diffusivities and Pr evolution is left for future work.

In general, the evolutionary trends shown in Fig. 5.10 are maintained regardless of Pr and Pm. Rm decreases similarly for all set runs. Finally, $E_{\rm mag}/E_{\rm kin}$, $\Lambda$, and $f_{\rm ohm}$ show a different behavior depending on their dipolarity. The two most strongly dipolar solutions ($\rm Pm = 1, Pr = 2$ and $\rm Pm = 2, Pr = 2$) consistently show the same behavior as noted above, that is, an increase in $E_{\rm mag}/E_{\rm kin}$ and $f_{\rm ohm}$ in time, and a plateau or mild decrease in $\Lambda$. In contrast, the most multipolar set of Prandtl numbers ($\rm Pm = 4, Pr = 1$) decreases in $f_{\rm ohm}$ and $\Lambda$ and increases slightly less in $E_{\rm mag}/E_{\rm kin}$. The other sets of runs are consistent with a transition from multipolar to dipolar, similar to Fig. 5.10.

### 5.2.10 Scaling laws between dynamo numbers

Yadav et al. (2013) used a large set of anelastic dynamo numerical solutions to derive scaling laws by relating several representative dimensionless diagnostic parameters. Their dynamos covered a large space of parameters: $0 \leq N_\rho \lesssim 5.5$, $0.1 \leq \eta \leq 0.75$, $0.3 \leq \rm Pr \leq 10$, $0.2 \leq \rm Pm \leq 20$, $10^{-6} \leq \rm E \leq 10^{-3}$, and $2.5 \cdot 10^5 \leq \rm Ra \leq 2.5 \cdot 10^9$. The scattering plots shown in Fig. 5.14 and 5.15 superimpose the results of Yadav et al. (2013) with the data representing our models.

To compare our results, we used the inverse rotation frequency $\Omega^{-1}$ as the time unit and the magnetic field units of $\Omega D\sqrt{\mu_o \rho_o}$. The buoyancy power $P$ is

$$P = \frac{\rm RaE^3}{\rm Pr}\frac{\langle \tilde{\alpha}\tilde{T}\tilde{g}s' u_r\rangle}{M} \ , \tag{5.15}$$

where $M$ is the dimensionless mass of the shell and is listed in Table E.1 for our models. In Fig. 5.14 we show Ro as a function of $P/\rm Pm^{13/45}$. For the runs of Yadav et al. (2013), there is a clear separation between the models with constant $\sigma$, which lie close to the best-fitting power law (gray line), and the runs with a decaying $\sigma$ profile near the surface, which lie slightly above but parallel to the trend. The reason is that for a decaying $\sigma$, the strong jets that appear in the external nonconductive layer tend to increase the total kinetic energy, and thus, Ro for the same amount of available $P$. Our definition of Ro differs by a factor of $1/\tilde{\lambda}$, which erases the outer jet contribution. Our runs, therefore, mostly lie above the scaling law itself. Our models represent different evolutionary stages of planetary dynamos that move through this dimensionless space. This progression is parallel to the power law $\rm Ro = 2.47\mathit{P}^{0.45}Pm^{-0.13}$ and reaches from higher to lower values: For a single model, Ro decreases by about half an order of magnitude, while $P/\rm Pm^{13/45}$ decreases by one order of magnitude (which corresponds to the 0.45 exponent).

Physically, this evolution should be positioned orders of magnitude away, but if the scaling law holds, so should this trend.

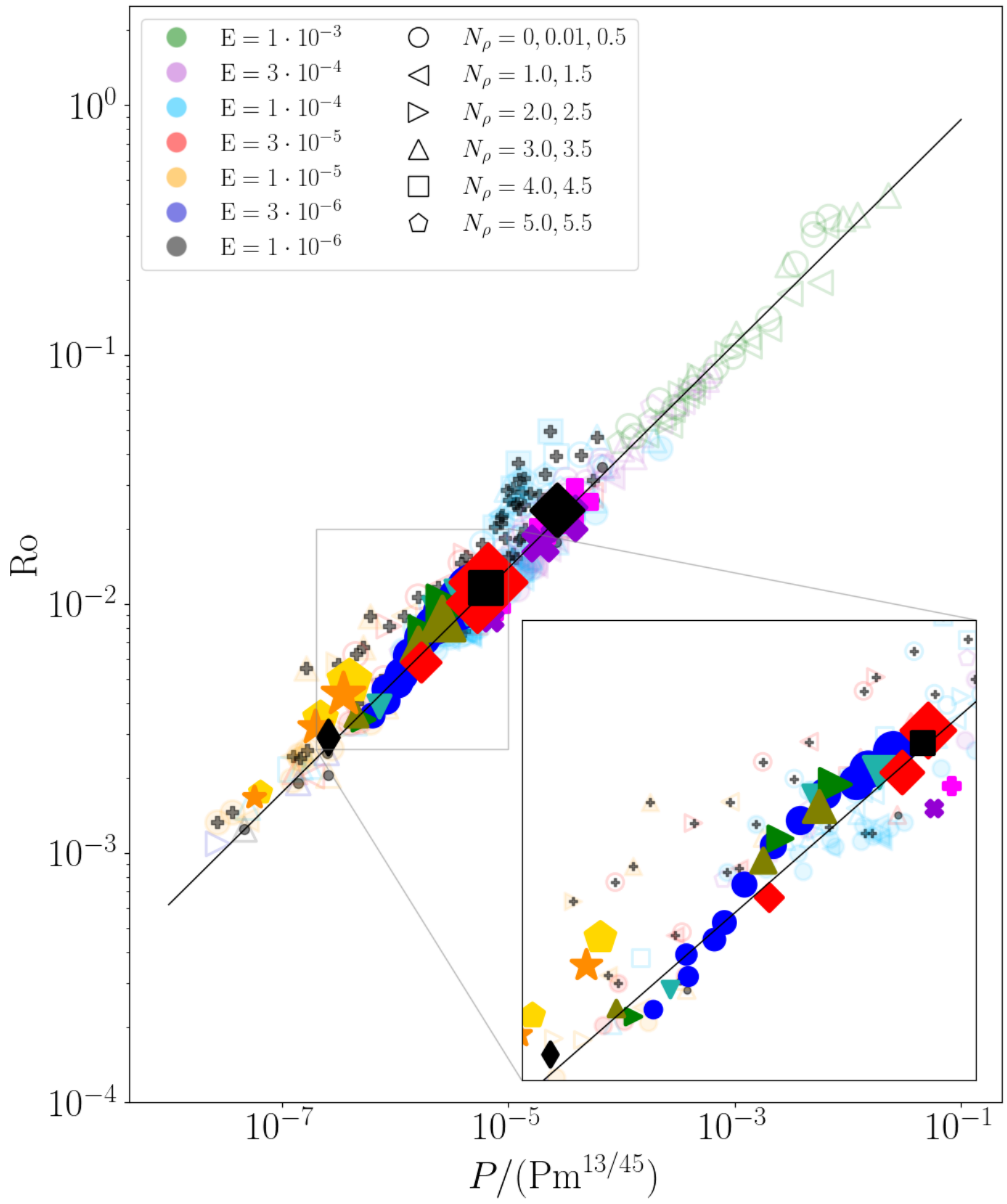

Figure 5.14: Rossby number as a function of a combination of nondimensional buoyancy power and magnetic Prandtl number. The solid line corresponds to the power law $\mathrm{Ro} = 2.47P^{0.45}\mathrm{Pm}^{-0.13}$. The semitransparent data are taken from Yadav et al. (2013) and are plotted with a similar symbol and color scheme (the legend is different from our runs). Filled (empty) symbols correspond to dipolar (multipolar) dynamos, and the authors define dipolar as $f_{\mathrm{dip,l<12}}^{\mathrm{axi,surf}} > 0.3$. The Ekman number is color-coded, and the marker shape indicates the degree of density stratification $N_\rho$. Symbols containing a plus have an exponentially decaying conductivity as Eq. (5.7), and those with a dot have a moderate outward decay of $\sigma$, $\nu$, and $\kappa$, all proportional to $\rho(r)$. The complete input and output set can be found in the additional data of the original paper. The results of this work are superposed with the same legend as in Fig. 5.10, and the size of the marker denotes the approximate age and mass of the planet.

We also defined the Lorentz number, Lo, as the nondimensional magnetic field strength in $\Omega^{-1}$ time units per unit of mass. In this case, $B$ is in units of $\Omega d\sqrt{\rho_o\mu_o}$, and we relate the definition of Lo with $\Lambda$, E and Pm,

$$\mathrm{Lo} = \frac{B_{rms}}{\Omega d\sqrt{\rho_o\mu_o}}\frac{1}{\sqrt{M}} = \sqrt{\frac{2E_{\mathrm{mag}}\mathrm{E}^2}{M}}\,. \tag{5.16}$$

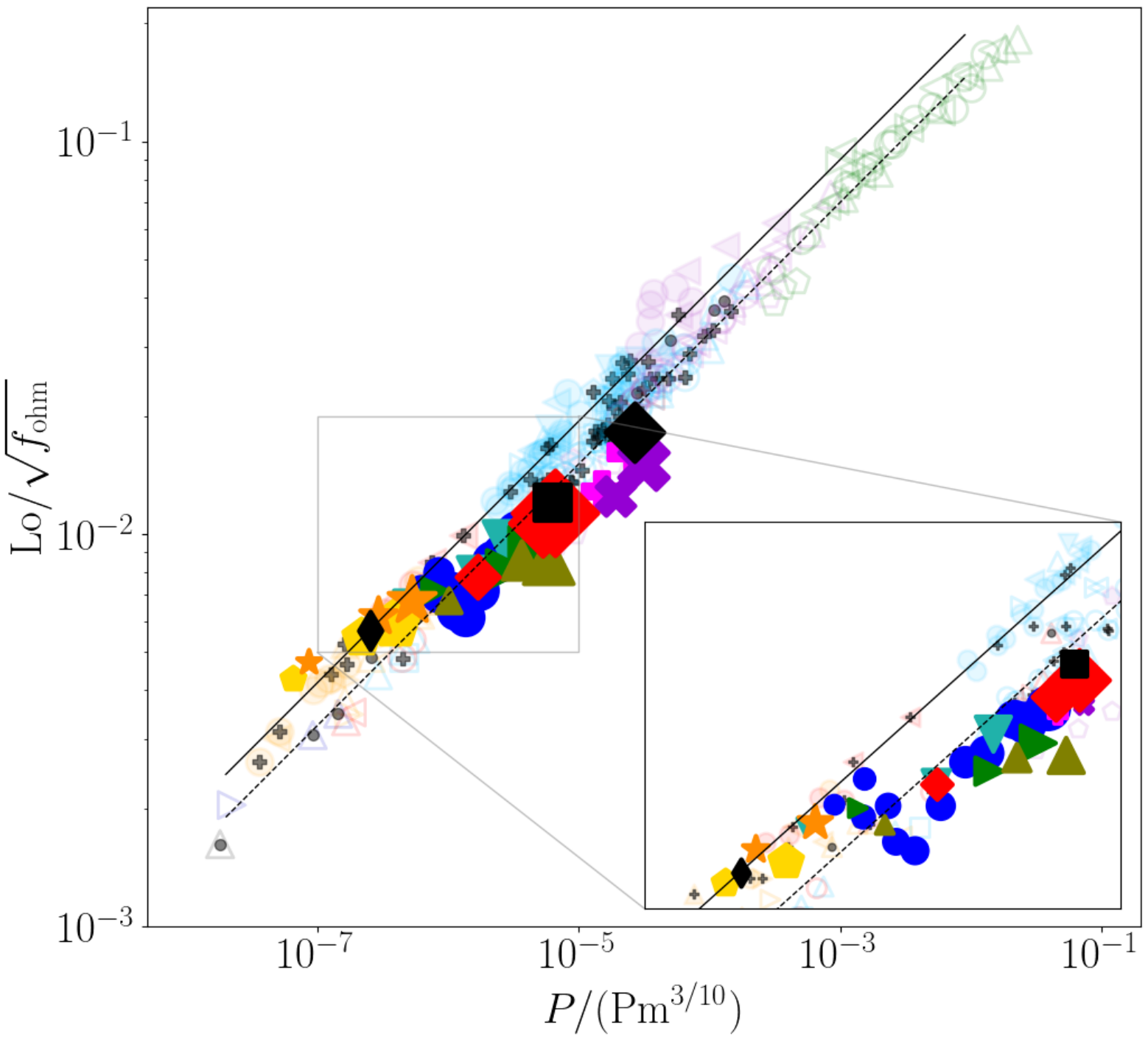

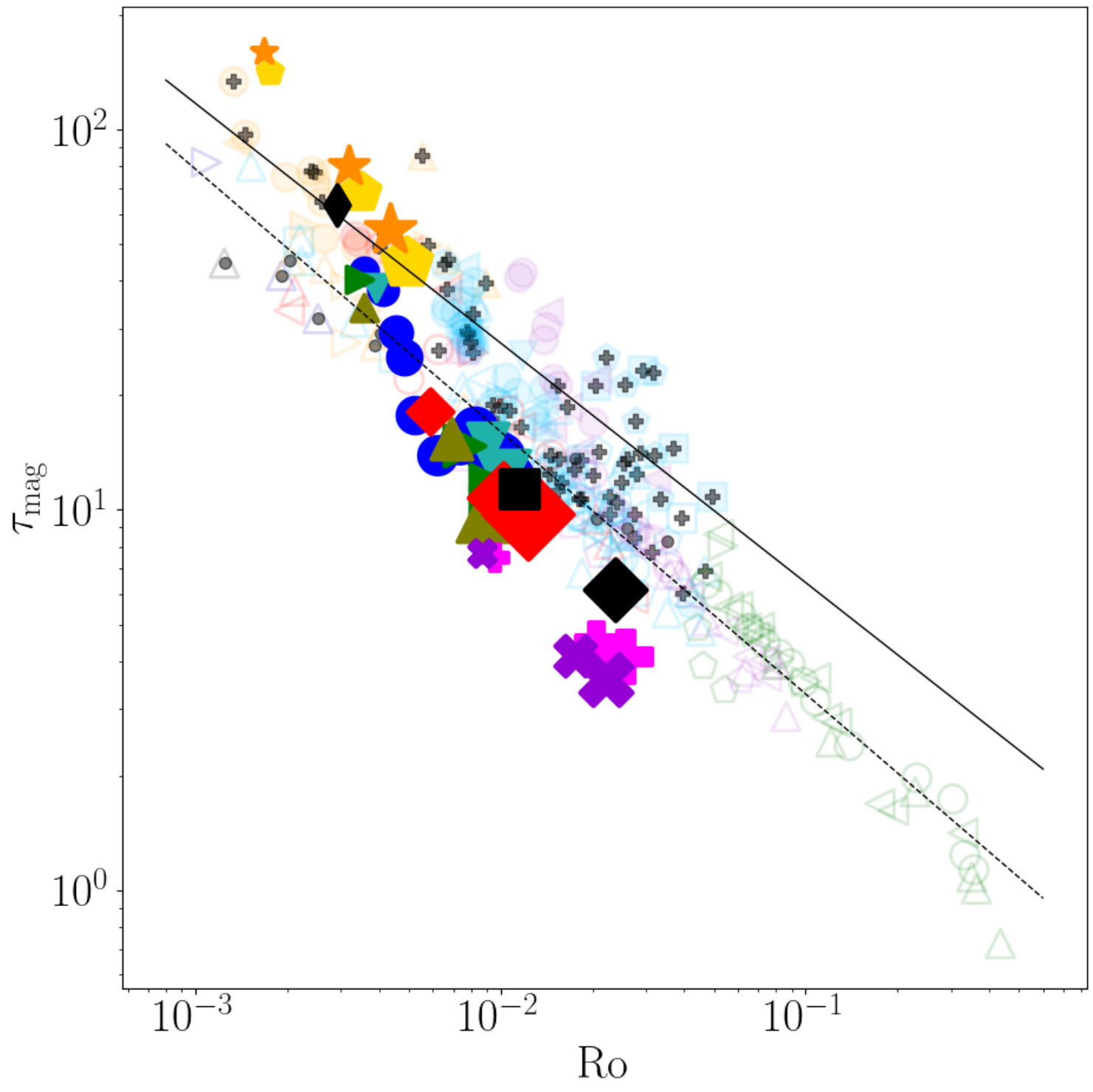

Figure 5.15: Similar to Fig. 5.14 (same legend) for magnetically related quantities. On the left: Lorentz number corrected for the fraction of Ohmic dissipation as a function of a combination of nondimensional buoyancy power and magnetic Prandtl number. The scaling relations are $\mathrm{Lo} f_{\mathrm{ohm}}^{-1/2} = A P^{\frac{1}{3}} \mathrm{Pm}^{\frac{1}{10}}$, where $A$ is 0.9 or 0.7 for dipolar and multipolar dynamos, respectively. Yadav et al. (2013) found that the value $f_{\mathrm{dip,l<12}}^{\mathrm{axi,surf}} > 0.3$ divides the data into dipolar and multipolar, and these two types of runs are best fit separately. On the right: Characteristic timescale of the magnetic energy dissipation as a function of Rossby number. The scaling relations are $\tau_{\mathrm{mag,dip}} = 1.51\ \mathrm{Ro}^{-0.63}$ and $\tau_{\mathrm{mag,mulip}} = 0.67\ \mathrm{Ro}^{-0.69}$.

The other magnetically related scaling law involves the characteristic timescale of magnetic energy dissipation $\tau_{\rm mag}$, which is defined as the nondimensional magnetic energy divided by the joule heat dissipation (all in units of $\Omega^{-1}$),

$$\tau_{\rm mag} = \frac{E_{\rm mag}{\rm E}^2}{P f_{\rm ohm} M} \ . \tag{5.17}$$

In Fig. 5.15 we show the scaling laws for Lo and $\tau_{\rm mag}$ with the same legend as in Fig. 5.14. Yadav et al. (2013) reported that dipolar- and multipolar-dominated solutions take similar but parallel trends. We overplot our runs with the joined dipolar and multipolar branches for the two scaling laws (they show them separately). As suspected from Sect. 5.2.6, our series evolves from the multipolar to the dipolar branch in both diagrams, but this is more clearly visible in the $\tau_{\rm mag}$ plot. This is also an argument in favor for this possible multipolar to dipolar transition.

The power-law relations shown above are purely fits obtained from Yadav et al. (2013). The velocity scaling of ${\rm Ro} \propto P^\alpha$ with $\alpha$ somewhat larger than 0.4 was theoretically justified by force balances by some authors (Aubert et al., 2001; Davidson, 2013; Starchenko and Jones, 2002), but none derived a Pm dependence. Similarly, $\tau_{\rm mag} \propto {\rm Ro}^\alpha$ was also discussed, where $\alpha \lesssim -1$ (Christensen and Tilgner, 2004; Stelzer and Jackson, 2013) or $\alpha \approx -0.75$ (Davidson, 2013). Finally, if the magnetic field is only a function of power, dimensional arguments dictate that it must depend on the cubic root of the power, that is, ${\rm Lo} \propto P^{1/3}$ (Kunnen et al., 2010; Christensen et al., 2009; Davidson, 2013).

### 5.2.11 Force decomposition

We now examine the spectral force decomposition for our dynamo solutions. We follow the same analysis introduced by Aubert et al. (2017); Schwaiger et al. (2019) (see Sect. 4.6.1). As a general example, we show in Fig. 5.16 the time-averaged force balance for the 0.2 Gyr run over the whole simulation domain. This spectrum is similar to Gastine and Wicht (2021), where they show the force balances. The geostrophic balance dominates throughout all $\ell$. Our runs are similar to Gastine and Wicht (2021) (except for the SSL); thus, a MAC balance dominates inside the conductive region, and inertia plays little role. Instead, an IAC balance is found in the non-conductive outer region. Overall, as Fig. 5.16 shows, the sum above and below $r_{\rm dyn}$, the ageostrophic component balances with buoyancy at the largest scales and Lorentz and inertia at the lowest scales. Viscosity remains the lowest contribution.

The force balance spectra for the different evolutionary models are shown in Fig. 5.17. For all the runs in the evolutionary series, we find a dominating geostrophic balance even at the largest $\ell$, and low contributions from viscosity. The overall trend for inertial forces is that they lose importance in the evolutionary track. Moreover, $\ell_{\rm IA}$ (the multipole where inertial and buoyant forces equilibrate), increases with age starting from $\ell_{\rm IA,\,2\,Gyr} \sim 10$ to $\ell_{\rm IA,\,10\,Gyr} \sim 90$. Instead, the magnetic force increases its relative importance with age, and the peak of magnetic energy as well as $\ell_{\rm MA}$ both tend to evolve to lower multiples with age, although not with such a clear trend as $\ell_{\rm IA}$. This is compatible with a decrease of equipartition level (i.e., increase in magnetic energy relative to the kinetic energy) shown in Fig. 5.9, as well as the decrease in the relative ratio of internal and magnetic forces (i.e., $\mathcal{F}_{\rm I}/\mathcal{F}_{\rm L}$) as seen by Zaire et al. (2022).

### 5.2.12 Observational consequences

The NASA exoplanet archive currently (March 2025) lists more than 650 confirmed gas-giant candidates with $M\sin(i) > 0.2\,M_J$ and $P_{orb} > 200$ days, for which irradiation and tidal synchro-nization are negligible. Only four exoplanets under 10 pc have an estimate for the Solar System age. The closest and youngest exoplanet (0.6 ± 0.2 Gyr) is $\epsilon$ Eridani b (Hatzes et al., 2000), with a mass of $0.66^{0.12}_{-0.09}$ $M_J$, which is a good candidate for an intense multipolar-dominated dynamo. The other three candidates, Gliese 832 b (Bailey et al., 2009), HD 219134 h (Motalebi et al., 2015; Vogt et al., 2015), and GJ 3512 b (Morales et al., 2019), with predicted ages >5 Gyr, are likely to instead host Jupiter-like dipolar dynamos. If their magnetic field is intense enough to produce electron-cyclotron maser emission that is detectable from ground (with an associated gyro-frequency, $\nu \simeq 2.8\ B[\mathrm{G}]$ MHz, higher than the ionospheric ≈10 MHz cutoff; Zarka 1998), current (LOFAR) and next-generation (SKA-low) low-frequency radio interferometers might eventually detect their exoplanetary radio emission. This might provide indications of the intensity of the magnetic fields and, possibly, their morphology. The overall conclusions for this chapter are summarized in Chapter 8.

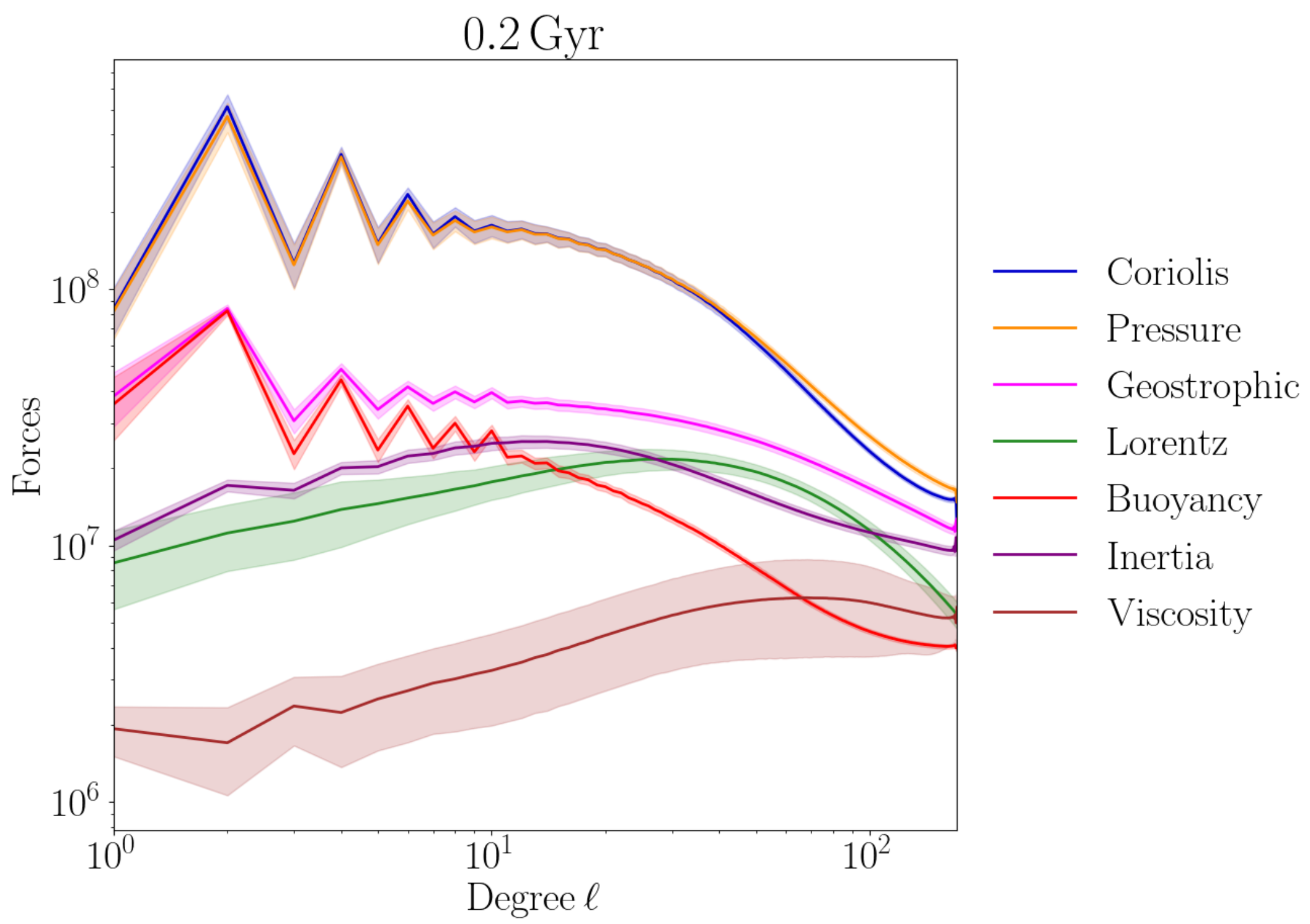

Figure 5.16: Time average force decomposition obtained with Eq. 4.19, with standard deviations depicted with shaded regions.

## 5.3 Summary

- We obtained 3D saturated dynamo solutions with numerical shell dynamo models with *MagIC*, incorporating the thermodynamic radial profiles from *MESA* at different evolutionary stages as background. We considered planets in the mass range $0.3\,M_J \leq M_P \leq 4\,M_J$ and ages from 0.2 to 10 Gyr. We assumed fixed Pr and Pm; changes in Ra and E result from evolving $\Delta T$, $T_o$, and shell thickness. We interpret saturated dynamo solutions as snapshots of long-term dynamo evolution.

- We identified a transition from a multipolar to dipolar dynamo regime as the planet cools. We observed decreasing trends in Rm, $P$, and Ro, with increasing surface and volumetric dipolarities. For multipolar dynamos, $\Lambda$, $f_{\mathrm{ohm}}$, and $E_{\mathrm{mag}}/E_{\mathrm{kin}}$ decrease with time. For dipo-lar dynamos, the same quantities increase with time. Trends are consistent across different Prandtl numbers and 4 $M_J$ models.

- Magnetic field decay at the dynamo surface matches scaling law expectations (Reiners et al., 2009). The evolutionary tracks trace a path through parameter space consistent with previously known anelastic scaling laws (Yadav et al., 2013).

- If we calibrate our results with the observed fields on Jupiter and Saturn, evolved gas-giant dynamos are likely dipolar-dominated. Dynamos may be born in a dipolar state, or alternatively, initial multipolar dynamos may evolve into dipolar regimes. Trends are expected to apply to mildly irradiated gas giants and brown dwarfs, but not to main-sequence stars.

- Over 650 cold Jupiter analogs have been confirmed, only four of which are within 10 pc and have age estimates. Among them, $\epsilon$ Eridani b is the best candidate to have multipole-dominated magnetic fields. This could potentially leave imprints on its magnetically powered ECM radio emission, if detected in the future.

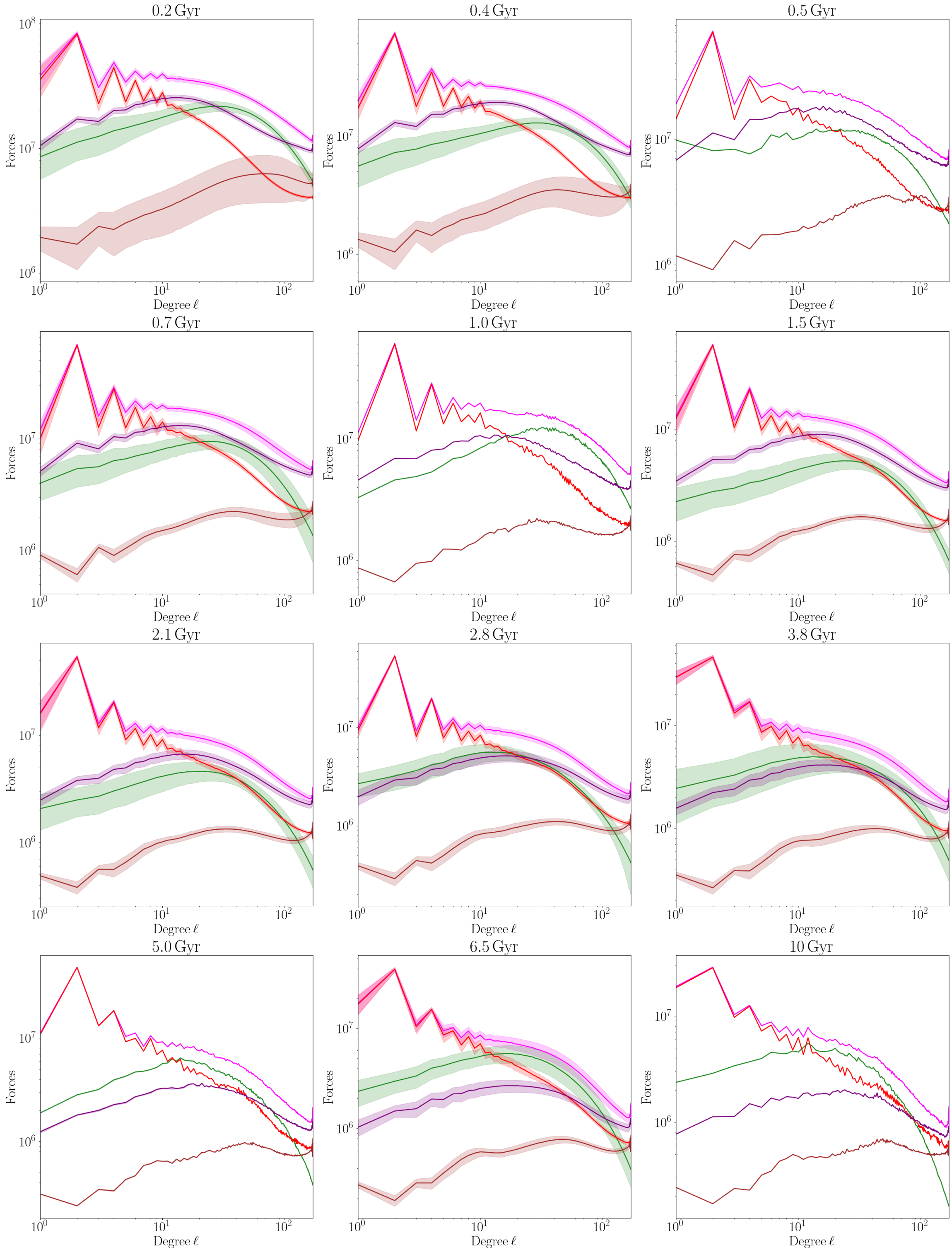

Figure 5.17: Time average force decomposition at different evolutionary stages. The Coriolis and pressure forces are not shown, but the geostrophic component is shown. Standard deviations are also depicted with shaded regions, except for the 0.5, 1, 5, and 10 Gyr models, where the spectra have been obtained from only one snapshot.

# 6

# Rossby number, convection suppression and magnetism in inflated hot Jupiters

Hot Jupiters (HJs) are gaseous giant planets on close-in orbits, characterized by orbital periods of just a few days and separations $\lesssim 0.1$ AU (Fortney et al., 2021). Due to the proximity to their host stars, they are highly irradiated and tidally locked, with equilibrium temperatures up to 3000 K. This extreme external heating creates strong thermal gradients between their day and night sides, driving strong atmospheric flows and equatorial zonal jets that shape their atmospheric dynamics (e.g., Showman and Guillot, 2002; Showman et al., 2009; Heng and Showman, 2015). Currently, several hundred HJs have been confirmed, enabling extensive population studies. One of the most intriguing findings is the so-called HJ radius anomaly (see Thorngren (2024) for a review), which remains one of the longest-standing open questions in exoplanetary science: observations show that HJs systematically exhibit inflated radii, significantly larger (up to twice the Jovian radius) than those predicted by standard evolutionary models. There is a clear trend between the degree of radius inflation and stellar irradiation, once the planetary equilibrium temperature exceeds $T_{\rm eq} \gtrsim 1000$ K (Weiss et al., 2013; Sestovic et al., 2018), a fundamental hint to the underlying physical mechanism. However, the irradiation alone can contribute to radius inflation only up to approximately $\sim 1.3\ R_J$ at Gyr ages, depending also on the mass and the amount of heavy elements (Guillot et al., 1996; Guillot and Showman, 2002; Arras and Bildsten, 2006; Burrows et al., 2007; Fortney et al., 2007). Several mechanisms have been proposed to delay the shrinking and cooling (Spiegel and Burrows, 2013; Thorngren, 2024): enhanced opacities (Burrows et al., 2007), inhibition of large-scale convection (Chabrier and Baraffe, 2007), hydrodynamic effects leading to dissipation in deep layers (Showman and Guillot, 2002; Showman et al., 2009; Li and Goodman, 2010; Youdin and Mitchell, 2010), and Ohmic dissipation of atmospherically induced magnetic field. Among these proposed mechanisms, the latter has received considerable attention, with several quantitative studies (e.g., Batygin and Stevenson, 2010; Batygin et al., 2011; Perna et al., 2010b; Wu and Lithwick, 2013; Ginzburg and Sari, 2016; Knierim et al., 2022; Soriano-Guerrero et al., 2023; Soriano-Guerrero et al., 2025; Viganò et al., 2025). This heating mechanism is caused by the hot ionized atmospheric flows that twist the magnetic field lines of the internal dynamo.

This magnetic induction happens in the very outermost layers of the planet, that is, for pressure $p \lesssim 10$ bar. But multiple mechanisms may operate simultaneously (Sarkis et al., 2021).

One of the most elusive characteristics of exoplanets is their magnetism. Observational estimates of magnetic fields in some HJs, with considerable uncertainties, have been proposed so far in very few works, via star-planet interaction (SPI) models, used to interpret the observed trends of X-ray luminosities vs presence of short-orbit planets (Scharf, 2010), or the modulation of Ca II K lines with HJ orbital periods (Cauley et al., 2019).

As mentioned in the previous chapter, self-sustained dynamo simulations for the generic planetary scenario have been used to derive scaling laws that relate various dimensionless parameters (Christensen and Aubert, 2006; Yadav et al., 2013), which have been validated by the comparison with the observed magnetic fields on Earth, Jupiter, and in fast-rotating low-mass stars (Christensen et al., 2009; Reiners and Christensen, 2010). These scaling laws state that, in the fast rotator regime, convective heat flux determines the magnetic field strength. The application of these scaling laws to the HJ context is not trivial, since the energy budget is different: the presence of irradiation and the inflated radii are not taken into account in the original laws. Nevertheless, existing applications of the heat-flux scaling laws generally found an increase by up to $\sim 1$ order of magnitude of the inferred magnetic fields compared to non-inflated gas giants, if the extra heat needed to explain the inflation is assumed to take the role of the heat convective flux in the scaling laws (Yadav and Thorngren, 2017; Kilmetis et al., 2024). Note that the latter is an implicit, but strong and not necessarily motivated assumption, since the scaling laws consider the heat flux in the highly conductive regions only, while the extra heat could be deposited elsewhere.

Additionally, as explained above, slowly rotating cool stars observationally show a correlation of the magnetic activity indicators with the rotation rate (Reiners et al., 2014). The transition from a rotation-dominated (slow rotators) to a convective-heat-flux-dominated (fast rotators) dynamo strength is seen to be fairly sharply defined by the Rossby number Ro $\sim 0.12$. In this sense, planetary scaling laws applicable to slow rotators assumed a proportionality with the rotation rate rather than the heat flux (e.g., Sánchez-Lavega, 2004; Stevens, 2005). A primary aim of this work is to shed light on which regime applies to HJs, considering the observed range of their relevant parameters: planetary mass, stellar type, and star-planet separation.

Moreover, since virtually all the existing scaling laws predict a correlation between magnetic field intensity and planetary mass, giant exoplanets, and HJs in particular, have been historically considered the leading candidates for the still elusive quest of Jovian-like coherent radio emission (e.g., Lazio (2024); Narang et al. (2024) and references within), in the form of electron cyclotron maser (Dulk, 1985). The latter has been detected from all magnetized planets in the solar system (Zarka, 1998), but can be detected from the ground only if the cyclotron frequency (ECM), $\nu_c \simeq 2.8\, B$ [G] MHz, is larger than the $\sim 10$ MHz ionospheric cut-off. This implies that, to be detected by ground-based radio interferometers, the magnetic field needs to be at least as intense as the Jovian one. In this sense, the second main aim of this work is to explore the trends of the estimates of magnetic field intensity at the dynamo region and the planetary surface, considering their Rossby number regime and applying the appropriate scaling law at a given age, with the internal structure and (convective) luminosity given by long-term evolutionary models. Understanding the magneto-thermal dependence on orbital distance and age can help guide future observations toward the most promising targets.

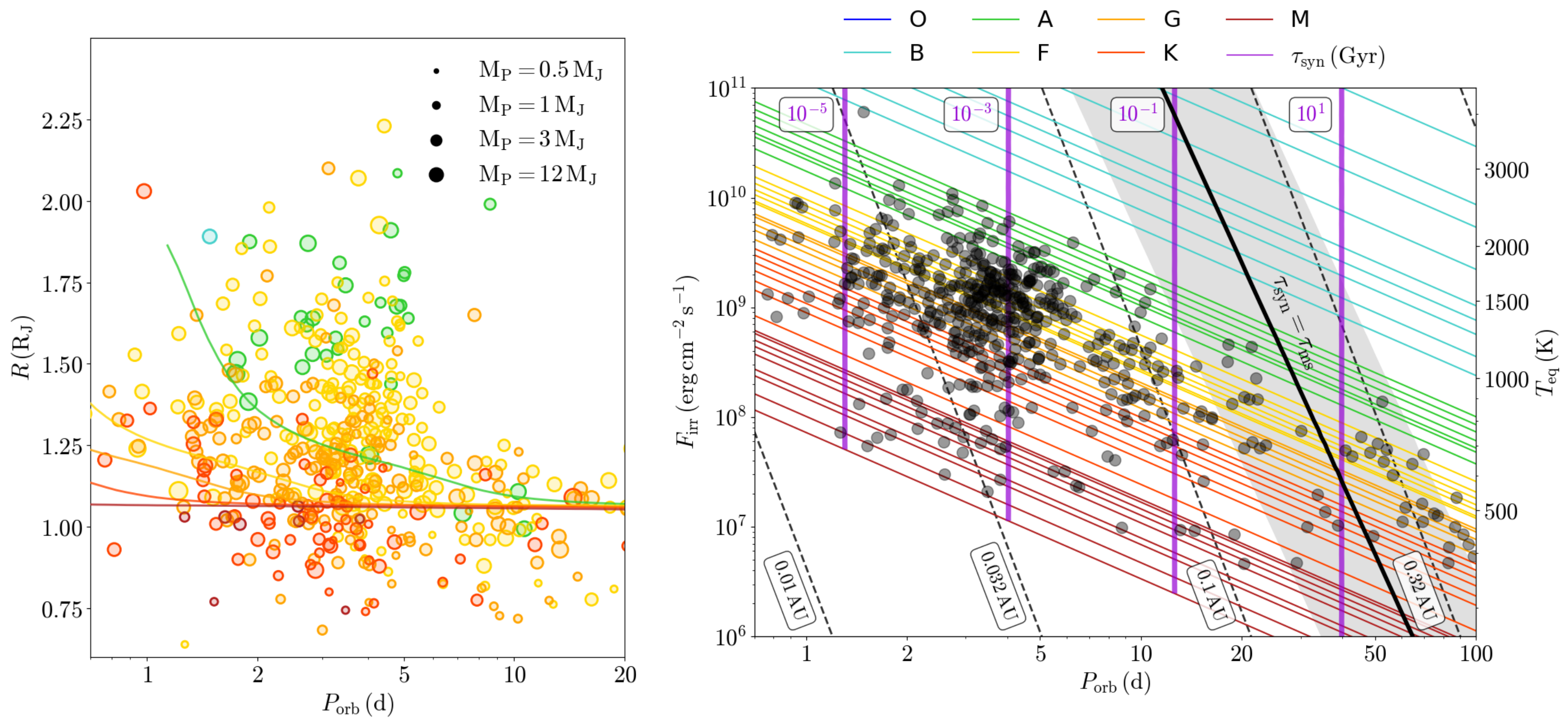

Figure 6.1: Population properties for the confirmed gas giant exoplanet population with mass $M_P > 0.2$ $M_J$, considering the host star effective temperature and radius to obtain the irradiation flux $F_{irr}$, and the corresponding equilibrium temperature $T_{eq}$. Data is extracted from the NASA exoplanet archive (Christiansen et al. 2025, https://exoplanetarchive.ipac.caltech.edu/). Colors indicate the main-sequence stellar type of the host star. Errors are not shown, for the sake of visibility. *Left:* planetary radius as a function of orbital period. Marker size denotes the planetary mass. Right: $F_{irr}$ as a function of planetary orbital period for the same exoplanetary systems. The dashed black lines mark constant separation $a$. The purple vertical lines correspond to different synchronization timescales, $\tau_{syn}$ in Gyr, obtained with Eq. 6.4, assuming $M_P = 1$ $M_J$, $Q'_P = 5 \cdot 10^5$, $\alpha = 2/5$, and $\omega_i = 2\pi/5\,h^{-1}$ (see text). The solid black line indicates where the main-sequence lifetime equals the synchronization timescale.

This chapter is organized as follows: in Sect. 6.1 we summarize the irradiation values and tidal locking timescales, based on the HJ population. Sect. 6.2 introduces the evolutionary models and the amount of additional internal heat as a function of irradiation, based on the statistical study Thorngren and Fortney (2018). We also discuss how the injection and irradiation affect the internal convective layers. In Sect. 6.3, we obtain internal profiles of the local Rossby number and discuss how it depends on age, star-planet separation, planetary mass, and stellar type. In Sect. 6.4 we estimate the dynamo-generated magnetic fields, and the corresponding values at the planetary surfaces, via the scaling laws mentioned above. We also discuss in Sect. 6.4 the star planet interaction and radio emission consequences for the magnetic field estimations.

## 6.1 Irradiation and tidal synchronization

The irradiation flux received by a planet in a specific circular orbit of radius $a$ (hereafter, separation) around its host star is:

$$F_{\rm irr} = \frac{L_\star}{4\pi a^2} = \sigma_{\rm SB} T_{\rm eff}^4 \frac{R_\star^2}{a^2} \quad ; \quad \sigma_B = \frac{2\pi^5 k_b^4}{15 c^2 h^3}, \tag{6.1}$$

where $R_\star$, $L_\star$, and $T_{\rm eff}$ are the stellar radius, luminosity, and effective temperature, respectively. $\sigma_{\rm SB}$ is the Stefan-Boltzmann constant, $k_b$ is the Boltzmann constant, $h$ is Plank's constant and $c$ is the speed of light. Stellar luminosities are taken from tabulated main-sequence star values[1] (i.e., obtained from mass-luminosity relations, $L \propto M^a$, where $a$ depends on stellar mass). The resulting irradiation fluxes for HJs typically range from $10^8$ to $10^{10}$ erg cm$^{-2}$ s$^{-1}$, orders of magnitude larger than the fluxes reaching Earth and Jupiter ($1.4 \cdot 10^6$ and $5 \cdot 10^4$ erg cm$^{-2}$ s$^{-1}$, respectively). The corresponding planetary equilibrium temperature is:

$$T_{\rm eq} = \left( \frac{F_{\rm irr}(1 - A_B)}{4\sigma_{\rm SB}} \right)^{1/4} = (1 - A_B)^{1/4} \left( \frac{R_\star}{2a} \right)^{1/2} T_{\rm eff} \,, \tag{6.2}$$

where $A_B$ is the planetary albedo (which we consider zero for simplicity).

The left panel of Fig. 6.1 shows planetary radii, $R_{\rm P}$, as a function of the orbital period $P_{\rm orb}$ for the currently known HJ population with masses $M_P > 0.2\ M_{\rm J}$. In the right panel, we also show the irradiation flux $F_{\rm irr}$ and equilibrium temperature $T_{\rm eq}$, as a function of $P_{\rm orb}$, for the same HJ population. Colors correspond to the spectral type of the host star. The bulk of known HJs orbit F, G, and K-type stars, with orbital distances lower than 0.1 AU, i.e., periods $P_{\rm orb} \lesssim 10$ days. The lines in the left panel correspond to the radii of standard planetary evolutionary models with 1 $\rm M_J$ masses, at 5 Gyr, considering only irradiation and no internal extra heat injection (see Sect. 6.2.1) around different stars (see Table 6.1 for details). Note that most planets are above these lines, showing signs of inflation and the need for a heat injection mechanism. The HJ population below these lines is mostly small planets compatible with high metallicities or large rocky cores (Laughlin et al., 2011), possibly related to early severe mass losses (Sestovic et al., 2018; Lazovik, 2023), which are factors that we do not consider for our evolutionary models, since we aim at studying the average properties of HJs, i.e. with inflated radii. Note that, as mentioned before, the clear correlation is between $R_{\rm P}$ and $F_{\rm irr}$ (or $T_{\rm eq}$). However, as we use $P_{\rm orb}$ to define the Rossby number and dynamo regime (see Sect. 6.3), we choose it as a key variable, instead of the stellar irradiation $F_{\rm irr}$. Due to the degeneracy between stellar mass and $P_{\rm orb}$ in defining $F_{\rm irr}$, the correlation $R_P$ vs $P_{\rm orb}$ (left panel) is essentially lost, in agreement with e.g. Weiss et al. (2013).

[1] In this work, we use the tables of https://sites.uni.edu/morgans/astro/course/Notes/section2/spectralmasses.html

Observed HJs are commonly seen to sit on low-eccentricity orbits (Zink and Howard, 2023), and assumed to be tidally locked, meaning that their rotational and orbital periods $P_{\rm orb}$ are identical. Secondary transits finding azimuthal displacements of the hottest points compared to the substellar points to the East (e.g., Knutson et al., 2007) or West (e.g., Dang et al., 2018), are usually interpreted in terms of atmospheric circulation effects, rather than a deviation from the tidal locking regime. Here, we revisit with some detail the quantitative assessment of the timescale $\tau_{\rm syn}$ required to achieve tidal synchronization, by calculating the rate of change of the planetary rotation $\omega$ under the assumption of circular orbits and zero obliquity:

$$\begin{aligned} \frac{d\omega}{dt} &= \frac{9}{4}\frac{1}{\alpha Q'_P}\frac{GM_{\rm P}}{R_{\rm P}^3}\left(\frac{M_\star}{M_{\rm P}}\right)^2\left(\frac{R_{\rm P}}{a}\right)^6 \\ \text{where} \quad \alpha &= \frac{I}{M_{\rm P}R_{\rm P}^2} \quad , \quad Q'_{\rm P} = \frac{3Q_{\rm P}}{2k_{2,\rm P}} \, , \end{aligned} \tag{6.3}$$

$M_{\rm P}$ and $R_{\rm P}$ are the planetary mass and radius, respectively, $I$ is the planetary moment of inertia, $Q_{\rm P}$ is the dissipation factor, $k_{2,\rm P}$ is the Love number of the planet and $G$ is the gravitational constant (e.g. Goldreich and Soter, 1966; Murray and Dermott, 1999; Grießmeier et al., 2007). The synchronization timescale is then given by:

$$\tau_{\rm syn} \approx \frac{\Delta\omega}{d\omega/dt} = \frac{4}{9}\frac{a^6 I Q'_P \Delta\omega}{GM_*^2 R_{\rm P}^5} \simeq \frac{G\alpha}{36\pi^4}\frac{M_{\rm P}Q'_P P^4\omega_i}{R_{\rm P}^3} \, , \tag{6.4}$$

where $P$ is the orbital period, and $\Delta\omega = \omega_i - \omega_f \sim \omega_i$, since $\omega_i \gg \omega_f$, the difference between initial and final rotation. On the right panel of Fig. 6.1 we show in purple different lines of constant $\tau_{\rm syn}$ for a $M_{\rm P} = 1$ $\rm M_J$ planet, assuming $\alpha = 2/5$ (homogeneous sphere), $R_{\rm P} = 1.5$ $\rm R_J$, $Q'_P = 5 \cdot 10^5$ as in Grießmeier et al. (2007), and an initial spin period of 5 hours typical of known fast-rotating sub-stellar objects, $\omega_i = (2\pi/5\ \rm h)$, and conservatively larger than the breakup values which can be used as initial condition (Batygin, 2018). Most of the parameters adopted in the above expression evolve significantly over time, particularly at early stages ($t \lesssim 10$ Myr) when planetary contraction proceeds rapidly, and orbital migration as well as interactions with the protoplanetary disk are important. For this reason, we adopted reference values that yield systematically conservative estimates, overestimating $\tau_{\rm syn}$ by factors of a few. Specifically, realistic adjustments would include considering a radially dependent density profile (moment of inertia factor $\alpha < 2/5$), larger radii of very young gas giants depending on their formation entropy (e.g., Fortney et al., 2007), and shorter initial spin periods, which can be as brief as about 2 hours for newborn brown dwarfs and low-mass stars.

We show in Fig. 6.1 the location where $\tau_{\rm syn}$ equals the main sequence lifetime of the host star as a black line. We calculated the duration of the main sequence as $\tau_{\rm ms} = 10$ Gyr $(M_\star/M_\odot)^{-2/5}$ (Carroll and Ostlie, 2017, chap. 13). Planets lying to the left of this line are expected to be tidally locked, or to reach that state within the stellar lifetime. To evaluate how planetary mass affects $\tau_{\rm syn}$, we observe that for a 0.2 $\rm M_J$, 2 $R_J$ planet, $\tau_{\rm syn}$ becomes a factor of 10 smaller. On the other hand, a massive $M_{\rm P} = 10$ $\rm M_J$ takes about one order of magnitude longer to synchronize. The gray shading shows this range of values around the $\tau_{\rm syn} = \tau_{\rm ms}$ (and each $\tau_{\rm syn}$ line has a similar uncertainty). From Fig. 6.1, it is clear that the HJs population with $T_{\rm eq} > 1000$ K should be tidally locked, and only warm Jupiters (having $P_{\rm orb} \gtrsim 20 - 30$ d) could still preserve their initial faster rotation.

Note that, recently, Wazny and Menou (2025) used global circulation models to show that the presence of a non-negligible magnetic torque in atmospheres permeated by thermally driven winds could lead to substantial spin-orbit asynchronization, particularly in planets with high equilibrium temperatures and strong magnetic fields. These effects are beyond the scope of this paper, and thus, we adopt the assumption of tidal synchronization throughout our analysis.

## 6.2 Planetary evolution

### 6.2.1 Cooling model

Similarly to Chapter 5, we use the public code *MESA*, specifically its version 24.08.1, to evolve irradiated gas giants, allowing for additional internal heat deposition. The equations include mass conservation, hydrostatic equilibrium, energy conservation, and energy transport. They are the same as Eqs. 5.1-5.4, except for the energy conservation, which now reads:

$$\frac{dL}{dm} = -T\frac{ds}{dt} + \epsilon_{\rm irr} + \epsilon_{\rm heat} \ , \tag{6.5}$$

The right-hand side includes the specific heating/cooling rates: gravitational contraction, stellar irradiation $\epsilon_{\rm irr}$ (Guillot et al., 1996), and any internal heat source $\epsilon_{\rm heat}$. The latter has been included in several previous works using *MESA* to consider the effect of continuous heat deposition and inflation in HJs (Komacek and Youdin, 2017; Thorngren and Fortney, 2018; Komacek et al., 2020).

The atmospheric irradiation is implemented in a simplified way, as in previous *MESA* HJ works (Paxton et al., 2013; Komacek and Youdin, 2017; Komacek et al., 2020): the specific energy absorption rate is assumed to be $\epsilon_{\rm irr} = F_{\rm irr}/(4\Sigma_\star)$, i.e. a uniform deposition of heat through the outermost mass column $m(r) \leq \Sigma_\star$, where $\Sigma_\star$ parametrises how deep is the column which atmospheric absorption occurs, without a more detailed modeling of the atmospheric composition and opacities. We fix it to 200 g cm$^{-2}$, which approximately corresponds to a grey opacity $\kappa = 5 \times 10^{-3}$ cm$^2$ g$^{-1}$, with a maximum depth of a fraction of a bar. We consider a constant-in-time value of the stellar irradiation $F_{\rm irr}$, thus neglecting the slight increase in stellar luminosity $L_\star(t)$ throughout the main sequence phase. We refer to Lopez and Fortney (2016); Komacek et al. (2020) for works including this effect, important especially at the end of the main sequence and beyond it. As commonly done in similar previous works, we assume here an inert homogeneous rocky core of mass 10 $M_\oplus$ and homogeneous density $\rho_c = 10$ g cm$^{-3}$ (which together set the core size $R_{\rm core}$), and a fixed solar composition for the envelope. We don't discuss here the effects of the total metallicity, and the comparison between a diluted or solid core, since they have been shown in detailed cooling studies for HJs (e.g., Burrows et al., 2007; Thorngren et al., 2016) and Solar planets (e.g., Wahl et al., 2017; Yıldız et al., 2024).

Generally speaking, during the initial stages ($t \lesssim 10$ Myr), the evolution of the planetary radius depends on the assumed initial condition (i.e., the internal entropy at formation). At later times, instead, models are insensitive to it, and the planet shrinks during its long-term evolution due to its cooling (e.g., Burrows et al., 2007; Paxton et al., 2013). Therefore, the detailed value of the initial condition is not relevant. To help *MESA* numerical convergence at early ages for extremely irradiated models, we also use a short relaxation phase at the beginning of evolution, after applying the irradiation.

We denote the depth at which irradiation reaches $R_{\rm irr}$, and the radiative-convective boundary as $R_{\rm RCB}$. We also define the outer radius of the dynamo region, $R_{\rm dyn}$, by the pressure above which hydrogen becomes metallic, which we take as $P_{\rm dyn} = 1$ Mbar (French et al., 2012). Note that, strictly speaking, the transition to the metallic hydrogen is not sharp and depends on the temperature, as well (Bonitz et al., 2024). However, for simplicity, and given the high theoretical uncertainties about the exact location of the phase transition, we consider a single value of $P_{\rm dyn}$ as a first approximation. In the metallic hydrogen region, the electrical conductivity is $\sigma \sim 10^6$ S m$^{-1}$ (French et al., 2012; Bonitz et al., 2024), orders of magnitude larger than in the molecule-dominated envelope, where the conductivity is dominated by the thermal ionization of alkali metals (e.g., Kumar et al., 2021). Considering a typical convective velocity $v_{\rm conv} \sim 0.1$ m s$^{-1}$ (Fuentes et al., 2023), a Jupiter-like shell thickness $L \sim 5 \cdot 10^7$ m, the resulting magnetic Reynolds number in the dynamo region is ${\rm Rm} = v_{\rm conv}\, L/\eta = \mu_0\, \sigma\, v_{\rm conv}\, L \sim 10^6 - 10^7$ (where $\eta$ is the magnetic diffusivity and $\mu_0$ is the magnetic permeability in vacuum), well above the minimum values commonly thought to be needed for dynamo, ${\rm Rm} \gtrsim 50$ (Christensen and Aubert, 2006).

Table 6.1: Characteristics of the stars considered throughout the text (specifically the planetary irradiated models in Fig. 6.1 and in App. 6.3.1.

| Stellar type | $T_\star$ [K] | $L_\star$ [L$_\odot$] | $M_\star$ [M$_\odot$] |
|---|---|---|---|
| AV3 | 8750 | 12 | 1.86 |
| FV5 | 6700 | 2.4 | 1.25 |
| GV2 | 5800 | 1 | 1.00 |
| KV3 | 4800 | 0.31 | 0.746 |
| MV3 | 3500 | 0.046 | 0.463 |

Hereafter, we consider planets with 1 d $\lesssim P_{\rm orb} \lesssim$ 30 d orbiting a representative sample of five main-sequence stars, listed in Table 6.1. We test planetary masses from 0.5 to 12 M$_{\rm J}$, sampling the whole inflated HJ population in period and flux as seen on the right of Fig. 6.1. We restrict to planets $M_P \geq 0.5 M_{\rm J}$ due to the above-mentioned contamination of partially evaporated, high-metallicity planets, as discussed also in e.g. Sestovic et al. (2018).

### 6.2.2 Extra heat

Since the radius inflation, which calls for continuous internal heat deposition, correlates with the irradiation flux (Weiss et al., 2013), a commonly used parameter is the heating efficiency $\epsilon$, defined as the ratio between the total deposited heat rate, $Q_{\rm dep} = \int_M \epsilon_{\rm heat}\, {\rm d}m$, and the irradiation flux integrated over the planetary surface:

$$\epsilon = \frac{Q_{\rm dep}}{\pi R^2 F_{\rm irr}} \ . \tag{6.6}$$

Values of efficiency of a few percent or less are enough to inflate planets to the observed radii, if the heat is deposited in the convection region (e.g., Batygin and Stevenson, 2010; Batygin et al., 2011; Wu and Lithwick, 2013; Komacek and Youdin, 2017). A detailed statistical study by Thorngren and Fortney (2018) derived an analytical expression for the amount of deposited heat fraction as a function of the incident flux, $\epsilon(F_{\rm irr})$, that best fits the trend $R_{\rm P}(F_{\rm irr})$ seen in the entire HJ population:

$$\epsilon = \left(2.37^{+1.3}_{-0.26}\%\right) \exp\left[ - \frac{\left(\log_{10}(F_{\rm irr}/F_0) - 0.14^{+0.060}_{-0.069}\right)^2}{2 \cdot \left(0.37^{+0.038}_{-0.059}\right)^2} \right] \ , \tag{6.7}$$

where $F_0 = 10^9$ erg s$^{-1}$cm$^{-2}$. This inferred efficiency shows a strong, non-monotonic dependence on the stellar irradiation: the maximum heating efficiency grows with $T_{\rm eq}$ to reach a maximum of $\sim 2.5\%$ at $T_{\rm eq} \approx 1500$ K and then decreases, having $\sim 0.2\%$ at $T_{\rm eq} \approx 2500$ K. Such a non-monotonic trend fits very well with the Ohmic models, for which, as the induced field increases with $T_{\rm eq}$, the Lorentz forces exert an effective larger drag on the thermal winds (e.g., Perna et al., 2010a; Menou, 2012). As a consequence, very hot Jupiters might have slower winds, which limit the available energy for the magnetic induction and decrease the overall heating efficiency.

Note that the expression above is purely phenomenological and shows high standard deviations, reflecting the large HJ radius dispersion at a given $T_{\rm eq}$. In a similar work, Sarkis et al. (2021) obtain a similar fit with a Gaussian function for the flux, but with the efficiency peaking at $T_{\rm eq} \approx 1850\,K$. Moreover, both Thorngren and Fortney (2018) and Sarkis et al. (2021) note that this heating efficiency is degenerate with heavy mass fraction. Despite these uncertainties and caveats, the analytical function above, which we use in this study, offers an easy way to explore the representative internal heating we can infer for a given planet and star.

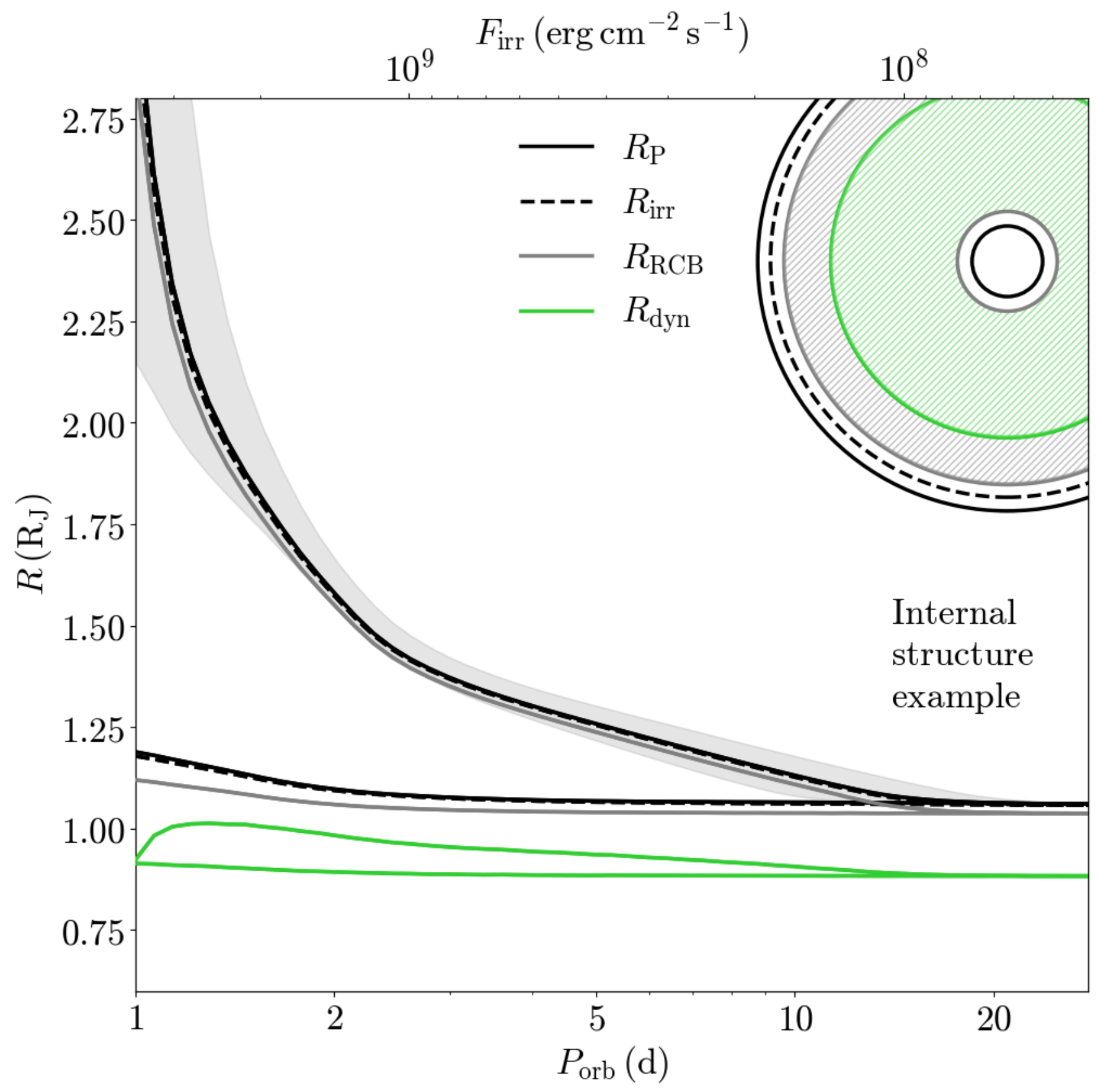

Figure 6.2: Dynamo outer surface radius $R_{\rm dyn}$ (green), RCB radius $R_{\rm rcb}$ (gray), bottom of the irradiated region $R_{\rm irr}$ (dashed), and planetary radius $R_{\rm P}$ (solid black), for a 1 $\rm M_J$ planet orbiting a 1 $\rm M_\odot$ star, at 5 Gyr, as a function of orbital period, for both heated (upper lines) and non-heated models (lower lines). The gray area shows the outer radius uncertainty for the heated models, accounting for the range of parameters quoted in Eq. (6.7). In the upper right corner, we show a sketch of the relative position, not at scale, of the different radii, including the internal core $R_{\rm core}$ (solid black) and a possible stratified layer close to the core (gray).

The efficiency is not the only parameter to consider: it is also important to specify where the heat is deposited, i.e., the $\epsilon_{\rm heat}(r)$ profile (Ginzburg and Sari, 2015; Komacek and Youdin, 2017). Since we are not assuming any particular mechanism responsible for the extra heat, for simplicity we inject a uniform specific heat rate, $\epsilon_{\rm heat} = Q_{\rm dep}/M_{\rm shell}$ (heat per unit time per unit mass), over a given shell with mass $M_{\rm shell}$, delimited by two radii, $r_{\rm bot}$ and $r_{\rm top}$, for which we consider two

simple scenarios:

1. Extended deposition in the dynamo region ($\epsilon_{\rm heat,extended}$): $r_{\rm bot} = R_{\rm core}$; $r_{\rm top} = R_{\rm dyn}$.
2. Deposition in the outer convective region, outside the dynamo region ($\epsilon_{\rm heat,outer}$): $r_{\rm bot} = R_{\rm dyn}$; $r_{\rm top} = R_{\rm irr}$.

Note that if one considers heating up to the irradiated layer ($r_{\rm bot} = R_{\rm core}$; $r_{\rm top} = R_{\rm irr}$), i.e. including part of the stratified, radiative layers, the results are almost indistinguishable results from the first one. As noticed in previous works (Komacek and Youdin, 2017), heating the radiative region is much less effective in terms of inflated radii, because the mass in the radiative region is very small and the heat mostly escapes outwards. Moreover, for very high values of irradiation and heating, including the outermost layers, the code can fail to converge more easily.

The second type of heating injection (i.e., above the dynamo radius) is inspired by the fact that most heating mechanisms deposit the bulk of energy in low-density regions. Physical models such as ohmic, tidal, or turbulent heating have different heating distributions (Batygin and Stevenson, 2010; Batygin et al., 2011; Wu and Lithwick, 2013; Ginzburg and Sari, 2015, 2016), but all of them deposit most of the heat in the outer layers of the planet, well outside the metallic dynamo region.

As we will show in the Sect. 6.2.4 and below, heat deposition above the dynamo region leads to significant differences. Further exploration of the heating radial profile has been conducted in detail in several other works, e.g., different power laws (Ginzburg and Sari, 2015), or Gaussian profiles at different depths (Komacek and Youdin, 2017; Komacek et al., 2020), and is not a matter of this study, where we only consider the two simplified scenarios above. The inlists used to run our models can be found on Zenodo[2].

### 6.2.3 Extended heat cases: general properties

In Fig. 6.2 we show the evolution of the planetary radius of a 1 $M_J$ planet orbiting a 1 $M_\odot$ star at 5 Gyr, as a function of the orbital period. In the plot, we compare irradiated models and heated models with the extended type of injection. Planets heated as shown in Eq. (6.7) (Thorngren and Fortney, 2018) inflate considerably and reach the range of observed HJ radii. The uncertainty for the outer radius of internally heated models reflects different values for $\epsilon_{\rm heat}(F_{\rm irr})$, considering the uncertainty ranges of the parameters in Eq. (6.7). We also show a not-to-scale sketch of the different radii for illustration purposes. The order $R_{\rm P} > R_{\rm irr} > R_{\rm RCB} > R_{\rm dyn}$ always holds for both heated and non-heated models. There is only an exception for the highest irradiated cases, in which an additional convective outer layer above $R_{\rm irr}$ appears. More considerations regarding the behavior of convective regions for highly irradiated and heated models are found below.

In Fig. 6.3 we show the internal profiles for some chosen periods in Fig. 6.2. Specifically, we show the temperature $T$, the density $\rho$ the density scale height $H_\rho(r) = P/\rho g$, the convective velocity $v_{\rm conv}$, and the convective flux $q_{\rm c}$. The convective velocity comes from mixing length theory (MLT) based on the model in Kuhfuss (1986), which reduces to the expression given by (Cox and Giuli, 1968, chap. 14), in the limit of long time steps (see Paxton et al. (2011); Jermyn et al. (2023) for details). We define $q_{\rm c}$ as the convective heat flux in the dynamo-generating region as follows:

$$q_{\rm c} = \frac{2\, c_P\, T\, \rho^2\, v_{\rm conv}^3}{P\, \delta} \, , \tag{6.8}$$

where $\delta = -(\partial \ln \rho / \partial \ln T)_P$.

[2] https://doi.org/10.5281/zenodo.15800305

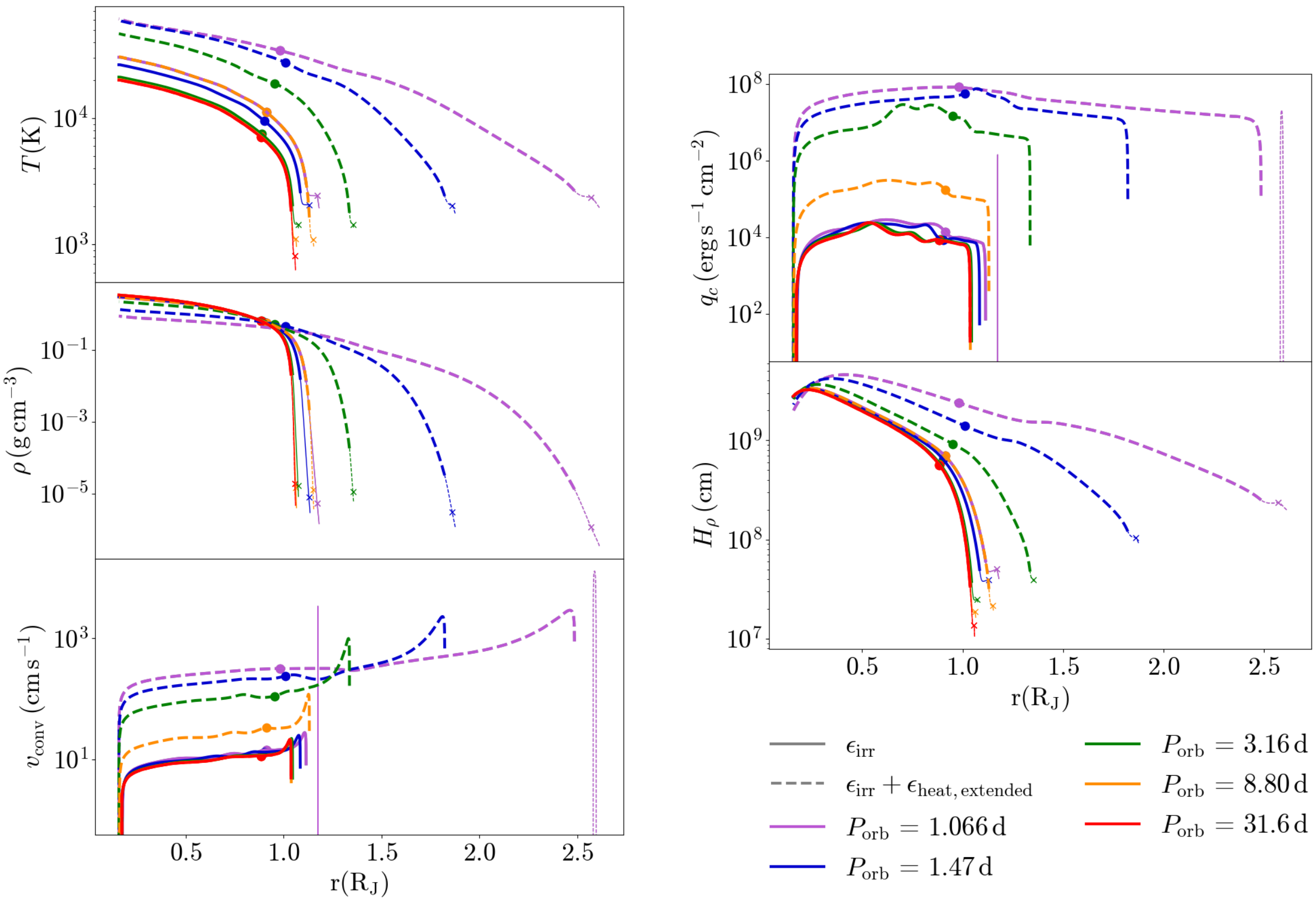

Figure 6.3: Radial profiles of temperature, density, convective velocity, heat convective flux, and density scale height, for 1 $M_J$ planets orbiting a 1 $M_\odot$ star at 5 Gyr with different $P_{\mathrm{orb}}$. Models without and with extra heating (below $R_{\mathrm{dyn}}$) are shown by solid and dashed lines, respectively. $R_{\mathrm{irr}}$ is marked with a cross, $R_{\mathrm{RCB}}$ is the transition from thick to thin lines, and $R_{\mathrm{dyn}}$ is marked with a dot.

When the planets are far from the heating efficiency peak, there is almost no difference between heated and non-heated models. Instead, inflation implies very different radial profiles: $T$, $H_\rho$, $v_{\mathrm{conv}}$, and $q_c$ substantially increase for a given model and age. At the same time, the density in the innermost parts decreases due to the higher internal temperature (i.e., inflation and smaller mass/volume ratio).

Note that we analyze models at 5 Gyr, which are representative of the bulk of HJs. Most evolution models of irradiated, internally heated HJs show that at ∼Gyr ages (corresponding to the vast majority of known HJs), the shrinking and cooling stalls, due to the essential balance between the internal heating and the long-term cooling (e.g., Komacek and Youdin, 2017). Therefore, the internal structure doesn't change notably. In reality, this behaviour relies on the assumption that both the stellar luminosity and the internal heat are constant in time, then neglecting the evolution of stellar luminosity (see Lopez and Fortney (2016); Komacek et al. (2020) for studies of re-inflation in this context), and the possible change in the heating efficiency of the underlying mechanism (see Viganò et al. (2025) for Ohmic models with evolving heating rates).

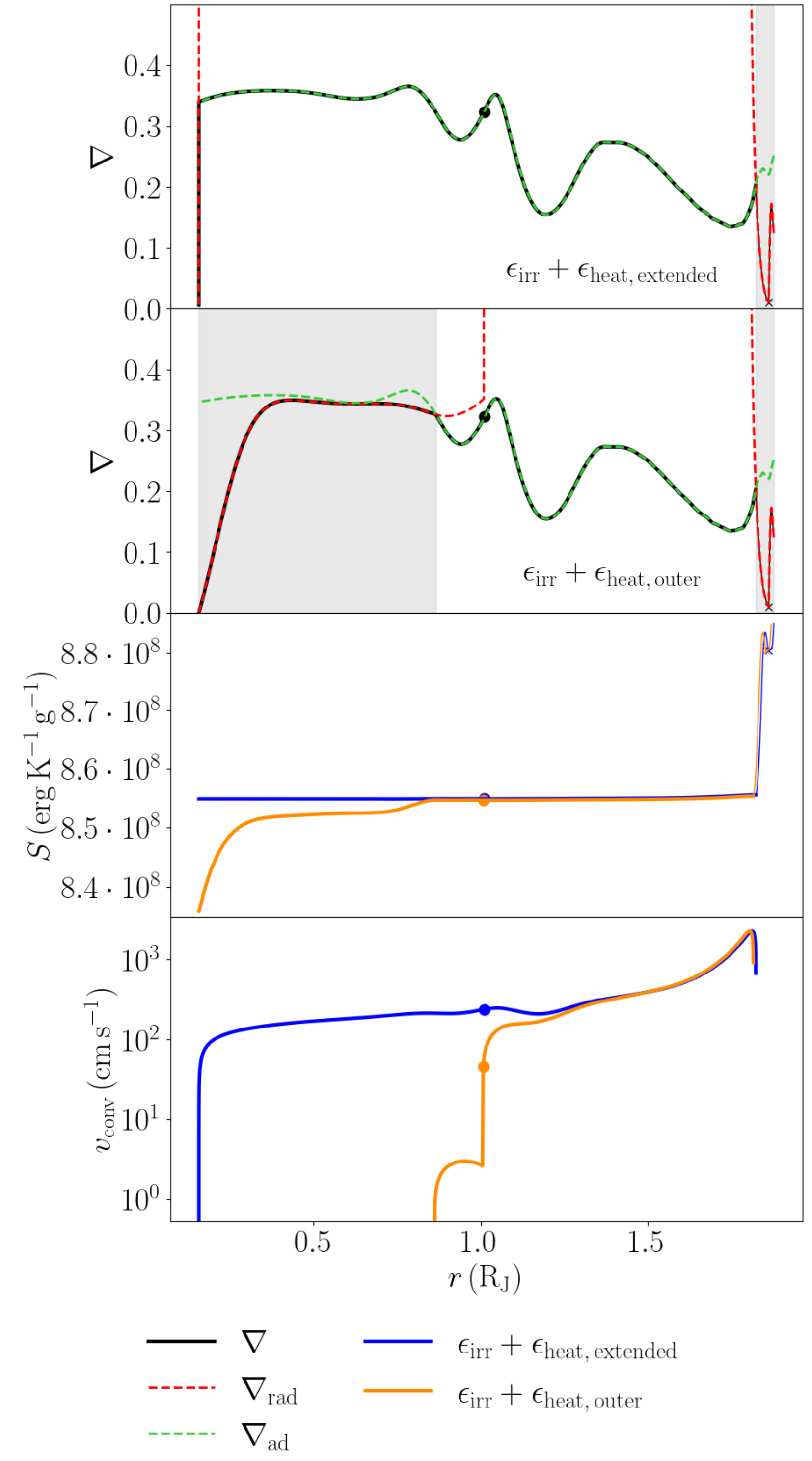

Figure 6.4: Radial profiles of extended and outer heating models for a 1 $M_J$ planet orbiting a 1 $M_\odot$ star at 5 Gyr, with $P_{\rm orb}$= 1.47 d. *Top panels*: Total logarithmic temperature gradient $\nabla$ (black solid line), with the adiabatic (green dashed line) and radiative (red dashed line) temperature gradients, $\nabla_{\rm ad}$ and $\nabla_{\rm rad}$, respectively. Gray zones mark stably stratified layers, i.e., $\nabla_{\rm rad} < \nabla_{\rm ad}$. *Bottom panels*: Specific entropy and convective velocity for the same models.

### 6.2.4 Heat deposited in the outer region: suppression of convection

Substantial differences appear with the outer type of heat injection. The heat is deposited in a significantly shallower layer, which, for a high enough deposited heat rate, leads to internal convection suppression. This is physically justified by comparing the logarithmic temperature radiative gradient ($\nabla_{\rm rad} = 3\kappa LP/64\pi\sigma GmT^4$) and convective/adiabatic gradient ($\nabla_{\rm ad}$), which MESA determines using MLT. Following the Schwarzchild criterion, the actual temperature gradient $\nabla$ is set to the smaller of the adiabatic or radiative, i.e., $\nabla = d\ln T/d\ln P = \min(\nabla_{\rm rad}, \nabla_{\rm ad})$. Therefore, convection develops where $\nabla_{\rm ad}$ is lower than $\nabla_{\rm rad}$, and on the other hand, stably stratified layers arise when $\nabla_{\rm rad}$ is lower than $\nabla_{\rm ad}$. Similar to Komacek and Youdin (2017), we plot $\nabla, \nabla_{\rm rad}$ and $\nabla_{\rm ad}$ in the two top panels of Fig. 6.4 for the same 1 $M_J$ planet with a $P_{\rm orb}$= 1.47 d. The gradient inversion is always present in the outermost part of the planets, covering at least the irradiation depth. For the model with outer heating, the inversion also happens in a substantial part of the interior near the core, creating an additional internal stratified layer.

This behavior can also be seen with the specific entropy gradient: the extended heated model (blue lines) remains isentropic over a large volume, whereas the outer heating model produces a positive entropy gradient in correspondence with the layers just below where the heat is deposited. With the outer heating case, $v_{\rm conv}$, and thus $q_{\rm c}$, significantly reduces and even vanishes below $R_{\rm dyn}$. The reduction of convective region size has been briefly mentioned in previous works where intense heating was deposited at specific interior locations (e.g., Komacek and Youdin 2017; Komacek et al. 2020). Importantly, it has relevant consequences for the internal dynamo (see Sect. 6.4.1).

Note that, on the contrary, planetary radius, external luminosity (set by $R_{\rm P}$ and $F_{\rm irr}$), and the profiles of $T$, $\rho$, $P$, $g$ are almost indistinguishable between the two heating types. This is because the layers responsive to heating are the external ones, due to the lower density: they are the ones responsible for the inflation, as seen by comparing how the density profiles extend, in Fig. 6.3. Therefore, depositing the heat over most of the planet, or only in the outer convective region, provides a very similar evolution of thermodynamic profiles.

Additionally, in our most irradiated models, a shallow convective layer arises in the outermost regions of the planet. This outer, shallow layer would be conductive due to alkali thermal ionization (Kumar et al., 2021) ($T_{\rm eq} > 1000$ K) and thus could interfere with the dynamo-generated magnetic field. But this layer is cut off if the internally generated magnetic field is strong enough. We use the critical value as given by (Jermyn and Cantiello, 2020, Eq. 9):

$$B_{\rm crit,\, r} = \sqrt{\frac{4\pi\rho c_s^2 Q(\nabla_{\rm rad} - \nabla_{\rm ad})}{1 - Q(\nabla_{\rm rad} - \nabla_{\rm ad}) + d\ln\Gamma_1/d\ln p}} \tag{6.9}$$

where $c_s$ is the sound speed, $\nabla_{\rm rad} = 3\kappa L/64\pi GM\sigma T^4$ it the radiative temperature gradient, $\nabla_{\rm ad} = \partial\ln T/\partial\ln P|_s$ it the adiabatic temperature gradient, $\Gamma_1 = \partial\ln P/\partial\ln\rho|_s$ it the usual adiabatic index, and

$$Q = 1 + 4\frac{P_{rad}}{P_{gas}}$$

For the models shown in Fig. 6.3 and 6.2, only the planet with $P_{\rm orb} = 1.066$ d models shows this shallow convective outer layer. Its associated $B_{\rm crit,\, r}$ corresponds to about 100 G. As seen in Sect. 6.4.1 the field near the planetary surface does not reach such high values for $P_{\rm orb} = 1.066$, not suppressing this possible convective layer. These results are sensitive to external boundary conditions and the heat deposition $\epsilon_{\rm irr}(P)$. Furthermore, the surface deposition for HJs is far from uniform, which would lead to a stronger convection layer in the dayside and hamper its existence in the nightside. Global circulation three-dimensional models are needed to address the possible existence of such a layer and their interaction with the thermal winds.

## 6.3 Rossby number estimation

The Rossby number, Ro, is the ratio between inertial forces (dominated by convection) and the Coriolis force. Most planets in the solar system are fast rotators, i.e. they are in a low Ro regime: defining $\mathrm{Ro} = P_{\mathrm{rot}}/\tau_{\mathrm{turn}}$, i.e. the ratio of rotation period to convection turnover timescales, Earth and Jupiter have $\mathrm{Ro} \sim 10^{-5}$ and $10^{-4}$, respectively. In this regime, the widely used dynamo scaling laws by Christensen et al. (2009), explained below, apply. In HJs, as seen in Sect. 6.1, tidal locking implies $P_{\mathrm{rot}} = P_{\mathrm{orb}}$. We obtain the convective timescale, $\tau_{\mathrm{turn}}(r)$, from the evolutionary models, by using the radial profiles of density scale height, $H_\rho(r)$, and convective velocity, $v_{\mathrm{conv}}(r)$, so that

$$\mathrm{Ro}(r) = \frac{P_{\mathrm{orb}}\, v_{\mathrm{conv}}(r)}{H_\rho(r)} \, . \tag{6.10}$$

In Fig. 6.5 we show $\mathrm{Ro}(r)$, for 50 different values of $P_{\mathrm{orb}} \in [1, 25]$ d, considering the representative case of a 1 $\mathrm{M_J}$ orbiting a 1 $\mathrm{M}_\odot$ star, at 5 Gyr. We compare purely irradiated cases with the irradiated plus extended heated ones. Even though the extra heat mostly inflates the outer layers, it leads to Ro increasing by about 1 order of magnitude in the inner layers, due to the higher internal entropy. Still, the dynamo region maintains $\mathrm{Ro} < 0.1$, regardless of the stellar irradiation, so that we can infer that HJs remain in the fast-rotation regime. Since the evolutionary models depend on the irradiation alone, the results for Ro are the same for other stellar types, but shifted in orbital periods, as shown in App. 6.3.1 for MV3, KV3, and FV5 stars. HJs around the most massive star lead to larger inflation and higher values for Ro, but the dynamo region still remains lower than 0.1 in all cases. The sensitivity of results with planetary masses is discussed below.

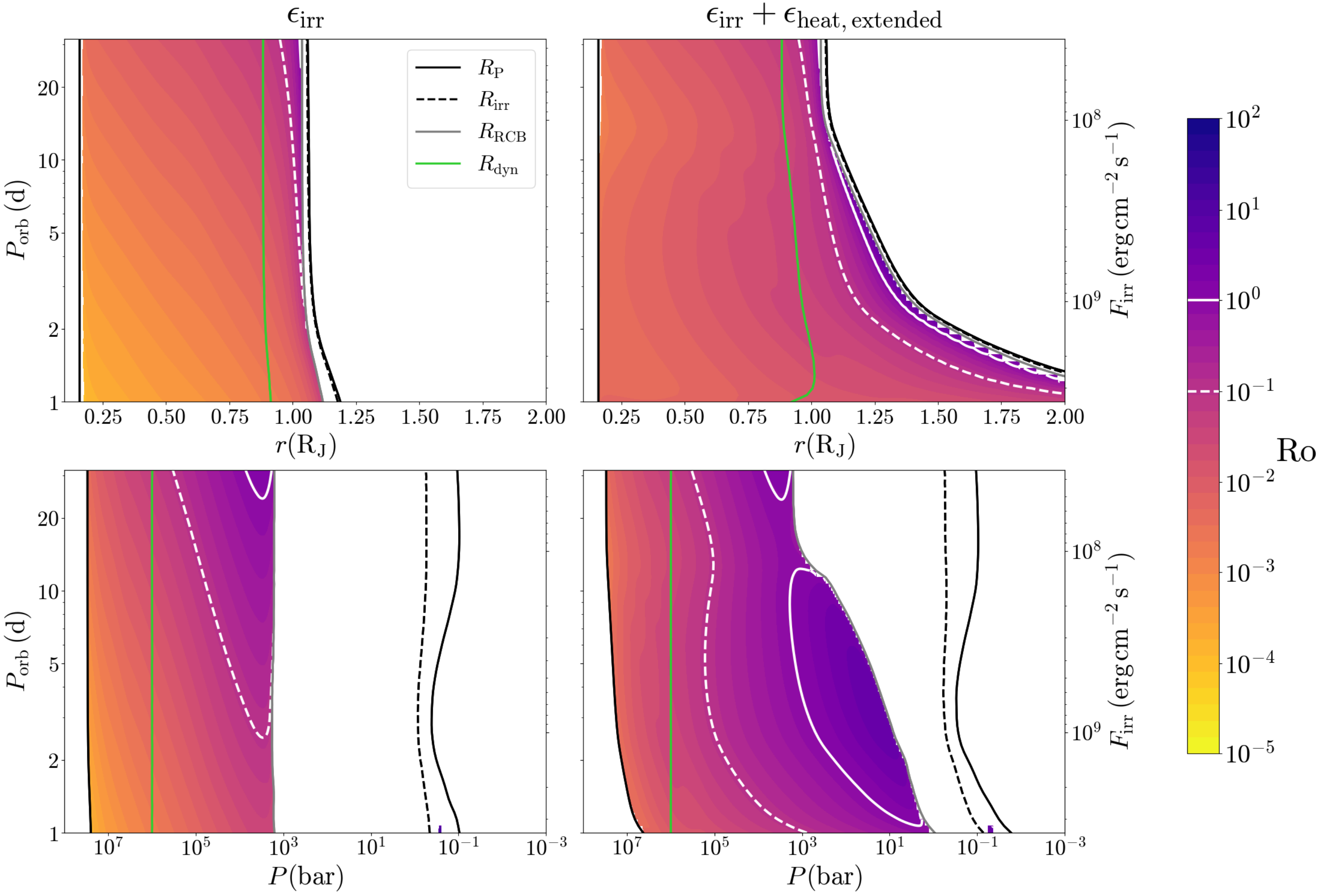

Figure 6.5: Rossby number as a function of radius or pressure for 1 $M_J$ planets orbiting a solar-like G2 star, comparing heated (left panels) and non-heated (right panels) models. The different radii introduced in Fig. 6.2 are shown as in the legend, while the white lines mark Ro=0.1 (dashed) and Ro=1 (solid).

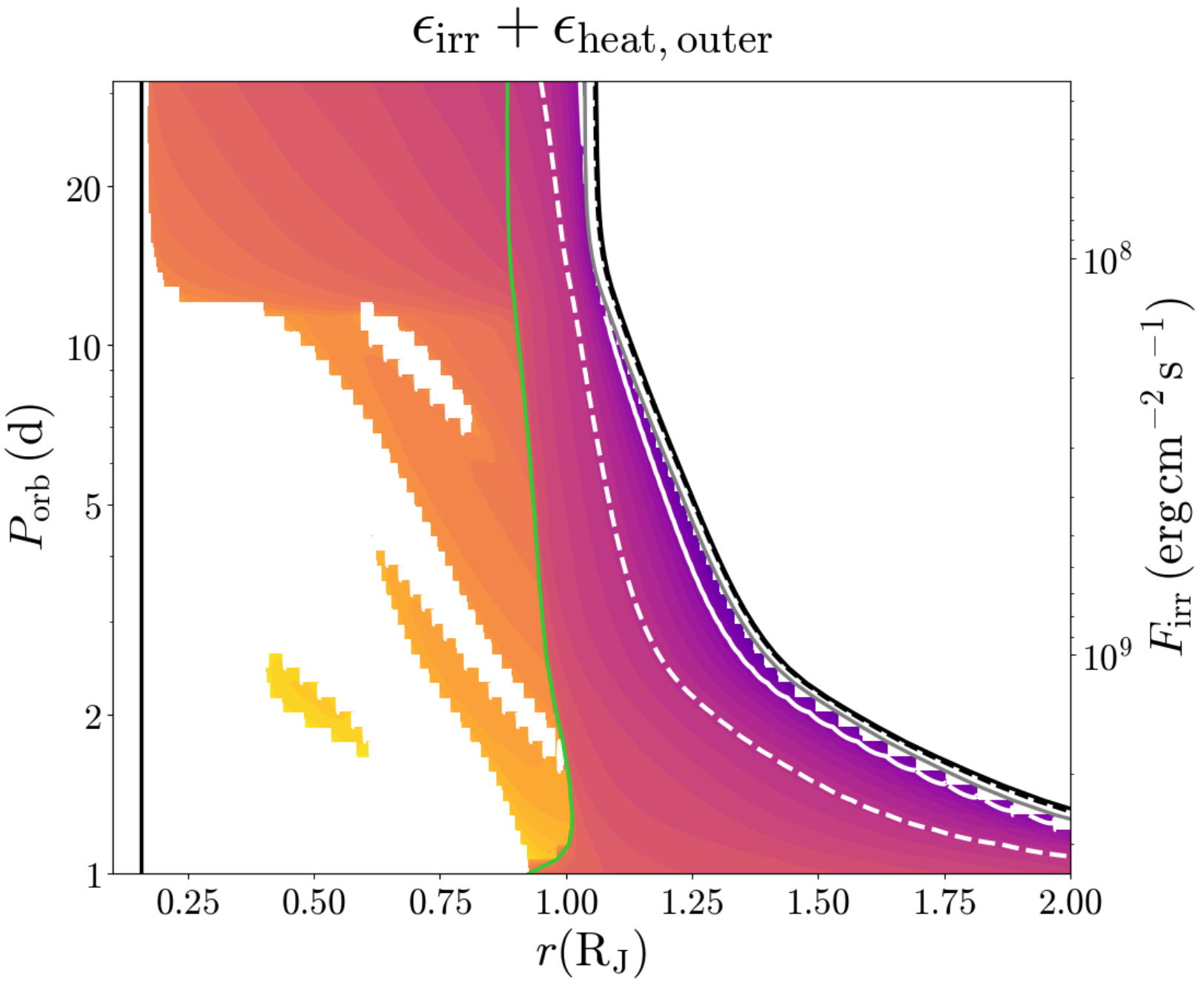

Figure 6.6: Similar to Fig. 6.5, but for the outer deposition of heat, i.e., only above the dynamo region.

In Fig. 6.6 we show the same upper right panel of Fig. 6.5 for the case with heat injection above the dynamo region. As expected from Fig. 6.4, there are large non-convective areas, where Ro is formally zero, in the metallic hydrogen region.

In Fig. 6.7 we show Ro for several different mass planets, for two representative orbital periods $P_{\rm orb} = 2.45$ and 31.6 d, for both heated and non-heated models. We show four planetary masses, which well represent the monotonic trend with $M_{\rm P}$. Heated and non-heated models are indistinguishable for long periods, as the $\epsilon(F_{\rm irr})$ is negligible. As seen in the top panel of Fig. 6.7, inflation has minimal effect for planets above 8 $\mathrm{M_J}$, with only a slight increase of Ro with a factor less than 2. Instead, a low-mass planet of 0.5 $\mathrm{M_J}$ experiences a more than one order of magnitude increase in Ro, although it remains in the fast rotator regime. This still has consequences for the interior dynamics, as an increase in Ro may lead to more multipolar but weak dynamo-generated magnetic fields (Christensen and Aubert, 2006; Jones, 2011; Davidson, 2013).

Planets with longer $P_{\rm orb}$ but still tidally locked, as the ones shown on the top panel of Fig. 6.7, have the highest Ro. The most conductive layers of low-mass planets are generally well within the fast-rotation regime, but for masses higher than 4 $\mathrm{M_J}$, a large fraction of their metallic hydrogen regions have Ro $\gtrsim 0.1$, being the only exceptions to the overall finding of this study. Note that for longer period planets, $P_{\rm orb} > 15$ d, the heating mechanism is low and there is no inflation independent of the host star stellar type (see section below). Therefore, the argument for convection suppression would not apply to high-mass long-period planets.

A NASA archive search reveals only three targets with 4 $\mathrm{M_J}$ closer than 100 pc and 15 d$<$ $P_{\rm orb}$ <40 d, all orbiting G-type stars. The most promising candidate is GJ 86 b (Stassun et al., 2017), which has a 15.8 d orbit, a mass of 4.4 $\mathrm{M_J}$, and is located only at 10.8 pc. The other two planets are farther away: HD 72892 b (39.4 d, 5.5 $\mathrm{M_J}$, 69.7 pc, Feng et al. (2022)) and TIC 393818343 b (16.2 d, 4.3 $\mathrm{M_J}$, 93.7 pc, Sgro et al. (2024)). Usual definitions for HJs require $P_{\rm orb} <$ 10 d or $T_{\rm eq} >$ 1000 K (see Gan et al. (2023) and within), then we can safely say that the vast majority of confirmed HJs are in the low-Ro regime.

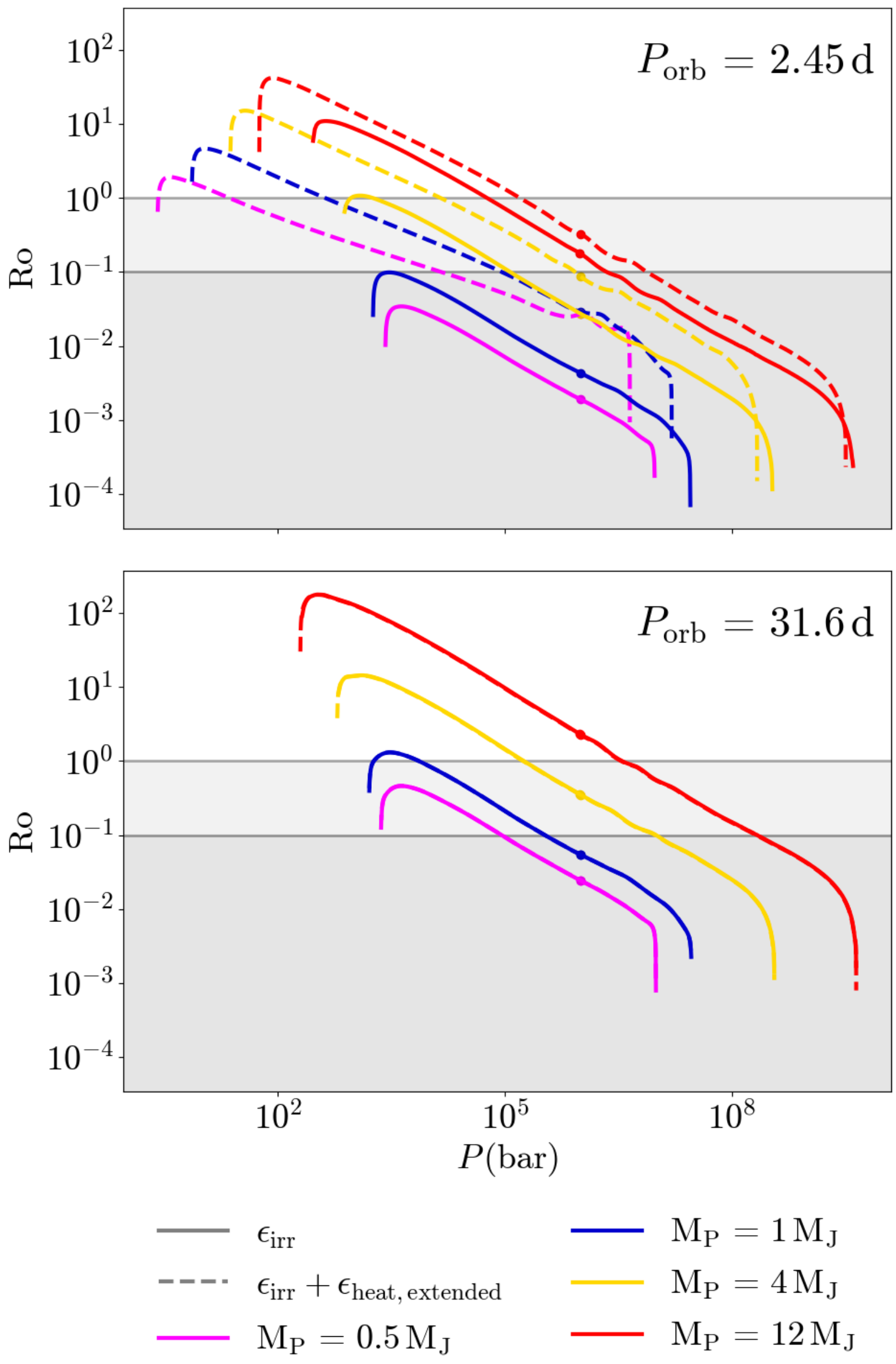

Figure 6.7: Rossby number as a function of pressure for different mass planets orbiting a G2 star with two different periods, comparing models with the extended heating (dashed) with only irradiated models (solid).

### 6.3.1 Rossby number dependence on stellar type

To evaluate the dependence of Ro on stellar type, we use the properties of the specific stars shown in Table 6.1. Note that, by construction, the evolutionary models are the same but shifted in orbital period, since the relevant parameter is the stellar irradiation, which depends on $P_{\rm orb}$ and stellar mass. The Rossby profiles are instead different, since the Ro definition includes a further $P_{\rm orb}$ dependency. Similarly to Fig. 6.5, in Fig. 6.8 we show Ro as a function of planetary depth and orbital period for an M3, K3, and F5 stars. As expected, the more massive the star, the more inflated the planets are, which leads to larger values of Ro, though they remain smaller than 0.1 for the F5 models.

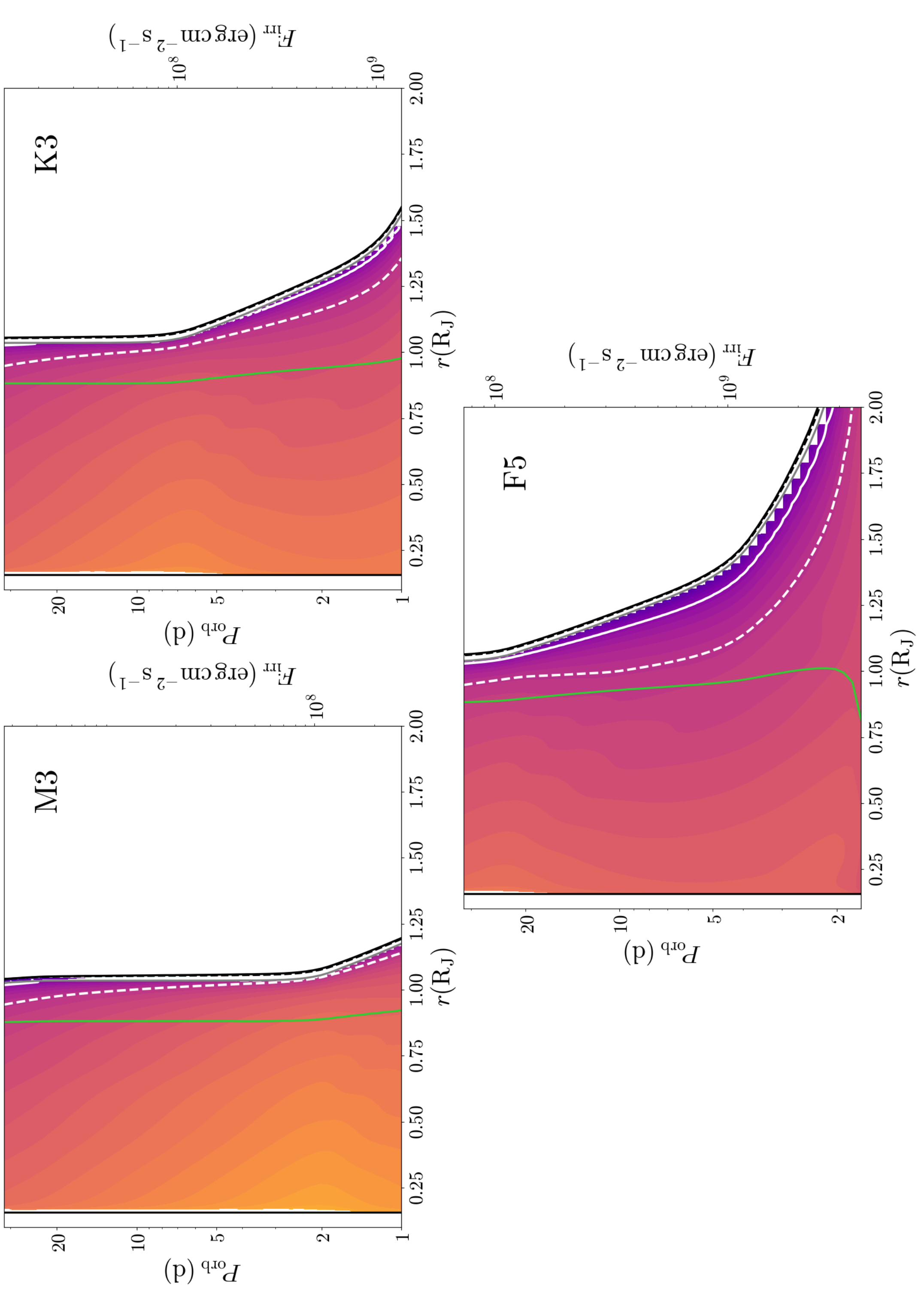

Figure 6.8: Similar to Fig. 6.5: Ro as a function of $r$ and $a$ for a 1 $M_J$ planet orbiting the different stars shown in Table 6.1. Only the inflated models with heat injection below the irradiated layer are shown.

## 6.4 Observational consequences

### 6.4.1 Magnetic field scaling laws

Combining energy equipartition arguments, the results for a large set of dynamo simulations (Christensen and Aubert, 2006) and magnetic field measurements, Christensen et al. (2009) argued that for the Earth, Jupiter, and fully-convective fast-rotating main-sequence stars, the convective heat flux is the key factor determining the order of magnitude of the dynamo-generated magnetic field strength:

$$\frac{B^2}{2\mu_0} \propto f_{\rm ohm}\, \rho^{1/3} \left(\frac{q_{\rm c} L}{H_T}\right)^{2/3} , \tag{6.11}$$

where $\mu_0$ is the magnetic permeability, $f_{\rm ohm} \leq 1$ is the ratio of ohmic to total dissipation, $L$ is the length scale of the largest convective structures and $H_T = P/(\rho g \nabla_{\rm ad})$ is the temperature scale height, with $c_p$ the heat capacity, $\alpha$ the thermal expansion coefficient, $g$ is the gravitational acceleration, and $q_{\rm c}$ is the convective heat flux as previously defined. As in Christensen et al. (2009), we hereafter assume $f_{\rm ohm} = 1$ for simplicity. Eq. (6.11) can be integrated in the spherically symmetric shell of volume $V$ (between $R_{\rm core}$ and $R_{\rm dyn}$) and define the root-mean-square value of the magnetic field in the dynamo region as:

$$\frac{B_{\rm dyn}^2}{2\mu_0} := \frac{\langle B\rangle^2}{2\mu_0} = c\, f_{\rm ohm}\, \langle\rho\rangle^{1/3}\, (F\, q_o)^{2/3} , \tag{6.12}$$

where brackets indicate volume averages, $q_0$ is a reference convective flux, which we take as the value of $q_{\rm c}$ at $R_{\rm dyn}$, $c$ is a proportionality constant, calibrated as $c = 0.63$ by Christensen et al. (2009). The factor $F$ includes the radial profile variations as follows:

$$\begin{aligned} F^{2/3} &= \frac{1}{V}\int_{R_{\rm core}}^{R_{\rm dyn}} \left(\frac{q_{\rm c}(r)}{q_0}\frac{L(r)}{H_T(r)}\right)^{2/3} \left(\frac{\rho(r)}{\langle\rho\rangle}\right)^{1/3} 4\pi r^2\, dr = \\ &= \frac{1}{V}\int_{Q_{\rm core}}^{Q_{\rm dyn}} \left(\frac{q_{\rm c}(r)}{q_0}\frac{L(r)}{H_T(r)}\right)^{2/3} \left(\frac{\rho(r)}{\langle\rho\rangle}\right)^{1/3} \frac{1}{\rho(r)} dQ \end{aligned} \tag{6.13}$$

where $L = \min(D, H_\rho)$. In the context of *MESA* variables, it is more natural to use the second expression and integrate in $Q$, where $dQ = dm/M$. We have used $dm/dr = 4\pi r^2 \rho$. All these quantities are obtained for the evolutionary models at each timestep. This implementation is similar to other planetary studies (e.g. Hori, 2021; Kilmetis et al., 2024).

Additionally, we can estimate the dipolar component at the planetary surface at the magnetic equator, $B_{\rm dip,surf}$, defined as:

$$B_{\rm dip,surf} = \frac{1}{2\sqrt{2}} \left(\frac{R_{\rm dyn}}{R_{\rm P}}\right)^3 B_{\rm dyn} , \tag{6.14}$$

where for simplicity we use the same factor $1/(2\sqrt{2})$ as in Reiners and Christensen (2010), which comes from the assumption that the rms dipole field strength is half of the rms and that we consider the value at the magnetic equator, which is $1/\sqrt{2}$ of the average dipole field. Note also that the scaling law suffers from considerable uncertainties: (i) Reiners et al. (2009); Reiners and Christensen (2010) reformulate the original scaling law in the way illustrated in Sect. 5.2.8, and evaluate a $\sim 60\%$ relative dispersion around that formula from data comparison; (ii) several factors (like the dipolar fraction and $f_{\rm ohm}$) which may be in reality systematically shifted, or vary from case to case; (iii) the calibration of the scaling law has been originally limited to the Earth, Jupiter and low-mass main-sequence stars, excluding other planets, with brown dwarfs found later to be surprising outliers (Kao et al., 2016, 2018), and with no available comparison with magnetic field measurements for exoplanets, except a few indirect estimates by (Cauley et al., 2019). Therefore, the absolute values should be taken with care. Nevertheless, our main focus is to show the trends of the predicted field, mainly as a function of age, $P_{\rm orb}$, and the type of heating.

Note also that, for each model, we show the evolution from 10 Myr to 10 Gyr. At early ages, convection is more vigorous, but more effects related to disk-planet interaction and migration could occur, which we do not account for. Moreover, as discussed above, the internal structure predicted by cooling models still depends on the specific initial entropy, while at later ages, $\gtrsim 0.1$ Gyr, the evolution has lost memory of the initial condition, so that results are more reliable in this sense. Since the bulk of detected HJs are $\mathcal{O}$(Gyr)-old, hereafter we will not focus on the $t \lesssim 100$ Myr part of the plots.

In Fig. 6.9, we show $B_{\rm dyn}$ (top) and $B_{\rm dip,surf}$ (bottom) for the same 1 $\rm M_J$ models shown in Fig. 6.5 and 6.6. Models with only irradiation (left panels) lead to a magnetic field decay in time ($B \propto t^{-0.3}$, where $B$ can be $B_{\rm dip,surf}$ or $B_{\rm dyn}$) proportional to the gradual cooling of the planet and independent of the orbital period, compatible with Reiners and Christensen (2010) and Elias-López et al. (2025b). Models with internal heat lead instead to much higher $B_{\rm dip,surf}$, since the shrinking is stalled, with an equilibrium between the long-term cooling and the internal heating. These values are compatible with Yadav and Thorngren (2017), which assumes that all the heat necessary to inflate the planet is involved with the magnetic field generation (see Viganò et al. (2025) for a discussion about the assumptions in the use of the scaling laws). As seen in Fig. 6.2 and Fig. 6.3, inflation primarily affects the outer layers of the planet, and the dynamo region does not expand nearly as much. The consequence is that, even though $B_{\rm dyn}$ plateaus at very high values $\sim 900$ G for $P_{\rm orb} \lesssim 2$ d, $B_{\rm dip,surf}$ has much lower values, since the layers between the outer dynamo region $R_{\rm dyn}$ and the planetary surface $R_{\rm P}$ are inflated, and the relative decay of the dipolar field, cubic in radius, is relevant (higher multipoles decay even more). The non-heated models shown in Kilmetis et al. (2024) have a similar trend where the magnetic field estimation decays for short $P_{\rm orb}$.

The predictions for models with the outer heating, for which convection can be suppressed, are very different, as shown in the right panels of Fig. 6.9. Whenever inflation becomes noticeable, the magnetic field strength dramatically decreases to values even lower than the Jovian values of $\sim 10$ G. The cases with the lowest predicted values of $B_{\rm dyn}$ have thin layers of convection below $R_{\rm dyn}$. This implies that the integral, Eq. (6.13), is very sensitive to both the assumed definition of $R_{\rm dyn}$, and the radial interval used in the outer heating. Moreover, it might not be accurate to estimate the magnetic field strength created in these shallower convective layers with the same scaling laws. Therefore, the values shown on the left panel of Fig. 6.9 could be even lower or potentially vanish for the convection-suppressed cases.

This fact leads us to draw a parallel with Venus. The most accepted argument for the lack of an active Venusian dynamo is the presence of a stagnant lid. The absence of an active tectonic plate system inhibits efficient cooling of its core, leading to steeper entropy gradients (Nimmo, 2002; Jacobson et al., 2017). The argument works similarly for HJs: if the heating necessary to reach the observed HJ radii is deposited in the outer layers, convection (and thus dynamo) is suppressed, essentially due to a blanketing effect.

### 6.4.2 Planetary coherent radio emission

One of the main observational consequences of exoplanetary magnetic fields is the possible coherent radio emission, as seen in all magnetized planets in the Solar system (Zarka, 1998). The mechanism behind this emission is the above-mentioned ECM (Dulk, 1985), which requires the presence of a source of keV electrons (arguably coming from the stellar wind) and a planetary magnetic field. As the charged particles from the stellar wind come into the vicinity of the planetary magnetic field, they start following helical trajectories around the magnetic field lines and get bunched together in phase. This results in a coherent, circularly polarized radio emission from the planet in a hollow cone, at the Larmor frequency of the local magnetic field $B$, $\nu_c$ [MHz] $= 2.8B$ [G]. Therefore, the detected emission traces the intensity of the magnetic field, with a frequency cut-off at the Larmor frequency of the largest value of the field in the emitting region, i.e., close to the surface. As seen in the bottom panel of Fig. 6.9, irradiated and extended heated models have $B_{\rm dip,surf}$ estimations that surpass the $\sim$ 10 MHz ionospheric cut-off, which corresponds to a minimum magnetic field strength of $\sim$ 3.5 G. On the contrary, when HJ are inflated with the outer heating, their $B_{\rm dip,surf}$ usually are below the ECM cut-off, suppressing any detectable radio emission.

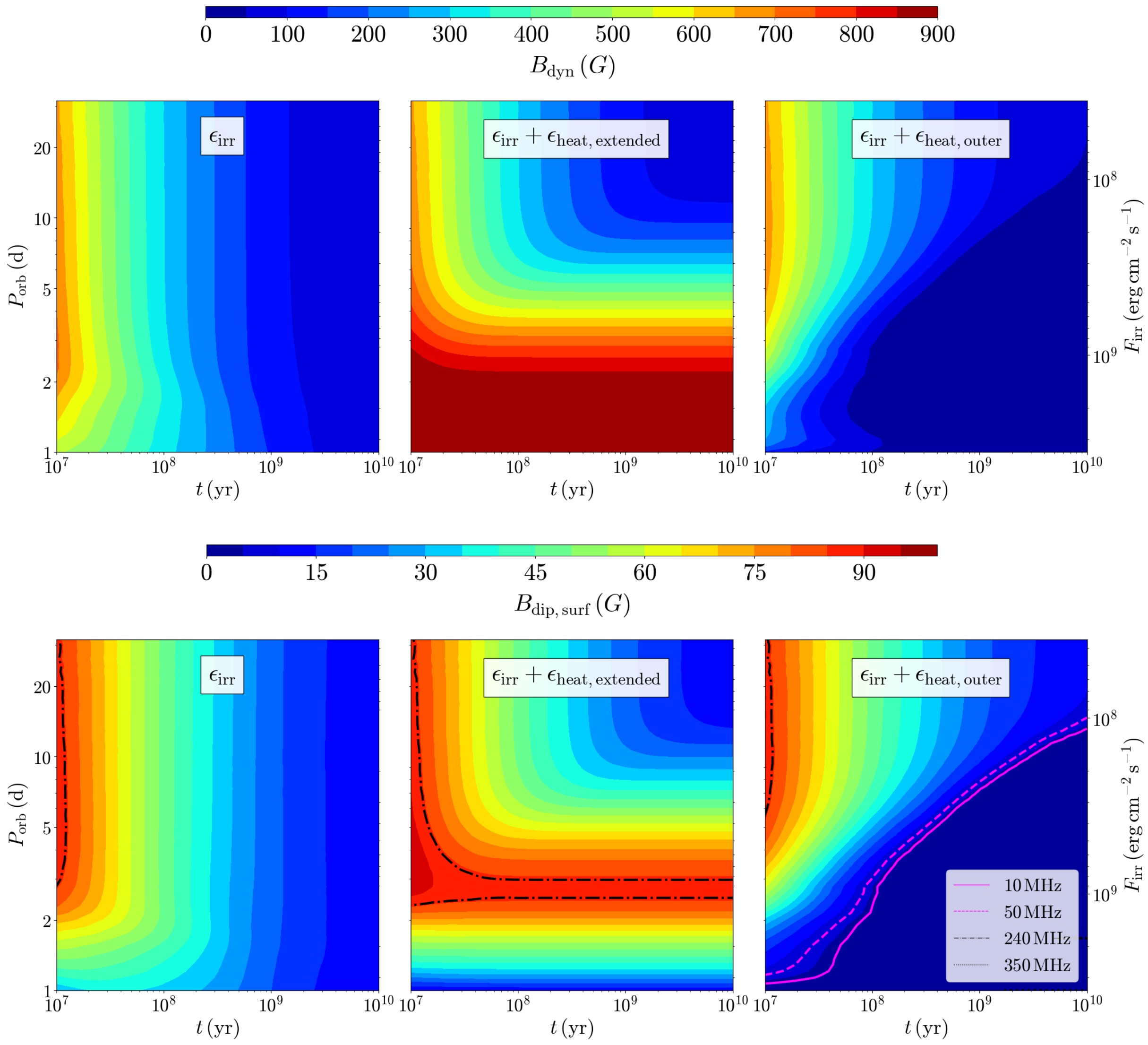

Figure 6.9: Magnetic field estimates for 1 $M_J$ planets orbiting a 1 $M_\odot$ main-sequence star (same models as Fig. 6.5 and 6.6). *Top panels*: average magnetic field strength at the dynamo surface obtained with the scaling law discussed in the text. *Bottom panels*: inferred dipole component at the planetary surface equator. Several of the most interesting frequency limits are included: 10 MHz, the ionospheric cut-off and lower range of LOFAR; 50 MHz, the lower range of SKA-low; 240 MHz, the upper range of LOFAR; and 350 MHz, the upper range of SKA-low.

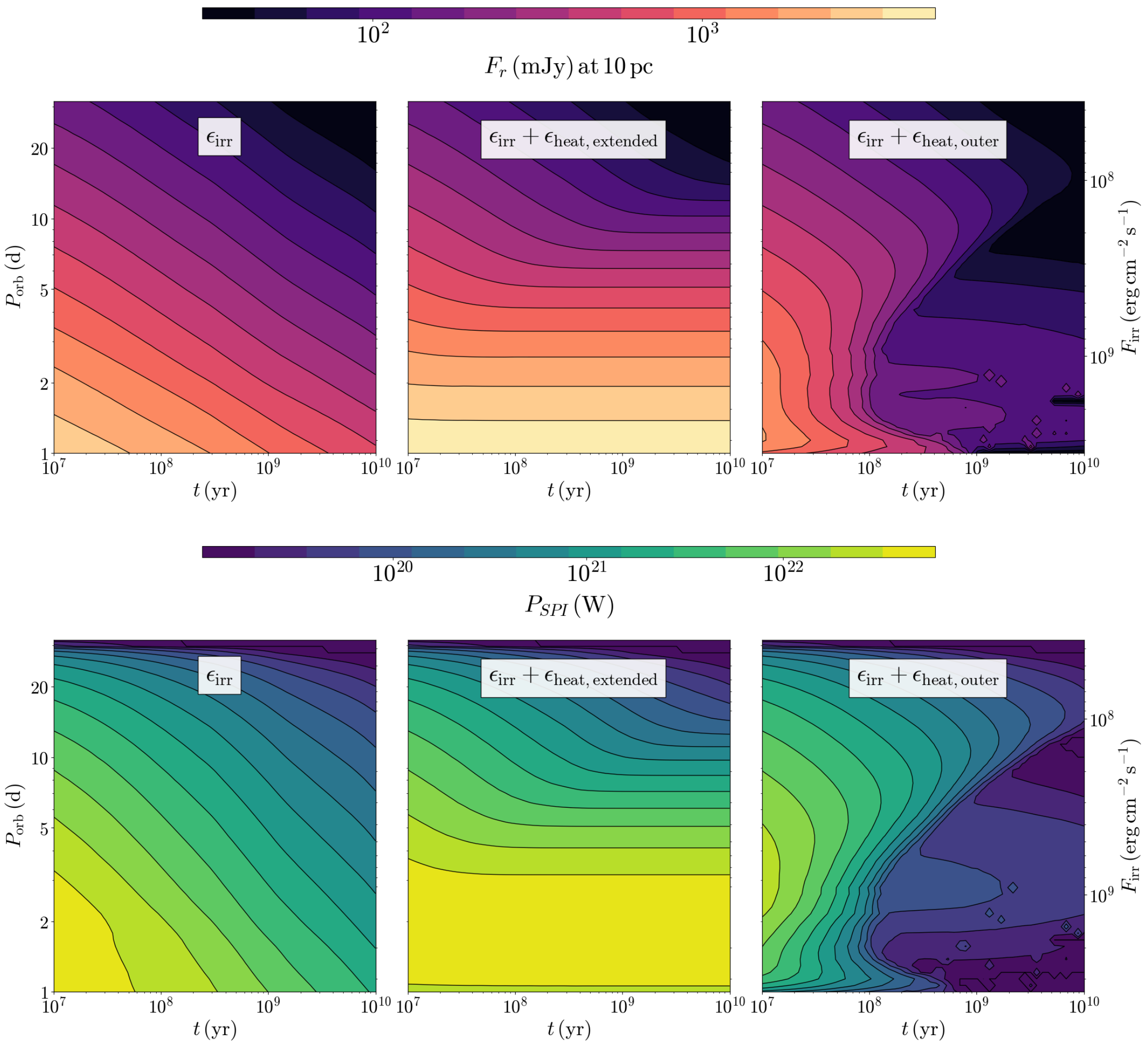

Figure 6.10: Estimates for Jovian-like coherent radio flux (top, for a 10 pc away HJ) and SPI available power (bottom), as a function of age and $P_{\rm orb}$, for the same models shown in Fig. 6.9.

Given that the Jovian case remains the only firm observable constraint for gas giants' ECM emission, there are huge uncertainties on how bright the expected radio emission should be. Simple estimates assume that the emitted radio power is proportional to the kinetic flux intercepted by the planetary magnetosphere and is given by the phenomenological radiometric Bode's law, which states the the total irradiated power is $P_r \propto \dot{M}_\star^{2/3} \cdot M_{\rm dip}^{2/3} a^{-4/3} V_W^{5/3}$, with an added factor for the distance to the source of $d^{-2}$ to recover the radio flux. This fits the well-characterized radio luminosity of Jupiter and that of the other solar planets, which are less studied, mainly coming from limited in-situ Cassini and Voyager measurements. The corresponding specific flux, scaled by the measured typical Jovian values, reads (Bastian et al., 2000; Zarka et al., 2001; Stevens, 2005):

$$F_r = 2.35 \cdot 10^{-2}\,{\rm mJy} \cdot \left(\frac{\dot{M}_\star}{\dot{M}_\odot}\right)^{2/3} \cdot \left(\frac{M_{\rm dip}}{M_{\rm dip,J}}\right)^{2/3} \cdot \\ \cdot \left(\frac{a}{5\,{\rm AU}}\right)^{-4/3} \cdot \left(\frac{V_W}{400\,{\rm km\,s^{-1}}}\right)^{5/3} \cdot \left(\frac{d}{10\,{\rm pc}}\right)^{-2} \tag{6.15}$$

where $\dot{M}_\star$ is the stellar mass loss rate, $M_{\rm dip} = B_{\rm dip,surf}\,R_{\rm P}^3$ is the planetary magnetic dipolar moment, and $d$ is the distance to the source. Note that in Stevens (2005) they use the same expression with a source distance of $d = 1$ AU, which corresponds to a pre-factor of $10^{11}$ mJy. The stellar wind properties may vary from case to case and can be inferred from UV and X-ray observations (see e.g. (Wood et al., 2021) for M dwarfs), or some models (e.g., Johnstone et al., 2015). Given the intrinsic uncertainties, here we simply assume a representative solar value, $\dot{M}_\star = \dot{M}_\odot \simeq 400$ km s$^{-1}$ (Wood et al., 2005), thus neglecting the large spatial and time variability measured for the Sun (e.g., Fig. 1 of Johnstone et al. 2015), and its dependence on $a$ from measurements or from e.g. the Parker (1965) model.

In the top panel of Fig. 6.10, we show the resulting estimates for the ECM radio flux, for the same models in Fig. 6.9. HJs with extended heating have very high flux predictions, overcoming 1 Jy, which is incompatible with much lower upper limits reported in literature for tens of HJs (e.g. Narang et al. (2024) and references within). Models with external heating have instead flux estimates which are orders of magnitude smaller, especially for the hottest cases at Gyr ages, where the convection and dynamo suppression bring the most relevant effects. Moreover, they are mostly below the 10 MHz ionospheric cut-off shown in Fig. 6.9 and cannot be detected by ground-based radio interferometers.

Note that the absolute values of the estimated flux for the purely irradiated case are compatible with the cases proposed in the original works by e.g. Stevens (2005), and fluxes $F_r \gtrsim 0.1$ mJy should be easily detectable by targeted observations and/or current wide-field radio surveys at low frequencies, LOFAR Stokes V survey V-LoTSS in particular (Callingham et al., 2023). However, after two decades of radio campaigns targeting tens of theoretically promising HJs have led to no confirmed planetary radio emission (see e.g. Fig. 6 of Narang et al. (2024), and references within), with a claim for one case (Turner et al., 2021), not confirmed by an extended follow-up (Turner et al., 2024). Therefore, the models are probably very optimistic, in part because they are often calibrated to the peak values of the Jovian short bursts, rather than the average one. In any case, the purpose of this work is to discuss the trends with irradiation and heating models, rather than the absolute values, which are arguably plagued by huge intrinsic uncertainties.

### 6.4.3 Star-planet interaction available power

Another potentially observable signature of planetary magnetism is the effect of magnetic SPI. One of the possible SPI mechanisms for HJ systems can arise when a close-in giant planet moves through its host star's magnetic field, enabling energy transfer due to the stretching of magnetic flux tubes that connect the planetary and stellar magnetospheres (Lanza, 2013). This interaction can produce observable signatures, such as spectroscopically detectable modulation with the orbital period of chromospheric activity (in particular, the emission line Ca II K, Cauley et al. 2019), or coherent emission at radio frequencies close to the stellar surface (e.g., Saur et al., 2013; Pérez-Torres et al., 2021). Here we don't enter into detail of the specific manifestation of the SPI at different wavelengths, but we simply follow Lanza (2013) to derive the total power available:

$$P_{\rm SPI} \simeq \frac{2\pi}{\mu_o} f_{\rm AP}\, R_{\rm P}^2\, (2B_{\rm dip,surf})^2\, v_{\rm rel}, \tag{6.16}$$

where $v_{\rm rel} = 2\pi a/P_{\rm rel}$ is the relative velocity between the planet and the stellar magnetic field lines (therefore, $P_{\rm rel} = 2\pi/(\omega_{orb} - \omega_\star)$, where $\omega_i$ are orbital frequencies), $2B_{\rm dip,surf}$ is the polar magnetic field (twice its equatorial value defined in Eq. 6.14), and $f_{\rm AP}$ is the fraction of the planetary surface covered by stellar magnetic flux tubes:

$$f_{\rm AP} = 1 - \left(1 - \frac{3\xi^{1/3}}{2+\xi}\right)$$

where $\xi = B_\star(a)/(2B_{\rm dip,surf})$. $B_\star(a)$ is the stellar magnetic field at the given orbital separation, which we derive assuming a solar-like value at the stellar surface for simplicity, $B_\star = 10$ G. Note that this underestimates the stellar field for at least M dwarfs, which, being fully convective, usually show hundreds to thousands gauss fields (Reiners et al., 2022), and for the highly magnetic, chemically peculiar Ap and Bp stars. However, the bulk of HJs orbit Sun-like stars whose spectral type is usually associated with (at most) Sun-like large-scale magnetic field values (e.g., Seach et al., 2020), therefore justifying our simple assumption.

Eq. (6.16) estimates the Poynting flux generated by the continuous stretching of magnetic field lines due to the planet's motion and was also used in Cauley et al. (2019) to estimate the magnetic field of four HJs by observing periodic enhancements in the Ca II K chromospheric emission line. In the same work, they show that other SPI estimates, like the pioneering reconnection model (Cuntz et al., 2000) and the Alfvén wing scenario (Saur et al., 2013; Strugarek, 2016), have different dependencies with the planetary and stellar magnetic fields, and typically provide less available power, incompatible with the observed activity modulation signal (Cauley et al., 2019), so we don't discuss them here. In the bottom panels of Fig. 6.10, we show the resulting available SPI power, a fraction of which (e.g., $\lesssim 1\%$ in Cauley et al. 2019) can become visible as orbit-modulation of the activity indicators. Similar to Fig. 6.9, only the extended heat scenario, for the hottest cases, provide optimistic predictions for inflated HJs, $P_{\rm SPI} \gtrsim 10^{22}$ W, potentially able then to power the observationally inferred values of $10^{20}$ W in Ca II K line modulation, assuming reasonable $\lesssim 1\%$ conversion efficiency, Cauley et al. (2019). When the heat is applied only above the dynamo, the available power is typically reduced by orders of magnitude.

Note that an essential requirement for SPI is that the local wind velocity should be less than the Alfvén velocity (i.e., the Alfvén Mach number $M_A = v_W/v_{alf} < 1$, where $v_{alf}$ is the Alfvén speed). This intrinsically depends on the individual stellar properties, and we cannot assess it here. However, it can further limit the detectability of SPI signals. Therefore, again, the values here indicated should be regarded as upper limits and be used only to infer the trends with the HJ properties (depending ultimately on irradiation and heating).

## 6.5 Summary

- We study the evolutionary internal properties of inflated HJs using the *MESA* code under assumptions of tidal locking and stellar irradiation. We investigate the influence of planetary and stellar masses, orbital separation, and internal heating location on planetary structure and magnetic field estimates.

- Heat injection is guided by irradiation-efficiency relations derived by (Thorngren and Fortney, 2018), with a focus on trends rather than precise inflation magnitudes. Two heating scenarios are tested: heat injected within the dynamo region ($P \gtrsim 10^6$ bar) and external to it (as in Ohmic heating models). We reproduce the radius inflation of previous studies, and we focus on the internal properties. Structural differences between the two heating types are minimal.

- Rossby number (Ro) profiles reveal that most models have Ro $\lesssim 0.1$, consistent with fast-rotating convection. Only very massive, long-period (15–40 days), tidally locked planets may develop higher Ro and potentially different dynamo regimes. The innermost heat injection type tends to increase Ro with more visible effects in low-mass planets, compared to the outer heating, but minimal changes are seen for high-mass planets. Moreover, outer heating leads to convection suppression via entropy inversions and reduces internal heat transport, making Ro non-defined.

- The convection suppression has deep implications for the expected magnetic fields. Using magnetic field scaling laws (Christensen et al., 2009), we predict that if the heat is mostly deposited in the dynamo regions, the surface magnetic fields are up to ∼100 G, consistent with Yadav and Thorngren (2017). On the other hand, external heating results in weaker convective power and much smaller magnetic field strengths, similar to Jupiter. Unlike Jupiter-mass HJs, massive HJs are largely unaffected by inflation; thus, they are the only cases with outer heating that could still host strong magnetic fields.

- Despite having encouraging scaling law flux predictions, SPI signals like coherent radio emission and orbital modulation of activity indicators have remained so far very scarce (Cauley et al., 2019). Based on the magnetic fields derived by scaling laws, we give observational predictions of radio emission and SPI available power, comparing different periods for the irradiated-only (no extra heating), inner heat, and outer heat models. Only the inner heat model has a promising prediction, but the outer heated model is much more likely, physically speaking (Batygin et al., 2011; Ginzburg and Sari, 2015, 2016; Viganò et al., 2025). Therefore, we conclude that the expectations of highly magnetized HJs are probably too optimistic. This can explain why HJ magnetism remains elusive despite two decades of radio campaigns aimed at it. However, other important biases, such as beaming, variability, or distance, certainly contribute negatively to the detection chances.

# 7

# Hot Jupiters dynamo simulations and on-going applications

This chapter presents some preliminary results from dynamo simulations of HJs. I follow the same methodology used in Chapter 5, with the difference that I use background thermodynamic profiles coming from the evolution of highly irradiated planets. Sect. 7.1, related to the curvature and the assessment of the electrical current intensity, is my contribution to a study of the Ohmic model for HJ radii inflation, currently under the final stages of revision (Viganò et al., 2025). Sect. 7.2 represents an ongoing work which, at the moment of finalizing this thesis, is being shaped as an article, first-authored by me. This chapter, far from being self-contained, is meant to provide some details of the current and future developments and applications of the work presented in the rest of the thesis.

## 7.1 Assessing the magnitude of electrical currents in HJs

As mentioned in previous chapters, the HJ inflated radii issue (Laughlin et al., 2011; Sestovic et al., 2018) is often understood within the Ohmic models, first proposed in the early 2010s (Perna et al., 2010b; Batygin and Stevenson, 2010; Batygin et al., 2011). The basic principle relies on the irradiation causing strong thermal winds which, due to the partial thermal ionization of the alkali metals (Kumar et al., 2021), carry electrical currents and locally modify the background magnetic field, generated much deeper, in the dynamo region. Such currents, due to fundamental electrodynamic principles and the non-zero conductivity, penetrate the interior. Therefore, their Ohmic dissipation is not limited only to the shallow outermost radiative and highly irradiated layers where winds circulate, but concerns also the deeper, convective layers, where the deposited heat can effectively slow down (Komacek and Youdin, 2017) or even revert (Komacek et al., 2020) the long-term cooling and shrinking, thus providing substantially higher values of the measured planetary radii. Previous works, for simplicity, have considered extremely simple and unrealistic geometries for the dynamo magnetic fields, like dipoles (Batygin and Stevenson, 2010; Batygin et al., 2011; Wu and Lithwick, 2013; Knierim et al., 2022). Such studies extended the calculations of the induced currents to the dynamo region, but they neglected how induced currents can affect the dynamo currents, and vice versa. In Viganò et al. (2025) we have refined such models, by three main novelties: (i) considering the coupling with an evolving internal field via scaling laws (Christensen et al., 2009) presented in the previous chapters; (ii) allowing for any arbitrary combination of multipoles for the background dynamo-generated field; (iii) including the latest theoretically and experimentally constrained values for the conductivity, considering both the thermal ionization (Kumar et al., 2021) and the high-density and degeneracy effects (Bonitz et al., 2024), to derive the radial profiles of the electrical currents. It turns out that the conductivity profile is the main factor shaping the radial profile of electrical currents. The latter vary by less than one order of magnitude across the planet, in contrast with the broad range of values (6-7 orders of magnitude) spanned by the conductivity, which steeply increases as one goes closer to the dynamo region, $p \sim 10^6$ bar. Typically, the magnitude of the currents can reach $\lesssim 10^{-2}$ A m$^{-2}$, with the hottest planets having the strongest currents. In Viganò et al. (2025), we show representative examples of the radial profiles of the conductivity, induced currents, and specific Ohmic heat.

Here, without entering into the details of the Ohmic models, we compare those values with what we can expect from dynamo simulations, which has been my practical contribution to Viganò et al. (2025). As explained through this thesis, in dynamo simulations, it is not possible to directly predict the absolute value of the magnetic field and currents. The reason is that they depend on Ra and E, which are computationally constrained to be many orders of magnitude away from the realistic ones, due to the impossibility of including all relevant scales. However, we can use the information on the typical topology from dynamo simulations, by looking at the ratio between currents and magnetic fields, and calibrate them by using the values measured in the solar planets, shown in Chapter 4. In particular, here we make use of the observations of the Jovian field, the one reconstructed with the highest fidelity among gas giants. Although this approach has several caveats and the numbers are not precise assessments of what can be expected, we aim to evaluate the order of magnitude and make a rough comparison of the intensity of the two electrical current systems in HJs.

### 7.1.1 Jovian magnetic field curvature

As mentioned in Chapter 4, the Jovian magnetic field produced by the internal dynamo has been accurately measured and can be reconstructed in terms of multipolar decomposition of the poloidal and toroidal components, as given by the best-fit model to the data (Connerney et al., 2018, 2022). With the set of these best-fitting weights, one can describe the potential magnetic field in the entire volume where there are no currents, down to the radius $R_{\rm dyn}$, below which the dynamo starts to be active and the field is no longer potential.

We are interested in estimating the intensity of electrical currents $|\mathbf{J}| = |\nabla \times \mathbf{B}|/\mu_0$, which dimensionally can be thought of as a magnetic field divided by a physical length scale $L_B$, $J \sim B/(L_B \mu_0)$. A natural proxy for this length-scale is the inverse of the magnetic field curvature, which is defined as $\boldsymbol{\kappa} = (\boldsymbol{b} \cdot \boldsymbol{\nabla})\boldsymbol{b}$, where $\boldsymbol{b} = \boldsymbol{B}/|\boldsymbol{B}|$ is the unit vector field. Using the Jovian solution, we compute the local values of the curvature and the average of its modulus over a sphere at a given radius, with the following expression:

$$\langle|\boldsymbol{\kappa}(r = r_\kappa)|\rangle = \frac{\int |\boldsymbol{\kappa}(\theta, \phi; r = r_\kappa)| \sin\theta \; d\theta \; d\phi}{\int \sin\theta \; d\theta \; d\phi} \,. \tag{7.1}$$

The full expressions used to obtain the components and mean magnetic field curvature are shown in App. C.2.

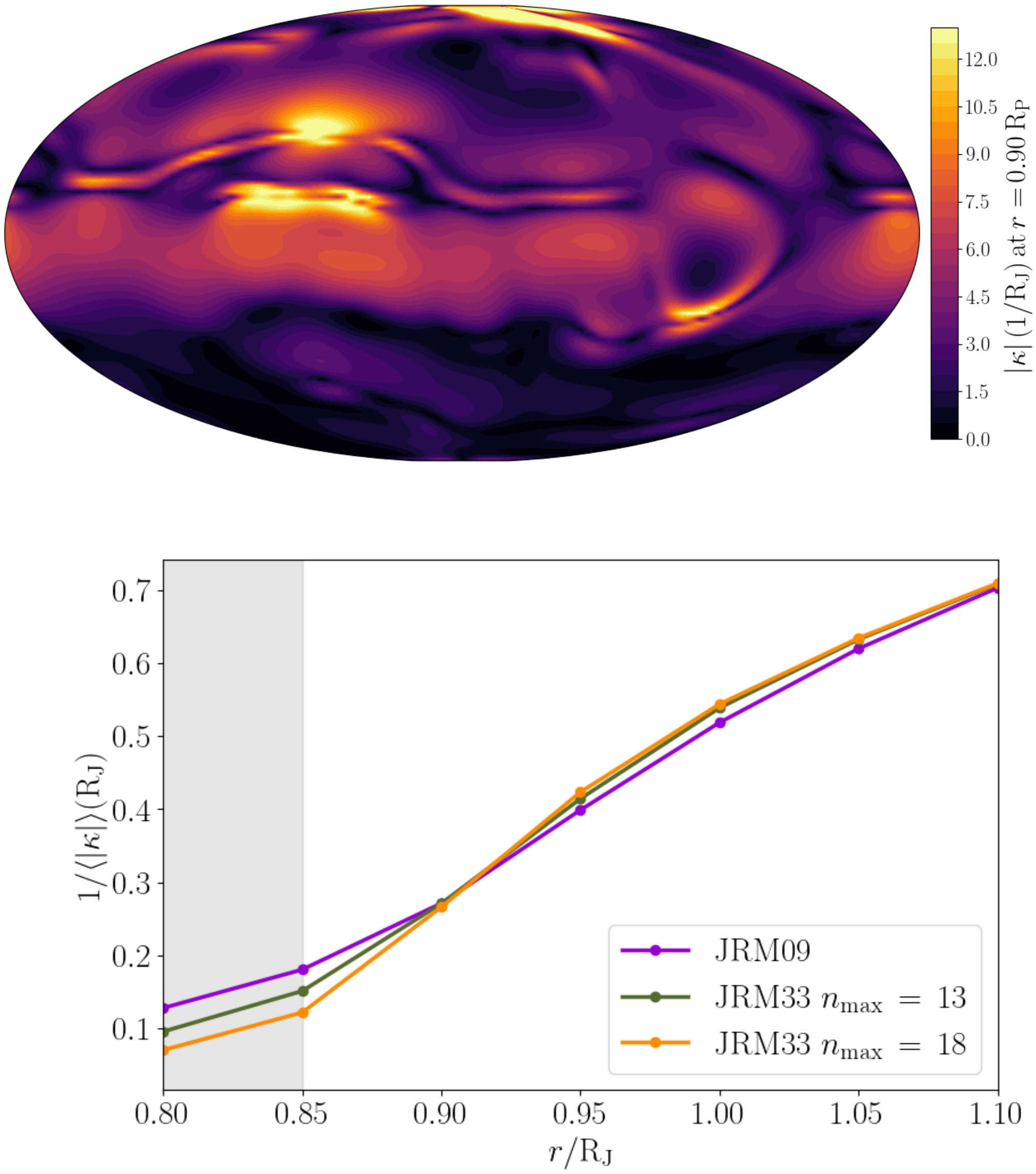

Figure 7.1: Top: Curvature modulus of Jupiter's magnetic field at $r = 0.9\,R_\mathrm{j}$ in the Mollweide projection, reconstructed with the model JRM33 (Connerney et al., 2022) with $n_{\mathrm{max}} = 13$. Bottom: Inverse of the mean curvature modulus for different Jupiter magnetic field models as a function of radius, obtained using a discretized version of Eq. 7.1. The curvature for the JRM33 models does not significantly change if the external multipoles are included.

In the bottom panel of Fig. 7.1, I show the inverse of the mean curvature at different radii for both models, including two different cut-offs for JRM33. The gray zone approximately marks the possible start of the conductive zone, where the gradual appearance of metallic hydrogen would prevent the validity of the potential solution. The inverse curvature has a mean value of the order of $\bar{\kappa} \sim (0.07 - 0.3\ \mathrm{R_J})^{-1}$ at $r \sim (0.8 - 0.9)\ \mathrm{R_J}$ (around where the dynamo surface should be), which can be taken as a lower limit of the magnetic field curvature inside the Jovian dynamo and a proxy for the electric currents. Below, we adopt this as a lower limit for $\bar{\kappa}$, since higher values are expected in deeper regions, as a consequence of the turbulent environment in which the magnetic field is continuously regenerated (see bottom panel of Fig. 7.2).

### 7.1.2 Magnetic field curvature estimation in the internal dynamo region

We can estimate the electric currents from planetary numerical simulations as a function of depth, translating the numerical results into physical quantities by using the values of $\kappa$ obtained above and the typical values of $B$. For these purposes, we performed MHD dynamo simulations that allowed us to estimate the values of the currents, and to validate at the same time the use of the magnetospheric value $\bar{\kappa}$ as a proxy for the internal $J/B$ ratio. Similarly to Chapter 5, *MagIC* has been used to solve the MHD equations in a spherical shell under the anelastic approximation with the *MESA* radial structure as background profiles. As explained in detail in Sect. 5.1.1, we implement the radial dependence of density, temperature, gravity, thermal expansion coefficient, and the Grüneisen parameter with a very high degree polynomial, fitting them better than $\lesssim 1\%$. We assume constant thermal and viscous diffusivities and we adopt the conductivity profile (and thus a corresponding magnetic diffusivity) first defined in Gómez-Pérez et al. (2010) which consists of an approximately constant conductivity in the hydrogen metallic region joined (at $r = r_m$ with $\sigma(r) = \sigma_m$) via a polynomial to an exponentially decaying outer molecular region (see Eq. 5.7). For these models, we have arbitrarily chosen $\sigma_m = 0.6$ and $a = 11$, and we fix $r_m = R_{\mathrm{dyn}}$, i.e., the radius of $10^6$ bar. Due to numerical limitations, dynamo models cannot cover the entire density range of the dynamo region, and we have to consider a more limited density contrast (ratio between the innermost and outermost densities) $N_\rho = \ln(\rho_i/\rho_o) \lesssim 4.6$. We externally cut the *MESA* radial profiles to have a density contrast $N_\rho$ of either ∼3.0 or 4.6. We have also used a variety of dynamo parameters and boundary conditions.

We show the results in Fig. 7.2, where only some representative models producing a dynamo are shown. The background profiles correspond to an irradiated $1\mathrm{M_J}$ planet with internal Ohmic heating at 10 Gyr (similar to the models shown in Viganò et al. (2025)). At that time, the simulated planetary radius is 1.218 $\mathrm{R_J}$, which after the outer cuts at $N_\rho \sim 3.0, 4.6$ leads to external pressures of 64 and 5.4 kbar, respectively. We used resolutions of $(N_r, N_\theta, N_\phi) = (193, 256, 512)$ for the cases with $N_\rho = 3.0$, and $(N_r, N_\theta, N_\phi) = (241, 320, 640)$ for the cases with $N_\rho = 4.6$. To explore how sensitive the magnitudes in Fig. 7.2 are, we vary the Pr, Pm, Ra, $N_\rho$, and the boundary conditions. Generally speaking, the models shown above reach a steady state dynamo with equatorial jets. More details and the main results of these simulations will be presented in an upcoming paper dedicated to the dynamo in gas giants.

Despite some minor differences among the various simulations, we obtain radial profiles with the same trend and order of magnitude: $|\boldsymbol{\nabla} \times \boldsymbol{B}|/(\bar{\kappa}B) \sim$ 1-3 and $1/\bar{\kappa} \sim (0.01\text{-}0.1)\ \mathrm{R_J}$. For the former (top panel), it confirms that the ratio between the magnetic field intensity and the mean radius of curvature, $B\bar{\kappa}$, is a very good tracer of the current density, $J$. Regarding $\bar{\kappa}$ itself (bottom panel), we obtain values similar to the ones inferred from Juno data (see section above).

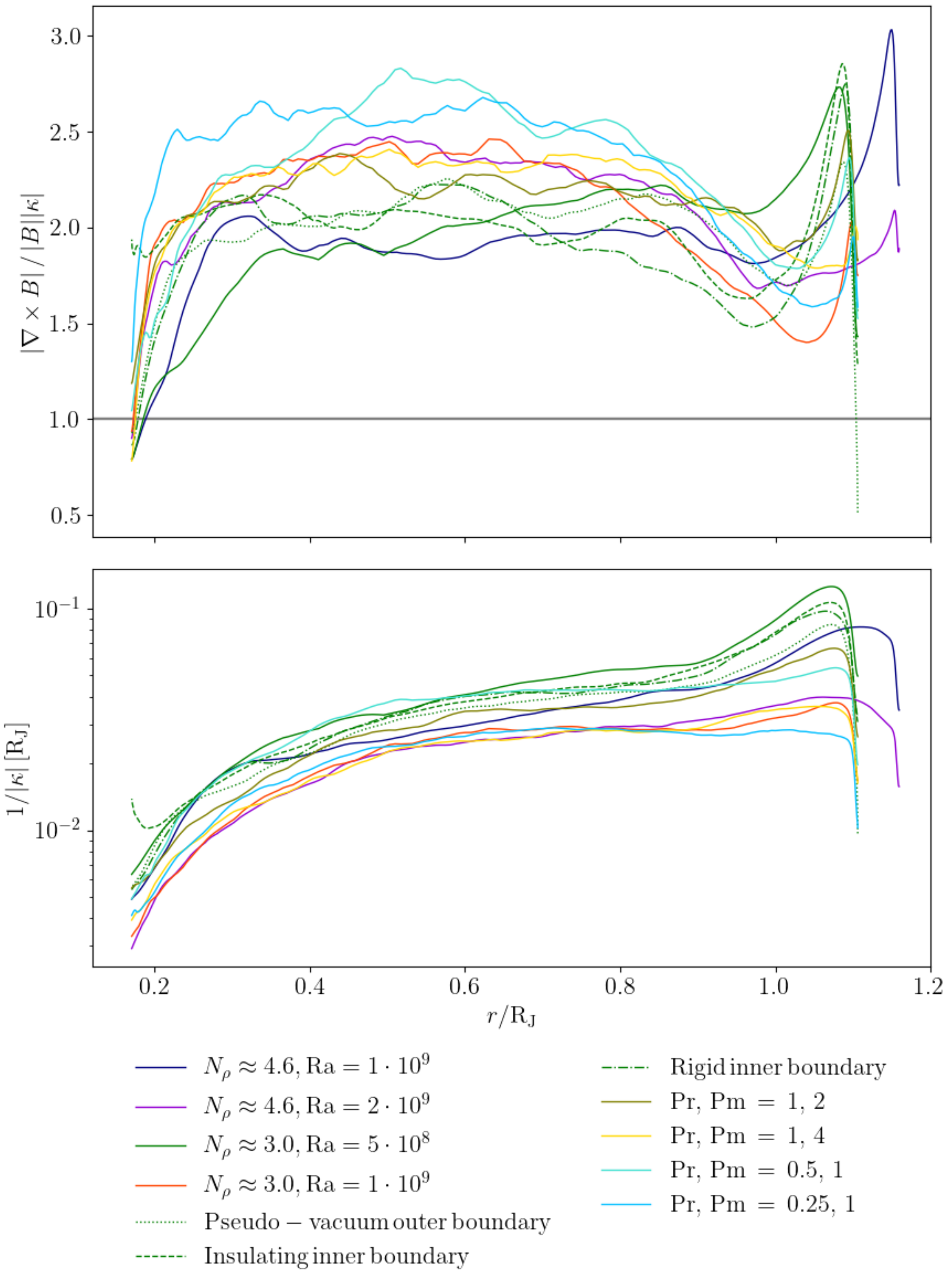

Figure 7.2: Typical radial profiles of $|\boldsymbol{\nabla} \times \boldsymbol{B}|/(\bar{\kappa}B)$ (top) and $1/\bar{\kappa}$ (bottom), averaged over the spherical surface at each radius, from 3D dynamo simulations with the code *MagIC*. To explore the sensitivity to the free parameters. All models have the same boundary conditions (i.e. stress-free for the fluid and insulating for the magnetic fields outside, stress-free and perfectly conducting inside) except for the cases marked in the legend. The Ekman number is $10^{-5}$, and both Prandtl numbers are set to 1 (except for those also marked in the legend, which have $N_\rho \approx 3.0$ and Ra $= 5\times10^8$. Note that $1/\bar{\kappa}$ becomes much smaller close to the external boundary, to match the potential configuration constraint. See Chapter 5 for more details.

Therefore, taking typical values $B_{\rm dyn} \sim 10$ G and $\kappa \sim 10^{-6}$ m$^{-1}$, one obtains $J \sim B\kappa/\mu_0 \sim 10^{-3}$ A/m$^2$. This range is the same as (or slightly lower than) the atmospherically-induced currents at $r \sim R_{\rm dyn}$. Therefore, they could be a significant factor to be taken into account to set the boundary conditions in dynamo simulations, which are instead almost always set to be a current-free solution. Since the geometry of the atmospherically induced currents depends on the wind and background field configurations themselves (Batygin et al., 2011; Viganò et al., 2025), this opens up the remarkable possibility of atmospheric processes affecting the deep dynamo solution. The implementation of atmospherically induced currents as top boundary conditions for the highly-conductive region would open a new type of dynamo simulations, so far non-existent, which would be specifically applicable for HJs only.

## 7.2 Hot Jupiter dynamo simulations: preliminary results

A natural, ongoing work which follows Chapters 5 and 6 is to obtain 3D MHD HJ saturated dynamo solutions at different orbital distances (i.e., irradiation levels). The main idea is to take the interior planetary structures of several HJs models shown in Chapter 6, without or with internal heating, and repeat the same procedure performed in Chapter 5. In this section, I include the preliminary results for such numerical experiments.

We first take several heated and non-heated planetary structure models, all at the same age, which we fix at 5 Gyr, and implement the background profiles as seen in Chapter 5, choosing a density contrast $N_\rho \approx 3$. We again assume that the diffusivities at a given radius remain constant for all models. This implies that we considered the same values of Pr and Pm along the orbital distance sequence, which we assume to be Pm = Pr = 1. It also means that the relative change in E comes from both the level of HJ inflation and the rotation rate $\Omega = 2\pi/P_{\rm orb}$:

$$\mathrm{E}' = \mathrm{E}_0 \frac{d_0^2}{d'^2}\frac{\Omega_0}{\Omega'} \;, \qquad \mathrm{Ra}' = \mathrm{Ra}_0 \frac{d'^3 \Delta T'}{d_0^3 \Delta T_0} \;.$$

We have considered a non-inflated 1 $\mathrm{M_J}$ HJs series with $P_{\rm orb}$ from 0.528 to 31.6 d around a solar-like star, logarithmically spaced. For comparison, we have added two internally heated models, with $P_{\rm orb}$ being 1.47 d and 4.08 d, with corresponding moderate heating efficiencies (Eq. 6.7) of $\epsilon \sim 0.6, 0.3\,\%$, respectively. In Table 7.1, we summarize the most important models and their corresponding input parameters.

We employ the same analytical conductivity expression (i.e., Eq. 5.7), taking $\sigma_m$= 0.1 and $r_m$ defined for each model corresponding to the radius where pressure reaches 1 Mbar. Instead of assuming a constant value for $a$, we now choose different values depending on the orbital separation (see Table 7.1). There are two reasons for this: a physical one and a numerical one. The outer layers of HJ are known to be more conductive than in cold Jupiters due to the high outer temperatures. However, the conductivity is still orders of magnitude lower than that of metallic hydrogen (Kumar et al., 2021; Bonitz et al., 2024). Therefore, the smaller $P_{\rm orb}$, the higher the surface temperature and the higher the conductivity, leading to an overall less steep conductivity decay. The numerical reason is similar to the 0.3 $\mathrm{M_J}$ model of Chapter 5: for heated models, the outer non-conductive layer is so thick that the decaying exponential leads to low numerical values of $\tilde{\sigma}$. We can avoid this problem by having not-so-steep conductivity drops for the most inflated hotter HJ models.

An important fact is how the E and Ra change within the HJ series: E decreases by two orders of magnitude while Ra only increases by a factor of 3. This change is far from the criticality relation, i.e., $\mathrm{Ra_c} \propto \mathrm{E}^{-4/3}$. Therefore, this will influence the trends in $P_{\rm orb}$, as the rotational effects are much more different within the HJ series than the inertial (buoyant) effects. This can already be seen by Ro, as it varies for more than 3 orders of magnitude. Notice that the last four longest-period non-inflated models are not in the low-Ro regime, for we expect a very different behavior.

| $P_{\rm orb}$ | $r_o(\mathrm{R_J})$ | $M$ | $V$ | E | Ra | $a$ | $r_m$ | Ro |
|---|---|---|---|---|---|---|---|---|
| 0.528 | 1.1135 | 50.5 | 7.63 | $3.709{\cdot}10^{-6}$ | $1.488{\cdot}10^{9}$ | 3.0 | 0.8359113 | 0.00115 |
| 0.880 | 1.0762 | 51.4 | 7.72 | $6.670{\cdot}10^{-6}$ | $1.133{\cdot}10^{9}$ | 3.5 | 0.8518471 | 0.00285 |
| 1.47 | 1.0426 | 53.8 | 7.94 | $1.210{\cdot}10^{-5}$ | $8.165{\cdot}10^{8}$ | 4.0 | 0.8644453 | 0.00651 |
| 2.45 | 1.0171 | 55.1 | 8.01 | $2.134{\cdot}10^{-5}$ | $6.222{\cdot}10^{8}$ | 4.5 | 0.8734342 | 0.0149 |
| 4.08 | 1.0103 | 56.8 | 8.20 | $3.659{\cdot}10^{-5}$ | $5.603{\cdot}10^{8}$ | 5.0 | 0.8757581 | 0.0354 |
| 6.81 | 1.0080 | 56.7 | 8.15 | $6.108{\cdot}10^{-5}$ | $5.496{\cdot}10^{8}$ | 5.5 | 0.8764779 | 0.19 |
| 11.4 | 1.0070 | 56.6 | 8.15 | $1.025{\cdot}10^{-4}$ | $5.435{\cdot}10^{8}$ | 6.0 | 0.8765579 | 0.444 |
| 19.0 | 1.0067 | 56.6 | 8.15 | $1.709{\cdot}10^{-4}$ | $5.417{\cdot}10^{8}$ | 6.5 | 0.8765823 | 0.447 |
| 31.6 | 1.0066 | 56.8 | 8.15 | $2.843{\cdot}10^{-4}$ | $5.403{\cdot}10^{8}$ | 7.0 | 0.8772614 | 0.671 |
| 1.47 | 1.2910 | 53.7500 | 4.0000 | $8.048{\cdot}10^{-6}$ | $2.863{\cdot}10^{9}$ | 4.0 | 0.7595887 | 0.0136 |
| 4.08 | 1.1107 | 59.4600 | 5.0000 | $3.172{\cdot}10^{-5}$ | $1.211{\cdot}10^{9}$ | 5.0 | 0.8359263 | 0.0696 |

Table 7.1: Similar to Table E.1: input parameters for a series of 1 $\mathrm{M_J}$ HJ models orbiting a solar-like star at 5 Gyr introduced in Chapter 6, and one output parameter, i.e., Ro. The top models are non-heated HJ, and the two bottom models are for the inflated counterpart with a moderate heat injection corresponding to an Ohmic efficiency of 0.6% and 0.3%, respectively, defined similarly as in Chapter 6.

The dependence on $P_{\rm orb}$ of the most important dimensionless dynamo parameters is shown in Fig. 7.3. Within the series that remains in the low-Ro regime (i.e., $P_{\rm orb} > 4.08$ d), we observe a steady growth in Rm and $\Lambda$ with $P_{\rm orb}$, which is compatible with the sterep increase in E. A less pronounced growth is observed for $E_{\rm mag}/E_{\rm kin}$ and $f_{\rm ohm}$. Both the overall dipolarity $f_{\rm dip}$ and at the dynamo surface $f_{\rm dyn,dip}$ (obtained from the 5 % integral below $r_m$, as explained in Chapter 5), decrease with $P_{\rm orb}$ from a dipolar regime and then they fall to the multipolar dominated regime at 4.08 d (which remains in the low-Ro regime). The four models with Ro > 0.1 show reversed trends, there is a decrease of Rm, $\Lambda$, $E_{\rm mag}/E_{\rm kin}$, and $f_{\rm ohm}$ with $P_{\rm orb}$. However, in comparison with the low-Ro runs, Rm values are much larger; $E_{\rm mag}/E_{\rm kin}$ and $f_{\rm ohm}$ take much lower values; and $\Lambda$ shows similar values with a smooth transition between the Ro > 0.1 and Ro < 0.1 regimes. Both dipolarity measures show a decrease with $P_{\rm orb}$, corresponding to typical multipolar dynamos. Note that typical values are not as high as the ones in Chapter 5, this is due to the more demanding change in parameter space for the HJ series.

Both internally heated models here considered fall in the low-Ro regime. Comparing the non-heated with the heated models, we can see that the latter have higher values for Rm and $\Lambda$, lower values for $E_{\rm mag}/E_{\rm kin}$, $f_{\rm dip}$ and $f_{\rm dyn,dip}$, and similar values for $f_{\rm ohm}$. This is due to the fact that heated HJs have more energetic convection, leading to higher internal velocities and magnetic field predictions, but at the same time, more multipolar fields with lower equipartition levels. With only two data points, we cannot make any strong statements about the trends, but we can see that there is still a dependence of Rm and $\Lambda$ with $P_{\rm orb}$, although less pronounced than the non-inflated series. The loss of dipolarity with larger $P_{\rm orb}$ is still maintained.

We show three corresponding spectral and radial energy distributions in Fig. 7.4. The main characteristics are similar to Chapter 5: the kinetic spectra are dominated by the equatorial jet at low multipoles; the magnetic spectra are mostly flat; and they show a similar diffusion range at about $\ell \sim 50 - 60$. By looking at surface spectra like the ones shown in Fig. 5.6, we can affirm that some of these runs are somewhat underresolved, and we are now in the process of running high-resolution models as well as changing to a less demanding parameter space.

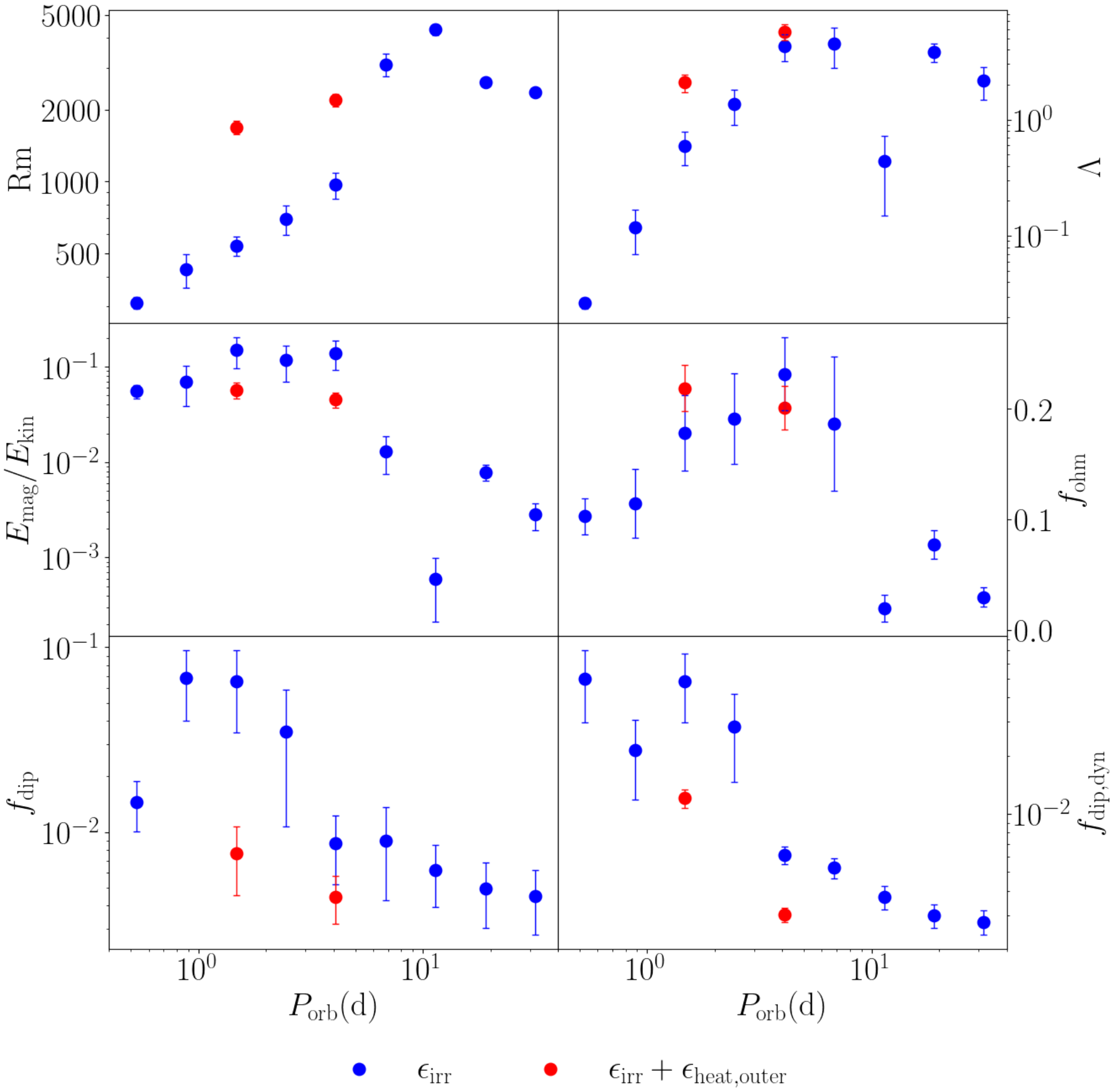

Figure 7.3: Similar to Fig. 5.10: diagnostics as a function of orbital distances. From left to right and top to bottom, the magnetic Reynolds number, the Elsasser number, the magnetic-to-kinetic energy ratio, the Ohmic fraction, total dipolarity, and dipolarity at the dynamo surface.

Finally, we plot this preliminary HJ sequence on top of the anelastic scaling laws of Yadav et al. (2013). Similarly to the evolutionary trends shown in Figs. 5.14 and 5.15, in Fig. 7.5 we show HJ trends in the dimensionless parameter space. In the Ro vs $P/(\mathrm{Pm}^{13/45})$ plot, we can see that HJs orbiting closer to their host star have lower Ro. This could already have been predicted: the smaller the $P_{\rm orb}$, internal convection is dominated by rotational effects (see Chapter 6). As noted above, the four longest-period planets are not in the low-Ro regime, and they tend to fall farther apart from the scaling law. What might seem surprising is the decrease of buoyant power per unit of mass $P$ for more irradiated planets. But it can be justified again by noting that rotational effects hinder convection processes (E decreases much more than Ra increases to maintain the same criticality). Both inflated models move up the scaling law (i.e., both Ra and $P$ increase) in comparison with their unheated counterparts. Similarly, we plot the HJ sequence on top of the scaling laws related to magnetic outputs. Again, the Ro $> 0.1$ do not exactly fall within the scaling law trends. The low-Ro regime models are spread throughout the scaling law down to very low values for Lo and high $\tau_{\rm mag}$. The transition from the multipolar to the dipolar dynamo branches as $P_{rmorb}$ gets smaller is not as clear as in Chapter 5. However, it can be slightly appreciated in the $\tau_{\rm mag}$ scaling law.

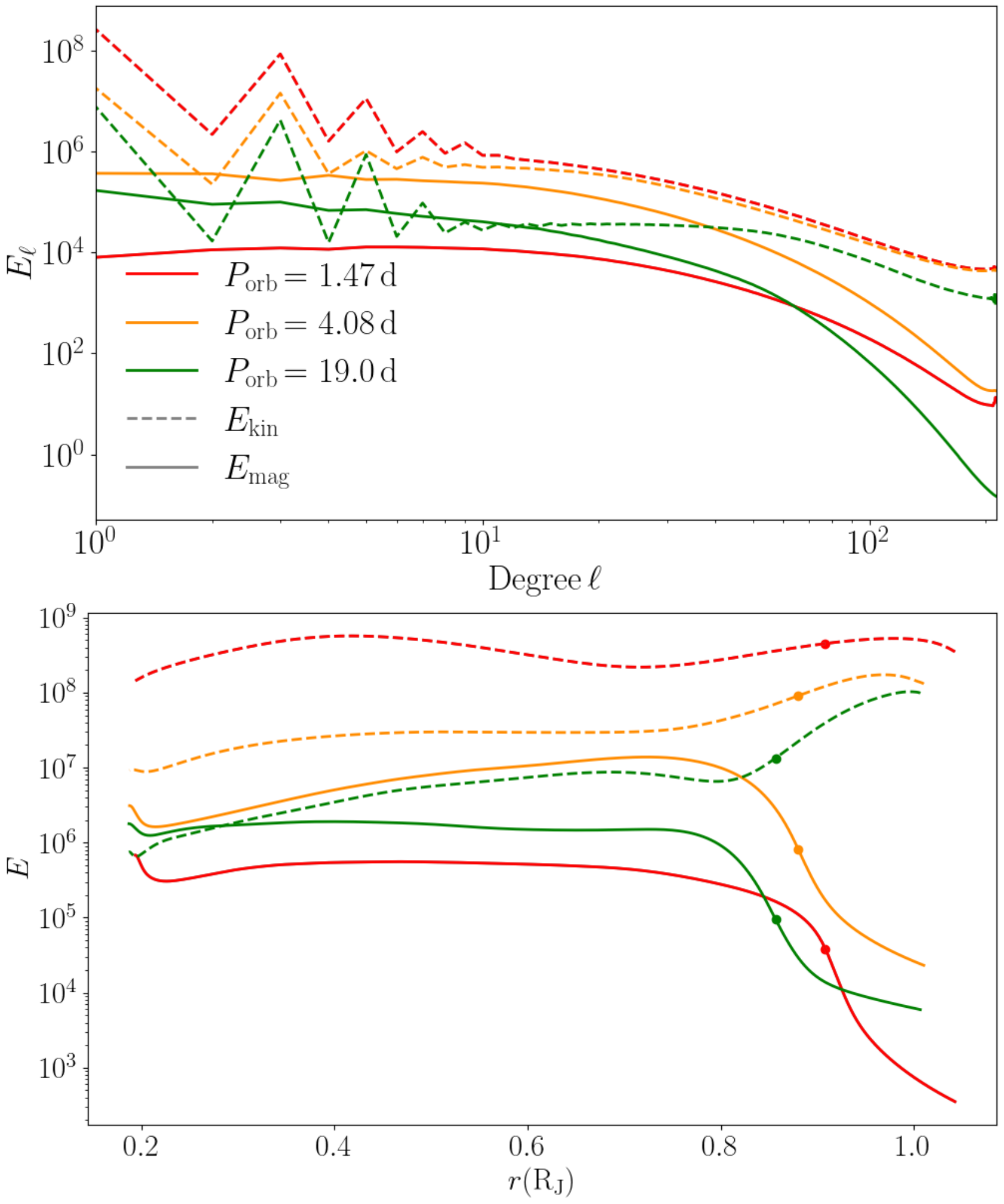

Figure 7.4: Similar to Fig. 5.5. Magnetic (solid) and kinetic (dashes) energy distribution over the multipole degrees $l$ (up), and the radius (down), for three models representing the same 1 $M_J$ planet at different orbital distances.

After the exploratory simulations summarized here, we are currently about to perform the production runs for an article, which will be the first one specifically dedicated to dynamos in HJs, to the best of our knowledge. We first need to increase the resolution so that $f_P = |\overline{P_\nu} - \overline{D_{\rm visc}} - \overline{D_{\rm ohm}}|/\overline{P_\nu}$ is no greater than 1 %. We will also restrict the $P_{\rm orb}$ interval so we are not pushing simulations to such a demanding low E regime (so that the runs in the Yadav et al. (2013) scaling laws are not so spread), and at the same time recover a low Ro regime for a full series of HJs. Lastly, we will use the inflation prescription by Thorngren and Fortney (2018), already used in Chap. 6. The HJ dynamo trends shown above will be confirmed or refuted after these improvements are made, but we expect the trends to be only marginally adjusted.

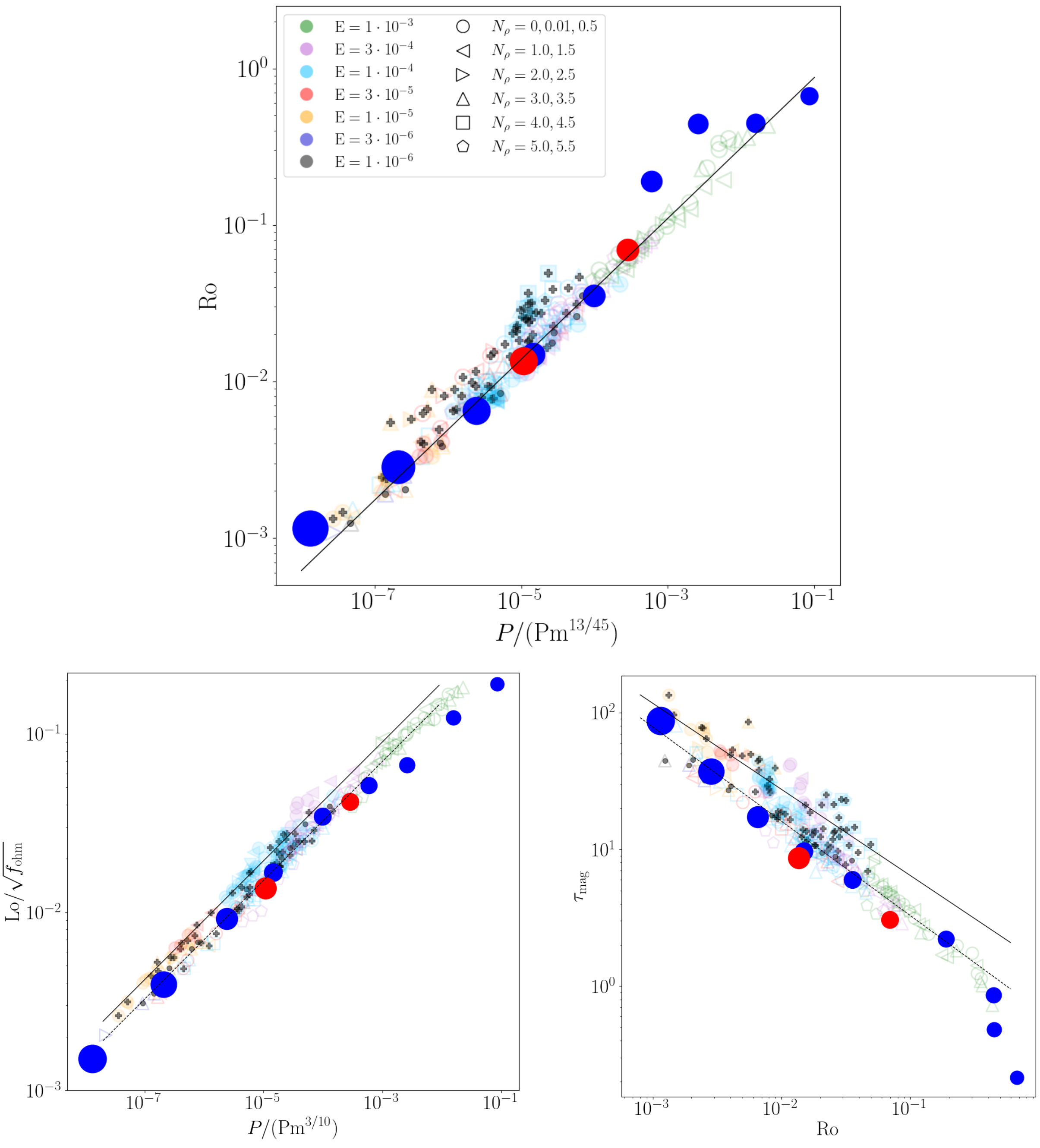

Figure 7.5: Similar to Fig. 5.14 and Fig. 5.15: scaling laws from Yadav et al. (2013) plotted alongside the HJ sequence. The size of the markers denotes stellar irradiation, not the inflation level.

# 8

# Conclusions

Here I summarize the conclusions from all the projects I have been involved with, which have been extensively presented in Chapters 3, 5, 6 and 7.

In Chapter 3, I explore the possibility of having vorticity production and dynamo action driven by a purely curl-free forcing of the velocity field. I ran numerical models that employed an irrotational forcing, which does not lead to any growth of the vorticity, neither in the HD nor in the MHD scenarios, if it is acting independently. However, a vortical flow is produced when the forcing interacts with a solid body rotation, in the presence of baroclinicity, or of a background shear. In the case of a rotating and baroclinic system, I have not found any dynamo action within the explored parameter space, neither with an initial random seed magnetic field, nor with an initial uniform field. This result, therefore, seems not to be dependent on the topology of the initial field.

Instead, in the presence of a background, sinusoidal shearing velocity, I could observe an amplification of the initial seed magnetic field as a consequence of a dynamo process, both in the barotropic and in the baroclinic cases. The main novelty in the presented work is to show how such a dynamo phase occurs after the onset of a hydrodynamical instability, driving an exponential growth of the vorticity. After the vorticity has grown, the magnetic energy is amplified too, approaching equipartition with the turbulent kinetic energy. This exponential amplification of vorticity occurs both in the purely HD case and in the MHD one, starting from small scales and then spreading up to the scale of the box. Similarly, the magnetic field is amplified first on small scales, but the inverse cascade makes the large scales dominate during the saturation phase of the simulations. Such a large-scale field is subject to a winding process, which I can identify after the system reaches equipartition between kinetic and magnetic energy. This winding enhances the field in the shearing direction, but the turbulent vorticity and velocity remain isotropic.

We also explored the role played by the scale on which the irrotational forcing is acting. I find that, independent of the scale of this forcing, no dynamo instability develops for systems that do not include any shear. Also, while the root mean square values of vorticity weakly depend on the forcing scale, in no case do I observe an exponential amplification of vorticity. Again, when shear is added to the picture, the vorticity is always exponentially amplified after a transient time if the kinematic viscosity is low enough, except in the case of a very small forcing scale, which does not lead to any growth over thousands of turnover times. Then it is followed by a dynamo instability if the magnetic diffusivity is not too high. By analyzing kinematic spectra, I see how the typical scale of the system is provided by the forcing scale of the turbulence before the vorticity is amplified and, conversely, by the scale of the shear, in the saturation phase. Based on that, I observe that the growth rate of the dynamo depends on the scale of the expansion waves if the growth rate is calculated using the turnover time as the time unit. Still, it exhibits no dependence on the forcing lengthscale if the time unit is the shear timescale. The scale of the forcing also sets the time needed for the instability to develop. Models with a larger forcing scale amplify vorticity and magnetic fields after shorter times.

If the baroclinic term is at work, I noticed how the vorticity growth takes place at a slightly later time, with a growth rate similar to the barotropic case. Instead, the magnetic field grows at a bit faster rate. The only models that immediately develop the instability are those that include both the baroclinic term and a cooling function. I observe an increase of one order of magnitude for the growth rate with the magnetic Prandtl number, which I varied between 0.1 and 10. I extrapolated a critical value for the magnetic Reynolds number of slightly less than 20.

Our results cannot firmly exclude that a dynamo is produced with irrotational forcing, rotation, and no shear, as, for example, seen by Seta and Federrath (2022) using a different Fourier-based forcing in a multi-phase MHD setup with the *Flash* code. However, in our setup, such a dynamo could only be produced for Re higher than what was tested in the chapter (up to a few hundred). Another possible source for this discrepancy is the difference in physical and numerical viscosity schemes between *Flash* and the*Pencil Code*, which may have a significant effect. In any case, the critical threshold would therefore be much larger than what is seen in the shear case ($\mathrm{Re}_{crit} \sim 50$).

We conclude that the presence of shear remains the basic ingredient for triggering a dynamo instability when subsonic turbulence is driven by spherical expansion waves, even if the curl-free acceleration assumption is relaxed and density fluctuations are taken into account. These results hold when solid rigid rotation, high Reynolds number, or more complex EoS are taken into consideration. Future work could take into account turbulence forced on more than one scale at the same time, as well as the role played by plane waves, before moving toward more complex models that include density stratification as well as shocks and supersonic flows.

When non-isothermality, rotation, shear, or density-dependent forcing are included, they contribute to increasing the vorticity. I find that turbulence driven by subsonic expansion waves can amplify the vorticity and magnetic field only in the presence of a background shearing profile. The presence of a cooling function makes the instability occur on a shorter timescale. I estimate critical Reynolds and magnetic Reynolds numbers of 40 and 20, respectively.

The second line of research was related to the study of dynamo solutions across the lifetime of Jupiter-like planets. For this project, outlined in Chapter 5, I used radial profiles taken from gas giant evolutionary models to obtain sequences of 3D MHD spherical shell dynamo models. Using the public code *MESA*, I obtained the radial hydrostatic profiles for planets with different masses at different stages of evolution: $0.3\,M_J \leq M_P \leq 4\,M_J$ and 0.2 Gyr $\leq t \leq$ 10 Gyr, respectively. From the evolutionary tracks, I derived the trends for the dynamo parameters. Using the radial profiles as the background state, I solved the resistive MHD equations under the anelastic approximation with the pseudospectral spherical shell code *MagIC*. I obtained saturated dynamo solutions and interpreted them as different snapshots of planetary dynamos during their long-term evolution.

For our longest set of runs that represents different evolutionary times of a 1 $\mathrm{M_J}$ planet, I find a transition from a multipolar- to a dipolar-dominated dynamo regime. Within the dipolar or multipolar branch, a few snapshots are enough to generally assess the behavior that cannot be straightforwardly derived from scaling laws. As the planet evolves and cools, I obtain a steady decrease for Rm, $P$, and Ro as well as an increase in the volumetric and surface dipolarities. For multipolar dynamo solutions $\Lambda$, $f_{\rm ohm}$, and $E_{\rm mag}/E_{\rm kin}$ decrease with time, whereas they increase for dipolar dynamos. These quantities are a proxy for the magnetic field energy, the dynamics of the power dissipation, and the energy ratio. These trends hold for different Prandtl numbers and for the 4 $M_J$ models. Future studies are needed to confirm that these trends are also observed at lower Ekman numbers because the physical nondimensional parameter space is currently computationally inaccessible.

The decay of the magnetic field strength at the dynamo surface (Fig. 5.12) is roughly compatible with existing scaling-law estimates. Our models are based on a sequence of realistic backgrounds, and thus, they can be representative of the real long-term dynamo evolution. The trends in mass and age are also expected to hold for the realistic but computationally unfeasible range of nondimensional numbers (which is the usual intrinsic caveat of any dynamo study). I compared our results with the anelastic scaling laws of Yadav et al. (2013) between nonlinear combinations of dimensionless diagnostics, and I showed that the long-term evolution of a cold Jupiter dynamo evolves in this parameter space. Consistent with the diagnostics on the morphology described above, some age sequences transition from the multipolar-dominated to the dipolar-dominated family of solutions, which are separated in this parameter space, as shown by Yadav et al. (2013).

Additionally, in the sequences, I considered fixed values of Pr and Pm, and the values of Ra and E changed only due to the evolving values of $\Delta T$, $T_o$, and the shell thickness. Taking into account the likely slight decrease in Pr (due to the long-term cooling, which decreases the thermal conductivity), I would expect a slightly steeper decrease in the magnetic field at the dynamo surface. In the presence of non-negligible external torques that would spin down the planet (e.g., tidal frictions with large satellites), E would (slightly) increase, which, for the trends seen in our sets, would lead to a (slight) enhancement of the slow magnetic decay. I leave this fine-tuned exploration for a follow-up work.

Taking Jupiter and Saturn as prototypes, I expect gas-giant dynamos to be dipolar-dominated at an advanced evolutionary age. However, according to our results, multipolar gas-giant dynamos could exist in the early stages of planetary dynamos, which would then evolve into a dipolar regime. Another possibility is that gas-giant dynamos are already born in a dipole-dominated parameter space region and remain so. These arguments can also be applied to mildly irradiated gas giants and brown dwarfs because they have similar low-Ro dynamos that follow the scaling laws of Christensen et al. (2009). After all, the orbital distance is large enough to prevent tidal interactions or inflation from dominating their energy budget. In contrast, these trends cannot be applied to rapidly rotating main-sequence stars as their evolution is dictated by hydrogen burning and not by slow cooling.

We also address the observational consequences regarding the radio emission produced by ECM emission that is detectable from ground (with an associated gyro-frequency, $\nu \simeq 2.8\ B[\mathrm{G}]$ MHz, higher than the ionospheric ≈10 MHz cutoff; Zarka (1998)), current (LOFAR), and next-generation (SKA-low) low-frequency radio interferometers might eventually detect their exoplanetary radio emission. With more than 650 confirmed cold Jupiter analogs ($M\sin(i) > 0.2\,M_J$ and $P_{\rm orb} > 200$ days), there are only four exoplanets under 10 pc with an age estimate. $\epsilon$ Eridani b ($0.66^{0.12}_{-0.09}\ M_J$, $0.6 \pm 0.2$ Gyr, Hatzes et al. (2000)) is the only suitable candidate for an early multipolar dynamo evolution.

In Chapter 6, I explore the interior evolutionary tracks of inflated hot Jupiters using the one-dimensional evolutionary code *MESA*. I assume tidal locking and irradiation from main-sequence host stars, and I explore dependencies on orbital distance, planetary and stellar mass, and the type of heat injection. Guided by observationally constrained flux-heating efficiency relations (Thorn-gren and Fortney, 2018), I inject heat into the internal layers of the planet to reproduce the observed radii. This relation and others in the literature (e.g., Sarkis et al., 2021) exhibit significant dispersions around the average values of inflated radii for a given irradiation. Therefore, this study only addresses the trends with the planetary properties (mass, separation, type of internal heating), keeping the parameters involving the amount of deposited heat fixed.

For a given stellar type, orbital period and planetary mass, I compare the purely irradiated case with two simplified scenarios of regions where extra heat is continuously deposited: extended (injected primarily on the dynamo region, $P \gtrsim 10^6$ bar), or outside the dynamo region, which is what is expected in most models, the Ohmic one especially (Batygin and Stevenson, 2010; Batygin et al., 2011; Thorngren, 2024; Viganò et al., 2025). Whether heat is injected uniformly, centrally, or throughout the convective region, the structural differences are minimal. Planets typically exhibit a shallow stratified outer layer, which fully contains the irradiation zone, followed by a deep convective interior where pressure is sufficient for hydrogen metallization and possible dynamo action.

However, important differences are seen in the Rossby number as a function of depth. The vast majority of models yield Ro $\lesssim$ 0.1, indicating a fast-rotating convection regime. Defining HJs as having $P_{\rm orb} < 10$ d or $T_{\rm eq} > 1000$ K (see Gan et al. (2023) and within), I can safely say that the vast majority of confirmed HJs are expected to be in the low-Ro regime. Only the most massive, distantly orbiting (yet still tidally locked) planets exceed Ro $\gtrsim$ 0.1 over significant interior regions, potentially altering the dynamo regime. Thus, massive HJs with orbital periods beyond 15-20 days may host low-Rossby-number dynamos that generate weaker, more multipolar magnetic fields despite similar convective power. Within the known HJs, only three targets with $M_P \gtrsim 4\ M_J$ closer than 100 pc have 15 d$<P_{\rm orb}<$40 d, all orbiting G-type stars.

The extra heat affects the outer layers and consistently increases Ro. While this effect is negligible for planets above 8 $M_J$ (especially those already exceeding Ro $\gtrsim$ 0.1), which can hardly get inflated due to their higher gravity, lower-mass planets experience a roughly one order of magnitude increase in Ro, which has important interior dynamic consequences even it they remain within the low-Ro regime. When the heat injection is localized outside the dynamo region, the temperature gradients are reduced or even inverted, as already briefly mentioned by e.g. (Komacek and Youdin, 2017; Komacek et al., 2020). The internal heat transport is significantly reduced, leading to positive entropy gradients that suppress convection.

The relevant consequence of this comes when one applies the widely-used, observationally constrained magnetic scaling laws from Christensen et al. (2009), suited for fast rotators. When heat is deposited in an extended way, I recover surface magnetic field strengths around 100 G for the most inflated planets, an order of magnitude higher than Jupiter. These estimates are compatible with the work of Yadav and Thorngren (2017), who assume that the extra heat gives the convective heat flux. This is conceptually similar to our extended heat case, where heat is essentially deposited in the dynamo region. In contrast, a more realistic external heating substantially reduces convective power and yields weaker magnetic fields.

This result is particularly interesting in the context of the Ohmic dissipation models (Batygin and Stevenson, 2010; Batygin et al., 2011; Perna et al., 2010b; Wu and Lithwick, 2013; Ginzburg and Sari, 2016; Knierim et al., 2022; Viganò et al., 2025). Essentially, it relies on the presence of conducting outermost layers due to thermal ionization (e.g., Kumar et al., 2021; Dietrich et al., 2022), and on the circulation of strong, mainly zonal winds which twist and stretch the background field coming from the deep-seated dynamo. The inducted currents also propagate and dissipate into the interior due to the finite conductivity values in the outer convective region (much smaller than in the dynamo region, but not zero). Combining estimation for the conductivity and the induced currents, models predict that the Ohmic heating profile steeply decays towards the interior, so that the bulk of it is deposited well above the dynamo region (Batygin et al., 2011; Ginzburg and Sari, 2016; Knierim et al., 2022; Viganò et al., 2025). Since the atmospheric induction depends also on the intensity of the background magnetic field (linearly only for magnetic Reynolds numbers Rm $\lesssim$ 1), there could be an interesting, non-trivial coupling between the dynamo and the atmospheric induction: if the Ohmic heating is significant, it can suppress convection and the underlying dynamo field, which would in turn decrease the Ohmic heating rate, causing the convection to be restored. This process might only lead to a less effective Ohmic heating mechanism or, potentially, show an oscillatory behavior, as proposed by Viganò et al. (2025).

The considerations on the magnetic field strengths are of direct interest for the observational detection of SPI signals, e.g., in terms of activity indicator orbital modulations (Cauley et al., 2019), and/or planetary Jovian-like coherent radio emission, which remains elusive, despite optimistic predictions (Stevens, 2005) and tens of observational campaigns (e.g., Narang et al., 2024, and references within). Several potential causes can account for the lack of detection, including beaming effects, large distances, intrinsic variability, and limited observational coverage. However, one cannot neglect the possibility that the typical estimates of HJ magnetic fields of the order 100 G might be too optimistic (Yadav and Thorngren, 2017; Kilmetis et al., 2024). In this sense, our results indicate that very massive HJs might offer an exception, since their internal structure is less sensitive to the combined effects of irradiation and heating, and could still maintain large magnetic fields. For Jupiter-like masses, if the external heating model is assumed, there is little hope of having a magnetic field much larger than Jupiter's.

Finally, in Chapter 7, I introduce the work on HJ dynamos, which naturally follows the previous chapters. On one side, I show my contribution to (Viganò et al., 2025), currently under review, where I compare the magnetic field curvature estimates for Jupiter's magnetic data (Connerney et al., 2018, 2022) with the numerical calculations obtained from spherical shell saturated dynamo solutions. Downward interpolation predicts a mean inverse of curvature (i.e., the typical length-scale of magnetic field structures) between 0.1 and 0.2 $R_J$ just above the dynamo surface. These estimates are compatible with the inverse of curvature obtained from dynamo simulations of HJs, $\lesssim$ 0.1 $R_J$. I also show that predictions for the electric currents in the dynamo region can be estimated with the curvature by a factor between 1.5 and 2.5 between $|\mathbf{J}|$ and $|\mathbf{B}||\kappa|/\mu_0$. These currents are comparable to or less than the ones induced by the atmosphere for Ohmic models (Batygin and Stevenson, 2010; Viganò et al., 2025). Therefore, one should be aware that more realistic HJ dynamo boundary conditions can lead to substantial current exchange with the interior. Additionally, I add the preliminary results of saturated dynamo simulations of different HJ interiors at several orbital periods. Overall, we obtain models with several similarities to the cold Jupiter cases, shown in Chapter 5, in terms of time series, spectra, and radial distributions. The main difference is that there is a strong influence on orbital period due to the tidal locking constraint, which leads to a 3-order of magnitude difference in Ro between the closest and farthest HJ.

## 8.1 Future prospects

- Regarding the vorticity project, I am considering several follow-up studies. In particular, I would like to study other types of forcing, such as the one employed by Federrath et al. (2011); Seta and Federrath (2022), which considers a purely rotational flow, similar to our work, but defined in Fourier space. Another possibility is to reach the supersonic regime by using shock viscosities, and possibly adding the Hall and ambipolar terms to the induction equation.

- As shown in the preliminary results of Chapter 7, we are implementing the 1D to 3D strategy described for cold gas-giants in Chapter 5 to the HJ *MESA* models of Chapter 6. We are improving the resolution and using the correct prescription for inflated models.

- I intend to make the 3D planetary dynamo projects more robust, by considering the expected temperature dependencies to Pr and Pm as described by Bonitz et al. (2024), and its evolution over the lifetime of planets. This could affect the dynamo trends observed in time (Chapter 5) and orbital distance (Chapter 7).

- I also might explore a new aspect of the suite of gas giant simulations presented in this thesis by analyzing the zonal winds that naturally develop in the upper layers of the 3D models. Achieving a realistic representation of the outer atmospheric spectrum will require higher-resolution grids. For cold Jupiters undergoing long-term cooling, incorporating SSL will be essential to reproducing more Jovian-like wind patterns. In contrast, for HJs, implementing outer boundary conditions that mimic the contrasting day and night sides will be necessary to capture their distinctive atmospheric dynamics.

- The preliminary results shown in Chapter 7 will represent the first dedicated HJ dynamo simulations, and can potentially open an entirely new line of research within the dynamo community. In particular, I will focus on the expected magnetic intensity configuration at the surface of HJs, since it can significantly affect the global circulation models and the atmospheric induction, which is in turn likely a key factor in explaining the HJ observed radii inflation.

- Related to the last points, the planetary dynamo studies finally aim at characterizing better the properties that make gas giants (cold or hot) more likely to observationally show signatures of SPI and/or coherent radio emission.

# A

# Magnetohydrodynamics: a basic derivation

For completeness, I add a short derivation of the four fundamental equations of the non-relativistic classical MHD. The applicability of the MHD equations is discussed in Chapter 2, together with arguments for the fluid approximation and the fully collisional limit. Let's consider a fluid element delimited by a surface $S$ enclosing a volume $V$. The integral form of the conservation principle for a general quantity $q$ can be expressed as:

$$\frac{d}{dt}\int_V q\, dV = -\int_{\partial V} \mathbf{F_q} \cdot d\mathbf{S} + \int_V S_q\, dV\,,$$

where $\mathbf{F_q}$ is the flux, $S_q$ are the sources or sinks and $d\mathbf{S} = \mathbf{n}\, dS$, and $\hat{\mathbf{n}}$ is the unit vector normal to the surface. This equation is equivalent to its differential form:

$$\frac{dq}{dt} + \boldsymbol{\nabla} \cdot \mathbf{F_q} = S_q\,.$$

When the quantity $q$ is mass, momentum, or energy, we recover the three hydrodynamic equations, i.e., the continuity equation, the Navier-Stokes equation, and the energy conservation equation. The induction equation is derived from Maxwell's equations, closed with the specific Ohm's law that describes the electrically conducting fluid.

## A.1 Mass continuity equation

Assuming no particle creation or annihilation, any fluid element can see an increase or decrease in the mass it contains, $m$, only as a consequence of fluid flow through its surface $S$: the mass change must be exactly equal to the net mass flux ($\rho\mathbf{u}$). This notion can be used to express mass conservation mathematically:

$$\frac{dm}{dt} = \frac{d}{dt}\int_V \rho\; dV = \int_V \frac{\partial \rho}{\partial t}\; dV = -\oint_S \rho\mathbf{u} \cdot d\mathbf{S}\,,$$

where $d\mathbf{S}$ is the vector representing the surface infinitesimal, pointing perpendicular and outward to it. We have imposed that the imaginary surface does not move or change in time, so time derivatives commute with the integral. The divergence theorem applied to the mass flux reads:

$$\oint_S \rho\mathbf{u} \cdot d\mathbf{S} = \int_V \boldsymbol{\nabla} \cdot (\rho\mathbf{u})\; dV\;.$$

Using this identity, we obtain a relation where both sides are integrated in the same volume. This equality must hold independently of the chosen fluid element; thus, the integrands must be the same:

$$\int_V \frac{\partial \rho}{\partial t}\, dV = -\int_V \boldsymbol{\nabla}\cdot(\rho\mathbf{u})\, dV \qquad \Rightarrow \qquad \frac{\partial \rho}{\partial t} + \boldsymbol{\nabla}\cdot(\rho\mathbf{u}) = 0\,.$$

This is known as the continuity equation, in its differential form, and it is simply the statement of mass conservation applied at any point in the fluid domain. Another slightly different version of this equation, used by the *Pencil code*, can be obtained by rearranging factors of $\rho$ and defining the material derivative as $D/Dt = \partial/\partial t + \mathbf{u}\cdot\boldsymbol{\nabla}$:

$$\frac{D\ln\rho}{Dt} = -\boldsymbol{\nabla}\cdot\mathbf{u}\,.$$

This is useful with compressible MHD codes, because using $\ln\rho$ rather than $\rho$ makes density fluctuations numerically more stable.

## A.2 The momentum equation

We can apply the same conservation principle to the momentum within a fluid element to write the equivalent of Newton's second law for fluids. In contrast with mass conservation, the change of momentum $\mathbf{P}$ in a fluid parcel is not only equal to the flux through the imaginary surface, but any force acting on the fluid has to be taken into account as a source or sink:

$$\frac{d\mathbf{P}}{dt} = \frac{d}{dt}\int_V \rho\mathbf{u}\, dV = -\oint_S (\rho\mathbf{u})\otimes\mathbf{u}\cdot d\mathbf{S} + \mathbf{F}\ ,$$

where we have introduced the outer product: $(\mathbf{u}\otimes\mathbf{u})_{ij} = u_i u_j$. The general force $\mathbf{F}$ is applied to the fluid element, e.g., gravity, viscous forces, or magnetic forces, among others. An alternative way to express the law is using subscript notation, where repeated indices are used to indicate the sum over them:

$$\int_V \frac{\partial}{\partial t}(\rho u_i)\, dV + \oint_S \rho u_i u_j\, dS_j = F_i\ .$$

Similarly, we can apply the divergence theorem for the momentum flux; thus, the surface term becomes a volume integral, so that:

$$\int_V \frac{\partial}{\partial t}(\rho u_i)\, dV + \int_V \frac{\partial}{\partial x^j}(\rho u_i u_j)\, dV = F_i\ .$$

We can now join both integrands and develop their derivatives. Using the continuity equation (A.1), we can further simplify this equation and use the vector notation:

$$\int_V \left(\rho\frac{\partial}{\partial t}u_i + u_i\frac{\partial}{\partial t}\rho + \rho u_j\frac{\partial}{\partial x^j}u_i + u_i\frac{\partial}{\partial x^j}(\rho u_j)\right) dV =$$
$$= \int_V \left(\rho\frac{\partial}{\partial t}u_i + \rho u_j\frac{\partial}{\partial x^j}u_i\right) dV = F_i \qquad \Rightarrow$$
$$\Rightarrow \qquad \int_V \rho\left(\frac{\partial}{\partial t}\mathbf{u} + (\mathbf{u}\cdot\boldsymbol{\nabla})\mathbf{u}\right) dV = \mathbf{F}\ .$$

Pressure is present in all fluids, and the total pressure force that a given volume element will feel will be the surface integral of the contributions coming from the neighbouring pressure over each portion of the surface, $-p\, d\mathbf{S}$, which by definition points inwards. This is mathematically expressed as:

$$\mathbf{F}_{\mathrm{p}} = -\oint_S p\, d\mathbf{S} = -\int_V \boldsymbol{\nabla}p\, dV\,.$$

If we then express all other possible forces as a force $\mathbf{f}$ per unit mass acting throughout the volume:

$$\int_V \rho \left( \frac{\partial}{\partial t}\mathbf{u} + (\mathbf{u}\cdot\boldsymbol{\nabla})\mathbf{u} \right) \, dV = -\int_V \boldsymbol{\nabla} p \, dV + \int_V \rho\mathbf{f} \, dV \ .$$

Finally, as this equality must hold regardless of the chosen volume, the integrands must equate: the differential form reads

$$\rho \left( \frac{\partial}{\partial t} + \mathbf{u}\cdot\boldsymbol{\nabla} \right) \mathbf{u} = -\boldsymbol{\nabla} p + \rho\mathbf{f} \, .$$

If there is no viscous term, the momentum equation is known as the Euler equation, which describes inviscid fluids, the movement of which can be mathematically expressed as potential flows.

However, in reality, for most fluids, one should take into account viscosity, which has a microphysical origin and leads to internal friction forces acting within the fluid elements as they rub together. Usually, one assumes that the viscous forces are linear with velocity, which is known as the Newtonian approximation. By logic, one can expect it must be proportional to the gradient of velocity and not to velocity itself. The basic assumption is that the Laplacian of velocity can be written as $\mu\nabla^2\mathbf{u}$, where $\mu$ is a proportionality constant known as dynamic viscosity.

A real derivation involves looking at all possible forces a fluid can have on itself, i.e., the stress tensor $\sigma_{ij}$. This tensor must be symmetric and can be divided into perpendicular and parallel components to the surface of any fluid element. The perpendicular components are directly interpreted as pressure, whereas the parallel forces constitute the shear tensor, $\tau_{ij}$, which contains the viscous forces per unit area:

$$\sigma_{ij} = -p\delta_{ij} + \tau_{ij} = -p\delta_{ij} + 2\rho\nu S_{ij} = -p\delta_{ij} + 2\mu \left( \frac{1}{2}\frac{\partial u_i}{\partial x_j} + \frac{1}{2}\frac{\partial u_j}{\partial x_i} - \frac{1}{3}\boldsymbol{\nabla}\cdot\mathbf{u}\delta_{ij} \right) \, .$$

The expansion defining the rate-of-strain tensor, $S_{ij}$, is obtained by taking the most general traceless symmetric tensor containing first derivatives of $\mathbf{u}$. Taking the gradient of $\sigma_{ij}$ will give the terms in the momentum equation. The pressure term recovers the term mentioned above, $-\boldsymbol{\nabla} p$. The shear tensor term requires more work. We first introduce the kinematic viscosity $\nu = \mu/\rho$. It is usually assumed that $\nu$ is constant in space, i.e., $\nu$ depends on $\rho$ much less than $\mu$. They both still vary with $T$, as microscopical processes (the ones dictating the viscous interactions) are highly temperature dependent. If we multiply $1/\rho$ to the total viscous force (i.e., consider the force per unit mass), the resulting expression is:

$$\mathbf{F}_\nu = \frac{1}{\rho}\boldsymbol{\nabla}\cdot\tau = \frac{1}{\rho}\boldsymbol{\nabla}\cdot(2\rho\nu\boldsymbol{S}) = \frac{2\nu}{\rho}\boldsymbol{S}\cdot\boldsymbol{\nabla}\rho + 2\nu\boldsymbol{\nabla}\cdot\boldsymbol{S} = 2\nu\boldsymbol{S}\cdot\boldsymbol{\nabla}\ln\rho + 2\nu\boldsymbol{\nabla}\cdot\boldsymbol{S} \, .$$

We can develop the gradient of the strain tensor easily by using the repeated index sum notation:

$$\partial_i S_{ij} = \partial_i \left( \frac{1}{2}\partial_j u_i + \frac{1}{2}\partial_i u_j - \frac{1}{3}\delta_{ij}\partial_k u_k \right) = \frac{1}{2}\partial_j(\partial_i u_i) + \frac{1}{2}(\partial_i\partial_i)u_j - \frac{1}{3}\partial_j(\partial_k u_k) =$$

$$= \frac{1}{6}\partial_j(\partial_i u_i) + \frac{1}{2}(\partial_i\partial_i)u_j \quad \Rightarrow \quad \boldsymbol{\nabla}\cdot\boldsymbol{S} = \frac{1}{6}\boldsymbol{\nabla}(\boldsymbol{\nabla}\cdot\mathbf{u}) + \frac{1}{2}\nabla^2\mathbf{u} \, .$$

If we also consider a non-uniform kinematic viscosity $\nu$, the general viscous force reads:

$$\mathbf{F}_\nu = 2\nu\boldsymbol{S}\cdot\boldsymbol{\nabla}ln\rho + 2\boldsymbol{S}\cdot\boldsymbol{\nabla}\nu + \nu\nabla^2\mathbf{u} + \frac{1}{3}\nu\boldsymbol{\nabla}(\boldsymbol{\nabla}\cdot\mathbf{u}) \, .$$

For an incompressible fluid, i.e., $\rho$ constant and $\boldsymbol{\nabla}\cdot\mathbf{u} = 0$, with constant $\nu$, the total viscous force is $\mathbf{F}_\nu = \nu\nabla^2\mathbf{u}$ as the simplified case before.

Another common force, present in electrically conductive fluids with a non-zero magnetic field (external or internally induced), is the Lorentz force. As both the free-flowing positive and negative charges carrying an electrical current in the fluid are displaced alongside the flow through a magnetic field, they feel the magnetic forces. Mathematically, the magnetic force per unit volume is the cross-product of the total electrical current and the magnetic field:

$$\mathbf{F}_L = \mathbf{J} \times \mathbf{B} = \frac{1}{\mu_0}(\boldsymbol{\nabla} \times \mathbf{B}) \times \mathbf{B} \ .$$

Therefore, the full MHD momentum equation is:

$$\rho\left(\frac{\partial}{\partial t} + \mathbf{u} \cdot \boldsymbol{\nabla}\right)\mathbf{u} = -\boldsymbol{\nabla} p - \frac{1}{\mu_0}\mathbf{B} \times \boldsymbol{\nabla} \times \mathbf{B} + \rho \mathbf{F}_\nu + \rho \mathbf{f} \ .$$

The conservative form of the equation can be written by rearranging the Lorentz and pressure terms to the left-hand side:

$$\frac{\partial}{\partial t}(\rho \mathbf{u}) + \nabla \cdot \left[\rho \mathbf{u} \otimes \mathbf{u} - \frac{1}{\mu_0}\mathbf{B} \otimes \mathbf{B} + \left(p + \frac{B^2}{2\mu_0}\right)\mathbf{I}\right] = \nabla \mathbf{S} + \rho \mathbf{f} \ ,$$

where $\mathbf{I}$ is the identity tensor, and the Lorentz force has been decomposed into two terms: the magnetic tension $\nabla \mathbf{B} \otimes \mathbf{B}$ which tries to straighten field lines (it is directed radially inward to the curvature), and the so-called magnetic pressure term $B^2/(2\mu_0)$, which exerts a force towards regions of lower magnetic field intensity. The term $\rho \mathbf{f}$ includes any other force, e.g., gravity, besides viscosity and Lorentz force.

## A.3 The energy equation

### A.3.1 Energy form

The most straightforward way to obtain the MHD energy equation is to apply a conservation law to the total energy per unit volume: $\mathcal{E} = \rho e + \rho u^2/2 + B^2/2\mu_0$, where $e$ is the specific internal energy, which for an ideal gas is $e = p/(\gamma - 1)$. A strategy to write the evolution equation in its conservative form is to obtain conservation laws for each of the internal, kinetic, and magnetic energies separately.

The first law of thermodynamics for the internal energy of a fluid element leads to the following expression:

$$\frac{De}{Dt} = \frac{1}{\rho}\left(-p \nabla \cdot \mathbf{u} + \Phi_\nu + \nabla \cdot (\kappa \nabla T)\right) \ ,$$

where $D/Dt = \partial/\partial t + \mathbf{u} \cdot \nabla$ is the material derivative, $\Phi_\nu$ is the viscous dissipation rate, $\kappa$ is the thermal conductivity, and $T$ is the temperature. This equation accounts for the work done by pressure forces, viscous heating, and heat conduction (from Fourier's law, i.e., $\nabla \cdot (\kappa \nabla T)$), respectively. The viscous dissipation term is proportional to the rate of strain tensor $\mathbf{S}$, i.e., $\Phi_\nu = 2\rho\nu \mathbf{S}^2$. This expression can be expanded by:

$$\mathbf{S}^2 = S_{ij}S_{ij} = \left(e_{ij} - \frac{1}{3}\delta_{ij}(\boldsymbol{\nabla} \cdot \mathbf{u})\right)\left(e_{ij} - \frac{1}{3}\delta_{ij}(\boldsymbol{\nabla} \cdot \mathbf{u})\right) =$$

$$= e_{ij}e_{ij} - \frac{2}{3}e_{ij}\delta_{ij}(\boldsymbol{\nabla} \cdot \mathbf{u}) + \frac{1}{9}\delta_{ij}\delta_{ij}(\boldsymbol{\nabla} \cdot \mathbf{u})^2 = e_{ij}e_{ij} - \frac{1}{3}(\boldsymbol{\nabla} \cdot \mathbf{u})^2 \ .$$

Multiplying both sides by $\rho$, we get the internal energy conservation:

$$\rho\left(\frac{\partial}{\partial t} + \mathbf{u} \cdot \nabla\right)e = -p \nabla \cdot \mathbf{u} + \Phi_\nu + \nabla \cdot (\kappa \nabla T) \ .$$

For the kinetic energy conservation, we can use the identity $\mathbf{u} \times (\boldsymbol{\nabla} \times \mathbf{u}) = \boldsymbol{\nabla}(\mathbf{u} \cdot \mathbf{u})/2 - (\mathbf{u} \cdot \boldsymbol{\nabla})\mathbf{u}$ in the momentum equation:

$$\rho \left( \frac{\partial}{\partial t} + \mathbf{u} \cdot \boldsymbol{\nabla} \right) \mathbf{u} = -\boldsymbol{\nabla} p + \rho \mathbf{f} \quad \Rightarrow \quad \rho \left( \frac{\partial \mathbf{u}}{\partial t} + \frac{1}{2} \boldsymbol{\nabla} \mathbf{u}^2 - \mathbf{u} \times (\boldsymbol{\nabla} \times \mathbf{u}) \right) = -\boldsymbol{\nabla} p + \rho \mathbf{f} \, .$$

Now we can take the scalar product of both sides with $\mathbf{u}$ and rearrange the terms:

$$\rho \left( \frac{\partial}{\partial t} \left( \frac{1}{2} u^2 \right) + \mathbf{u} \cdot \nabla \left( \frac{1}{2} u^2 \right) \right) = -\mathbf{u} \cdot \nabla p + \rho \mathbf{u} \cdot \mathbf{f} \, .$$

Note that the viscous force (i.e., $\rho \mathbf{u} \cdot \mathbf{F}_\nu$) is not included, as it is already taken into account by the internal energy conservation. The Lorentz force term (i.e., $-\mathbf{u} \cdot (\mathbf{B} \times (\boldsymbol{\nabla} \times \mathbf{B}))/\mu_0$) is not added either, as it naturally arises from magnetic energy conservation.

We can develop the magnetic energy starting from the induction equation and apply the identity $\mathbf{b} \cdot (\boldsymbol{\nabla} \times \mathbf{a}) = \mathbf{a} \cdot (\boldsymbol{\nabla} \times \mathbf{b}) + \boldsymbol{\nabla} \cdot (\mathbf{a} \times \mathbf{b})$ for both terms:

$$\begin{aligned} \frac{\partial}{\partial t} \left( \frac{B^2}{2\mu_0} \right) &= \frac{1}{\mu_0} \mathbf{B} \cdot [\nabla \times (\mathbf{u} \times \mathbf{B}) + \nabla \times (\lambda \nabla \times \mathbf{B})] = \\ &= \frac{1}{\mu_0} (\boldsymbol{\nabla} \times \mathbf{B}) \cdot (\mathbf{u} \times \mathbf{B}) - \frac{1}{\mu_0} \boldsymbol{\nabla} \cdot (\mathbf{B} \times (\mathbf{u} \times \mathbf{B})) + \\ &\quad + \frac{\lambda}{\mu_0} (\boldsymbol{\nabla} \times \mathbf{B})^2 + \frac{1}{\mu_0} \boldsymbol{\nabla} \cdot (\mathbf{B} \times (\lambda \boldsymbol{\nabla} \times \mathbf{B})) \, . \end{aligned}$$

This expression can be reshuffled into a magnetic conservation law, the first term is equivalent to $\mathbf{u} \cdot (\mathbf{B} \times (\boldsymbol{\nabla} \times \mathbf{B}))/\mu_0$, the second and fourth terms can be rearranged to the electric field: $\mathbf{E} = \mathbf{u} \times \mathbf{B} - \lambda \boldsymbol{\nabla} \times \mathbf{B}$ (see the following section for details), and the fourth term is the

$$\frac{\partial}{\partial t} \left( \frac{B^2}{2\mu_0} \right) + \boldsymbol{\nabla} \cdot \left( \frac{1}{\mu_0} \mathbf{E} \times \mathbf{B} \right) = -\frac{1}{\mu_0} \mathbf{u} \cdot (\mathbf{B} \times (\boldsymbol{\nabla} \times \mathbf{B})) + \frac{\lambda}{\mu_0} (\nabla \times \mathbf{B})^2$$

Finally, adding the three contributions, we obtain the MHD total energy conservation equation:

$$\begin{aligned} &\rho \left( \frac{\partial}{\partial t} + \mathbf{u} \cdot \nabla \right) \left( e + \frac{1}{2} u^2 \right) + \frac{\partial}{\partial t} \left( \frac{B^2}{2\mu_0} \right) + \boldsymbol{\nabla} \cdot \left( \frac{1}{\mu_0} \mathbf{E} \times \mathbf{B} \right) = \\ &= \Phi_\nu + \nabla \cdot (\kappa \nabla T) + \frac{\lambda}{\mu_0} (\nabla \times \mathbf{B})^2 - \nabla \cdot \left( p\mathbf{u} + \frac{\mathbf{B} \times (\mathbf{u} \times \mathbf{B})}{\mu_0} \right) \, . \end{aligned}$$

### A.3.2 Entropy form

We can also recast the energy conservation equation using the entropy as the conserved variable. We begin with the conservation equation for internal energy with magnetic effects included:

$$\rho \left( \frac{\partial}{\partial t} + \mathbf{u} \cdot \nabla \right) e = -p \nabla \cdot \mathbf{u} + \Phi_\nu + \nabla \cdot (\kappa \nabla T) + \frac{\lambda}{\mu_0} (\nabla \times \mathbf{B})^2 \, ,$$

From thermodynamics, the change in entropy per unit mass is related to internal energy and volume by:

$$T \frac{Ds}{Dt} = \frac{De}{Dt} + p \frac{D}{Dt} \left( \frac{1}{\rho} \right) \, ,$$

where $s$ is the specific entropy. The second term can be developed by using the continuity equation:

$$\frac{D}{Dt} \left( \frac{1}{\rho} \right) = -\frac{1}{\rho^2} \frac{D\rho}{Dt} = \frac{1}{\rho} \nabla \cdot \mathbf{u} \quad \Rightarrow \quad \rho T \frac{Ds}{Dt} = \rho \frac{De}{Dt} + p \nabla \cdot \mathbf{u} \, .$$

where a factor of $\rho$ has been added. Finally, the expression for $\rho De/Dt$ can be substituted in the internal energy equation, canceling the pressure-divergence terms to give:

$$\rho T \left( \frac{\partial}{\partial t} + \mathbf{u} \cdot \nabla \right) s = \Phi_\nu + \nabla \cdot (\kappa \nabla T) + \frac{\lambda}{\mu_0} (\nabla \times \mathbf{B})^2 \, .$$

This is the resistive MHD entropy equation found throughout this thesis and used by *MagIC* and *Pencil*.

## A.4 The induction equation

Maxwell's equations completely describe the electric $\mathbf{E}$ and magnetic $\mathbf{B}$ fields by defining their divergence and curl through space. The fundamental limits employed in MHD are that the fluid is to neglect the displacement current in Ampere's law and charge neutrality

$$\begin{cases} \nabla \cdot \mathbf{E} = \dfrac{\rho_{\mathrm{q}}}{\varepsilon_0} \\ \nabla \cdot \mathbf{B} = 0 \\ \nabla \times \mathbf{E} = -\dfrac{\partial \mathbf{B}}{\partial t} \\ \nabla \times \mathbf{B} = \mu_0 \left( \mathbf{J} + \varepsilon_0 \dfrac{\partial \mathbf{E}}{\partial t} \right) \end{cases} \Rightarrow \begin{cases} \nabla \cdot \mathbf{E} = 0 \\ \nabla \cdot \mathbf{B} = 0 \\ \nabla \times \mathbf{E} = -\dfrac{\partial \mathbf{B}}{\partial t} \\ \nabla \times \mathbf{B} = \mu_0 \mathbf{J} \end{cases}$$

The first approximation if $|\partial \mathbf{E}/\partial t| \ll c^2 |\nabla \times \mathbf{B}|$ . This holds if the fluid is conductive enough so that the timescales of the $\mathbf{E}$ variation can be considered instantaneous compared to the other dynamic timescales, and the displacement currents, $\varepsilon_0 \partial \mathbf{E}/\partial t$, become negligible. The other assumption is that the charge density $\rho_q$ is assumed to be zero based on the quasi-neutrality assumption, i.e., the forces associated with any unbalanced charges imply a potential energy per particle much greater than the mean thermal energy per particle.

Rearranging the last equations, we have the electromagnetic form of the MHD equations, which are fulfilled in any reference frame, specifically for both the observer ($_{obs}$) and the fluid ($_{fl}$) frames.

$$\begin{cases} \nabla \cdot \mathbf{B}_{\mathrm{fl}} = 0 \\ \dfrac{\partial \mathbf{B}_{\mathrm{fl}}}{\partial t} = -\nabla \times \mathbf{E}_{\mathrm{fl}} \\ \mathbf{J}_{\mathrm{fl}} = \dfrac{1}{\mu_0} \nabla \times \mathbf{B}_{\mathrm{fl}} \end{cases} ; \quad \begin{cases} \nabla \cdot \mathbf{B}_{\mathrm{obs}} = 0 \\ \dfrac{\partial \mathbf{B}_{\mathrm{obs}}}{\partial t} = -\nabla \times \mathbf{E}_{\mathrm{obs}} \\ \mathbf{J}_{\mathrm{obs}} = \dfrac{1}{\mu_0} \nabla \times \mathbf{B}_{\mathrm{obs}} \end{cases}$$

There is a boost in the relationship between the two sets of systems. We are interested in the observer frame, as it will be the reference for measuring velocities and magnetic fields. But it is in the fluid frame that we know an expression for $\mathbf{E}$: either the fluid is a perfect conductor, i.e., $\mathbf{E} = 0$, or takes some value using Ohm's law in a finitely conducting case. Naively, one would assume that $\mathbf{E}$ and $\mathbf{B}$ do not change under a boost transform in the classical approximation ($v \ll c$), but that is not the case. The general Lorentz transform for the fields can be divided into the boosted parallel and perpendicular components:

$$\begin{cases} \mathbf{E}_{\|}{}' &= \mathbf{E}_{\|} \\ \mathbf{B}_{\|}{}' &= \mathbf{B}_{\|} \\ \mathbf{E}_{\perp}{}' &= \gamma \left( \mathbf{E}_{\perp} + \mathbf{v} \times \mathbf{B} \right) \\ \mathbf{B}_{\perp}{}' &= \gamma \left( \mathbf{B}_{\perp} - \dfrac{1}{c^2} \mathbf{v} \times \mathbf{E} \right) \end{cases}$$

where $\gamma$ is the Lorentz factor, i.e. $1/\sqrt{1 - v^2/c^2}$, and $\mathbf{v}$ is the velocity difference between both reference systems. These expressions can be joined in the following way:

$$\mathbf{E}' = \gamma \left( \mathbf{E} + \mathbf{v} \times \mathbf{B} \right) - (\gamma - 1) \left( \mathbf{E} \cdot \mathbf{n} \right) \mathbf{n}$$
$$\mathbf{B}' = \gamma \left( \mathbf{B} - \frac{\mathbf{v} \times \mathbf{E}}{c^2} \right) - (\gamma - 1) \left( \mathbf{B} \cdot \mathbf{n} \right) \mathbf{n}$$

where $\mathbf{n} = \mathbf{v}/|\mathbf{v}|$. Taking the classical limit $\gamma \to 1$, we obtain the following relation:

$$\mathbf{E}' = \mathbf{E} + \mathbf{v} \times \mathbf{B}\,,$$
$$\mathbf{B}' = \mathbf{B}\,.$$

Thus, $\mathbf{B}$ is independent of reference frame, but the $\mathbf{E}$ changes from frame to frame depending on $\mathbf{B}$ itself. These relations can more easily and more directly be obtained from equating the total Lorentz force, i.e., the sum of magnetic and electric, that a charge sees for two observers, given that one observer is in the charge's frame, but the derivation above is more general. Thus, in our case, for a comoving observer with the fluid and one external observer:

$$\mathbf{E}_{\text{fl}} = \mathbf{E}_{\text{obs}} + \mathbf{v}_{\text{obs}} \times \mathbf{B}_{\text{obs}}$$
$$\mathbf{B}_{\text{fl}} = \mathbf{B}_{\text{obs}}$$

We now assume that the fluid is not perfectly conducting and fulfills the simplest Ohm's law, i.e., $\mathbf{E}_{\text{fl}} = \mathbf{J}_{\text{fl}}/\sigma$, where $\mathbf{J}_{\text{fl}}$ is the current seen in the fluid frame and $\sigma$ is the finite conductivity. Thus, the electric field transformation is:

$$\mathbf{E}_{\text{obs}} = \mathbf{E}_{\text{fl}} - \mathbf{v}_{\text{obs}} \times \mathbf{B}_{\text{obs}} = \frac{1}{\sigma}\mathbf{J}_{\text{fl}} - \mathbf{v}_{\text{obs}} \times \mathbf{B}_{\text{obs}} = \frac{1}{\mu_0 \sigma}\boldsymbol{\nabla} \times \mathbf{B}_{\text{fl}} - \mathbf{v}_{\text{obs}} \times \mathbf{B}_{\text{obs}}$$

We can define the magnetic diffusivity as $\lambda = 1/(\mu_0\sigma)$, use the fact that $\mathbf{B}$ is frame independent, and plug this expression into the magnetic field evolution for the observable we get:

$$\frac{\partial \mathbf{B}_{\text{obs}}}{\partial t} = \boldsymbol{\nabla} \times (\mathbf{u}_{\text{obs}} \times \mathbf{B}_{\text{obs}}) - \boldsymbol{\nabla} \times (\lambda \boldsymbol{\nabla} \times \mathbf{B}_{\text{obs}})$$

where the fluid velocity as seen by the observer has been changed to $u$, as it is much more common in literature, other than here, the frame sub-indices are dropped, as we are only interested in the observer's frame. If the fluid is perfectly conductive, i.e., the magnetic diffusivity $\lambda$ is 0, the second term vanishes.

### A.4.1 Extended Ohm's law

There is a more complete Ohm's law for partially ionized plasmas and conducting fluids (e.g., Pandey and Wardle, 2008; Koskinen et al., 2014):

$$\mathbf{E} = \frac{\mathbf{J}}{\sigma} + \frac{\mathbf{J} \times \mathbf{B}}{ne} - \frac{(\mathbf{J} \times \mathbf{B}) \times \mathbf{B}}{\gamma_d \rho_i \rho_n}\,, \tag{A.1}$$

where $n$ is the electron number density, $e$ is the elementary charge, $\rho_i$ and $\rho_n$ are ion and neutral mass densities, and $\gamma_d$ is the ion-neutral collision drag coefficient. The first new term (i.e., $\propto \mathbf{J} \times \mathbf{B}$) is known as the Hall term and arises from the different reactions of electrons and ions in response to the same electric and magnetic fields due to their different masses. In other words, electrons drift faster from ions, creating a separation that causes an additional electric field in the rest frame of the fluid known as the Hall electric field. The ambipolar term is the second new term (i.e., $\propto (\mathbf{J} \times \mathbf{B}) \times \mathbf{B}$). It arises in partially ionized plasmas, where only ions feel the Lorentz force. As the neutral particles do not couple to the magnetic field, they can only be dragged along magnetic field lines via ion-neutral collisions. The ambipolar diffusion is the fact that magnetic fields can "slip through" the neutral component of the gas. Note that both terms rely on a low collisional rate (i.e., low densities): when interactions between particles are infrequent, both terms have greater effects.

Substituting these new terms into Faraday's law, we obtain:

$$\begin{aligned}\frac{\partial \mathbf{B}}{\partial t} &= -\boldsymbol{\nabla} \times (\mathbf{E} - \mathbf{u} \times \mathbf{B}) \\ &= \boldsymbol{\nabla} \times \left( \mathbf{u} \times \mathbf{B} - \lambda \mathbf{J} - \frac{\mathbf{J} \times \mathbf{B}}{ne} + \frac{(\mathbf{J} \times \mathbf{B}) \times \mathbf{B}}{\gamma_d \rho_i \rho_n} \right) .\end{aligned}$$

The first two terms are the ideal MHD advection and the Ohmic diffusion, as seen previously. Due to its mathematical behavior, the Hall term is non-dissipative but redistributes the magnetic field lines, i.e., the topology, and can lead to whistler waves and magnetic reconnection events. Instead, the ambipolar effect is dissipative by nature (but quadratically in $B$ instead of linearly) and converts magnetic energy into heat via friction between ions and neutrals. Note that they become negligible for high plasma or conducting densities (i.e., high $n$, $\rho_i$, $\rho_n$). In the context of the dense planetary interiors explored in Chapters 5 and 7, these two terms are negligible. However, they can play a role in the ISM, so that including it could represent a future extension of the studies illustrated in Chapter 3.

## A.5 Frozen-in flux theorem

The *frozen-in flux theorem*, also known as *Alfvén's theorem*, states the following:

> *In the limit of a perfectly conducting fluid (ideal MHD), the magnetic flux through any material surface moving with the fluid remains constant in time.*

The first assumption is that the induction equation takes its ideal form. No magnetic diffusion is taken into account: $\partial \mathbf{B}/\partial t = \nabla \times (\mathbf{u} \times \mathbf{B})$ (see Sec 2.2.2). We also need to define the magnetic flux across a surface $S(t)$ that moves along the fluid flow:

$$\Phi_B(t) = \int_{S(t)} \mathbf{B} \cdot d\mathbf{S}.$$

Then the *frozen-in flux theorem* is mathematically expressed as:

$$\frac{D\Phi_B}{Dt} = \frac{\partial \Phi_B}{\partial t} + (\mathbf{u} \cdot \nabla)\Phi_B = 0,$$

where $D/Dt$ is the advective (material) derivative already introduced above. It has a rather short proof, as follows.

Consider an arbitrary, orientable, and open advected surface (i.e., it moves with the fluid) in a magnetized moving fluid. We can define two different infinitesimally close times, $t$ and $t + \delta t$, and their corresponding advected surfaces $S_1$ and $S_2$. The rate of change of magnetic flux through the surface can be obtained by taking the limit $\delta t \to 0$ of the difference between these fluxes:

$$\frac{D\Phi_B}{Dt} = \lim_{\delta t \to 0} \frac{1}{\delta t} \left( \int_{S_2} \mathbf{B}(t + \delta t) \cdot d\mathbf{S}_2 - \int_{S_1} \mathbf{B}(t) \cdot d\mathbf{S}_1 \right) .$$

Let us consider a closed surface defined by $S_1$, $S_2$, and $S_3$, where $S_3$ is the surface connecting the boundaries of $S_1$ and $S_2$. By Gauss' law, $\nabla \cdot \mathbf{B} = 0$, the total magnetic flux through any closed surface vanishes, so that:

$$0 = -\int_{S_1} \mathbf{B}(t + \delta t) \cdot d\mathbf{S}_1 + \int_{S_2} \mathbf{B}(t + \delta t) \cdot d\mathbf{S}_2 + \int_{S_3} \mathbf{B}(t + \delta t) \cdot d\mathbf{S}_3 \, .$$

Thus, for the previous integral to vanish, $d\mathbf{S}_1$ needs to have a change of sign. The mathematical trick for integrating over surface $S_3$ is to construct its differential area element by taking the vector product between an element of length of the boundary $\partial S_1$, $d\mathbf{l}$, and the distance traveled by this same element in the time interval $\delta t$:

$$d\mathbf{S}_3 = d\mathbf{l} \times \mathbf{u}\,\delta t \ .$$

Note that the surface integral is then converted to a line integral around $S_1$. We can solve for the flux through surface $S_2$:

$$\int_{S_2} \mathbf{B}(t+\delta t)\cdot d\mathbf{S}_2 = \int_{S_1} \mathbf{B}(t+\delta t)\cdot d\mathbf{S}_1 - \int_{\partial S_1} (\mathbf{u}\times\mathbf{B}(t+\delta t))\cdot d\mathbf{l}\,\delta t \ ,$$

and substitute it into $D\Phi_B/Dt$:

$$\begin{aligned} \frac{D\Phi_B}{Dt} &= \lim_{\delta t\to 0}\frac{1}{\delta t}\left(\int_{S_1} [\mathbf{B}(t+\delta t)-\mathbf{B}(t)]\cdot d\mathbf{S}_1 - \int_{\partial S_1} (\mathbf{u}\times\mathbf{B}(t+\delta t))\cdot d\mathbf{l}\,\delta t\right) = \\ &= \int_{S_1} \frac{\partial \mathbf{B}}{\partial t}\cdot d\mathbf{S}_1 - \int_{\partial S_1} (\mathbf{u}\times\mathbf{B})\cdot d\mathbf{l} \ . \end{aligned}$$

For the final step, we can use Stokes' theorem to convert the line integral back to a surface integral:

$$\int_{\partial S_1} (\mathbf{u}\times\mathbf{B})\cdot d\mathbf{l} = \int_{S_1} \nabla\times(\mathbf{u}\times\mathbf{B})\cdot d\mathbf{S}_1 \ .$$

Now the total flux derivative is expressed only in terms of the integral over $S_1$:

$$\frac{D\Phi_B}{Dt} = \int_{S_1} \left(\frac{\partial \mathbf{B}}{\partial t} - \nabla\times(\mathbf{u}\times\mathbf{B})\right)\cdot d\mathbf{S}_1 \ ,$$

in which the integrand, in ideal MHD, exactly vanishes:

$$\frac{D\Phi_B}{Dt} = 0 \, .$$

Thus, if the magnetic flux through any material surface moving with the fluid is conserved, then the magnetic field lines must be frozen into the fluid moving along with it.

This theorem has profound effects, since the fluid movements can advect, stretch, and twist the magnetic field lines, implying the possibility of changing the direction and intensity of the local magnetic field (despite maintaining the same mathematical topology). Simple examples are stellar collapses, where the magnetic field lines are advected inward, causing a higher average intensity by flux conservation, or different types of shear (e.g., differential rotation), which can amplify the magnetic field lines in the direction of the flow (see Sect. 2.3.3). In this sense, any kind of dynamo mechanism relies on the advective effects to transfer kinetic energy to magnetic energy. On the contrary, when one goes beyond ideal MHD and considers resistivity, magnetic field dissipation, and reconnections can happen, which generally decrease the field intensity and imply topological simplifications.

Another consequence of the ideal MHD limit is the conservation of magnetic helicity, $H_\mathrm{M}$ :

$$H_\mathrm{M} \equiv \int_V \mathbf{A}\cdot\mathbf{B}\,dV \quad \Rightarrow \quad \frac{\partial H_\mathrm{M}}{\partial t} = 0 \, , \tag{A.2}$$

where $\mathbf{A}$ is the vector potential (i.e., $\mathbf{B} = \boldsymbol{\nabla}\times\mathbf{A}$). This statement is proven by using $\partial\mathbf{B}/\partial t = \nabla\times(\mathbf{u}\times\mathbf{B})$, and the gauge condition is chosen so that $\boldsymbol{\nabla}\Phi = 0$.

## A.6 The vorticity equation

In Chapter 3, I study the vorticity evolution equation, and specifically its sources or dissipative terms. In this section, I derive the diagnostic quantities related to each source or sink term for the vorticity. I start with the vorticity evolution equation:

$$\begin{aligned}\frac{\partial \boldsymbol{\omega}}{\partial t} &= \boldsymbol{\nabla}\times(\mathbf{u}\times\boldsymbol{\omega}) + \boldsymbol{\nabla}\times\mathbf{F}_\nu + \frac{\boldsymbol{\nabla}\rho\times\boldsymbol{\nabla}p}{\rho^2} + \\ &+\boldsymbol{\nabla}\times\frac{\mathbf{J}\times\mathbf{B}}{\rho} - 2\boldsymbol{\nabla}\times(\boldsymbol{\Omega}\times\mathbf{u}) + \boldsymbol{\nabla}\times\mathbf{f}_{shear} + \boldsymbol{\nabla}\times\mathbf{f}\ ,\end{aligned} \tag{A.3}$$

Here the first term on the right-hand side is similar to the amplification term in the induction equation, the second term is about the viscous forces acting on the system, the third is the baroclinic term, related to the equation of state, the forth is the effect of the Lorentz force, the fifth appears if the system is rotating, the sixth if some large-scale shear acts on the system, and the seventh is due to the effect of a generic forcing. The fundamental thermodynamics quantities used in *Pencil Code* are not ($\rho$, p, e) but ($\rho$, T, s), thus the baroclinic term must be changed into $-\boldsymbol{\nabla}T\times\boldsymbol{\nabla}s$ (derived in A.6.1).

If we use the definition of viscous force for a constant kinematic viscosity, then the curl of it is:

$$\mathbf{F}_\nu = \frac{1}{\rho}\boldsymbol{\nabla}\cdot(2\rho\nu\boldsymbol{S}) = 2\nu\boldsymbol{S}\cdot\boldsymbol{\nabla}ln\rho + \nu\nabla^2\mathbf{u} + \frac{1}{3}\nu\boldsymbol{\nabla}(\boldsymbol{\nabla}\cdot\mathbf{u})$$

$$\boldsymbol{\nabla}\times\mathbf{F}_\nu = 2\nu\boldsymbol{\nabla}\times(\boldsymbol{S}\cdot\boldsymbol{\nabla}ln\rho) + \nu\nabla^2\boldsymbol{\nabla}\times\mathbf{u} + \frac{1}{3}\nu\cancelto{0}{\boldsymbol{\nabla}\times\boldsymbol{\nabla}(\boldsymbol{\nabla}\cdot\mathbf{u})}$$

These terms are rather complex to analyze in time and space, and the easiest way to determine their relative importance is to use some scalar quantity averaged over all simulated space. An option is to take the dot product with $\boldsymbol{\omega}$ and integrate over the periodic rigid volume domain. Using the divergence theorem and the following vector identities: $(\boldsymbol{\nabla}\times\mathbf{a})\cdot\mathbf{b} = \boldsymbol{\nabla}\cdot(\mathbf{a}\times\mathbf{b}) + \mathbf{a}\cdot(\boldsymbol{\nabla}\times\mathbf{b})$; $\nabla^2\mathbf{a} = \boldsymbol{\nabla}(\boldsymbol{\nabla}\cdot\mathbf{a}) - \boldsymbol{\nabla}\times(\boldsymbol{\nabla}\times\mathbf{a})$ we reach:

$$\begin{aligned}\int\frac{\partial\boldsymbol{\omega}}{\partial t}\cdot\boldsymbol{\omega}dV = \int\Bigg(&\boldsymbol{\nabla}\times(\mathbf{u}\times\boldsymbol{\omega}) + \nu\nabla^2\boldsymbol{\omega} + 2\nu\boldsymbol{\nabla}\times(\mathbf{S}\boldsymbol{\nabla}ln\rho) + \boldsymbol{\nabla}T\times\boldsymbol{\nabla}s + \\ &+\boldsymbol{\nabla}\times\frac{\mathbf{J}\times\mathbf{B}}{\rho} - 2\boldsymbol{\nabla}\times(\boldsymbol{\Omega}\times\mathbf{u}) + \boldsymbol{\nabla}\times\frac{1}{\tau}(u_y^S - u_y)\hat{\mathbf{y}} + \boldsymbol{\nabla}\times\mathbf{f}\Bigg)\cdot\boldsymbol{\omega}dV\end{aligned}$$

$$\begin{aligned}\frac{1}{2}\frac{\partial}{\partial t}\int\boldsymbol{\omega}^2dV &= \int(\mathbf{u}\times\boldsymbol{\omega})\cdot\boldsymbol{\nabla}\times\boldsymbol{\omega}dV + \cancelto{0}{\int\boldsymbol{\nabla}\cdot((\mathbf{u}\times\boldsymbol{\omega})\times\boldsymbol{\omega})dV} + \\ &+\nu\cancelto{0}{\int\boldsymbol{\nabla}(\boldsymbol{\nabla}\cdot\boldsymbol{\omega})\cdot\boldsymbol{\omega}dV} - \nu\int\boldsymbol{\nabla}\times\boldsymbol{\omega}\cdot\boldsymbol{\nabla}\times\boldsymbol{\omega}dV - \nu\cancelto{0}{\int\boldsymbol{\nabla}\cdot((\boldsymbol{\nabla}\times\boldsymbol{\omega})\times\boldsymbol{\omega})dV} \\ &+2\nu\int(\mathbf{S}\boldsymbol{\nabla}ln\rho)\cdot\boldsymbol{\nabla}\times\boldsymbol{\omega}dV + 2\nu\cancelto{0}{\int\boldsymbol{\nabla}\cdot((\mathbf{S}\boldsymbol{\nabla}ln\rho)\times\boldsymbol{\omega})dV} - \\ &+\int(\boldsymbol{\nabla}T\times\boldsymbol{\nabla}s)\cdot\boldsymbol{\omega}dV + \int\left(\frac{\mathbf{J}\times\mathbf{B}}{\rho}\right)\cdot\boldsymbol{\nabla}\times\boldsymbol{\omega}dV + \cancelto{0}{\int\boldsymbol{\nabla}\cdot\left(\frac{\mathbf{J}\times\mathbf{B}}{\rho}\times\boldsymbol{\omega}\right)dV} - \\ &-2\Omega\int(\boldsymbol{e_z}\times\mathbf{u})\cdot\boldsymbol{\nabla}\times\boldsymbol{\omega}dV - 2\Omega\cancelto{0}{\int\boldsymbol{\nabla}\cdot((\boldsymbol{e_z}\times\mathbf{u})\times\boldsymbol{\omega})dV} + \\ &+\frac{1}{\tau}\int(u_y^S - u_y)\hat{\mathbf{y}}\cdot\boldsymbol{\nabla}\times\boldsymbol{\omega}dV + \frac{1}{\tau}\cancelto{0}{\int\boldsymbol{\nabla}\cdot((u_y^S - u_y)\hat{\mathbf{y}}\times\boldsymbol{\omega})dV} + \\ &+\int\mathbf{f}\cdot\boldsymbol{\nabla}\times\boldsymbol{\omega}dV + \cancelto{0}{\int\boldsymbol{\nabla}\cdot(\mathbf{f}\times\boldsymbol{\omega})dV}\end{aligned}$$

Where we have assumed that the quantities are well behaved inside the volume, thus their surface integrals vanish because of the periodic rigid volume. With the usual definition of enstrophy of $\boldsymbol{\omega}\cdot\boldsymbol{\omega}/2$, we can then finally reach an expression for the mean enstrophy evolution:

$$\begin{aligned}\frac{1}{2}\frac{\partial}{\partial t}\langle\boldsymbol{\omega}^2\rangle = \langle(\mathbf{u}\times\boldsymbol{\omega})\cdot\mathbf{q}\rangle - \nu\langle\mathbf{q}^2\rangle + 2\nu\langle\mathbf{S}\boldsymbol{\nabla}ln\rho\cdot\mathbf{q}\rangle + \langle\boldsymbol{\nabla}T\times\boldsymbol{\nabla}s\cdot\boldsymbol{\omega}\rangle + \\ +\langle\frac{\mathbf{J}\times\mathbf{B}}{\rho}\cdot\mathbf{q}\rangle - 2\Omega\langle(\boldsymbol{e_z}\times\mathbf{u})\cdot\mathbf{q}\rangle + \frac{1}{\tau}\langle(u_y^S - u_y)\hat{\mathbf{y}}\cdot\mathbf{q}\rangle + \langle\mathbf{f}\cdot\mathbf{q}\rangle\end{aligned}, \tag{A.4}$$

where we have defined $\mathbf{q} = \boldsymbol{\nabla}\times\boldsymbol{\omega}$. The *Pencil Code* tools did not have all these diagnostic terms implemented, so we had to add the baroclinic ($4^{th}$ term in the right hand side), the Lorentz force ($5^{th}$), rotation ($6^{th}$), and shear terms ($7^{th}$) to the analysis.

### A.6.1 Baroclinic term in the vorticity equation

If we apply the first law of thermodynamics to a general fluid element and use the intensive thermodynamic variables $e = E/m = E/V\rho$, $s = S/m$:

$$dE = dQ - pdV = TdS - pdV \quad \Rightarrow \quad de = Tds - pd(\frac{1}{\rho}) \ .$$

Using the definition of enthalpy, we get the relation:

$$H = E + pV \qquad \Rightarrow \quad h = e + \frac{p}{\rho} \quad \Rightarrow \qquad \frac{p}{\rho} = h - e \ .$$

The differential of any scalar quantity ($A$) can be expressed as:

$$dA = \frac{\partial A}{\partial x}dx + \frac{\partial A}{\partial y}dy + \frac{\partial A}{\partial z}dz = \boldsymbol{\nabla}A\cdot\boldsymbol{dx} \ .$$

Applying this to the energy equation:

$$\boldsymbol{\nabla}e\cdot\boldsymbol{dx} = T\boldsymbol{\nabla}s\cdot\boldsymbol{dx} - pd\boldsymbol{\nabla}\left(\frac{1}{\rho}\right)\cdot\boldsymbol{dx} \quad \Rightarrow \quad \boldsymbol{\nabla}e = T\boldsymbol{\nabla}s - p\boldsymbol{\nabla}(\frac{1}{\rho})$$

Taking the gradient of $h - e$ and using the expression just derived for $\boldsymbol{\nabla}e$:

$$\begin{aligned}\boldsymbol{\nabla}(h-e) = \boldsymbol{\nabla}h - \boldsymbol{\nabla}e = \boldsymbol{\nabla}h - T\boldsymbol{\nabla}s + p\boldsymbol{\nabla}(\frac{1}{\rho}) \\ \boldsymbol{\nabla}(\frac{p}{\rho}) = p\boldsymbol{\nabla}(\frac{1}{\rho}) + \frac{1}{\rho}\boldsymbol{\nabla}p \\ \Rightarrow \frac{1}{\rho}\boldsymbol{\nabla}p = \boldsymbol{\nabla}h - T\boldsymbol{\nabla}s\end{aligned}$$

This term appears on the RHS of the momentum MHD equation, which, when taken the curl, leads to the baroclinic term of the vorticity equation.

$$\frac{\partial\boldsymbol{\omega}}{\partial t} = -\boldsymbol{\nabla}\times\left(\frac{1}{\rho}\boldsymbol{\nabla}p\right) + ... = \frac{1}{\rho^2}\boldsymbol{\nabla}\rho\times\boldsymbol{\nabla}p - \cancelto{0}{\frac{1}{\rho}\boldsymbol{\nabla}\times\boldsymbol{\nabla}p} + ...$$

$$\begin{aligned}\frac{\partial\boldsymbol{\omega}}{\partial t} = -\boldsymbol{\nabla}\times\Big(\boldsymbol{\nabla}h - T\boldsymbol{\nabla}s\Big) + ... = \\ -\cancelto{0}{\boldsymbol{\nabla}\times\boldsymbol{\nabla}h} + \boldsymbol{\nabla}T\times\boldsymbol{\nabla}s + \cancelto{0}{T\boldsymbol{\nabla}\times\boldsymbol{\nabla}s} + ...\end{aligned}$$

Therefore, the baroclínic term can be calculated either from the $\boldsymbol{\nabla}\rho\times\boldsymbol{\nabla}p$ term or $\boldsymbol{\nabla}T\times\boldsymbol{\nabla}s$. In Chapter 3, it is more natural to use the second option, as the *Pencil Code* uses entropy as one of its fundamental thermodynamic quantities.

## A.7 Reynolds rules for scale decomposition

Both theoretical and numerical research rely on Reynolds decomposition, a mathematical technique that separates the expectation value of a quantity from its fluctuations. This assumption is widely accepted in both HD and MHD. For example, given a steady state fluid flow on top of which turbulence occurs, the velocity field $\boldsymbol{u}(x, y, z, t)$ is naturally decomposed by:

$$\boldsymbol{u}(x, y, z, t) = \overline{\boldsymbol{u}(x, y, z, t)} + \boldsymbol{u}'(x, y, z)$$

where $\overline{\boldsymbol{u}}$ is the expectation value and $\mathbf{u}'$ are the fluctuation. The fluctuations can be defined as the quantity subtracted from $\boldsymbol{u}$ so that $\partial(\boldsymbol{u} - \boldsymbol{u}')/\partial t = 0$. The procedure to *bar* a variable can be understood as an averaging procedure that eliminates the fluctuations. Given two decomposed time and space dependent fields, $F = \overline{F} + f$ and $G = \overline{G} + g$, the set of identities that must be accomplished:

$$\overline{\overline{F}} = \overline{F} \quad ; \quad \overline{f} = 0 \quad ; \quad \overline{F + G} = \overline{F} + \overline{G} \quad ; \quad \overline{\overline{F}\,\overline{G}} = \overline{F}\,\overline{G} \quad ; \quad \overline{\overline{G}f} = \overline{G} \tag{A.5}$$

The small- and large-scale dynamo theories use this decomposition of the induction equation to derive possible growth for the magnetic field. The basic assumption is that both the magnetic and velocity fields can be separated into their large and small-scale components. This procedure is used in Chapter 3 to separate the turbulent velocity and vorticity set by the forcing from the large-scale pattern set by the imposed sinusoidal shearing profile.

## A.8 Helmholtz decomposition for vectorial fields

Helmholtz showed that any vector field $\mathbf{F}$ which vanishes suitably quickly at infinity can be decomposed into irrotational (longitudinal, purely divergent) and solenoidal (transverse) components. Therefore, it can be expressed as the sum of the gradient of a scalar potential $\Phi$ and the curl of a vector potential $\mathbf{A}$:

$$\mathbf{F} = \mathbf{F}_{pot} + \mathbf{F}_{rot} = -\nabla\phi + \nabla \times \mathbf{A} \ . \tag{A.6}$$

Aided by the *Pencil code* available tools, we have used this decomposition in Chapter 3 to evaluate how much of the flow is irrotational, with no associated vorticity, and how much is rotational. To numerically obtain such decomposition we made use of the 3D Fourier transform to obtain the flow components in Fourier space: $\tilde{\mathbf{u}}_{pot}(\mathbf{k}) = \tilde{\mathbf{u}}(\mathbf{k}) \cdot \mathbf{k}$, $\tilde{\mathbf{u}}_{rot}(\mathbf{k}) = \tilde{\mathbf{u}}(\mathbf{k}) \times \mathbf{k}$. If we inverse transform these functions, we can obtain the decomposition. We can check both the percentage of the rotational flow and the error of the numerical procedure by comparing the squared volume quantities with the relation:

$$\mathbf{u}(\mathbf{r}) = \mathbf{u}_{pot}(\mathbf{r}) + \mathbf{u}_{rot}(\mathbf{r}) \quad \rightarrow \quad \langle|\mathbf{u}|^2\rangle = \langle|\mathbf{u}_{pot}|^2\rangle + \langle|\mathbf{u}_{rot}|^2\rangle. \tag{A.7}$$

## A.9 Energy spectra

Originally introduced by Isaac Newton in the $17^{\text{th}}$ century within the field of optics, the term *spectrum* is now widely used across physics to refer to the decomposition or analysis of a signal as a function of a variable, which may be continuous or discrete. It often also refers to the graphical representation of that signal concerning the variable in question. A typical example is the light spectrum, typically represented as a plot of energy versus wavelength or frequency, more precisely known as the spectral density. Other wave phenomena, such as sound or surface water waves, are similarly characterized by their respective spectra (e.g., noise spectra, sea wave spectra). The concept has since been generalized to more abstract quantities, such as particle counts or intensities in spectroscopy (as a function of particle energy), or kinetic energy density in hydrodynamics (as a function of wavelength). We are interested in the latter, which describes how kinetic energy is spatially distributed across flow structures of different scales.

Spectra of a continuous variable $\mathbf{u}$ over a three-dimensional domain can be obtained by computing its Fourier transform, provided an analytical expression is available. In practice, however, numerical simulations based on finite-difference or finite-volume methods yield only discrete values on a computational grid. Consequently, a numerical approach must be employed to compute the spectrum from the discretized data. The *Pencil Code*, for example, automatically obtains the 3D spectra by computing:

$$\begin{aligned}\tilde{\mathbf{u}}(\mathbf{k}) = \frac{1}{(2\pi)^3}\int \mathbf{u}(\mathbf{r})e^{-i\mathbf{k}\cdot\mathbf{r}}dr^3 \approx \\ \approx \frac{1}{N_x N_y N_z}\sum_{p,q,r} u(x_p, y_q, z_r)e^{-ik_x x_p}e^{-ik_y y_q}e^{-ik_z z_r} \ ,\end{aligned} \tag{A.8}$$

where $N_x$, $N_y$, and $N_z$ are the resolutions of the box, and we have used that $L_x = L_y = L_z = 2\pi$. Notice that the wave numbers can only reasonably reach up to the $\pi N/L$, that is $|k_x| < \pi N_x/L_x$, $|k_y| < \pi N_y/L_y$, and $|k_x| < \pi N_z/L_z$. Then the three-dimensional spectrum is defined as:

$$P(k) = \frac{1}{2}\tilde{\mathbf{u}}(\mathbf{k})\tilde{\mathbf{u}}^*(\mathbf{k}) \qquad \text{where} \qquad k = \sqrt{k_x^2 + k_y^2 + k_z^2} \ , \tag{A.9}$$

where the three-dimensional dependence of the spectrum on $\mathbf{k}$ is reduced to a function of the modulus $k$ alone, by rebinning in the Fourier space. Python libraries with multidimensional Fourier transforms make these calculations readily available.

In Chapter 3, this has been used for the velocity (kinetic energy spectra, neglecting the small density deviations from unity), magnetic field (magnetic energy spectra), and vorticity (enstrophy spectra). All the spectra shown in Chapter 3 have units of the largest $k$ fitting the whole box. If we use cubic periodic boxes, the expression simplifies to:

$$\begin{aligned}\tilde{\mathbf{u}}(\mathbf{k}) = \frac{1}{(2\pi)^3}\int_0^L\int_0^L\int_0^L \mathbf{u}(\mathbf{r})e^{-i\mathbf{k}\cdot\mathbf{r}}dx\ dy\ dz \approx \\ \approx \frac{1}{N^3}\sum_{p=0}^{N-1}\sum_{q=0}^{N-1}\sum_{r=0}^{N-1} u(x_p, y_q, z_r)e^{-ik_x x_p}e^{-ik_y y_q}e^{-ik_z z_r} \ .\end{aligned} \tag{A.10}$$

Spectra for finite-difference or finite-volume methods need to compute the discretized integrals every time step that a spectrum is required.

One of the advantages of spectral methods is that spectra are readily available. The orthonormal set of functions always has an associated length scale related to the indices describing such a set. All relevant dynamical variables are expressed as the sum of weights associated with each function; thus, one can directly recover many quantities as a function of the indices. For a cubic or rectangular domain, the usual strategy is to use sinusoidal functions in all directions as orthonormal bases. Then the direct sum of weights for each mode $k$ will automatically lead to the same Fourier spectra as in Eq. A.8. For spherical shell MHD codes, such as *MagIC*, it is more natural to use spherical harmonics in the $\theta, \phi$ direction.

# B

# Vorticity and dynamo from expansion Gaussian waves: inputs and diagnostics

The following tables contain all the simulations included in Elias-López et al. (2023) and Elias-López et al. (2024). Note that some runs from the first article do not have all diagnostics. For all models, we used $B_0 = 10^{-6}$ as the seed field, $\Delta t = 0.02$ as the interval between two different explosions, $c_{s0}^2$=1 in the case of non-isothermal runs, and an amplitude of the shearing profile $A = 0.2$ in the shearing cases. We indicate the initial amplitude of the magnetic field $B_0$ along with ('G') for random values for the potential vector components or ('U$_i$') for uniform distribution in a given direction $i = \{x, y, z\}$. The included diagnostic magnitudes are: turnover time, vorticity proxy $k_\omega/k_f$, Re (and Rm if Pm $\neq$ 1), and Re$_\omega$. The last column corresponds to the rotational flow contribution obtained using the Helmholtz decomposition, only for some specific runs. The blue highlighted values are used to perform the linear fit $k_\omega/k_f(\Omega)$ discussed in the text. The four simulations marked with † became numerically unstable before reaching a fully steady saturated state: we still indicate their diagnostics, even if it is not completely comparable to the others. Simulation marked with † became unstable before reaching a fully steady saturated state. Nevertheless, we still calculate their diagnostics, although they cannot be robustly compared to the other runs.

Table B.1: HD simulations with different rigid rotation rates, and no shear.

| $256^3$ | $\nu$ | $\chi$ | $\Omega$ | $\phi_0$ | $\Delta$t | $R_f$ |
|---|---|---|---|---|---|---|
| H_0s | $2{\cdot}10^{-4}$ | - | 0 | 1 | 1 | 0.2 |
| H_2s | $2{\cdot}10^{-4}$ | - | 2 | 1 | 1 | 0.2 |
| H_0 | $2{\cdot}10^{-4}$ | - | 0 | 1 | 0.02 | 0.2 |
| H_2 | $2{\cdot}10^{-4}$ | - | 2 | 1 | 0.02 | 0.2 |
| H_0c | $2{\cdot}10^{-4}$ | - | 0 | 1 | $\delta t$ | 0.2 |
| H_2c | $2{\cdot}10^{-4}$ | - | 2 | 1 | $\delta t$ | 0.2 |
| H_0cW1 | $2{\cdot}10^{-3}$ | - | 0 | 1 | $\delta t$ | 1 |
| H_2cW1 | $2{\cdot}10^{-3}$ | - | 2 | 1 | $\delta t$ | 1 |
| HB_0 | $2{\cdot}10^{-4}$ | $2{\cdot}10^{-4}$ | 0 | 1 | 0.02 | 0.2 |
| HB_2 | $2{\cdot}10^{-4}$ | $2{\cdot}10^{-4}$ | 2 | 1 | 0.02 | 0.2 |

Table B.2: Barotropic (i.e., isothermal) runs without shear.

| $256^3$ | $\nu$ | $\eta$ | $\Omega$ | $\phi_0$ | $R_f$ | $t_{tot}$ | $t_{turn}$ | $k_\omega/k_f$ | Re (Rm) | $Re_\omega$ | $u_{rms}$ | $u_{rot}/u_{tot}$ | r ($t_{turn}^{-1}$) | $r_\omega$ ($t_{turn}^{-1}$) |
|---|---|---|---|---|---|---|---|---|---|---|---|---|---|---|
| M_0W0.10 | $2\cdot10^{-4}$ | $2\cdot10^{-4}$ | 0 | 2.6 | 0.10 | 282.77 | 0.77 | 0.04158 | 16.15 | 0.671 | $(6.458 \pm 0.042)\cdot10^{-2}$ | 0.019 | - | - |
| M_0W0.20 | $2\cdot10^{-4}$ | $2\cdot10^{-4}$ | 0 | 1 | 0.20 | 1266.21 | 1.42 | 0.01823 | 35.33 | 0.644 | $(7.067 \pm 0.047)\cdot10^{-2}$ | 0.014 | - | - |
| M_0W0.30 | $2\cdot10^{-4}$ | $2\cdot10^{-4}$ | 0 | 0.52 | 0.30 | 1304.95 | 2.16 | 0.01749 | 52.12 | 0.911 | $(6.949 \pm 0.072)\cdot10^{-2}$ | 0.011 | - | - |
| M_0W0.40 | $2\cdot10^{-4}$ | $2\cdot10^{-4}$ | 0 | 0.36 | 0.40 | 1293.14 | 2.78 | 0.02089 | 72.06 | 1.505 | $(7.21 \pm 0.12)\cdot10^{-2}$ | 0.0085 | - | - |
| M_0W0.50 | $2\cdot10^{-4}$ | $2\cdot10^{-4}$ | 0 | 0.24 | 0.50 | 1313.37 | 3.56 | 0.01908 | 86.91 | 1.658 | $(6.95 \pm 0.14)\cdot10^{-2}$ | 0.0072 | - | - |
| M_0W0.60 | $2\cdot10^{-4}$ | $2\cdot10^{-4}$ | 0 | 0.175 | 0.60 | 1233.70 | 4.39 | 0.01845 | 102.5 | 1.890 | $(6.83 \pm 0.20)\cdot10^{-2}$ | 0.0051 | - | - |
| M_0W0.80 | $2\cdot10^{-4}$ | $2\cdot10^{-4}$ | 0 | 0.118 | 0.80 | 1329.22 | 5.74 | 0.02050 | 139.9 | 2.861 | $(6.98 \pm 0.29)\cdot10^{-2}$ | 0.0044 | - | - |
| M_0W1.00 | $2\cdot10^{-4}$ | $2\cdot10^{-4}$ | 0 | 0.087 | 1.00 | 1339.16 | 7.03 | 0.02186 | 178.5 | 3.901 | $(7.14 \pm 0.43)\cdot10^{-2}$ | 0.0041 | - | - |
| M_0W1.50 | $2\cdot10^{-4}$ | $2\cdot10^{-4}$ | 0 | 0.048 | 1.50 | 1433.08 | 11.2 | 0.01528 | 253.7 | 3.890 | $(6.77 \pm 0.73)\cdot10^{-2}$ | 0.0031 | - | - |
| M_0W2.00 | $2\cdot10^{-4}$ | $2\cdot10^{-4}$ | 0 | 0.031 | 2.00 | 1489.96 | 16.7 | 0.01550 | 304.5 | 4.713 | $(6.09 \pm 0.81)\cdot10^{-2}$ | 0.0022 | - | - |
| M_2W0.10 | $2\cdot10^{-4}$ | $2\cdot10^{-4}$ | 2 | 2.0 | 0.10 | 1028.6 | 0.84 | 0.1666 | 14.82 | 2.469 | $(5.927 \pm 0.021)\cdot10^{-2}$ | 0.11 | - | - |
| M_2W0.10 | $2\cdot10^{-4}$ | $2\cdot10^{-4}$ | 2 | 1.5 | 0.10 | 1114.23 | 1.03 | 0.1640 | 12.13 | 1.990 | $(4.854 \pm 0.020)\cdot10^{-2}$ | 0.11 | - | - |
| M_2W0.20 | $2\cdot10^{-4}$ | $2\cdot10^{-4}$ | 2 | 1 | 0.20 | 1276.66 | 0.85 | 0.3822 | 59.07 | 22.57 | $(11.81 \pm 0.12)\cdot10^{-2}$ | 0.29 | - | - |
| M_2W0.30 | $2\cdot10^{-4}$ | $2\cdot10^{-4}$ | 2 | 0.52 | 0.30 | 1237.98 | 1.08 | 0.6644 | 104.5 | 69.41 | $(13.94 \pm 0.29)\cdot10^{-2}$ | 0.42 | - | - |
| M_2W0.40 | $2\cdot10^{-4}$ | $2\cdot10^{-4}$ | 2 | 0.36 | 0.40 | 1225.51 | 1.28 | 0.8895 | 157.0 | 139.6 | $(15.70 \pm 0.50)\cdot10^{-2}$ | 0.50 | - | - |
| M_2W0.50 | $2\cdot10^{-4}$ | $2\cdot10^{-4}$ | 2 | 0.24 | 0.50 | 1250.13 | 1.65 | 1.030 | 190.1 | 195.7 | $(15.21 \pm 0.45)\cdot10^{-2}$ | 0.54 | - | - |
| M_2W0.60 | $2\cdot10^{-4}$ | $2\cdot10^{-4}$ | 2 | 0.175 | 0.60 | 1178.19 | 1.93 | 1.105 | 233.5 | 257.7 | $(15.57 \pm 0.54)\cdot10^{-2}$ | 0.52 | - | - |
| M_2W0.80 | $2\cdot10^{-4}$ | $2\cdot10^{-4}$ | 2 | 0.118 | 0.80 | 1252.24 | 2.36 | 1.247 | 339.3 | 422.9 | $(16.96 \pm 0.39)\cdot10^{-2}$ | 0.57 | - | - |
| M_2W1.00 | $2\cdot10^{-4}$ | $2\cdot10^{-4}$ | 2 | 0.087 | 1.00 | 1261.90 | 3.03 | 1.424 | 413.3 | 588.4 | $(16.53 \pm 0.38)\cdot10^{-2}$ | 0.57 | - | - |
| M_2W1.50 | $2\cdot10^{-4}$ | $2\cdot10^{-4}$ | 2 | 0.048 | 1.50 | 1285.2 | 4.52 | 1.983 | 625.0 | 1246 | $(16.7 \pm 1.2)\cdot10^{-2}$ | 0.55 | - | - |
| M_2W2.00 | $2\cdot10^{-4}$ | $2\cdot10^{-4}$ | 2 | 0.031 | 2.00 | 1329.72 | 6.47 | 1.845 | 790.1 | 1450 | $(15.8 \pm 2.3)\cdot10^{-2}$ | 0.44 | - | - |
| M_0W0.20_Pm0.25 | $2\cdot10^{-4}$ | $8\cdot10^{-4}$ | 0 | 1.0 | 0.20 | 1288.08 | 1.42 | 0.01828 | 35.1 (8.776) | 0.6417 | $(7.021 \pm 0.044)\cdot10^{-2}$ | 0.014 | - | - |
| M_0W0.20_Pm4 | $8\cdot10^{-4}$ | $2\cdot10^{-4}$ | 0 | 1.0 | 0.20 | 1305.14 | 1.62 | 0.00620 | 7.729 (30.92) | 0.04792 | $(6.184 \pm 0.049)\cdot10^{-2}$ | 0.013 | - | - |
| M_2W0.20_Pm0.25 | $2\cdot10^{-4}$ | $8\cdot10^{-4}$ | 2 | 1.0 | 0,20 | 1241.23 | 0.85 | 0.3814 | 59.30 (14.80) | 22.39 | $(11.840 \pm 0.098)\cdot10^{-2}$ | 0.29 | - | - |
| M_2W0.20_Pm4 | $8\cdot10^{-4}$ | $2\cdot10^{-4}$ | 2 | 1.0 | 0.20 | 1295.54 | 1.30 | 0.3036 | 9.566 (38.27) | 29.04 | $(7.653 \pm 0.051)\cdot10^{-2}$ | 0.23 | - | - |

| $512^3$ | $\nu$ | $\eta$ | $\Omega$ | $\phi_0$ | $R_f$ | $t_{tot}$ | $t_{turn}$ | $k_\omega/k_f$ | Re (Rm) | $Re_\omega$ | $u_{rms}$ | $u_{rot}/u_{tot}$ | r ($t_{turn}^{-1}$) | $r_\omega$ ($t_{turn}^{-1}$) |
|---|---|---|---|---|---|---|---|---|---|---|---|---|---|---|
| M_0W0.20_512 | $2\cdot10^{-4}$ | $2\cdot10^{-4}$ | 0 | 1 | 0.20 | 47.34 | 1.47 | 0.00581 | 34.02 | 0.1975 | $(6.803 \pm 0.041)\cdot10^{-2}$ | 0.012 | - | - |
| M_0W0.60_512 | $2\cdot10^{-4}$ | $2\cdot10^{-4}$ | 0 | 0.175 | 0.60 | 598.85 | 4.531 | 0.01148 | 99.41 | 1.141 | $(6.627 \pm 0.19)\cdot10^{-2}$ | - | - | - |
| M_2W0.60_512 | $2\cdot10^{-4}$ | $2\cdot10^{-4}$ | 2 | 0.175 | 0.60 | 339.57 | 2.13 | 1.174 | 211.0 | 247.8 | $(14.07 \pm 0.22)\cdot10^{-2}$ | - | - | - |

Table B.2: (continuation)

| $128^3$ | $\nu$ | $\eta$ | $\Omega$ | $\phi_0$ | $\Delta$t | $R_f$ | $t_{turn}$ | $k_\omega/k_f$ | Re (Rm) | $Re_\omega$ | $u_{rot}/u_{tot}$ | r ($t_{turn}^{-1}$) | $r_\omega$ ($t_{turn}^{-1}$) |
|---|---|---|---|---|---|---|---|---|---|---|---|---|---|
| M_0_128 | $2{\cdot}10^{-4}$ | $2{\cdot}10^{-4}$ | 0 | 1 | 0.02 | 0.2 | 1.32 | 0.04693 | 38.0 | 1.78 | 0.039 | - | - |
| M_1_128 | $2{\cdot}10^{-4}$ | $2{\cdot}10^{-4}$ | 1 | 1 | 0.02 | 0.2 | 1.12 | 0.190 | 44.8 | 8.53 | 0.131 | - | - |
| M_2_128 | $2{\cdot}10^{-4}$ | $2{\cdot}10^{-4}$ | 2 | 1 | 0.02 | 0.2 | 0.84 | 0.385 | 59.6 | 22.9 | 0.284 | - | - |
| M_3_128 | $2{\cdot}10^{-4}$ | $2{\cdot}10^{-4}$ | 3 | 1 | 0.02 | 0.2 | 0.77 | 0.543 | 64.6 | 35.1 | 0.417 | - | - |
| M_4_128 | $2{\cdot}10^{-4}$ | $2{\cdot}10^{-4}$ | 4 | 1 | 0.02 | 0.2 | 0.76 | 0.606 | 65.6 | 39.7 | 0.481 | - | - |
| M_5_128 | $2{\cdot}10^{-4}$ | $2{\cdot}10^{-4}$ | 5 | 1 | 0.02 | 0.2 | 0.78 | 0.642 | 64.0 | 41.1 | 0.524 | - | - |
| M_0_128_Pr | $2{\cdot}10^{-3}$ | $2{\cdot}10^{-5}$ | 0 | 1 | 0.02 | 0.2 | 1.93 | 0.00565 | 2.6 (259.4) | 0.01 | - | - | - |
| M_2_128_Pr | $2{\cdot}10^{-3}$ | $2{\cdot}10^{-5}$ | 2 | 1 | 0.02 | 0.2 | 1.86 | 0.29462 | 2.7 (269.2) | 0.79 | - | - | - |

| $256^3$ | $\nu$ | $\eta$ | $\Omega$ | $\phi_0$ | $\Delta$t | $R_f$ | $t_{turn}$ | $k_\omega/k_f$ | Re (Rm) | $Re_\omega$ | $u_{rot}/u_{tot}$ | r ($t_{turn}^{-1}$) | $r_\omega$ ($t_{turn}^{-1}$) |
|---|---|---|---|---|---|---|---|---|---|---|---|---|---|
| M_0 | $2{\cdot}10^{-4}$ | $2{\cdot}10^{-4}$ | 0 | 1 | 0.02 | 0.2 | 1.42 | 0.01820 | 35.3 | 0.64 | 0.014 | - | - |
| M_0s | $2{\cdot}10^{-4}$ | $2{\cdot}10^{-4}$ | 0 | 1 | 1 | 0.2 | 4.32 | 0.00061 | 11.6 | 0.007 | - | - | - |
| M_0c | $2{\cdot}10^{-4}$ | $2{\cdot}10^{-4}$ | 0 | 1 | $\delta t$ | 0.2 | 1.46 | 0.01609 | 34.3 | 0.55 | - | - | - |
| M_0low† | $2{\cdot}10^{-5}$ | $2{\cdot}10^{-5}$ | 0 | 1 | 0.02 | 0.2 | 3.05 | 0.5913 | 163.8 | 97.03 | - | - | - |
| M_0low_F2† | $2{\cdot}10^{-5}$ | $2{\cdot}10^{-5}$ | 0 | 2 | 0.02 | 0.2 | 1.03 | 1.4231 | 486.9 | 708.34 | - | - | - |
| M_0lowc | $2{\cdot}10^{-5}$ | $2{\cdot}10^{-5}$ | 0 | 1 | $\delta t$ | 0.2 | 1.38 | 0.08582 | 362.6 | 31.12 | - | - | - |
| M_0highcF10 | $2{\cdot}10^{-2}$ | $2{\cdot}10^{-2}$ | 0 | 10 | $\delta t$ | 0.2 | 0.61 | 0.03443 | 0.81 | 0.028 | - | - | - |
| M_2 | $2{\cdot}10^{-4}$ | $2{\cdot}10^{-4}$ | 2 | 1 | 0.02 | 0.2 | 0.84 | 0.3809 | 59.2 | 22.55 | 0.292 | - | - |
| M_2c | $2{\cdot}10^{-4}$ | $2{\cdot}10^{-4}$ | 2 | 1 | $\delta t$ | 0.2 | 0.85 | 0.38179 | 58.5 | 22.35 | - | - | - |
| M_2W0.5† | $2{\cdot}10^{-4}$ | $2{\cdot}10^{-4}$ | 2 | 1 | $\delta t$ | 0.5 | 0.29 | 0.64222 | 171.7 | 110.24 | - | - | - |
| M_2low† | $2{\cdot}10^{-5}$ | $2{\cdot}10^{-5}$ | 2 | 1 | 0.02 | 0.2 | 2.91 | 1.60296 | 171.8 | 276.5 | - | - | - |
| M_0W0.1 | $2{\cdot}10^{-2}$ | $2{\cdot}10^{-2}$ | 0 | 1 | 0.02 | 0.1 | 13.25 | 0.00073 | 0.0094 | 0.00001 | - | - | - |
| M_0W0.2 | $2{\cdot}10^{-2}$ | $2{\cdot}10^{-2}$ | 0 | 1 | 0.02 | 0.2 | 6.51 | 0.00348 | 0.0077 | 0.00027 | - | - | - |
| M_0W0.5 | $2{\cdot}10^{-2}$ | $2{\cdot}10^{-2}$ | 0 | 1 | 0.02 | 0.5 | 2.79 | 0.00177 | 1.12 | 0.0198 | - | - | - |
| M_0W1 | $2{\cdot}10^{-2}$ | $2{\cdot}10^{-2}$ | 0 | 1 | 0.02 | 1.0 | 1.96 | 0.05571 | 6.37 | 0.355 | - | - | - |

| $256^3$ | $\nu$ | $\eta$ | $B_0$(U) | $\Omega$ | $\phi_0$ | $\Delta$t | $R_f$ | $t_{turn}$ | $k_\omega/k_f$ | Re (Rm) | $Re_\omega$ | $u_{rot}/u_{tot}$ | r ($t_{turn}^{-1}$) | $r_\omega$ ($t_{turn}^{-1}$) |
|---|---|---|---|---|---|---|---|---|---|---|---|---|---|---|
| M_0B | $2{\cdot}10^{-4}$ | $2{\cdot}10^{-4}$ | $10^{-2}$ | 0 | 1 | 0.02 | 0.2 | 1.42 | 0.01820 | 35.3 | 0.64 | - | - | - |
| M_2Bxs | $2{\cdot}10^{-4}$ | $2{\cdot}10^{-4}$ | $10^{-6}$ | 2 | 1 | 1 | 0.2 | 3.93 | 0.39751 | 12.7 | 5.05 | - | - | - |
| M_2Bx | $2{\cdot}10^{-4}$ | $2{\cdot}10^{-4}$ | $10^{-2}$ | 2 | 1 | 0.02 | 0.2 | 0.85 | 0.3809 | 59.2 | 22.54 | - | - | - |
| M_2By | $2{\cdot}10^{-4}$ | $2{\cdot}10^{-4}$ | $10^{-2}$ | 2 | 1 | 0.02 | 0.2 | 0.85 | 0.3808 | 59.2 | 22.53 | - | - | - |
| M_2Bz | $2{\cdot}10^{-4}$ | $2{\cdot}10^{-4}$ | $10^{-2}$ | 2 | 1 | 0.02 | 0.2 | 0.84 | 0.3821 | 59.2 | 22.63 | - | - | - |

Table B.3: Baroclinic (i.e., non-isothermal) runs without shear.

| $256^3$ | $\nu$ | $\chi$ | $\eta$ | $\tau_{cool}$ | $\Omega$ | $\phi_0$ | $R_f$ | $t_{tot}$ | $t_{turn}$ | $k_\omega/k_f$ | Re (Rm) | $Re_\omega$ | $u_{rms}$ | $u_{rot}/u_{tot}$ | r ($t_{turn}^{-1}$) | $r_\omega$ ($t_{turn}^{-1}$) |
|---|---|---|---|---|---|---|---|---|---|---|---|---|---|---|---|---|
| MB_0W0.20_notau | $2{\cdot}10^{-4}$ | $2{\cdot}10^{-4}$ | $2{\cdot}10^{-4}$ | 0.1 | 0 | 2 | 0.20 | 1015.53 | 2.25 | 0.0304 | 22.23 | 0.675 | $(4.445 \pm 0.037) \cdot 10^{-2}$ | 0.015 | - | - |
| MB_0W0.20 | $2{\cdot}10^{-4}$ | $2{\cdot}10^{-4}$ | $2{\cdot}10^{-4}$ | - | 0 | 1 | 0.20 | 905.98 | 2.69 | 0.1550 | 18.60 | 2.883 | $(3.720 \pm 0.038) \cdot 10^{-2}$ | 0.036 | - | - |
| MB_0W0.30_notau | $2{\cdot}10^{-4}$ | $2{\cdot}10^{-4}$ | $2{\cdot}10^{-4}$ | 0.1 | 0 | 1.04 | 0.30 | 970.32 | 2.52 | 0.0101 | 44.69 | 0.451 | $(5.959 \pm 0.067) \cdot 10^{-2}$ | 0.011 | - | - |
| MB_0W0.30 | $2{\cdot}10^{-4}$ | $2{\cdot}10^{-4}$ | $2{\cdot}10^{-4}$ | - | 0 | 0.52 | 0.30 | 1079.47 | 3.77 | 0.1748 | 29.84 | 5.215 | $(3.979 \pm 0.062) \cdot 10^{-2}$ | 0.040 | - | - |
| MB_0W0.40_notau | $2{\cdot}10^{-4}$ | $2{\cdot}10^{-4}$ | $2{\cdot}10^{-4}$ | 0.1 | 0 | 0.72 | 0.40 | 964.04 | 3.23 | 0.0137 | 61.88 | 0.847 | $(6.19 \pm 0.10) \cdot 10^{-2}$ | 0.0076 | - | - |
| MB_0W0.40 | $2{\cdot}10^{-4}$ | $2{\cdot}10^{-4}$ | $2{\cdot}10^{-4}$ | - | 0 | 0.36 | 0.40 | 1183.27 | 5.04 | 0.1744 | 39.77 | 6.930 | $(3.98 \pm 0.14) \cdot 10^{-2}$ | 0.037 | - | - |
| MB_0W0.50_notau | $2{\cdot}10^{-4}$ | $2{\cdot}10^{-4}$ | $2{\cdot}10^{-4}$ | 0.1 | 0 | 0.48 | 0.50 | 980.79 | 4.19 | 0.0142 | 74.68 | 1.060 | $(5.97 \pm 0.14) \cdot 10^{-2}$ | 0.0072 | - | - |
| MB_0W0.50 | $2{\cdot}10^{-4}$ | $2{\cdot}10^{-4}$ | $2{\cdot}10^{-4}$ | - | 0 | 0.24 | 0.50 | 1234.23 | 6.45 | 0.1655 | 48.50 | 8.018 | $(3.88 \pm 0.13) \cdot 10^{-2}$ | 0.030 | - | - |
| MB_0W0.60_notau | $2{\cdot}10^{-4}$ | $2{\cdot}10^{-4}$ | $2{\cdot}10^{-4}$ | 0.1 | 0 | 0.35 | 0.60 | 998.6 | 5.14 | 0.0142 | 87.66 | 1.248 | $(5.84 \pm 0.16) \cdot 10^{-2}$ | 0.0065 | - | - |
| MB_0W0.60 | $2{\cdot}10^{-4}$ | $2{\cdot}10^{-4}$ | $2{\cdot}10^{-4}$ | - | 0 | 0.175 | 0.60 | 1251.19 | 7.53 | 0.1633 | 59.99 | 9.777 | $(4.00 \pm 0.17) \cdot 10^{-2}$ | 0.024 | - | - |
| MB_0W0.80_notau | $2{\cdot}10^{-4}$ | $2{\cdot}10^{-4}$ | $2{\cdot}10^{-4}$ | 0.1 | 0 | 0.236 | 0.80 | 1009.31 | 6.73 | 0.0183 | 191.1 | 2.174 | $(5.95 \pm 0.25) \cdot 10^{-2}$ | 0.0050 | - | - |
| MB_0W0.80 | $2{\cdot}10^{-4}$ | $2{\cdot}10^{-4}$ | $2{\cdot}10^{-4}$ | - | 0 | 0.118 | 0.80 | 1252.50 | 8.69 | 0.1732 | 92.52 | 15.95 | $(4.63 \pm 0.31) \cdot 10^{-2}$ | 0.022 | - | - |
| MB_0W1.00_notau | $2{\cdot}10^{-4}$ | $2{\cdot}10^{-4}$ | $2{\cdot}10^{-4}$ | 0.1 | 0 | 0.174 | 1.00 | 1020.70 | 8.32 | 0.0216 | 150.6 | 3.247 | $(6.03 \pm 0.33) \cdot 10^{-2}$ | 0.0040 | - | - |
| MB_0W1.00 | $2{\cdot}10^{-4}$ | $2{\cdot}10^{-4}$ | $2{\cdot}10^{-4}$ | - | 0 | 0.087 | 1.00 | 1248.69 | 9.80 | 0.1828 | 128.6 | 23.31 | $(5.14 \pm 0.47) \cdot 10^{-2}$ | 0.022 | - | - |
| MB_0W1.50_notau | $2{\cdot}10^{-4}$ | $2{\cdot}10^{-4}$ | $2{\cdot}10^{-4}$ | 0.1 | 0 | 0.096 | 1.50 | 1077.97 | 12.4 | 0.0168 | 229.5 | 3.852 | $(6.12 \pm 0.65) \cdot 10^{-2}$ | 0.0029 | - | - |
| MB_0W1.50 | $2{\cdot}10^{-4}$ | $2{\cdot}10^{-4}$ | $2{\cdot}10^{-4}$ | - | 0 | 0.048 | 1.50 | 1248.71 | 13.4 | 0.2479 | 214.2 | 51.83 | $(5.71 \pm 0.87) \cdot 10^{-2}$ | 0.027 | - | - |
| MB_0W2.00_notau | $2{\cdot}10^{-4}$ | $2{\cdot}10^{-4}$ | $2{\cdot}10^{-4}$ | 0.1 | 0 | 0.062 | 2.00 | 1164.33 | 18.8 | 0.0146 | 272.3 | 4.028 | $(5.45 \pm 0.85) \cdot 10^{-2}$ | 0.0016 | - | - |
| MB_0W2.00 | $2{\cdot}10^{-4}$ | $2{\cdot}10^{-4}$ | $2{\cdot}10^{-4}$ | - | 0 | 0.031 | 2.00 | 1259.42 | 19.0 | 0.2886 | 273.4 | 75.96 | $(5.5 \pm 1.1) \cdot 10^{-2}$ | 0.020 | - | - |
| MB_0W0.60_tau001 | $2{\cdot}10^{-4}$ | $2{\cdot}10^{-4}$ | $2{\cdot}10^{-4}$ | 0.01 | 0 | 0.175 | 0.60 | crashed | - | - | - | - | - | - | - | - |
| MB_0W0.60_tau002 | $2{\cdot}10^{-4}$ | $2{\cdot}10^{-4}$ | $2{\cdot}10^{-4}$ | 0.02 | 0 | 0.175 | 0.60 | crashed | - | - | - | - | - | - | - | - |
| MB_0W0.60_tau005 | $2{\cdot}10^{-4}$ | $2{\cdot}10^{-4}$ | $2{\cdot}10^{-4}$ | 0.05 | 0 | 0.175 | 0.60 | 1224.46 | 5.71 | 0.1240 | 78.95 | 9.772 | $(5.26 \pm 0.22) \cdot 10^{-2}$ | 0.016 | - | - |
| MB_0W0.60_tau | $2{\cdot}10^{-4}$ | $2{\cdot}10^{-4}$ | $2{\cdot}10^{-4}$ | 0.1 | 0 | 0.175 | 0.60 | 1251.19 | 7.52 | 0.1633 | 59.99 | 9.777 | $(3.99 \pm 0.17) \cdot 10^{-2}$ | 0.024 | - | - |
| MB_0W0.60_tau02 | $2{\cdot}10^{-4}$ | $2{\cdot}10^{-4}$ | $2{\cdot}10^{-4}$ | 0.2 | 0 | 0.175 | 0.60 | 1264.19 | 9.56 | 0.1958 | 47.19 | 9.223 | $(3.14 \pm 0.13) \cdot 10^{-2}$ | 0.038 | - | - |
| MB_0W0.60_tau05 | $2{\cdot}10^{-4}$ | $2{\cdot}10^{-4}$ | $2{\cdot}10^{-4}$ | 0.5 | 0 | 0.175 | 0.60 | 1264.94 | 10.38 | 0.1919 | 43.42 | 8.324 | $(2.89 \pm 0.10) \cdot 10^{-2}$ | 0.039 | - | - |
| MB_0W0.60_tau1 | $2{\cdot}10^{-4}$ | $2{\cdot}10^{-4}$ | $2{\cdot}10^{-4}$ | 1.0 | 0 | 0.175 | 0.60 | 1248.29 | 8.80 | 0.1520 | 51.21 | 7.774 | $(3.41 \pm 0.11) \cdot 10^{-2}$ | 0.026 | - | - |
| MB_2W0.20 | $2{\cdot}10^{-4}$ | $2{\cdot}10^{-4}$ | $2{\cdot}10^{-4}$ | 0.1 | 2 | 2 | 0.20 | 891.15 | 2.10 | 0.4497 | 23.82 | 10.71 | $(4.764 \pm 0.035) \cdot 10^{-2}$ | 0.42 | - | - |
| MB_2W0.30 | $2{\cdot}10^{-4}$ | $2{\cdot}10^{-4}$ | $2{\cdot}10^{-4}$ | 0.1 | 2 | 1.04 | 0.30 | 1062.24 | 2.62 | 0.5963 | 42.92 | 25.59 | $(5.722 \pm 0.060) \cdot 10^{-2}$ | 0.54 | - | - |
| MB_2W0.40 | $2{\cdot}10^{-4}$ | $2{\cdot}10^{-4}$ | $2{\cdot}10^{-4}$ | 0.1 | 2 | 0.72 | 0.40 | 1108.78 | 2.82 | 0.6709 | 70.95 | 47.59 | $(7.095 \pm 0.11) \cdot 10^{-2}$ | 0.60 | - | - |
| MB_2W0.50 | $2{\cdot}10^{-4}$ | $2{\cdot}10^{-4}$ | $2{\cdot}10^{-4}$ | 0.1 | 2 | 0.48 | 0.50 | 1176.45 | 3.95 | 0.6898 | 79.28 | 54.66 | $(6.342 \pm 0.26) \cdot 10^{-2}$ | 0.60 | - | - |
| MB_2W0.60 | $2{\cdot}10^{-4}$ | $2{\cdot}10^{-4}$ | $2{\cdot}10^{-4}$ | 0.1 | 2 | 0.35 | 0.60 | 1190.03 | 4.70 | 0.6984 | 95.95 | 66.99 | $(6.397 \pm 0.20) \cdot 10^{-2}$ | 0.62 | - | - |
| MB_2W0.80 | $2{\cdot}10^{-4}$ | $2{\cdot}10^{-4}$ | $2{\cdot}10^{-4}$ | 0.1 | 2 | 0.236 | 0.80 | 1174.07 | 5.33 | 0.7139 | 150.8 | 107.5 | $(7.541 \pm 0.49) \cdot 10^{-2}$ | 0.57 | - | - |
| MB_2W1.00 | $2{\cdot}10^{-4}$ | $2{\cdot}10^{-4}$ | $2{\cdot}10^{-4}$ | 0.1 | 2 | 0.174 | 1.00 | 1160.27 | 5.65 | 0.7419 | 222.8 | 165.0 | $(8.912 \pm 0.77) \cdot 10^{-2}$ | 0.60 | - | - |
| MB_2W1.50 | $2{\cdot}10^{-4}$ | $2{\cdot}10^{-4}$ | $2{\cdot}10^{-4}$ | 0.1 | 2 | 0.096 | 1.50 | 1175.55 | 8.33 | 0.8140 | 343.3 | 278.5 | $(9.154 \pm 1.2) \cdot 10^{-2}$ | 0.54 | - | - |
| MB_2W2.00 | $2{\cdot}10^{-4}$ | $2{\cdot}10^{-4}$ | $2{\cdot}10^{-4}$ | 0.1 | 2 | 0.062 | 2.00 | 1194.15 | 12.6 | 0.8870 | 407.1 | 358.3 | $(8.142 \pm 1.2) \cdot 10^{-2}$ | 0.43 | - | - |

Table B.3: (continuation)

| | | | | | | | | | | | | | | | |
|---|---|---|---|---|---|---|---|---|---|---|---|---|---|---|---|
| MB_0W_Pm0.25 | $2\cdot10^{-4}$ | $8\cdot10^{-4}$ | $2\cdot10^{-4}$ | 0.01 | 0 | 2.00 | 0.20 | 1129.75 | 1.29 | 0.0762 | 38.78 (9.694) | 2.954 | $(7.755 \pm 0.074)\cdot10^{-2}$ | 0.019 | - | - |
| MB_0W_Pm4 | $8\cdot10^{-4}$ | $2\cdot10^{-4}$ | $2\cdot10^{-4}$ | 0.01 | 0 | 2.00 | 0.20 | 1154.10 | 2.77 | 0.0770 | 4.499 (18.00) | 0.346 | $(3.600 \pm 0.035)\cdot10^{-2}$ | 0.022 | - | - |
| MB_2W_Pm0.25 | $2\cdot10^{-4}$ | $8\cdot10^{-4}$ | $2\cdot10^{-4}$ | 0.01 | 2 | 2.00 | 0.20 | 1138.86 | 2.11 | 0.450 | 23.65 (5.914) | 10.65 | $(4.731 \pm 0.033)\cdot10^{-2}$ | 0.42 | - | - |
| MB_2W_Pm4 | $8\cdot10^{-4}$ | $2\cdot10^{-4}$ | $2\cdot10^{-4}$ | 0.01 | 2 | 2.00 | 0.20 | 1158.22 | 2.16 | 0.419 | 11.59 (23.17) | 4.858 | $(4.635 \pm 0.035)\cdot10^{-2}$ | 0.39 | - | - |

| $512^3$ | $\nu$ | $\chi$ | $\eta$ | $\tau_{cool}$ | $\Omega$ | $\phi_0$ | $R_f$ | $t_{tot}$ | $t_{turn}$ | $k_\omega/k_f$ | Re (Rm) | $Re_\omega$ | $u_{rms}$ | $u_{rot}/u_{tot}$ | r ($t_{turn}^{-1}$) | $r_\omega$ ($t_{turn}^{-1}$) |
|---|---|---|---|---|---|---|---|---|---|---|---|---|---|---|---|---|
| MB_0W0.20_512 | $2\cdot10^{-4}$ | $2\cdot10^{-4}$ | $2\cdot10^{-4}$ | - | 0 | 1 | 0.2 | 44.10 | 1.69 | 0.00717 | 29.6 | 0.21 | $(5.884 \pm 0.072)\cdot10^{-2}$ | - | - | - |
| MB_0W0.60_512 | $2\cdot10^{-4}$ | $2\cdot10^{-4}$ | $2\cdot10^{-4}$ | 0.1 | 0 | 0.35 | 0.60 | 199.97 | 7.10 | 0.09355 | 63.5 | 5.92 | $(4.23 \pm 0.20)\cdot10^{-2}$ | - | - | - |
| MB_2W0.60_512 | $2\cdot10^{-4}$ | $2\cdot10^{-4}$ | $2\cdot10^{-4}$ | 0.1 | 2 | 0.35 | 0.60 | 199.18 | 4.69 | 0.7196 | 96.1 | 69.1 | $(6.40 \pm 0.26)\cdot10^{-2}$ | - | - | - |

| $128^3$ | $\nu$ | $\chi$ | $\eta$ | $B_0$(G) | $\Omega$ | $\phi_0$ | $\Delta$t | $R_f$ | $t_{turn}$ | $k_\omega/k_f$ | Re (Rm) | $Re_\omega$ | $u_{rot}/u_{tot}$ | r ($t_{turn}^{-1}$) | $r_\omega$ ($t_{turn}^{-1}$) |
|---|---|---|---|---|---|---|---|---|---|---|---|---|---|---|---|
| MB_0_128 | $2\cdot10^{-4}$ | $2\cdot10^{-4}$ | $2\cdot10^{-4}$ | $10^{-6}$ | 0 | 1 | 0.02 | 0.2 | 1.59 | 0.0086 | 31.5 | 0.27 | - | - | - |
| MB_2_128 | $2\cdot10^{-4}$ | $2\cdot10^{-4}$ | $2\cdot10^{-4}$ | $10^{-6}$ | 2 | 1 | 0.02 | 0.2 | 1.20 | 0.195 | 41.8 | 8.19 | - | - | - |
| MB_0_128_Pr | $2\cdot10^{-3}$ | $2\cdot10^{-3}$ | $2\cdot10^{-5}$ | $10^{-6}$ | 0 | 1 | 0.02 | 0.2 | 2.50 | 0.00319 | 2.002 (200.2) | 0.0064 | - | - | - |
| MB_2_128_Pr | $2\cdot10^{-3}$ | $2\cdot10^{-3}$ | $2\cdot10^{-5}$ | $10^{-6}$ | 2 | 1 | 0.02 | 0.2 | 2.39 | 0.21101 | 2.095 (209.5) | 0.44 | - | - | - |

| $256^3$ | $\nu$ | $\chi$ | $\eta$ | $\tau_{cool}$ | $\Omega$ | $\phi_0$ | $\Delta$t | $R_f$ | $t_{tot}$ | $t_{turn}$ | $k_\omega/k_f$ | Re (Rm) | $Re_\omega$ | $u_{rot}/u_{tot}$ | r ($t_{turn}^{-1}$) | $r_\omega$ ($t_{turn}^{-1}$) |
|---|---|---|---|---|---|---|---|---|---|---|---|---|---|---|---|---|
| MB_0 | $2\cdot10^{-4}$ | $2\cdot10^{-4}$ | $2\cdot10^{-4}$ | - | 0 | 1 | 0.02 | 0.2 | - | 1.65 | 0.0878 | 30.3 | 0.27 | - | - | - |
| MB_0c | $2\cdot10^{-4}$ | $2\cdot10^{-4}$ | $2\cdot10^{-4}$ | - | 0 | 1 | $\delta t$ | 0.2 | - | 1.65 | 0.00876 | 30.3 | 0.27 | - | - | - |
| MB_0highcF10 | $2\cdot10^{-2}$ | $2\cdot10^{-4}$ | $2\cdot10^{-2}$ | - | 0 | 10 | $\delta t$ | 0.2 | - | 0.64 | 0.01376 | 0.78 | 0.0108 | - | - | - |
| MB_0lowc | $5\cdot10^{-5}$ | $2\cdot10^{-4}$ | $5\cdot10^{-5}$ | - | 0 | 1 | $\delta t$ | 0.2 | - | 1.61 | 0.02876 | 124.1 | 3.57 | - | - | - |
| MB_0low | $5\cdot10^{-5}$ | $2\cdot10^{-4}$ | $5\cdot10^{-5}$ | - | 0 | 1 | 0.02 | 0.2 | - | 1.61 | 0.03046 | 124.4 | 3.79 | - | - | - |
| MB_0low2c | $2\cdot10^{-5}$ | $2\cdot10^{-4}$ | $2\cdot10^{-5}$ | - | 0 | 1 | $\delta t$ | 0.2 | - | 1.60 | 0.0434 | 313.1 | 13.58 | - | - | - |
| MB_2 | $2\cdot10^{-4}$ | $2\cdot10^{-4}$ | $2\cdot10^{-4}$ | - | 2 | 1 | 0.02 | 0.2 | - | 1.08 | 0.340 | 46.5 | 15.83 | - | - | - |
| MB_2c | $2\cdot10^{-4}$ | $2\cdot10^{-4}$ | $2\cdot10^{-4}$ | - | 2 | 1 | $\delta t$ | 0.2 | - | 1.06 | 0.343 | 47.0 | 16.12 | - | - | - |
| MB_2low† | $2\cdot10^{-5}$ | $2\cdot10^{-4}$ | $2\cdot10^{-5}$ | - | 2 | 1 | 0.02 | 0.2 | - | 0.96 | 0.6172 | 522.2 | 323.0 | - | - | - |
| MB_0W0.1 | $2\cdot10^{-2}$ | $2\cdot10^{-2}$ | $2\cdot10^{-2}$ | - | 0 | 1 | 0.02 | 0.1 | - | 15.31 | 0.00076 | 0.0082 | 0.00001 | - | - | - |
| MB_0W0.2 | $2\cdot10^{-2}$ | $2\cdot10^{-2}$ | $2\cdot10^{-2}$ | - | 0 | 1 | 0.02 | 0.2 | - | 7.84 | 0.00419 | 0.0638 | 0.00027 | - | - | - |
| MB_0W0.5 | $2\cdot10^{-2}$ | $2\cdot10^{-2}$ | $2\cdot10^{-2}$ | - | 0 | 1 | 0.02 | 0.5 | - | 1.36 | 0.01923 | 0.9209 | 0.01772 | - | - | - |
| MB_0W1 | $2\cdot10^{-2}$ | $2\cdot10^{-2}$ | $2\cdot10^{-2}$ | - | 0 | 1 | 0.02 | 1.0 | - | 0.46 | 0.06042 | 5.399 | 0.3263 | - | - | - |

| $256^3$ | $\nu$ | $\chi$ | $\eta$ | $B_0$(U) | $\Omega$ | $\phi_0$ | $\Delta$t | $R_f$ | $t_{turn}$ | $k_\omega/k_f$ | Re (Rm) | $Re_\omega$ | $u_{rot}/u_{tot}$ | r ($t_{turn}^{-1}$) | $r_\omega$ ($t_{turn}^{-1}$) |
|---|---|---|---|---|---|---|---|---|---|---|---|---|---|---|---|
| MB_0B | $2\cdot10^{-4}$ | $2\cdot10^{-4}$ | $2\cdot10^{-4}$ | $10^{-2}$ | 0 | 1 | 0.02 | 0.2 | 1.65 | 0.00898 | 30.3 | 0.27 | - | - | - |
| MB_2Bx | $2\cdot10^{-4}$ | $2\cdot10^{-4}$ | $2\cdot10^{-4}$ | $10^{-2}$ | 2 | 1 | 0.02 | 0.2 | 1.05 | 0.30828 | 47.5 | 14.63 | - | - | - |
| MB_2By | $2\cdot10^{-4}$ | $2\cdot10^{-4}$ | $2\cdot10^{-4}$ | $10^{-2}$ | 2 | 1 | 0.02 | 0.2 | 1.05 | 0.30744 | 47.7 | 14.65 | - | - | - |
| MB_2Bz | $2\cdot10^{-4}$ | $2\cdot10^{-4}$ | $2\cdot10^{-4}$ | $10^{-2}$ | 2 | 1 | 0.02 | 0.2 | 1.05 | 0.30858 | 47.5 | 14.66 | - | - | - |

Table B.4: MHD simulations with different values for $\tau_{cool}$.

| $256^3$ | $\nu$ | $\chi$ | $\eta$ | $B_0$(G) | $\tau_{cool}$ | $\phi_0$ | $\Delta$t | $R_f$ | Re (Rm) |
|---|---|---|---|---|---|---|---|---|---|
| MB_t_1.57 | 0.1 | 0.1 | 0.1 | $10^{-6}$ | 1.5708 | 10 | 0.02 | 0.2 | 0.0980 |
| MB_t_3.14 | 0.1 | 0.1 | 0.1 | $10^{-6}$ | 3.1415 | 10 | 0.02 | 0.2 | 0.0953 |
| MB_t_6.28 | 0.1 | 0.1 | 0.1 | $10^{-6}$ | 6.2831 | 10 | 0.02 | 0.2 | 0.0870 |
| MB_t_31.4 | 0.1 | 0.1 | 0.1 | $10^{-6}$ | 31.415 | 10 | 0.02 | 0.2 | 0.0759 |
| MB_t_314 | 0.1 | 0.1 | 0.1 | $10^{-6}$ | 314.15 | 10 | 0.02 | 0.2 | 0.0726 |
| MB_nt | 0.1 | 0.1 | 0.1 | $10^{-6}$ | - | 10 | 0.02 | 0.2 | 0.0638 |

Table B.5: Simulations involving higher forcing and diffusivities, using neither shear nor rotation.

| | $\nu$ | $\chi$ | $\phi_0$ | $\Delta$t | $R_f$ | Re |
|---|---|---|---|---|---|---|
| M_0highc_F5 | $10^{-3}$ | - | 5 | $\delta t$ | 0.2 | 18.13 |
| M_0highc_F20 + | 0.1 | - | 10 | $\delta t$ | 0.2 | 0.1421 |
| MB_0highc_F50 | 0.1 | 0.1 | 50 | $\delta t$ | 0.2 | 0.3467 |
| MB_0high_F50 | 0.1 | 0.1 | 50 | 0.02 | 0.2 | 0.4319 |
| MB_0high2_F50 | 1 | 1 | 50 | 0.02 | 0.2 | 0.0089 |
| MB_0high_F100 | 1 | 1 | 100 | 0.02 | 0.2 | 0.0329 |
| MB_0high_F100_W1† | 1 | 1 | 100 | 0.02 | 1 | 0.3931 |
| MB_0high_F200 | 1 | 1 | 200 | 0.02 | 0.2 | 0.0657 |
| MB_0high_F200_W1† | 1 | 1 | 200 | 0.02 | 1 | 0.8864 |
| MB_0high_F500 | 1 | 1 | 500 | 0.02 | 0.2 | 0.0647 |
| MB_0high_F500_W1† | 1 | 1 | 500 | 0.02 | 1 | 3.4525 |

Table B.6: Models with sinusoidal shearing velocity profile. ND stands for the growth rates not determined, even though a dynamo was present.

| $256^3$ | $\nu$ | $\chi$ | $\eta$ | $\tau_{cool}$ | $A$ | $\phi_0$ | $R_f$ | $t_{tot}$ | $t_{turn}$ | $k_\omega/k_f$ | Re(Rm) | $Re_\omega$ | $u_{rot,0}/u_{tot}$ | $u_{rot}/u_{tot}$ | r ($t_{turn}^{-1}$) | $r_\omega$ ($t_{turn}^{-1}$) |
|---|---|---|---|---|---|---|---|---|---|---|---|---|---|---|---|---|
| M_S_Pm0.1 | $2{\cdot}10^{-4}$ | - | $20{\cdot}10^{-4}$ | - | 0.20 | 1 | 0.2 | 4189.19 | 0.637 | 0.0915 | 78.5 (7.85) | 7.200 | 0.819 | 0.899 | - | $1.052{\cdot}10^{-2}$ |
| M_S_Pm0.25 | $2{\cdot}10^{-4}$ | - | $8{\cdot}10^{-4}$ | - | 0.20 | 1 | 0.2 | 2277.66 | 0.635 | 0.0919 | 78.8 (19.7) | 7.296 | 0.811 | 0.894 | $3.261{\cdot}10^{-3}$ | $8.573{\cdot}10^{-3}$ |
| M_S_Pm0.5 | $2{\cdot}10^{-4}$ | - | $4{\cdot}10^{-4}$ | - | 0.20 | 1 | 0.2 | 3492.59 | 0.634 | 0.0912 | 78.9 (39.5) | 7.247 | 0.810 | 0.873 | $1.224{\cdot}10^{-2}$ | $9.333{\cdot}10^{-3}$ |
| M_S_Pm0.75 | $2{\cdot}10^{-4}$ | - | $2.667{\cdot}10^{-4}$ | - | 0.20 | 1 | 0.2 | 3927.51 | 0.634 | 0.0911 | 78.9 (59.2) | 7.221 | 0.809 | 0.867 | $2.254{\cdot}10^{-2}$ | $1.096{\cdot}10^{-2}$ |
| M_S_Pm1 | $2{\cdot}10^{-4}$ | - | $2{\cdot}10^{-4}$ | - | 0.20 | 1 | 0.2 | 3559.77 | 0.633 | 0.0910 | 78.9 (78.9) | 7.200 | 0.806 | 0.861 | $2.783{\cdot}10^{-2}$ | $1.022{\cdot}10^{-2}$ |
| M_S_Pm1.25 | $2.5{\cdot}10^{-4}$ | - | $2{\cdot}10^{-4}$ | - | 0.20 | 1 | 0.2 | 3965.97 | 0.636 | 0.0910 | 62.9 (78.6) | 5.751 | 0.825 | 0.852 | $2.861{\cdot}10^{-2}$ | $9.176{\cdot}10^{-3}$ |
| M_S_Pm1.5 | $3{\cdot}10^{-4}$ | - | $2{\cdot}10^{-4}$ | - | 0.20 | 1 | 0.2 | 3811.68 | 0.637 | 0.0910 | 52.3 (78.5) | 4.767 | 0.817 | 0.865 | $3.073{\cdot}10^{-2}$ | $1.001{\cdot}10^{-2}$ |
| M_S_Pm2 | $4{\cdot}10^{-4}$ | - | $2{\cdot}10^{-4}$ | - | 0.20 | 1 | 0.2 | 4765.49 | 0.640 | 0.0914 | 39.0 (78.1) | 3.575 | 0.824 | 0.872 | $3.438{\cdot}10^{-2}$ | $9.378{\cdot}10^{-3}$ |
| M_S_Pm4 | $8{\cdot}10^{-4}$ | - | $2{\cdot}10^{-4}$ | - | 0.20 | 1 | 0.2 | 4757.68 | 0.649 | 0.0922 | 19.3 (77.0) | 1.775 | 0.848 | - | - | - |
| M_S_Pm10 | $20{\cdot}10^{-4}$ | - | $2{\cdot}10^{-4}$ | - | 0.20 | 1 | 0.2 | 3340.23 | 0.672 | 0.0947 | 7.44 (74.4) | 0.705 | 0.903 | - | - | - |
| M_S_W0.10 | $2{\cdot}10^{-4}$ | - | $2{\cdot}10^{-4}$ | - | 0.20 | 2 | 0.10 | 3012.18 | 0.33 | 0.04733 | 37.5 | 1.775 | 0.888 | - | - | - |
| M_S_W0.20 | $2{\cdot}10^{-4}$ | - | $2{\cdot}10^{-4}$ | - | 0.20 | 1 | 0.20 | 3332.61 | 0.63 | 0.09071 | 78.8 | 7.103 | 0.809 | 0.855 | $2.616{\cdot}10^{-2}$ | $1.022{\cdot}10^{-2}$ |
| M_S_W0.30 | $2{\cdot}10^{-4}$ | - | $2{\cdot}10^{-4}$ | - | 0.20 | 0.52 | 0.30 | 4923.52 | 0.95 | 0.1372 | 118.2 | 15.98 | 0.818 | 0.861 | $3.997{\cdot}10^{-2}$ | $1.233{\cdot}10^{-2}$ |
| M_S_W0.40 | $2{\cdot}10^{-4}$ | - | $2{\cdot}10^{-4}$ | - | 0.20 | 0.36 | 0.40 | 4867.96 | 1.26 | 0.1819 | 159.0 | 28.42 | 0.807 | 0.851 | $5.017{\cdot}10^{-2}$ | $2.273{\cdot}10^{-2}$ |
| M_S_W0.50 | $2{\cdot}10^{-4}$ | - | $2{\cdot}10^{-4}$ | - | 0.20 | 0.24 | 0.50 | 4889.66 | 1.58 | 0.2275 | 197.5 | 44.37 | 0.818 | 0.861 | $7.514{\cdot}10^{-2}$ | $2.572{\cdot}10^{-2}$ |
| M_S_W0.60 | $2{\cdot}10^{-4}$ | - | $2{\cdot}10^{-4}$ | - | 0.20 | 0.175 | 0.60 | 4092.11 | 1.90 | 0.2738 | 236.5 | 63.89 | 0.812 | 0.862 | $8.791{\cdot}10^{-2}$ | $3.361{\cdot}10^{-2}$ |
| M_S_W0.80 | $2{\cdot}10^{-4}$ | - | $2{\cdot}10^{-4}$ | - | 0.20 | 0.118 | 0.80 | 4908.57 | 2.51 | 0.3607 | 319.3 | 113.5 | 0.812 | 0.846 | $1.190{\cdot}10^{-1}$ | $4.175{\cdot}10^{-2}$ |
| M_S_W1.00 | $2{\cdot}10^{-4}$ | - | $2{\cdot}10^{-4}$ | - | 0.20 | 0.087 | 1.00 | 4918.58 | 3.08 | 0.4455 | 404.7 | 177.5 | 0.805 | 0.847 | $1.444{\cdot}10^{-1}$ | $5.803{\cdot}10^{-2}$ |
| M_S_W1.50 | $2{\cdot}10^{-4}$ | - | $2{\cdot}10^{-4}$ | - | 0.20 | 0.048 | 1.50 | 1217.34 | 4.85 | 0.6884 | 579.9 | 399.1 | 0.804 | 0.840 | $1.729{\cdot}10^{-1}$ | $5.959{\cdot}10^{-2}$ |
| M_S_W2.00 | $2{\cdot}10^{-4}$ | - | $2{\cdot}10^{-4}$ | - | 0.20 | 0.031 | 2.00 | 1333.81 | 3.48 | 0.4909 | 359.4 | 176.4 | 0.942 | – | $1.476{\cdot}10^{-1}$ | $5.058{\cdot}10^{-2}$ |
| MB_S_W0.20 | $2{\cdot}10^{-4}$ | $2{\cdot}10^{-4}$ | $2{\cdot}10^{-4}$ | 0.1 | 0.20 | 2 | 0.20 | 1710.04 | 0.68 | 01.102 | 73.00 | 8.046 | 0.97 | 0.97 | $2.433{\cdot}10^{-2}$ | $8.832{\cdot}10^{-3}$ |
| MB_S_W0.30 | $2{\cdot}10^{-4}$ | $2{\cdot}10^{-4}$ | $2{\cdot}10^{-4}$ | 0.1 | 0.20 | 1.04 | 0.30 | 2307.15 | 1.02 | 01.628 | 110.2 | 17.82 | 0.96 | 0.96 | $3.377{\cdot}10^{-2}$ | $1.116{\cdot}10^{-2}$ |
| MB_S_W0.40 | $2{\cdot}10^{-4}$ | $2{\cdot}10^{-4}$ | $2{\cdot}10^{-4}$ | 0.1 | 0.20 | 0.72 | 0.40 | 2401.10 | 1.35 | 02.165 | 148.7 | 32.19 | 0.95 | 0.93 | $3.942{\cdot}10^{-2}$ | $1.556{\cdot}10^{-2}$ |
| MB_S_W0.50 | $2{\cdot}10^{-4}$ | $2{\cdot}10^{-4}$ | $2{\cdot}10^{-4}$ | 0.1 | 0.20 | 0.48 | 0.50 | 2444.66 | 1.68 | 02.590 | 185.9 | 48.15 | 0.95 | 0.92 | $5.656{\cdot}10^{-2}$ | $2.470{\cdot}10^{-2}$ |
| MB_S_W0.60 | $2{\cdot}10^{-4}$ | $2{\cdot}10^{-4}$ | $2{\cdot}10^{-4}$ | 0.1 | 0.20 | 0.35 | 0.60 | 2448.34 | 2.02 | 0.3010 | 223.3 | 67.21 | 0.96 | 0.93 | $6.921{\cdot}10^{-2}$ | $2.768{\cdot}10^{-2}$ |
| MB_S_W0.80 | $2{\cdot}10^{-4}$ | $2{\cdot}10^{-4}$ | $2{\cdot}10^{-4}$ | 0.1 | 0.20 | 0.236 | 0.80 | 2407.49 | 2.64 | 0.3889 | 302.9 | 117.8 | 0.94 | 0.92 | $9.382{\cdot}10^{-2}$ | $4.174{\cdot}10^{-2}$ |
| MB_S_W1.00 | $2{\cdot}10^{-4}$ | $2{\cdot}10^{-4}$ | $2{\cdot}10^{-4}$ | 0.1 | 0.20 | 0.174 | 1.00 | 2342.14 | 3.25 | 0.4744 | 384.4 | 182.3 | 0.93 | 0.91 | $1.202{\cdot}10^{-1}$ | $6.614{\cdot}10^{-2}$ |
| MB_S_W1.50 | $2{\cdot}10^{-4}$ | $2{\cdot}10^{-4}$ | $2{\cdot}10^{-4}$ | 0.1 | 0.20 | 0.096 | 1.50 | 2421.20 | 4.80 | 0.6864 | 586.3 | 402.2 | 0.92 | 0.88 | $1.516{\cdot}10^{-1}$ | $8.403{\cdot}10^{-2}$ |
| MB_S_W2.00 | $2{\cdot}10^{-4}$ | $2{\cdot}10^{-4}$ | $2{\cdot}10^{-4}$ | 0.1 | 0.20 | 0.062 | 2.00 | 2421.66 | 6.44 | 0.9191 | 777.1 | 713.7 | 0.95 | 0.87 | $2.245{\cdot}10^{-1}$ | $1.137{\cdot}10^{-1}$ |
| MB_S_W0.50_tau0005 | $2{\cdot}10^{-4}$ | $2{\cdot}10^{-4}$ | $2{\cdot}10^{-4}$ | 0.005 | 0.20 | 0.48 | 0.50 | 1245.53 | 1.58 | 0.2232 | 198.4 | 44.28 | 0.81 | 0.77 | $5.863{\cdot}10^{-2}$ | $3.137{\cdot}10^{-2}$ |
| MB_S_W0.50_tau001 | $2{\cdot}10^{-4}$ | $2{\cdot}10^{-4}$ | $2{\cdot}10^{-4}$ | 0.01 | 0.20 | 0.48 | 0.50 | 1277.6 | 1.59 | 0.2258 | 196.1 | 44.27 | 0.84 | 0.80 | $5.993{\cdot}10^{-2}$ | $2.776{\cdot}10^{-2}$ |
| MB_S_W0.50_tau05 | $2{\cdot}10^{-4}$ | $2{\cdot}10^{-4}$ | $2{\cdot}10^{-4}$ | 0.05 | 0.20 | 0.48 | 0.50 | 1297.98 | 1.68 | 0.2362 | 187.5 | 44.28 | 0.92 | 0.88 | $7.864{\cdot}10^{-2}$ | $3.164{\cdot}10^{-2}$ |
| MB_S_W0.50_tau01 | $2{\cdot}10^{-4}$ | $2{\cdot}10^{-4}$ | $2{\cdot}10^{-4}$ | 0.1 | 0.20 | 0.48 | 0.50 | 1297.98 | 1.70 | 0.2410 | 183.7 | 44.28 | 0.95 | 0.93 | $6.769{\cdot}10^{-2}$ | $3.251{\cdot}10^{-2}$ |
| MB_S_W0.50_notau | $2{\cdot}10^{-4}$ | $2{\cdot}10^{-4}$ | $2{\cdot}10^{-4}$ | - | 0.20 | 0.48 | 0.50 | 979.59 | 1.57 | 0.2225 | 199.0 | 44.29 | 0.79 | 0.81 | $7.601{\cdot}10^{-2}$ | $3.203{\cdot}10^{-2}$ |
| MB_S_W0.20_notau | $2{\cdot}10^{-4}$ | $2{\cdot}10^{-4}$ | $2{\cdot}10^{-4}$ | - | 0.20 | 1 | 0.20 | 1769.62 | 0.65 | 0.0925 | 76.75 | 7.100 | 0.85 | 0.87 | $3.255{\cdot}10^{-2}$ | $9.601{\cdot}10^{-3}$ |

Table B.6: (continuation)

| $512^3$ | $\nu$ | $\chi$ | $\eta$ | $\tau_{cool}$ | $A$ | $\phi_0$ | $R_f$ | $t_{tot}$ | $t_{turn}$ | $k_\omega/k_f$ | Re (Rm) | $Re_\omega$ | $u_{rot,0}/u_{tot}$ | $u_{rot}/u_{tot}$ | r ($t_{turn}^{-1}$) | $r_\omega$ ($t_{turn}^{-1}$) |
|---|---|---|---|---|---|---|---|---|---|---|---|---|---|---|---|---|
| M_0A020_512 | $2{\cdot}10^{-4}$ | - | $2{\cdot}10^{-4}$ | - | 0.20 | 1 | 0.2 | 1606.63 | 0.64 | 0.091 | 78.5 | 7.10 | 0.83 | - | 0.0264 | 0.0107 |
| M_0A020_512 | $1{\cdot}10^{-4}$ | - | $1{\cdot}10^{-4}$ | - | 0.20 | 1 | 0.2 | 1950.99 | 0.64 | 0.090 | 157.0 | 14.2 | - | - | 0.0348 | 0.0162 |
| M_0A060_512 | $2{\cdot}10^{-4}$ | - | $2{\cdot}10^{-4}$ | - | 0.60 | 1 | 0.2 | 1642.89 | 1.98 | 0.281 | 227.3 | 6.39 | - | - | 0.0995 | 0.0323 |
| $128^3$ | $\nu$ | $\chi$ | $\eta$ | $\tau_{cool}$ | $A$ | $\phi_0$ | $R_f$ | $t_{tot}$ | $t_{turn}$ | $k_\omega/k_f$ | Re (Rm) | $Re_\omega$ | $u_{rot,0}/u_{tot}$ | $u_{rot}/u_{tot}$ | r ($t_{turn}^{-1}$) | $r_\omega$ ($t_{turn}^{-1}$) |
| M_0A000_128 | $2{\cdot}10^{-4}$ | - | $2{\cdot}10^{-4}$ | - | 0 | 1 | 0.2 | - | - | - | 38.0 | - | - | - | - | - |
| M_0A002_128 | $2{\cdot}10^{-4}$ | - | $2{\cdot}10^{-4}$ | - | 0.02 | 1 | 0.2 | - | - | - | 38.6 | - | - | - | - | - |
| M_0A005_128 | $2{\cdot}10^{-4}$ | - | $2{\cdot}10^{-4}$ | - | 0.05 | 1 | 0.2 | - | - | - | 41.9 | - | - | - | - | - |
| M_0A010_128 | $2{\cdot}10^{-4}$ | - | $2{\cdot}10^{-4}$ | - | 0.10 | 1 | 0.2 | - | - | - | 52.1 | - | - | - | ND | ND |
| M_0A020_128 | $2{\cdot}10^{-4}$ | - | $2{\cdot}10^{-4}$ | - | 0.20 | 1 | 0.2 | - | 0.62 | 0.092 | 80.2 | 7.37 | 0.78 | 0.81 | 0.0296 | 0.00983 |
| M_0A050_128 | $2{\cdot}10^{-4}$ | - | $2{\cdot}10^{-4}$ | - | 0.50 | 1 | 0.2 | - | - | - | 180.3 | - | - | - | ND | ND |
| M_0A100_128 | $2{\cdot}10^{-4}$ | - | $2{\cdot}10^{-4}$ | - | 1.00 | 1 | 0.2 | - | - | - | 355.1 | - | - | - | ND | ND |
| H_0A020_128 | $2{\cdot}10^{-4}$ | - | - | - | 0.20 | 1 | 0.2 | - | - | - | 80.2 | - | - | - | - | - |
| H_0A020_F0_128 | $2{\cdot}10^{-4}$ | - | - | - | 0.20 | 0 | - | - | - | - | 70.5 | - | - | - | - | - |
| H_0A050_128 | $2{\cdot}10^{-4}$ | - | - | - | 0.50 | 1 | 0.2 | - | - | - | 180.5 | - | - | - | - | - |
| H_0A050_F0_128 | $2{\cdot}10^{-4}$ | - | - | - | 0.50 | 0 | - | - | - | - | 176.5 | - | - | - | - | - |
| M_0A020_W0.10_128 | $2{\cdot}10^{-4}$ | - | $2{\cdot}10^{-4}$ | - | 0.20 | 1 | 0.1 | - | - | - | 36.1 | - | - | - | - | - |
| M_0A020_W0.15_128 | $2{\cdot}10^{-4}$ | - | $2{\cdot}10^{-4}$ | - | 0.20 | 1 | 0.15 | - | 0.50 | 0.071 | 56.6 | 4.01 | 0.87 | 0.91 | 0.0286 | 0.00918 |
| M_0A020_W0.25_128 | $2{\cdot}10^{-4}$ | - | $2{\cdot}10^{-4}$ | - | 0.20 | 1 | 0.25 | - | 0.74 | 0.112 | 104.9 | 11.76 | 0.73 | 0.76 | 0.0235 | 0.00853 |
| M_0A020_F0_128 | $2{\cdot}10^{-4}$ | - | $2{\cdot}10^{-4}$ | - | 0.20 | 0 | - | - | - | - | 70.5 | - | - | - | - | - |
| M_0A020_F0.5_128 | $2{\cdot}10^{-4}$ | - | $2{\cdot}10^{-4}$ | - | 0.20 | 0.5 | 0.2 | - | 0.68 | 0.096 | 74.1 | 7.10 | 0.91 | 0.92 | 0.0290 | 0.00875 |
| M_0A020_F1.5_128 | $2{\cdot}10^{-4}$ | - | $2{\cdot}10^{-4}$ | - | 0.20 | 1.5 | 0.2 | - | 0.58 | 0.099 | 85.8 | 8.49 | 0.70 | 0.75 | 0.0239 | 0.00857 |
| $256^3$ | $\nu$ | $\chi$ | $\eta$ | $\tau_{cool}$ | $A$ | $\phi_0$ | $R_f$ | $t_{tot}$ | $t_{turn}$ | $k_\omega/k_f$ | Re (Rm) | $Re_\omega$ | $u_{rot,0}/u_{tot}$ | $u_{rot}/u_{tot}$ | r ($t_{turn}^{-1}$) | $r_\omega$ ($t_{turn}^{-1}$) |
| M_0A010 | $2{\cdot}10^{-4}$ | - | $2{\cdot}10^{-4}$ | - | 0.10 | 1 | 0.2 | - | 1.00 | 0.073 | 49.8 | 3.61 | 0.51 | 0.68 | 0.0155 | 0.00756 |
| M_0A015 | $2{\cdot}10^{-4}$ | - | $2{\cdot}10^{-4}$ | - | 0.15 | 1 | 0.2 | - | 0.79 | 0.085 | 63.5 | 5.40 | 0.70 | 0.76 | 0.0220 | 0.00944 |
| H_0A020 | $2{\cdot}10^{-4}$ | - | - | - | 0.20 | 1 | 0.2 | - | - | - | 78.9 | - | - | - | - | - |
| M_0A020 | $2{\cdot}10^{-4}$ | - | $2{\cdot}10^{-4}$ | - | 0.20 | 1 | 0.2 | - | 0.63 | 0.091 | 78.9 | 7.19 | 0.81 | 0.86 | 0.0262 | 0.00969 |
| MB_0A020 | $2{\cdot}10^{-4}$ | $2{\cdot}10^{-4}$ | $2{\cdot}10^{-4}$ | - | 0.50 | 1 | 0.2 | - | 0.65 | 0.093 | 76.9 | 7.15 | 0.85 | 0.88 | 0.0316 | 0.00925 |
| M_0A050 | $2{\cdot}10^{-4}$ | - | $2{\cdot}10^{-4}$ | - | 0.50 | 1 | 0.2 | - | 0.28 | 0.103 | 181.1 | 18.74 | 0.96 | 0.97 | 0.0510 | 0.0143 |

Table B.7: MHD simulations with linear shear and the barotropic EoS. For all runs, $B_0 = 10^{-6}$, $\phi_0 = 1$, $\Delta$t= 0.02, $R = 0.2$. The magnetic field growth (last column) seen in some of them is dominated by a spurious boundary effect. The lower part of the table contains the diagnostics for some of the cases with magnetic field growth.

| $128^3$ | $\nu$ | $\eta$ | $\Omega$ | $S$ | $B_{growth}$ |
|---|---|---|---|---|---|
| M_0S00_128 | $2{\cdot}10^{-3}$ | $2{\cdot}10^{-3}$ | 0 | 0 | No |
| M_0S01_128 | $2{\cdot}10^{-3}$ | $2{\cdot}10^{-3}$ | 0 | 0.01 | No |
| M_0S05_128 | $2{\cdot}10^{-3}$ | $2{\cdot}10^{-3}$ | 0 | 0.05 | No |
| M_0S10_128 | $2{\cdot}10^{-3}$ | $2{\cdot}10^{-3}$ | 0 | 0.1 | No |
| M_0S15_128 | $2{\cdot}10^{-3}$ | $2{\cdot}10^{-3}$ | 0 | 0.15 | No |
| M_0S20_128 | $2{\cdot}10^{-3}$ | $2{\cdot}10^{-3}$ | 0 | 0.2 | No |
| M_0S30_128 | $2{\cdot}10^{-3}$ | $2{\cdot}10^{-3}$ | 0 | 0.3 | No |
| M_0S40_128 | $2{\cdot}10^{-3}$ | $2{\cdot}10^{-3}$ | 0 | 0.4 | No |
| M_0S50_128 | $2{\cdot}10^{-3}$ | $2{\cdot}10^{-3}$ | 0 | 0.5 | No |
| M_0S00_Pm_128 | $2{\cdot}10^{-3}$ | $2{\cdot}10^{-5}$ | 0 | 0 | No |
| M_0S01_Pm_128 | $2{\cdot}10^{-3}$ | $2{\cdot}10^{-5}$ | 0 | 0.01 | Yes |
| M_0S05_Pm_128 | $2{\cdot}10^{-3}$ | $2{\cdot}10^{-5}$ | 0 | 0.05 | Yes |
| M_0S10_Pm_128 | $2{\cdot}10^{-3}$ | $2{\cdot}10^{-5}$ | 0 | 0.1 | Yes |
| M_0S15_Pm_128 | $2{\cdot}10^{-3}$ | $2{\cdot}10^{-5}$ | 0 | 0.15 | Yes |
| M_0S20_Pm_128 | $2{\cdot}10^{-3}$ | $2{\cdot}10^{-5}$ | 0 | 0.2 | Yes |
| M_0S30_Pm_128 | $2{\cdot}10^{-3}$ | $2{\cdot}10^{-5}$ | 0 | 0.3 | Yes |
| M_0S40_Pm_128 | $2{\cdot}10^{-3}$ | $2{\cdot}10^{-5}$ | 0 | 0.4 | Yes |
| M_0S50_Pm_128 | $2{\cdot}10^{-3}$ | $2{\cdot}10^{-5}$ | 0 | 0.5 | Yes |
| M_.2S10_Pm_128 | $2{\cdot}10^{-3}$ | $2{\cdot}10^{-5}$ | 0.2 | 0.1 | Yes |
| M_2S10_Pm_128 | $2{\cdot}10^{-3}$ | $2{\cdot}10^{-5}$ | 2 | 0.1 | Yes |
| M_.2S20_Pm_128 | $2{\cdot}10^{-3}$ | $2{\cdot}10^{-5}$ | 0.2 | 0.2 | Yes |
| M_2S20_Pm_128 | $2{\cdot}10^{-3}$ | $2{\cdot}10^{-5}$ | 2 | 0.2 | Yes |
| M_.2S30_Pm_128 | $2{\cdot}10^{-3}$ | $2{\cdot}10^{-5}$ | 0.2 | 0.3 | Yes |
| M_2S30_Pm_128 | $2{\cdot}10^{-3}$ | $2{\cdot}10^{-5}$ | 2 | 0.3 | Yes |
| | $S$ | $t_{turn}$ | $k_\omega/k_f$ | Re | $Re_\omega$ |
| M_0S00_Pm_128 | 0.00 | 1.99 | 0.00615 | 2.52 | 0.02 |
| M_0S01_Pm_128 | 0.01 | 1.99 | 0.11165 | 2.52 | 0.28 |
| M_0S05_Pm_128 | 0.05 | 2.00 | 0.11509 | 2.49 | 0.29 |
| M_0S10_Pm_128 | 0.10 | 2.03 | 0.12138 | 2.46 | 0.3 |
| M_0S15_Pm_128 | 0.15 | 2.06 | 0.12949 | 2.43 | 0.32 |
| M_0S20_Pm_128 | 0.20 | 2.11 | 0.13702 | 2.37 | 0.32 |
| M_0S30_Pm_128 | 0.30 | 2.17 | 0.15462 | 2.32 | 0.34 |
| M_0S40_Pm_128 | 0.40 | 2.29 | 0.16855 | 2.19 | 0.37 |
| M_0S50_Pm_128 | 0.50 | 2.34 | 0.18544 | 2.14 | 0.39 |

# C

# Potential magnetic fields: spherical harmonics formalism

Non-residual planetary magnetic fields arise from an active dynamo process in electrically conducting fluids. Observers located outside the dynamo region, where there are no free currents and thus there can be no magnetic field sources, a scalar potential can be used to describe the magnetic field. Starting from Maxwell's equations and assuming a non-relativistic, non-charged conducting fluid observed from a region with no free currents ($\mathbf{J} = 0$), the equations simply become:

$$\nabla \cdot \mathbf{B} = 0$$
$$\nabla \times \mathbf{B} = 0$$

These relations ensure that the magnetic field can be expressed as the gradient of a scalar potential. For planets, it is usually assumed that the only source of magnetic field is from an internal origin (magnetic intensity decays as r grows), but one can include external sources such as magnetospheres or magneto-disks.

$$\boldsymbol{B} = -\nabla V = -\nabla(V_{\text{int}} + V_{\text{ext}}) \tag{C.1}$$

The equations for the potential expansions shown in this section are taken from Davis (2004); a more extensive review is Winch et al. (2005), where they also include the external potential contribution. As planets are close to a sphere, the most logical strategy is to express the magnetic field as an expansion of spherical harmonics, as done by Gauss:

$$V_{\text{int}} = a \sum_{n=1}^{n_{max}} \left(\frac{a}{r}\right)^{n+1} \sum_{m=0}^{n} P_n^m(\cos\theta) \left[g_n^m \cos(m\phi) + h_n^m \sin(m\phi)\right] \ . \tag{C.2}$$

The decomposition is analogous to the one used by spectral methods (see Appendix D), but with another normalization and using real numbers only rather than complex. Note that here we indicate the degree of spherical harmonic is $n$ instead of $\ell$, for a simple notation consistent with most of the planetary magnetism literature. To obtain the magnetic field strength, one needs to use the gradient of $V$ in spherical coordinates:

$$B_{r,\text{int}} = -\frac{\partial V_{\text{int}}}{\partial r} =$$
$$= \sum_{n=1}^{n_{max}} \left(\frac{a}{r}\right)^{n+2} (n+1) \sum_{m=0}^{n} P_n^m(\cos\theta) \left[g_n^m \cos(m\phi) + h_n^m \sin(m\phi)\right]$$

$$B_{\theta,\text{int}} = -\frac{1}{r}\frac{\partial V_{\text{int}}}{\partial \theta} =$$
$$= -\sum_{n=1}^{n_{max}} \left(\frac{a}{r}\right)^{n+2} \sum_{m=0}^{n} \frac{\partial P_n^m(\cos\theta)}{\partial \theta} \left[g_n^m \cos(m\phi) + h_n^m \sin(m\phi)\right]$$

$$B_{\phi,\text{int}} = -\frac{1}{r\sin\theta}\frac{\partial V_{\text{int}}}{\partial \phi} =$$
$$= -\frac{1}{\sin\theta}\sum_{n=1}^{n_{max}} \left(\frac{a}{r}\right)^{n+2} \sum_{m=0}^{n} m P_n^m(\cos\theta) \left[-g_n^m \sin(m\phi) + h_n^m \cos(m\phi)\right]$$

The same can be repeated for the external sources:

$$V_{\text{ext}} = a\sum_{n=1}^{n_{max}} \left(\frac{r}{a}\right)^{n} \sum_{m=0}^{n} P_n^m(\cos\theta) \left[G_n^m \cos(m\phi) + H_n^m \sin(m\phi)\right] \tag{C.3}$$

$$B_{r,\text{ext}} = -\frac{\partial V_{\text{ext}}}{\partial r} =$$
$$= -\sum_{n=1}^{n_{max}} \left(\frac{r}{a}\right)^{n-1} n \sum_{m=0}^{n} P_n^m(\cos\theta) \left[G_n^m \cos(m\phi) + H_n^m \sin(m\phi)\right]$$

$$B_{\theta,\text{ext}} = -\frac{1}{r}\frac{\partial V_{\text{ext}}}{\partial \theta} =$$
$$= -\sum_{n=1}^{n_{max}} \left(\frac{r}{a}\right)^{n-1} \sum_{m=0}^{n} \frac{\partial P_n^m(\cos\theta)}{\partial \theta} \left[G_n^m \cos(m\phi) + H_n^m \sin(m\phi)\right]$$

$$B_{\phi,\text{ext}} = -\frac{1}{r\sin\theta}\frac{\partial V_{\text{ext}}}{\partial \phi} =$$
$$= -\frac{1}{\sin\theta}\sum_{n=1}^{n_{max}} \left(\frac{r}{a}\right)^{n-1} \sum_{m=0}^{n} m P_n^m(\cos\theta) \left[-G_n^m \sin(m\phi) + H_n^m \cos(m\phi)\right]$$

With the magnetic field components, the declination and inclination can be obtained. The former is the deviation angle from the true north, and the latter is the angle between the magnetic field and the planetary surface. Their respective mathematical expressions are:

$$D = \arctan\left(-\frac{B_\phi}{B_\theta}\right) \tag{C.4}$$

$$I = \arctan\left(-\frac{B_r}{\sqrt{B_\phi^2 + B_\theta^2}}\right) \tag{C.5}$$

## C.1 Schmidt quasi-normalized associated Legendre polynomials

To use the expressions above, the Schmidt quasi-normalized associated Legendre polynomials must be obtained. The original Legendre polynomials can be defined as a complete and orthogonal set of polynomials, or via a generating function, or the solution of a differential equation:

$$\int_{-1}^{1} P_m(x)P_n(x)\,dx = 0 \quad \text{if} \quad n \neq m$$

$$\frac{1}{\sqrt{1-2xt+t^2}} = \sum_{n=0}^{\infty} P_n(x)t^n$$

$$(1-x^2)P_n''(x) - 2xP_n'(x) + n(n+1)P_n(x) = 0$$

These implicit definitions are not useful for obtaining a given polynomial, but there is a compact expression known as Rodrigues' formula, which can be derived:

$$P_n(x) = \frac{1}{2^n n!}\frac{d^n}{dx^n}(x^2-1)^n \tag{C.6}$$

The associated Legendre polynomials are the canonical solutions of the general Legendre equation, which is a bit more complicated than the previous differential equations. These polynomials can be defined as:

$$P_{n,m}(x) = (x^2-1)^{1/2m}\frac{d^m}{dx^m}P_n(x) \tag{C.7}$$

Therefore, the associated Legendre polynomials are zero for all m greater than n. These expressions have not been normalized, and the most common normalization is the Gaussian normalization obtained by multiplying by a factor:

$$P^{n,m}(x) = \frac{2^n!(n-m)!}{(2n)!}P_{n,m}(x) \tag{C.8}$$

Then the Schmidt quasi-normalized associated Legendre polynomials can be obtained with another factor $S_{n,m}$:

$$P_n^m(x) = S_{n,m}P^{n,m}(x) = \left(\frac{(2-\delta_m^0)(n-m)!}{(n+m)!}\right)^{1/2}\frac{(2n-1)!!}{(n-m)!}P^{n,m}(x) \qquad m \neq 0 \tag{C.9}$$

Another way to go directly to the Schmidt polynomials is from the associated Legendre functions themselves as follows:

$$\begin{aligned} P_n^m(x) &= P_n(x) && m = 0 \\ P_n^m(x) &= \left(\frac{2(n-m)!}{(n+m)!}\right)^{1/2}P_{n,m}(x) && m \neq 0 \end{aligned} \tag{C.10}$$

$$P_n^m(x) = \left((2-\delta_m^0)\frac{(n-m)!}{(n+m)!}\right)^{1/2}P_{n,m}(x) \tag{C.11}$$

There are recurrent formulas for the original associated polynomials and for both normalizations. But for coding purposes, the easiest ones to implement are the Gaussian normalized associated Legendre polynomials. These recursive expressions are:

$$\begin{aligned} P^{0,0} &= 1 \\ P^{n,n} &= \sin\theta P^{n-1,n-1} \\ P^{n,m} &= \cos\theta P^{n-1,m} - K^{n,m}P^{n-2,m} \end{aligned} \tag{C.12}$$

$$\begin{aligned} K^{n,m} &= 0, && n = 1 \\ K^{n,m} &= \frac{(n-1)^2 - m^2}{(2n-1)(2n-3)}, && n > 1 \end{aligned} \tag{C.13}$$

The Schmidt quasi-normalization factors can be found recursively using the following formulas:

$$\begin{aligned} S_{0,0} &= 1 \\ S_{n,0} &= \left(\frac{2n-1}{n}\right)S_{n-1,0} \\ S_{n,m} &= \left(\frac{(n-m+1)(\delta_m^1+1)}{n+m}\right)^{1/2}S_{n,m-1} \end{aligned} \tag{C.14}$$

Now, to obtain $B_\theta$, we need the derivatives of the polynomials; thus, Eqs. C.15 are not enough. But they can be derived, and another set of recursive formulas can be used:

$$\begin{aligned}\frac{\partial P^{0,0}}{\partial\theta} &= 0\\ \frac{\partial P^{n,n}}{\partial\theta} &= \sin\theta\frac{\partial P^{n-1,n-1}}{\partial\theta} + \cos\theta P^{n-1,n-1}, \qquad n\geq 1\\ \frac{\partial P^{n,m}}{\partial\theta} &= \cos\theta\frac{\partial P^{n-1,m}}{\partial\theta} - \sin\theta P^{n-1,m} - K^{n,m}\frac{\partial P^{n-2,m}}{\partial\theta}\end{aligned} \tag{C.15}$$

All of the above recursive formulae have been used in Chapter 4 for a set of ($r$,$\theta$,$\phi$) points to recreate the magnetic field for all the dynamos for the planetary bodies in the solar system.

## C.2 Magnetic field curvature

The curvature of the magnetic field (or any vector field) is defined with its corresponding unitary vector field.

$$\boldsymbol{\kappa} = (\mathbf{b}\cdot\boldsymbol{\nabla})\mathbf{b} \qquad \text{where} \qquad \mathbf{b} = \frac{\mathbf{B}}{|\mathbf{B}|} \tag{C.16}$$

The magnetic unitary vector field and gradient in spherical coordinates are $\mathbf{b} = (b_r\mathbf{e}_r + b_\theta\mathbf{e}_\theta + b_\phi\mathbf{e}_\phi)$, $\boldsymbol{\nabla} = \partial_r\mathbf{e}_r + \partial_\theta\mathbf{e}_\theta/r + \partial_\phi\mathbf{e}_\phi/r\sin\theta$, respectively. Then the curvature is:

$$\boldsymbol{\kappa} = \left\{b_r\frac{\partial}{\partial r} + \frac{b_\theta}{r}\frac{\partial}{\partial\theta} + \frac{b_\phi}{r\sin\theta}\frac{\partial}{\partial\phi}\right\}(b_r\mathbf{e}_r + b_\theta\mathbf{e}_\theta + b_\phi\mathbf{e}_\phi) \qquad \Rightarrow$$

$$\begin{aligned}\kappa_r &= b_r\frac{\partial b_r}{\partial r} + \frac{b_\theta}{r}\frac{\partial b_r}{\partial\theta} + \frac{b_\phi}{r\sin\theta}\frac{\partial b_r}{\partial\phi} - \frac{b_\theta^2 + b_\phi^2}{r}\\ \kappa_\theta &= b_r\frac{\partial b_\theta}{\partial r} + \frac{b_\theta}{r}\frac{\partial b_\theta}{\partial\theta} + \frac{b_\phi}{r\sin\theta}\frac{\partial b_\theta}{\partial\phi} + \frac{b_\theta b_r}{r} - \frac{b_\phi^2}{rtan\theta}\\ \kappa_\phi &= b_r\frac{\partial b_\phi}{\partial r} + \frac{b_\theta}{r}\frac{\partial b_\phi}{\partial\theta} + \frac{b_\phi}{r\sin\theta}\frac{\partial b_\phi}{\partial\phi} + \frac{b_\phi b_r}{r} + \frac{b_\phi b_\theta}{rtan\theta}\end{aligned} \tag{C.17}$$

Note that the non-vanishing unit vector derivatives have been taken into account:

$$\begin{aligned}&\frac{\partial}{\partial r}\mathbf{e}_r = 0 \qquad \frac{\partial}{\partial\theta}\mathbf{e}_r = \mathbf{e}_\theta \qquad \frac{\partial}{\partial\phi}\mathbf{e}_r = \sin\theta\mathbf{e}_\phi\\ &\frac{\partial}{\partial r}\mathbf{e}_\theta = 0 \qquad \frac{\partial}{\partial\theta}\mathbf{e}_\theta = -\mathbf{e}_r \qquad \frac{\partial}{\partial\phi}\mathbf{e}_\theta = \cos\theta\mathbf{e}_\phi\\ &\frac{\partial}{\partial r}\mathbf{e}_\phi = 0 \qquad \frac{\partial}{\partial\theta}\mathbf{e}_\phi = 0 \qquad \frac{\partial}{\partial\phi}\mathbf{e}_\phi = -\sin\theta\mathbf{e}_r - \cos\theta\mathbf{e}_\theta\end{aligned}$$

If we want to obtain the mean curvature in a spherical surface of radius $r = r_\kappa$, we must integrate over the surface with the corresponding weight for the latitude:

$$\langle|\boldsymbol{\kappa}(r = r_\kappa)|\rangle = \frac{\int|\boldsymbol{\kappa}(\theta,\phi;r = r_\kappa)|\sin(\theta)d\theta d\phi}{\int\sin(\theta)d\theta d\phi} \tag{C.18}$$

For a given latitudinal-longitudinal resolution ($N_\theta \times N_\phi$) this becomes:

$$\langle|\boldsymbol{\kappa}(r = r_\kappa)|\rangle = \frac{\sum_{i=1}^{N_\theta}\sum_{j=1}^{N_\phi}|\boldsymbol{\kappa}(\theta_i,\phi_j;r = r_\kappa)|\sin(\theta_i)}{\sum_{i=1}^{N_\theta}\sum_{j=1}^{N_\phi}\sin(\theta_i)} = \frac{\sum_{i=1}^{N_\theta}\sum_{j=1}^{N_\phi}|\boldsymbol{\kappa}(\theta_i,\phi_j;r = r_\kappa)|\sin(\theta_i)}{N_\phi\sum_{i=1}^{N_\theta}\sin(\theta_i)} \tag{C.19}$$

# D

# *MagIC* numerical technique

This Appendix is largely based on two sections of the *MagIC* code manual, and we report the main points here for completeness. The numerical technique it employs was originally developed by P. Gilman and G. Glatzmaier, published first in the articles Glatzmaier (1984, 1985a,b). the method relies on spectral methods, which take advantage of the domain geometry (spherical shell) and a preferred direction (the radial one, along which thermodynamical gradients mostly lie).

Therefore, in this approach, the fields are expanded into complete sets of functions in radial and angular directions: Chebyshev polynomials or finite differences in the radial direction, and spherical harmonic functions over the tangential spheres (azimuthal and latitudinal directions). This allows one to express all partial derivatives analytically. Employing orthogonality relations of spherical harmonic functions and using collocation in radius then leads to algebraic equations that are integrated in time with a mixed implicit/explicit time stepping scheme. The nonlinear terms and the Coriolis force are evaluated in the physical (or grid) space rather than in spectral space. Although this approach requires costly numerical transformations between the two representations (from spatial to spectral using Legendre and Fourier transforms), the resulting decoupling of all spherical harmonic modes leads to a net gain in computational speed. Before explaining these methods in more detail, we introduce the poloidal/toroidal decomposition.

## D.1 Poloidal and toroidal decomposition

The poloidal–toroidal decomposition is a mathematical statement that can be applied to any divergence-free three-dimensional vector field $\mathbf{F}$:

$$\mathbf{F} = \mathbf{P} + \mathbf{T} = \boldsymbol{\nabla} \times (\boldsymbol{\nabla} \times \phi_{\rm pol}\, \mathbf{e_r}) + \boldsymbol{\nabla} \times \phi_{\rm tor}\, \mathbf{e_r} \ , \tag{D.1}$$

where $\phi_{\rm pol}$ and $\phi_{\rm tor}$ are the toroidal and poloidal potentials, respectively. This decomposition fulfills the solenoidal condition, i.e., $\boldsymbol{\nabla} \cdot \mathbf{F} = 0$, by construction. It is similar to the Helmholtz decomposition (see Sect. A.8) but more restricted in the sense that it further decomposes a solenoidal field, not any general vector field. In the spherical shell MHD anelastic approximation scenario one can avoid solving for the solenoidal condition of $\mathbf{B}$ and the continuity equation, $\boldsymbol{\nabla}(\tilde{\rho}\mathbf{u}) = 0$, if the corresponding potentials are evolved in time and not the vector fields themselves. In *MagIC*, these potentials are defined as:

$$\tilde{\rho}\mathbf{u} = \boldsymbol{\nabla} \times (\boldsymbol{\nabla} \times W\,\mathbf{e_r}) + \boldsymbol{\nabla} \times Z\,\mathbf{e_r}\,, \\ \mathbf{B} = \boldsymbol{\nabla} \times (\boldsymbol{\nabla} \times g\,\mathbf{e_r}) + \boldsymbol{\nabla} \times h\,\mathbf{e_r}\,. \tag{D.2}$$

Then one can recover $\mathbf{B}$ and $\tilde{\rho}\mathbf{u}$ in spherical coordinates:

$$\begin{aligned}\tilde{\rho}\mathbf{u} = &\left(-\frac{1}{r^2\sin\theta}\frac{\partial}{\partial\theta}\left(\sin\theta\frac{\partial W}{\partial\theta}\right) - \frac{1}{r^2\sin^2\theta}\frac{\partial^2 W}{\partial^2\phi}\right)\mathbf{e_r} + \\ &+\left(\frac{1}{r}\frac{\partial^2 W}{\partial r\partial\theta} + \frac{1}{r\sin\theta}\frac{\partial Z}{\partial\phi}\right)\mathbf{e}_\theta + \\ &+\left(\frac{1}{r\sin\theta}\frac{\partial^2 W}{\partial r\partial\phi} - \frac{1}{r}\frac{\partial Z}{\partial\theta}\right)\mathbf{e}_\phi\end{aligned} \tag{D.3}$$

$$\begin{aligned}\mathbf{B} = &\left(-\frac{1}{r^2\sin\theta}\frac{\partial}{\partial\theta}\left(\sin\theta\frac{\partial g}{\partial\theta}\right) - \frac{1}{r^2\sin^2\theta}\frac{\partial^2 g}{\partial^2\phi}\right)\mathbf{e_r} + \\ &+\left(\frac{1}{r}\frac{\partial^2 g}{\partial r\partial\theta} + \frac{1}{r\sin\theta}\frac{\partial h}{\partial\phi}\right)\mathbf{e}_\theta + \\ &+\left(\frac{1}{r\sin\theta}\frac{\partial^2 g}{\partial r\partial\phi} - \frac{1}{r}\frac{\partial h}{\partial\theta}\right)\mathbf{e}_\phi\end{aligned} \tag{D.4}$$

*MagIC* makes uses the horizontal part of the Laplacian, $\Delta_H$, and the horizontal part of the divergence, $\boldsymbol{\nabla}_H$:

$$\Delta_H = \frac{1}{r^2\sin\theta}\frac{\partial}{\partial\theta}\left(\sin\theta\frac{\partial}{\partial\theta}\right) + \frac{1}{r^2\sin^2\theta}\frac{\partial^2}{\partial^2\phi}.$$

$$\boldsymbol{\nabla}_H = \frac{1}{r\sin\theta}\frac{\partial}{\partial\theta}\sin\theta\,\boldsymbol{e_\theta} + \frac{1}{r\sin\theta}\frac{\partial}{\partial\phi}\,\boldsymbol{e_\phi}$$

With this operators, the expression for $\tilde{\rho}\mathbf{u}$ and $\mathbf{B}$ are shortened to:

$$\tilde{\rho}\mathbf{u} = -\Delta_H\,\boldsymbol{e_r}\,W + \boldsymbol{\nabla_H}\frac{\partial}{\partial r}W + \vec{\nabla}_H \times \boldsymbol{e_r}\,Z$$

$$\mathbf{B} = -\Delta_H\,\boldsymbol{e_r}\,g + \boldsymbol{\nabla_H}\frac{\partial}{\partial r}g + \vec{\nabla}_H \times \boldsymbol{e_r}\,h$$

The calculation of their curls then becomes fairly simple:

$$\nabla\times\tilde{\rho}\,\mathbf{u} = -\Delta_H\,\mathbf{e_r}\,Z + \boldsymbol{\nabla_H}\frac{\partial}{\partial r}Z - \boldsymbol{\nabla_H}\times\Delta_H\mathbf{e_r}\,W\,.$$

$$\nabla\times\,\mathbf{B} = -\Delta_H\,\mathbf{e_r}\,h + \boldsymbol{\nabla_H}\frac{\partial}{\partial r}h - \boldsymbol{\nabla_H}\times\Delta_H\mathbf{e_r}\,g\,.$$

## D.2 Spectral schemes in the horizontal and radial directions

A finite difference code would define a spherical grid in which the four potentials are defined and evolved in time within some numerical scheme (e.g., for the *Pencil Code*, fourth-order difference in space and third-order Runge-Kutta in time). A more accurate but more computationally expensive option is decomposing the potentials into a set of orthonormal functions. In this case, the time evolution is performed in the weights themselves. A natural option for the horizontal expansion in $\theta$-$\phi$ are the spherical harmonics functions $Y_l^m$:

$$g(r,\theta,\phi) = \sum_{\ell=0}^{\ell_{max}}\sum_{m=-\ell}^{\ell} g_{\ell m}(r)\,Y_\ell^m(\theta,\phi) \tag{D.5}$$

where $\ell$ and $m$ denote spherical harmonic degree and order, respectively. A finite difference or volume method needs to define a grid with some specific resolution, but a spectral method needs to truncate the infinite set of functions, which in this case is done at some degree and order $l_{max}$.

In *MagIC*, the radial dependence of the potential can be chosen to use a finite difference method or a spectral method which uses Chebyshev polynomials $\mathcal{T}$. In this thesis, we have only used the latter, which also needs to be truncated to some degree $N$:

$$g_{\ell m}(r) = \sum_{n=0}^{N} g_{\ell mn}\, \mathcal{T}_n(r) \tag{D.6}$$

So, finally, the complete decomposition of the magnetic poloidal potential is:

$$g(t, r, \theta, \phi) = \sum_{\ell=0}^{\ell_{max}} \sum_{m=-\ell}^{\ell} \sum_{n=0}^{N} g_{\ell mn}(t)\, \mathcal{T}_n(r)\, Y_\ell^m(\theta, \phi) \tag{D.7}$$

The set of weights $g_{\ell mn}$ and $h_{\ell mn}$ completely describes $\mathbf{B}$ up to the chosen resolution. In Eq. D.7 the time dependence has been included, showing that the weights themselves are the quantities being evolved in time by solving the MHD equations under your chosen approximation, i.e. in *MagIC* you can solve under either the Boussinesq or anelastic approximations depending if a density difference has been introduced. For simplicity, below this point, we omit the ranges of $l, m, n$ and shorten the triple sums under one summation symbol, as follows:

$$\begin{aligned}
g(t, r, \theta, \phi) &= \sum_{l,m,n} g_{\ell mn}(t)\, \mathcal{T}_n(r)\, Y_\ell^m(\theta, \phi) \\
h(t, r, \theta, \phi) &= \sum_{l,m,n} h_{\ell mn}(t)\, \mathcal{T}_n(r)\, Y_\ell^m(\theta, \phi) \\
W(t, r, \theta, \phi) &= \sum_{l,m,n} W_{\ell mn}(t)\, \mathcal{T}_n(r)\, Y_\ell^m(\theta, \phi) \\
Z(t, r, \theta, \phi) &= \sum_{l,m,n} Z_{\ell mn}(t)\, \mathcal{T}_n(r)\, Y_\ell^m(\theta, \phi)
\end{aligned} \tag{D.8}$$

The fields $\mathbf{B}$ and $\mathbf{u}$ can be recovered at any time-step by plugging the weights in Eq. D.2:

$$\begin{aligned}
\tilde{\rho}\mathbf{u} = &\left( \frac{1}{r^2} \sum_{l,m,n} l\,(l+1)\, W_{\ell mn}\, \mathcal{T}_n\, Y_\ell^m \right) \mathbf{e_r} + \\
+ &\left( \frac{1}{r} \sum_{l,m,n} W_{\ell mn} \frac{\partial \mathcal{T}_n}{\partial r} \frac{\partial Y_\ell^m}{\partial \theta} + \frac{1}{r \sin\theta} \sum_{l,m,n} Z_{\ell mn}\, \mathcal{T}_n \frac{\partial Y_\ell^m}{\partial \phi} \right) \mathbf{e}_\theta + \\
+ &\left( \frac{1}{r \sin\theta} \sum_{l,m,n} W_{\ell mn} \frac{\partial \mathcal{T}_n}{\partial r} \frac{\partial Y_\ell^m}{\partial \phi} - \frac{1}{r} \sum_{l,m,n} Z_{\ell mn}\, \mathcal{T}_n \frac{\partial Y_\ell^m}{\partial \theta} \right) \mathbf{e}_\phi
\end{aligned} \tag{D.9}$$

$$\begin{aligned}
\mathbf{B} = &\left( \frac{1}{r^2} \sum_{l,m,n} l\,(l+1)\, g_{\ell mn}\, \mathcal{T}_n\, Y_\ell^m \right) \mathbf{e_r} + \\
+ &\left( \frac{1}{r} \sum_{l,m,n} g_{\ell mn} \frac{\partial \mathcal{T}_n}{\partial r} \frac{\partial Y_\ell^m}{\partial \theta} + \frac{1}{r \sin\theta} \sum_{l,m,n} h_{\ell mn}\, \mathcal{T}_n \frac{\partial Y_\ell^m}{\partial \phi} \right) \mathbf{e}_\theta + \\
+ &\left( \frac{1}{r \sin\theta} \sum_{l,m,n} g_{\ell mn} \frac{\partial \mathcal{T}_n}{\partial r} \frac{\partial Y_\ell^m}{\partial \phi} - \frac{1}{r} \sum_{l,m,n} h_{\ell mn}\, \mathcal{T}_n \frac{\partial Y_\ell^m}{\partial \theta} \right) \mathbf{e}_\phi
\end{aligned} \tag{D.10}$$

To simplify the radial component, the following spherical harmonic identity has been used:

$$-\Delta_H Y_\ell^m = \frac{\ell(\ell+1)}{r^2}\, Y_\ell^m.$$

With these expressions, *MagIC* reconstructs the velocity and magnetic fields from the potentials at any given point in the simulation and creates the computer format outputs. Such files have been used to recreate Fig. 5.4 and other visuals.

Similar expansions are needed for pressure $p$, entropy or temperature $s$, and chemical composition $\xi$:

$$\begin{aligned} p(t,r,\theta,\phi) &= \sum_{l,m,n} p_{\ell mn}(t)\,\mathcal{T}_n(r)\,Y_\ell^m(\theta,\phi) \\ s(t,r,\theta,\phi) &= \sum_{l,m,n} s_{\ell mn}(t)\,\mathcal{T}_n(r)\,Y_\ell^m(\theta,\phi) \\ \xi(t,r,\theta,\phi) &= \sum_{l,m,n} \xi_{\ell mn}(t)\,\mathcal{T}_n(r)\,Y_\ell^m(\theta,\phi) \end{aligned} \tag{D.11}$$

These expansions are substituted in the MHD equations applied to convection in rotating spherical shells, shown in Chapter 4 either under the Boussinesq or anelastic approximation. Therefore the time evolution of the MHD system is given by the new equations for $g_{\ell mn}$, $h_{\ell mn}$, $W_{\ell mn}$, $Z_{\ell mn}$, $p_{\ell mn}$, $s_{\ell mn}$ and $\xi_{\ell mn}$. All terms are easily calculable except the non-linear terms (i.e., the advection, buoyancy, and Coriolis terms in the momentum equation, and the induction term). To avoid the complexity of computing convolutions in spectral space directly, the fields are transformed from spectral space to real space, where the non-linear products are calculated pointwise, and then transformed back to spectral space. This introduces numerical errors due to spurious high-wavenumber interactions known as aliasing errors, which MagIC prevents with dealias algorithms. See further details in https://magic-sph.github.io/.

## D.3 Spherical harmonics

A natural choice for the $\theta - \phi$ expansion is the spherical harmonics $Y_\ell^m(\theta,\phi) \propto P_\ell^m(\cos\theta)\,e^{im\phi}$, where $P_\ell^m$ are the associated Legendre polynomials[1]. *MagIC* uses the following normalization:

$$Y_\ell^m(\theta,\phi) = \left(\frac{(2\ell+1)}{4\pi}\frac{(\ell-|m|)!}{(\ell+|m|)!}\right)^{1/2} P_\ell^m(\cos\theta)\,e^{im\phi},$$

so that the orthogonality relation is:

$$\int_0^{2\pi} d\phi \int_0^{\pi} \sin\theta\, d\theta\; Y_\ell^m(\theta,\phi)\, Y_{\ell'}^{m'}(\theta,\phi) \;=\; \delta_{\ell\ell'}\delta^{mm'}.$$

Then any potential mentioned above is expanded and truncated at a specific degree $\ell_{max}$ as expressed in Eq. D.5. The coefficients $g_{\ell m}(r)$ for the magnetic poloidal potential $g(r,\theta,\phi)$ are computed with the following integrals:

$$\begin{aligned} g_{\ell m}(r) &= \frac{1}{\pi}\int_0^{\pi} d\theta \sin\theta\; g_{\text{mag}}(r,\theta)\; P_\ell^m(\cos\theta), \\ g_m(r,\theta) &= \frac{1}{2\pi}\int_0^{2\pi} d\phi\; g(r,\theta,\phi)\; e^{-im\phi}. \end{aligned}$$

Note that for the potentials to be real functions, each coefficient needs to equal its complex conjugate for an inverse value of $m$: $g^{\star}_{\ell m}(r) = g_{\ell,-m}(r)$. The same expressions are similar for all other potentials $h_{\ell m}(r)$, $W_{\ell m}(r)$ and $Z_{\ell m}(r)$), the pressure $p_{\ell m}(r)$, entropy or temperature $s_{\ell m}(r)$, and chemical composition $\xi_{\ell m}(r)$. These expressions provide the transforms from the longitude/latitude representation to the spherical harmonic representation and its inverse.

[1] Attention! They are not the same as the Schmidt semi-normalized associated Legendre polynomials shown in Chapter 4 and Appendix C. They are also defined in the same Appendix C, Eq. C.7.

## D.4 Chebyshev polynomials

Chebyshev polynomials are widely used as an orthogonal basis in a finite one-dimensional domain. The following relation is the most straightforward way to define the polynomials:

$$\mathcal{T}_n(\cos\theta) \equiv \cos(n\theta) = \sum_{i=0}^{n} a_i cos^i(\theta) \quad \Rightarrow \quad \mathcal{T}_n(x) \equiv \cos(n \text{ arccos}\, x)$$

Note that, by construction, $\mathcal{T}_n(x)$ is only defined from (-1 < x < 1). Some other interesting properties are that a Chebyshev polynomial $\mathcal{T}_n(x)$ has exactly $n$ distinct zeros; $\mathcal{T}_{even}(x)$ is an even function; $\mathcal{T}_{odd}(x)$ is an odd function; $|\mathcal{T}_n(x)|$ has $n+1$ number of maxima separated by zeros and they take the value of 1; a simple function composition relation, i.e., $\mathcal{T}_m(\mathcal{T}_n(x)) = \mathcal{T}_{n\cdot m}(x)$; and the leading coefficient of $\mathcal{T}_n(x)$ is $2^{n-1}$.

With some trigonometric properties, one can derive their recursive formula:

$$\mathcal{T}_n(x) = 2x\,\mathcal{T}_{n-1}(x) - \mathcal{T}_{n-2}(x)$$

Proof:

$$\mathcal{T}_{n+1}(x) = \cos((n+1)\theta) = \cos(n\theta + \theta) = \cos\, n\theta\, \cos\, \theta - \sin\, n\theta\, \sin\, \theta$$

$$\mathcal{T}_{n-1}(x) = \cos((n-1)\theta) = \cos(n\theta - \theta) = \cos\, n\theta\, \cos\, \theta + \sin\, n\theta\, \sin\, \theta$$

$$\mathcal{T}_{n+1}(x) + \mathcal{T}_{n-1}(x) = 2\cos\, n\theta\, \cos\, \theta = 2x\mathcal{T}_n(x)$$

Which can be used to construct the series of polynomials: $\mathcal{T}_0(x) = 1$, $\mathcal{T}_1(x) = x$, $\mathcal{T}_2(x) = 2x^2 - 1$, $\mathcal{T}_3(x) = 4x^3 - 3x$, etc. These polynomials are called Chebyshev polynomials of the first kind, which *MagIC* employs. Interestingly, they can also be similarly defined with hyperbolic trigonometric functions, i.e., $\mathcal{T}_n(\cosh\theta) \equiv \cosh(n\theta)$.

Any two $\mathcal{T}_n(x)$ and $\mathcal{T}_m(x)$ obey an "almost" orthogonal relation with the weight function $1/\sqrt{1-x^2}$:

$$\int_{-1}^{1} \mathcal{T}_n(x)\mathcal{T}_{\text{mag}}(x)\frac{1}{\sqrt{1-x^2}}dx = -\int_{\pi}^{0} \mathcal{T}_n(\cos\theta)\mathcal{T}_{\text{mag}}(\cos\theta)d\theta =$$
$$= \int_{0}^{\pi} \cos(n\theta)\cos(m\theta)d\theta = \frac{1}{2}\int_{0}^{\pi} [\cos((n-m)\theta) + \cos((n+m)\theta)]$$

$$\int_{-1}^{1} \mathcal{T}_n(x)\mathcal{T}_{\text{mag}}(x)\frac{\mathrm{d}x}{\sqrt{1-x^2}} = \begin{cases} 0, & \text{if } n \neq m, \\ \pi, & \text{if } n = m = 0, \\ \frac{\pi}{2}, & \text{if } n = m \neq 0. \end{cases}$$

The other kind of Chebyshev polynomials is defined by the relation:

$$U_{n-1}(\cos\theta)\sin\theta \equiv \sin(n\theta) \quad \rightarrow \quad U_n(x) = \frac{\sin((n+1)\text{ arccos}\, x)}{\sin(\text{arccos}\, x)}$$

Which leads to $U_0(x) = 1$, $U_1(x) = 2x$, $U_2(x) = 4x^2 - 1$. Both families of polynomials are the solution of the Chebyshev differential equation:

$$(1-x^2)\frac{d^2y}{dx^2} - x\frac{dy}{dx} + n^2y = 0$$

where $n$ is a positive integer. By using the substitution $x = cost$:

$$\frac{d^2y}{dt^2} + n^2y = 0$$

which has a general direct solution: $y(t) = A\ \cos(nt) + B\ \sin(nt)$ or equivalently $y(x) = A\ \cos(n\ \arccos x) + B\ \sin(n\ \arccos x)$. This is another valid definition for Chebyshev polynomials.

Similarly, they can be defined in a third way with a generating function:

$$\frac{1 - tx}{1 - 2tx - t^2} = \sum_{n=0}^{\infty} \mathcal{T}_n(x) t^n$$

As they are an infinite set of orthogonal functions, they are commonly used as a basis for a general single-valued continuous function defined in (-1 < x < 1):

$$f(x) = \sum_{n=0}^{\infty} a_n \mathcal{T}_n(x) \quad \text{with} \quad a_n = \frac{2 - \delta_{n0}}{\pi} \int_{-1}^{1} \frac{f(x)\mathcal{T}_n(x)dx}{\sqrt{1 - x^2}}$$

This decomposition is shown in D.6 for *MagIC*, which then to reconstruct each coefficient $g_{\ell mn}$ a similar expression is used:

$$g_{\ell mn} = \frac{2 - \delta_{n0}}{\pi} \int_{-1}^{1} \frac{dx\, g_{\ell m}(r(x))\ \mathcal{C}_n(x)}{\sqrt{1 - x^2}}$$

where the range (-1 < x < 1) needs to be linearly mapped to $(r_i \leq r \leq r_o)$ by

$$x(r) = 2\frac{r - r_i}{r_o - r_i} - 1.$$

Again, the same is repeated for the other potentials $h_{\ell mn}$, $W_{\ell mn}$ and $Z_{\ell mn}$, the pressure $p_{\ell mn}$, entropy or temperature $s_{\ell mn}$, and chemical composition $\xi_{\ell mn}$.

## D.5 Boundary conditions

As mentioned in Chapters 5 and 7, we use the *MagIC* code under a variety of boundary conditions. For the magnetic field, we try insulating, a perfect conductor, or pseudovacuum conditions. These mathematical statements can be translated into requirements for the constant $g_{\ell m}$ and $h_{\ell m}$ for the magnetic field, and $Z_{\ell m}$ and $W_{\ell m}$ for the velocity field:

Magnetic boundary conditions

Insulating: $$\frac{\partial g_{\ell m}}{\partial r} + \frac{\ell}{r} g_{\ell m} = 0 \quad ; \quad \frac{\partial h_{\ell m}}{\partial r} = 0$$

Perfect conductor: $$g_{\ell m} = \frac{\partial^2 g_{\ell m}}{\partial r^2} = 0 \qquad \frac{\partial h_{\ell m}}{\partial r} = 0$$

Pseudo vacuum: $$\frac{\partial g_{\ell m}}{\partial r} = 0 \qquad h_{\ell m} = 0$$

Mechanical boundary conditions

No radial velocity (always): $$W_{\ell m} = 0$$

No-slip: $$\frac{\partial W_{\ell m}}{\partial r} = 0 \quad ; \quad Z_{\ell m} = 0$$

Stress-free: $$\left(\frac{\partial^2}{\partial r^2} - \frac{2}{r}\frac{\partial}{\partial r}\right) W_{\ell m} = 0 \quad ; \quad \left(\frac{\partial}{\partial r} - \frac{2}{r}\right) Z_{\ell m} = 0$$

The thermal boundary conditions are only applied as constant fixed values at the boundaries: $s$ = const. at $r_i$ and $r_o$. There is the possibility of setting a fixed flux boundary condition by setting $\partial s/\partial r$ = const., but it is not used in Chapters 5 or 7.

## D.6 Background thermodynamical profiles in *MagIC*

*MagIC* gives different available options for the radial dependency of the gravity profile: either a constant, $\sim r$, $\sim 1/r^2$, or any linear combination. With a specific gravity expression together with a value for $m$, one can derive the specific analytical profiles readily available in *MagIC*. Apart from this radial implementation, there are other internal background profiles resembling particular planets or stars, some being isentropic and some not. By default, *MagIC* uses an ideal equation of state to give the radial dependence:

$$\frac{1}{\rho}\partial\rho = -\alpha\partial T + \beta\partial\rho - \delta\partial\xi$$

Where we have introduced the thermal expansivity, $\alpha$, compressibility coefficient, $\beta$ and composition parameter, $\delta$:

$$\alpha = -\frac{1}{\rho}\left(\frac{\partial\rho}{\partial T}\right)_{P,\xi} = \frac{1}{V}\left(\frac{\partial V}{\partial T}\right)_{P,\xi}$$

$$\beta = \frac{1}{\rho}\left(\frac{\partial\rho}{\partial P}\right)_{T,\xi} = -\frac{1}{V}\left(\frac{\partial V}{\partial P}\right)_{T,\xi}$$

$$\delta = -\frac{1}{\rho}\left(\frac{\partial\rho}{\partial\xi}\right)_{P,T}$$

Chemical buoyancy has not been considered at any point in this thesis; thus, all terms containing $\xi$ should be dropped. To derive the associated radial background:

$$\frac{\boldsymbol{\nabla}T}{T} = \frac{1}{T}\left(\frac{\partial T}{\partial p}\right)_s \boldsymbol{\nabla}p + \frac{1}{T}\left(\frac{\partial T}{\partial s}\right)_p \boldsymbol{\nabla}s = -\frac{\rho}{T}\left(\frac{\partial T}{\partial p}\right)_s \boldsymbol{g} + \frac{1}{T}\left(\frac{\partial T}{\partial s}\right)_p \boldsymbol{\nabla}s$$

We have used the hydrostatic equation $\boldsymbol{\nabla}p = -\rho\boldsymbol{g}$. We can make use of different thermodynamic identities to simplify the equation. The first is the usual definition of entropy, assuming heat transfer at constant pressure:

$$\partial s = \frac{\delta q}{T} = \frac{\partial H}{T} = \frac{\partial H}{\partial T}\frac{\partial T}{T} = c_p\frac{\partial T}{T} \quad \rightarrow \quad \left(\frac{\partial s}{\partial T}\right)_p = \frac{c_p}{T} \quad \rightarrow \quad \frac{1}{T}\left(\frac{\partial T}{\partial s}\right)_p = \frac{1}{c_p}$$

The other term is trickier:

$$\frac{\rho}{T}\left(\frac{\partial T}{\partial p}\right)_s = \frac{\rho}{T}\frac{1}{\left(\dfrac{\partial p}{\partial T}\right)_s} = \frac{\rho}{T}\frac{1}{\left(\dfrac{\partial s}{\partial T}\right)_p\left(-\dfrac{\partial p}{\partial s}\right)_T} = \frac{\rho}{T}\frac{1}{\dfrac{c_p}{T}\left(\dfrac{\partial T}{\partial V}\right)_P} = \frac{\rho}{c_p}\frac{1}{\dfrac{1}{\alpha V}} = \frac{\alpha}{c_p}$$

We have used a cyclic relation between partial differential thermodynamic variables and a Maxwell relation. The final equation for the temperature gradient is:

$$\frac{\boldsymbol{\nabla}T}{T} = -\frac{\alpha}{c_p}\boldsymbol{g} + \frac{1}{c_p}\boldsymbol{\nabla}s \tag{D.12}$$

Repeating the same process for the gradient of density:

$$\frac{\boldsymbol{\nabla}\rho}{\rho} = \frac{1}{\rho}\left(\frac{\partial\rho}{\partial p}\right)_s \boldsymbol{\nabla}p + \frac{1}{\rho}\left(\frac{\partial\rho}{\partial s}\right)_p \boldsymbol{\nabla}s = -\beta\rho\boldsymbol{g} + V\left(\frac{\partial(1/V)}{\partial s}\right)_p \boldsymbol{\nabla}s$$

Massaging the second term, again with both a cyclic and a Maxwell relation:

$$V\left(\frac{\partial(1/V)}{\partial s}\right)_p = -\frac{1}{V}\left(\frac{\partial V}{\partial s}\right)_p = -\frac{1}{V}\frac{1}{\left(\dfrac{\partial s}{\partial V}\right)_p} = -\frac{1}{V}\frac{1}{\left(\dfrac{\partial p}{\partial T}\right)_s} =$$

$$= \frac{1}{V} \frac{1}{\left(\frac{\partial s}{\partial T}\right)_p \left(\frac{\partial p}{\partial s}\right)_T} = \rho \frac{1}{\frac{c_p}{T}\frac{1}{\alpha V}} = \frac{\alpha T}{c_p}$$

Then the final equation for the density gradient reads:

$$\frac{\boldsymbol{\nabla}\rho}{\rho} = -\beta\rho\boldsymbol{g} - \frac{\alpha T}{c_p}\boldsymbol{\nabla}s \tag{D.13}$$

As spherical symmetry is assumed in *MagIC*, the background static quantities only depend on radius. Both equations become, respectively:

$$\frac{dT}{dr} = -\frac{\alpha T}{c_p}g + \frac{T}{c_p}\frac{ds}{dr} \quad ; \quad \frac{d\rho}{dr} = -\beta\rho^2 g - \frac{\alpha T\rho}{c_p}\frac{ds}{dr} \tag{D.14}$$

These equations are non-dimensionalized in the following way:

$$\frac{d\tilde{T}}{dr} = -Di\,\tilde{T}\tilde{\alpha}\tilde{g} + \epsilon_S\tilde{T}\frac{d\tilde{s}}{dr} \quad ; \quad \frac{d\tilde{\rho}}{dr} = -\frac{Di\,\tilde{\rho}\tilde{\alpha}\tilde{g}}{\Gamma_o\tilde{\Gamma}} - Co\,\epsilon_S\tilde{\rho}\tilde{\alpha}\tilde{T}\frac{d\tilde{s}}{dr},$$

where four dimensionless parameters are involved:

$$Di = \frac{\alpha_o g_o d}{c_p}, \quad Co = \alpha_o T_o, \quad \Gamma_o, \quad \epsilon_S = \frac{d}{c_p}\left|\frac{ds}{dr}\right|_{r_o}.$$

Another usual assumption is that convection is vigorous enough so that the entropy is constant.

$$\frac{d\tilde{T}}{dr} = -Di\,\tilde{T}\tilde{\alpha}\tilde{g} \quad ; \quad \frac{d\tilde{\rho}}{dr} = -\frac{Di\,\tilde{\rho}\tilde{\alpha}\tilde{g}}{\Gamma_o\tilde{\Gamma}}$$

Finally, assuming that the gas behaves ideally, i.e. $p = \rho T$ and $\alpha = 1/T$, we only need the temperature equation:

$$\frac{d\tilde{T}}{dr} = -Di\,\tilde{g}(r) \quad ; \quad \tilde{\rho} = \tilde{T}^m$$

where $m = 1/(\gamma - 1)$ and $\gamma = c_p/c_v$. From the classical equipartition theorem, the heat capacity ratio $\gamma$ is predicted to be $\gamma = 1 + 2/f$, where $f$ is the thermally accessible degrees of freedom. For an ideal gas, $f$ can only be as small as 3, with no upper limit. Therefore, the range for the adiabatic index is $1 < \gamma < 5/3$. Thus, the physically available range for the variable $m$ used in *MagIC* is $3/2 < m < \infty$.

At the beginning of the project shown in Chapter 5, we tried fitting the *MESA* thermodynamic properties to these expressions for $\tilde{T}$, $\tilde{\rho}$, and $\tilde{g}$. But due to the more complex *MESA* EoS, the hydrostatic profiles could not be fitted correctly. At the end, we opted for the procedure explained in Sect. 5.1.1: we locally expanded the *radial.f90* file to directly introduce as a high-dimensional Taylor series, any general background state from *MESA*, a one-dimensional evolutionary code.

## D.7 *MagIC* energy spectra

*MagIC* sums the potential weights over $\ell$ and $m$ and integrates over radius to calculate the total energy. For example, the kinetic energy:

$$E_{\text{kin}} = \frac{1}{2}\int_V \tilde{\rho}u^2\,\mathrm{d}V = E_{\text{k,pol}} + E_{k,tor}$$

$$E_{\text{kin,pol}} = \frac{1}{2}\sum_{\ell,m}\ell(\ell+1)\int_{r_i}^{r_o}\frac{1}{\tilde{\rho}}\left[\frac{\ell(\ell+1)}{r^2}|W_{\ell m}(r)|^2 + \left|\frac{\mathrm{d}W_{\ell m}(r)}{\mathrm{d}r}\right|^2\right]\mathrm{d}r$$

$$E_{\text{kin,tor}} = \frac{1}{2}\sum_{\ell,m}\ell(\ell+1)\int_{r_i}^{r_o}\frac{1}{\tilde{\rho}}|Z_{\ell m}(r)|^2\,\mathrm{d}r$$

This operation is done for each timestep that the user needs total averaged quantities.

As explained in Sect. A.9, the energy distribution describes how kinetic or magnetic energy is spatially distributed across flow structures of different scales. For a spectral code that uses spherical harmonics, the energy as a function of degree or order, i.e., the spectral distribution, comes for free from Eq. D.7 if one does not integrate radially. Computationally, the shortest way to reach an expression of the sort is to start with the energy defined at some radius $r_j$ in the radial grid and for some $\ell$ and $m$:

$$E_{\text{kin}}(r_j,\ell,m) = \frac{1}{2}\ell(\ell+1)\frac{1}{\tilde{\rho}}\left[\frac{\ell(\ell+1)}{r^2}|W_{\ell m}(r_j)|^2 + \left|\frac{\mathrm{d}W_{\ell m}}{\mathrm{d}r}\right|^2\right] + \frac{1}{2}\ell(\ell+1)\frac{1}{\tilde{\rho}}|Z_{\ell m}|^2$$

One can sum in $m$ ($\ell$) and obtain a spectra depending both on $r$ and $\ell$ ($m$):

$$E_{\text{kin}}(r_j,\ell) = \sum_m E_{\text{kin}}(r_j,\ell,m) \qquad E_{\text{kin}}(r_j,m) = \sum_\ell E_{\text{kin}}(r_j,\ell,m)$$

These 2D quantities are already spectrum-dependent on radius, and can be quite informative if there is strong radial dependence. These are used in Chapter 5 to help define the dynamo radius. The last step to obtain the usual $\ell$ or $m$ spectra is to numerically integrate in the radial direction:

$$E_{\text{kin}}(\ell) = \int_{r_i}^{r_o} E_{\text{kin}}(r,\ell)dr \quad ; \quad E_{\text{kin}}(m) = \int_{r_i}^{r_o} E_{\text{kin}}(r,m)dr$$

The integration can be done with more developed methods, and *MagIC*'s default option is Simpson's method. The same process can be repeated for the magnetic field, taking into account that there is no density factor:

$$E_{\text{mag}}(r_j,\ell,m) = \frac{1}{2}\ell(\ell+1)\left[\frac{\ell(\ell+1)}{r^2}|b_{\ell m}(r_j)|^2 + \left|\frac{\mathrm{d}b_{\ell m}}{\mathrm{d}r}\right|^2\right] + \frac{1}{2}\ell(\ell+1)|j_{\ell m}|^2$$

$$E_{\text{mag}}(r_j,\ell) = \sum_m E_{\text{mag}}(r_j,\ell,m) \qquad E_{\text{mag}}(r_j,m) = \sum_\ell E_{\text{mag}}(r_j,\ell,m)$$

$$E_{\text{mag}}(\ell) = \int_{r_i}^{r_o} E_{\text{mag}}(r,\ell)dr \quad ; \quad E_{\text{mag}}(m) = \int_{r_i}^{r_o} E_{\text{mag}}(r,m)dr$$

## D.8 Physical interpretation of the dimensionless units

As mentioned in Chapter 4, one needs to transition to dimensionless units to obtain the MHD equations under the anelastic approximation, which are controlled by the usual dimensionless dynamo parameters (e.g., Ra, E, Pr, and Pm). This transition sets the physical meaning for the new magnitudes for the code units. For example, *MagIC* uses:

$$\begin{gathered} r \to r\,d, \quad t \to t\,(d^2/\nu), \quad s \to s\,\Delta s, \quad \xi \to \xi\,\Delta\xi, \\ B \to B\,(\mu_0 f \lambda_i \tilde{\rho_o}\Omega)^{1/2}, \quad \rho \to \rho\,\rho_o, \quad p \to p\,(\rho_o\nu_o\Omega) \end{gathered} \tag{D.15}$$

which means in code units, a distance of 1 is equivalent to the shell thickness $d$; a unit of time time is equivalent to a viscous diffusion timescale $d^2/\nu$; entropy and chemical composition are measured in units of the total difference between the inner and outer boundaries ($\Delta s$, $\Delta\xi$); magnetic field with the factor $(\mu_0 f \lambda_i \tilde{\rho_o}\Omega)^{1/2}$; density in units of the outer density $\rho_o$; and pressure in units of $\rho_o\nu_o\Omega$.

Therefore, once the simulation is done, the most straightforward way to give physical meaning to *MagIC* units is to associate physical values for the shell thickness, $d$, and rotation frequency, $\Omega$:

$$[L] = d \qquad [t] = \frac{d^2}{\nu_o} = \frac{1}{\Omega \mathrm{E}} \qquad [u] = \frac{[L]}{[t]} = d\,\Omega\,\mathrm{E}$$

As we are interested in the magnetic field, it is more appropriate to adopt units of magnetic diffusion timescale, thus for the same corresponding physical number, there is a factor of $\nu/\mu$=Pm:

$$\begin{aligned} t_{\mathrm{phys}} &= \frac{d^2}{\nu} t_{\mathrm{visc,diff(code)}} = \frac{d^2}{\mu} \\ t_{\mathrm{mag,diff}} \quad \rightarrow \quad t_{\mathrm{magdiff}} &= \mathrm{Pm} t_{\mathrm{viscdiff(code)}} \end{aligned} \tag{D.16}$$

Then, to convert code units to physical units:

$$d_{\mathrm{phys}} = d\, d_{MagIC} \qquad t_{\mathrm{phys}} = \frac{t_{MagIC}}{\Omega \mathrm{E}} \qquad u_{\mathrm{phys}} = d\,\Omega\,\mathrm{E}\, u_{MagIC}$$

With these basic units, one can derive all other quantities. For example, the units of kinetic energy are:

$$E_{k,\mathrm{phys}} = \frac{1}{2} \rho_o u_{\mathrm{phys}}^2 d_{\mathrm{phys}}^3 E_{k,\mathrm{norm}} = \rho_o d^5 \mathrm{E}^2 \Omega^2 E_{k,MagIC}$$

Similarly, for magnetic energy:

$$E_{m,\mathrm{phys}} = \frac{B_{\mathrm{phys}}^2 d_{\mathrm{phys}}^3}{\mu_0} E_{m,\mathrm{norm}} = \rho_o \lambda_i \Omega d^3 E_{m,MagIC}$$

As both energies must have the same units, we can recover a relation between the dimensional energy factors, $E_{m,\mathrm{factor}}$ and $E_{k,\mathrm{factor}}$:

$$E_{m,\mathrm{factor}} = \rho_o \lambda_i \Omega d^3 = \rho_o \nu \Omega d^3 \frac{1}{\mathrm{Pm}} = \rho_o \frac{\mathrm{E}\Omega^2 d^5}{\mathrm{Pm}} = \frac{1}{\mathrm{PmE}} E_{k,\mathrm{factor}}$$

# E

# Planetary dynamos in evolving cold gas giants: inputs and diagnostics

Here we include the tables related to the simulations presented in Chapter 5. Table E.1 shows the input parameters for all cold Jupiter models. Table E.2 shows the output parameters used for Fig. 5.10 and 5.13.

Table E.1: Parameters of the 1D models *MESA* models for the $1M_J$ and $4M_J$ at different times and density cuts, and the corresponding values of $\eta = r_i/r_o$, $\chi_m = r_m/r_o$, the dimensionless mass of the shell $M = 4\pi \int_{r_i}^{r_o} r^2 \frac{\rho(r)}{\rho_o} dr$, and the dynamo parameters E, Ra (after calibrating them in one case as described in Sect. 5.1.2).

| Model | $N_\rho$ | $\rho_o$(g·cm$^{-3}$) | $\Delta T$ (K) | $T_o$ (K) | $P_o$ (kbar) | $r_o$(R$_J$) | $\eta$ | $M$ | $\chi_m$ | E | Ra |
|---|---|---|---|---|---|---|---|---|---|---|---|
| $1M_J$ 0.4 Gyr | 2.99 | 0.165 | 19782 | 4858 | 55.6 | 1.030 | 0.164 | 47.36 | 0.874 | $1.08\cdot10^{-5}$ | $1.27\cdot10^{9}$ |
| $1M_J$ 0.5 Gyr | 2.98 | 0.171 | 18847 | 4665 | 57.3 | 1.023 | 0.156 | 45.58 | 0.878 | $1.08\cdot10^{-5}$ | $1.22\cdot10^{9}$ |
| $1M_J$ 0.7 Gyr | 2.99 | 0.172 | 17345 | 4306 | 54.8 | 1.014 | 0.166 | 48.03 | 0.881 | $1.12\cdot10^{-5}$ | $1.05\cdot10^{9}$ |
| $1M_J$ 1 Gyr | 2.99 | 0.177 | 16158 | 4002 | 54.5 | 1.005 | 0.159 | 46.91 | 0.884 | $1.12\cdot10^{-5}$ | $9.81\cdot10^{8}$ |
| $1M_J$ 1.5 Gyr | 2.98 | 0.181 | 15004 | 3716 | 54.0 | 0.998 | 0.1628 | 47.47 | 0.884 | $1.15\cdot10^{-5}$ | $8.78\cdot10^{8}$ |
| $1M_J$ 2.1 Gyr | 2.99 | 0.181 | 13999 | 3428 | 51.0 | 0.992 | 0.169 | 49.61 | 0.887 | $1.18\cdot10^{-5}$ | $7.87\cdot10^{8}$ |
| $1M_J$ 2.8 Gyr | 2.99 | 0.184 | 13278 | 3220 | 50.0 | 0.987 | 0.170 | 49.77 | 0.890 | $1.20\cdot10^{-5}$ | $7.33\cdot10^{8}$ |
| $1M_J$ 3.5 Gyr | 2.99 | 0.187 | 12424 | 2983 | 49.2 | 0.981 | 0.171 | 50.07 | 0.889 | $1.21\cdot10^{-5}$ | $6.71\cdot10^{8}$ |
| $1M_J$ 5 Gyr | 2.99 | 0.190 | 11742 | 2779 | 48.3 | 0.975 | 0.172 | 52.29 | 0.892 | $1.23\cdot10^{-5}$ | $6.22\cdot10^{8}$ |
| $1M_J$ 6.5 Gyr | 2.98 | 0.192 | 11031 | 2591 | 47.3 | 0.970 | 0.183 | 53.35 | 0.892 | $1.28\cdot10^{-5}$ | $5.52\cdot10^{8}$ |
| $1M_J$ 10 Gyr | 3.00 | 0.193 | 10180 | 2327 | 44.7 | 0.964 | 0.184 | 53.35 | 0.894 | $1.30\cdot10^{-5}$ | $4.98\cdot10^{8}$ |
| $1M_J$ 1 Gyr | 2.30 | 0.351 | 14973 | 5220 | 240 | 0.967 | 0.165 | 26.48 | 0.919 | $1.23\cdot10^{-5}$ | $7.92\cdot10^{8}$ |
| $1M_J$ 1 Gyr | 3.68 | 0.0881 | 17093 | 3100 | 14.9 | 1.022 | 0.156 | 89.66 | 0.869 | $1.08\cdot10^{-5}$ | $1.10\cdot10^{9}$ |
| $1M_J$ 1 Gyr | 4.58 | 0.0358 | 17962 | 2231 | 3.46 | 1.034 | 0.155 | 214.0 | 0.860 | $1.05\cdot10^{-5}$ | $1.20\cdot10^{9}$ |
| $0.3M_J$ 5 Gyr | 1.10 | 0.545 | 3022 | 3754 | 563 | 0.650 | 0.275 | 16.89 | 0.879 | $3.61\cdot10^{-5}$ | $3.18\cdot10^{7}$ |
| $0.7M_J$ 5 Gyr | 2.98 | 0.144 | 8927 | 2243 | 25.1 | 0.944 | 0.190 | 53.69 | 0.843 | $1.37\cdot10^{-5}$ | $4.01\cdot10^{8}$ |
| $2M_J$ 5 Gyr | 2.98 | 0.338 | 20486 | 4195 | 198 | 1.015 | 0.167 | 50.13 | 0.955 | $1.12\cdot10^{-5}$ | $1.24\cdot10^{9}$ |
| $4M_J$ 5 Gyr | 2.98 | 0.671 | 38765 | 6762 | 1060 | 1.025 | 0.159 | 47.94 | 1.000 | $1.08\cdot10^{-5}$ | $2.49\cdot10^{9}$ |
| $4M_J$ 0.5 Gyr | 4.60 | 0.109 | 69086 | 5533 | 33.6 | 1.137 | 0.140 | 32.20 | 0.959 | $8.37\cdot10^{-6}$ | $6.50\cdot10^{9}$ |
| $4M_J$ 1 Gyr | 4.58 | 0.120 | 59273 | 4829 | 33.9 | 1.109 | 0.148 | 31.26 | 0.961 | $8.98\cdot10^{-5}$ | $5.02\cdot10^{9}$ |
| $4M_J$ 10 Gyr | 4.60 | 0.137 | 36481 | 2880 | 27.7 | 1.048 | 0.169 | 31.81 | 0.967 | $1.06\cdot10^{-5}$ | $2.42\cdot10^{9}$ |

Table E.2: Quantitative details of all models of Table E.1, which was focused on the background setup. Here we show, for all of them, the input non-dimensional numbers (E and Ra fixed as in Table E.1, and the Prandtl numbers Pm, Pr), and the diagnostics: $\Lambda$, $E_{\rm mag}$ (in code units), $E_{\rm kin}$ (in code units), dipolarity $f_{\rm dip}$, Ohmic fraction $f_{\rm ohm}$, buoyancy power $P_\nu$, power imbalance $f_P$. The values are time-averaged. The typical standard deviations, here omitted for the sake of space, are shown as error bars in Figs. 5.10, 5.13.

| Model | $N_\rho$ | E | Ra | Pm | Pr | Rm | $\Lambda$ | $E_{\rm mag}$ | $E_{\rm kin}$ | $f_{\rm dip}$ | $f_{\rm ohm}$ | $P_\nu$ | $f_P(\%)$ |
|---|---|---|---|---|---|---|---|---|---|---|---|---|---|
| $1M_J$ 0.2 Gyr | 2.99 | $1.01\cdot10^{-5}$ | $1.63\cdot10^{9}$ | 1 | 1 | 1107 | 1.38 | $5.42\cdot10^{6}$ | $3.44\cdot10^{7}$ | 0.0271 | 0.235 | $2.123\cdot10^{11}$ | 2.4 |
| $1M_J$ 0.4 Gyr | 2.99 | $1.08\cdot10^{-5}$ | $1.27\cdot10^{9}$ | 1 | 1 | 956 | 0.99 | $3.8\cdot10^{6}$ | $3.13\cdot10^{7}$ | 0.028 | 0.197 | $1.49\cdot10^{11}$ | 0.45 |
| $1M_J$ 0.5 Gyr | 2.98 | $1.08\cdot10^{-5}$ | $1.22\cdot10^{9}$ | 1 | 1 | 904 | 0.987 | $3.7\cdot10^{6}$ | $2.56\cdot10^{7}$ | 0.031 | 0.194 | $1.29\cdot10^{11}$ | 0.22 |
| $1M_J$ 0.7 Gyr | 2.99 | $1.12\cdot10^{-5}$ | $1.05\cdot10^{9}$ | 1 | 1 | 829 | 0.71 | $2.58\cdot10^{6}$ | $2.54\cdot10^{7}$ | 0.028 | 0.176 | $9.27\cdot10^{10}$ | 0.31 |
| $1M_J$ 1 Gyr | 2.99 | $1.12\cdot10^{-5}$ | $9.81\cdot10^{8}$ | 1 | 1 | 734 | 0.649 | $2.31\cdot10^{6}$ | $1.68\cdot10^{7}$ | 0.030 | 0.172 | $7.36\cdot10^{10}$ | 0.025 |
| $1M_J$ 1.5 Gyr | 2.98 | $1.15\cdot10^{-5}$ | $8.78\cdot10^{8}$ | 1 | 1 | 641 | 0.385 | $1.32\cdot10^{6}$ | $1.22\cdot10^{7}$ | 0.0267 | 0.141 | $5.42\cdot10^{10}$ | 0.062 |
| $1M_J$ 2.1 Gyr | 2.99 | $1.18\cdot10^{-5}$ | $7.87\cdot10^{8}$ | 1 | 1 | 525 | 0.44 | $1.29\cdot10^{6}$ | $8.61\cdot10^{6}$ | 0.045 | 0.191 | $4.10\cdot10^{10}$ | 0.066 |
| $1M_J$ 2.8 Gyr | 2.99 | $1.20\cdot10^{-5}$ | $7.33\cdot10^{8}$ | 1 | 1 | 439 | 0.490 | $1.52\cdot10^{6}$ | $7.1\cdot10^{6}$ | 0.104 | 0.216 | $3.32\cdot10^{10}$ | 0.18 |
| $1M_J$ 3.5 Gyr | 2.99 | $1.21\cdot10^{-5}$ | $6.71\cdot10^{8}$ | 1 | 1 | 400 | 0.39 | $1.85\cdot10^{6}$ | $6.16\cdot10^{6}$ | 0.114 | 0.207 | $2.93\cdot10^{10}$ | 0.26 |
| $1M_J$ 5 Gyr | 2.99 | $1.23\cdot10^{-5}$ | $6.22\cdot10^{8}$ | 1 | 1 | 370 | 0.51 | $1.94\cdot10^{6}$ | $4.89\cdot10^{6}$ | 0.130 | 0.232 | $2.32\cdot10^{10}$ | 0.20 |
| $1M_J$ 6.5 Gyr | 2.98 | $1.28\cdot10^{-5}$ | $5.52\cdot10^{8}$ | 1 | 1 | 323 | 0.53 | $2.57\cdot10^{6}$ | $4.40\cdot10^{6}$ | 0.114 | 0.247 | $2.15\cdot10^{10}$ | 0.079 |
| $1M_J$ 10 Gyr | 3.00 | $1.30\cdot10^{-5}$ | $4.97\cdot10^{8}$ | 1 | 1 | 274 | 0.558 | $2.08\cdot10^{6}$ | $3.65\cdot10^{6}$ | 0.130 | 0.250 | $1.51\cdot10^{10}$ | 0.043 |
| $1M_J$ 1 Gyr | 2.30 | $1.23\cdot10^{-5}$ | $7.92\cdot10^{8}$ | 1 | 1 | 1092 | 1.87 | $3.07\cdot10^{6}$ | $1.48\cdot10^{7}$ | 0.0057 | 0.283 | $8.74\cdot10^{10}$ | 0.18 |
| | 3.68 | $1.08\cdot10^{-5}$ | $1.10\cdot10^{9}$ | 1 | 1 | 461 | 0.172 | $1.38\cdot10^{6}$ | $1.54\cdot10^{7}$ | 0.053 | 0.101 | $5.10\cdot10^{10}$ | 0.31 |
| | 4.58 | $1.05\cdot10^{-5}$ | $1.20\cdot10^{9}$ | 1 | 1 | 301 | 0.072 | $1.37\cdot10^{6}$ | $1.58\cdot10^{7}$ | 0.041 | 0.066 | $3.83\cdot10^{10}$ | 0.43 |
| $0.3M_J$ 5 Gyr | 1.10 | $3.61\cdot10^{-5}$ | $3.18\cdot10^{7}$ | 1 | 1 | 213 | 0.48 | $1.49\cdot10^{5}$ | $8.3\cdot10^{5}$ | 0.030 | 0.28 | $9.9\cdot10^{8}$ | 4.4 |
| $0.7M_J$ 5 Gyr | 2.98 | $1.37\cdot10^{-5}$ | $4.01\cdot10^{8}$ | 1 | 1 | 213 | 0.300 | $1.41\cdot10^{6}$ | $2.55\cdot10^{6}$ | 0.131 | 0.305 | $5.31\cdot10^{9}$ | 0.033 |
| $2M_J$ 5 Gyr | 2.98 | $1.12\cdot10^{-5}$ | $1.24\cdot10^{9}$ | 1 | 1 | 1032 | 3.14 | $1.03\cdot10^{7}$ | $1.82\cdot10^{7}$ | 0.055 | 0.358 | $2.26\cdot10^{11}$ | 0.90 |
| $4M_J$ 5 Gyr | 2.98 | $1.08\cdot10^{-5}$ | $2.49\cdot10^{9}$ | 1 | 1 | 2203 | 14.0 | $3.57\cdot10^{7}$ | $5.98\cdot10^{7}$ | 0.0041 | 0.526 | $1.020\cdot10^{12}$ | 15 |

Table E.2: (continuation)

| Model | $N_\rho$ | E | Ra | Pm | Pr | Rm | $\Lambda$ | $E_{mag}$ | $E_{kin}$ | $f_{dip}$ | $f_{ohm}$ | $P_\nu$ | $f_P(\%)$ |
|---|---|---|---|---|---|---|---|---|---|---|---|---|---|
| $1M_J$ 0.5 Gyr | 2.98 | $1.08\cdot10^{-5}$ | $1.22\cdot10^{9}$ | 0.5 | 1 | 514 | 0.41 | $3.0\cdot10^{6}$ | $3.2\cdot10^{7}$ | 0.076 | 0.228 | $1.335\cdot10^{11}$ | 0.16 |
| | 2.98 | $1.08\cdot10^{-5}$ | $1.22\cdot10^{9}$ | 2 | 1 | 1790 | 1.93 | $3.50\cdot10^{6}$ | $2.47\cdot10^{7}$ | 0.0045 | 0.218 | $1.32\cdot10^{11}$ | 5.4 |
| | 2.98 | $1.08\cdot10^{-5}$ | $1.22\cdot10^{9}$ | 4 | 1 | 3250 | 5.30 | $4.86\cdot10^{6}$ | $2.27\cdot10^{7}$ | 0.00204 | 0.331 | $1.40\cdot10^{11}$ | 53 |
| | 2.98 | $1.08\cdot10^{-5}$ | $1.22\cdot10^{9}$ | 0.5 | 0.5 | 1190 | 2.11 | $1.50\cdot10^{7}$ | $8.10\cdot10^{7}$ | 0.020 | 0.292 | $5.78\cdot10^{11}$ | 6 |
| | 2.98 | $1.08\cdot10^{-5}$ | $1.22\cdot10^{9}$ | 1 | 0.5 | 2050 | 4.11 | $1.46\cdot10^{7}$ | $7.16\cdot10^{7}$ | 0.00934 | 0.332 | $5.66\cdot10^{11}$ | 11 |
| | 2.98 | $1.08\cdot10^{-5}$ | $1.22\cdot10^{9}$ | 1 | 2 | 454 | 0.213 | $8.7\cdot10^{5}$ | $7.62\cdot10^{6}$ | 0.073 | 0.127 | $2.84\cdot10^{10}$ | 0.086 |
| | 2.98 | $1.08\cdot10^{-5}$ | $1.22\cdot10^{9}$ | 2 | 2 | 803 | 0.803 | $1.63\cdot10^{6}$ | $6.12\cdot10^{6}$ | 0.054 | 0.180 | $3.07\cdot10^{10}$ | 0.22 |
| $1M_J$ 1 Gyr | 2.99 | $1.12\cdot10^{-5}$ | $9.81\cdot10^{8}$ | 0.5 | 1 | 440 | 0.25 | $1.81\cdot10^{6}$ | $2.4\cdot10^{7}$ | 0.093 | 0.153 | $7.57\cdot10^{10}$ | 0.061 |
| | 2.99 | $1.12\cdot10^{-5}$ | $9.81\cdot10^{8}$ | 2 | 1 | 1360 | 1.31 | $2.33\cdot10^{6}$ | $1.45\cdot10^{7}$ | 0.0088 | 0.186 | $7.57\cdot10^{10}$ | 1.36 |
| | 2.99 | $1.12\cdot10^{-5}$ | $9.81\cdot10^{8}$ | 4 | 1 | 2460 | 3.52 | $3.17\cdot10^{6}$ | $1.25\cdot10^{7}$ | 0.00236 | 0.230 | $7.98\cdot10^{10}$ | 20.9 |
| | 2.99 | $1.12\cdot10^{-5}$ | $9.81\cdot10^{8}$ | 0.5 | 0.5 | 918 | 0.95 | $6.6\cdot10^{6}$ | $8.7\cdot10^{7}$ | 0.0141 | 0.213 | $3.08\cdot10^{11}$ | 0.65 |
| | 2.99 | $1.12\cdot10^{-5}$ | $9.81\cdot10^{8}$ | 1 | 0.5 | 164 | 2.07 | $7.2\cdot10^{6}$ | $7.1\cdot10^{7}$ | 0.0081 | 0.245 | $3.18\cdot10^{11}$ | 1.54 |
| | 2.99 | $1.12\cdot10^{-5}$ | $9.81\cdot10^{8}$ | 1 | 2 | 310 | 0.221 | $9.1\cdot10^{5}$ | $3.83\cdot10^{6}$ | 0.124 | 0.162 | $1.461\cdot10^{10}$ | 0.037 |
| | 2.99 | $1.12\cdot10^{-5}$ | $9.81\cdot10^{8}$ | 2 | 2 | 570 | 0.73 | $1.45\cdot10^{6}$ | $3.13\cdot10^{6}$ | 0.073 | 0.203 | $1.593\cdot10^{10}$ | 0.084 |
| $1M_J$ 10 Gyr | 3.00 | $1.30\cdot10^{-5}$ | $4.97\cdot10^{8}$ | 0.5 | 1 | 157 | 0.252 | $2.01\cdot10^{6}$ | $5.0\cdot10^{6}$ | 0.333 | 0.219 | $1.89\cdot10^{10}$ | 0.024 |
| | 3.00 | $1.30\cdot10^{-5}$ | $4.97\cdot10^{8}$ | 2 | 1 | 532 | 1.11 | $2.04\cdot10^{6}$ | $3.30\cdot10^{6}$ | 0.062 | 0.250 | $1.55\cdot10^{10}$ | 0.070 |
| | 3.00 | $1.30\cdot10^{-5}$ | $4.97\cdot10^{8}$ | 4 | 1 | 1100 | 1.77 | $1.59\cdot10^{6}$ | $3.18\cdot10^{6}$ | 0.0139 | 0.220 | $1.63\cdot10^{10}$ | 3.87 |
| | 3.00 | $1.30\cdot10^{-5}$ | $4.97\cdot10^{8}$ | 0.5 | 0.5 | 368 | 0.634 | $4.29\cdot10^{6}$ | $1.87\cdot10^{7}$ | 0.088 | 0.271 | $8.113\cdot10^{10}$ | 0.23 |
| | 3.00 | $1.30\cdot10^{-5}$ | $4.97\cdot10^{8}$ | 1 | 0.5 | 699 | 1.40 | $4.78\cdot10^{6}$ | $1.71\cdot10^{7}$ | 0.0456 | 0.281 | $8.49\cdot10^{10}$ | 0.072 |
| | 3.00 | $1.30\cdot10^{-5}$ | $4.97\cdot10^{8}$ | 1 | 2 | 135 | 0.128 | $5.13\cdot10^{5}$ | $7.78\cdot10^{5}$ | 0.301 | 0.177 | $3.14\cdot10^{9}$ | 0.0068 |
| | 3.00 | $1.30\cdot10^{-5}$ | $4.97\cdot10^{8}$ | 2 | 2 | 257 | 0.473 | $8.8\cdot10^{5}$ | $6.76\cdot10^{5}$ | 0.085 | 0.251 | $3.36\cdot10^{9}$ | 0.069 |
| $4M_J$ 0.5 Gyr | 4.60 | $8.37\cdot10^{-}6$ | $6.50\cdot10^{9}$ | 1 | 1 | 1460 | 4.4 | $7.28\cdot10^{7}$ | $1.250\cdot10^{8}$ | 0.0058 | 0.374 | $2.39\cdot10^{12}$ | 3.6 |
| $4M_J$ 1 Gyr | 4.58 | $8.98\cdot10^{-5}$ | $5.02\cdot10^{9}$ | 1 | 1 | 1130 | 3.3 | $5.26\cdot10^{7}$ | $8.47\cdot10^{7}$ | 0.00768 | 0.353 | $1.542\cdot10^{12}$ | 0.63 |
| $4M_J$ 10 Gyr | 4.60 | $1.06\cdot10^{-5}$ | $2.42\cdot10^{9}$ | 1 | 1 | 554 | 1.1 | $1.61\cdot10^{7}$ | $2.48\cdot10^{7}$ | 0.0051 | 0.256 | $3.28\cdot10^{11}$ | 0.054 |